\documentclass[11pt,a4paper]{article}

\usepackage[a4paper,margin=1in]{geometry}
\usepackage[T1]{fontenc}
\usepackage[utf8]{inputenc}
\usepackage{lmodern}
\usepackage{microtype}
\usepackage{setspace}
\usepackage{amsmath}
\usepackage{amssymb}
\usepackage{tikz}
\usetikzlibrary{arrows.meta,decorations.pathreplacing,decorations.pathmorphing}
\definecolor{otocLight}{HTML}{8FB2D4}
\definecolor{otocDark}{HTML}{14416F}
\usepackage{amsmath,amssymb,amsfonts,amsthm}
\usepackage{mathtools}
\numberwithin{equation}{section}   
\usepackage{bm}

\usepackage{graphicx}
\usepackage{subcaption}
\usepackage{booktabs}

\usepackage[numbers,sort&compress]{natbib}
\usepackage[colorlinks=true,
            linkcolor=blue,
            citecolor=blue,
            urlcolor=blue]{hyperref}
\hypersetup{
  pdftitle={Classical and Quantum Chaos from Foundations to Holography},
  pdfauthor={Bhasker Shukla},
  pdfsubject={Lecture notes: classical chaos, quantum chaos, scrambling, and holography},
  pdfkeywords={chaos, Lyapunov exponent, KAM theorem, quantum chaos, random matrix theory,
               out-of-time-order correlators, holography, black holes, scrambling}
}

\usepackage{amsthm}

\theoremstyle{definition}
\newtheorem{exercise}{Exercise}[section]

\begin{document}

\begin{center}
\vspace*{1cm}

{\LARGE Classical and Quantum Chaos\\[2mm]
from Foundations to Holography\par}

\vspace{6mm}

{\large Bhasker Shukla$^{1}$\par}

\vspace{3mm}

{\small\itshape $^{1}$Department of Physics and Astronomy,
National Institute of Technology Rourkela, Odisha 769008, India\par}

\vspace{3mm}

{\small\texttt{bhaskarspace@outlook.com}\par}

\vspace{5mm}

{Lecture Notes from the ST4 Workshop,\\
Chennai Mathematical Institute (CMI), India\par}

\vspace{3mm}

{July 2026\par}

\end{center}

\vspace{4mm}

\begin{abstract}
\noindent
These notes develop the diagnostics of chaos from their classical definitions through to holography,
in six lectures given at the ST4 Workshop, Chennai Mathematical Institute, in 2026. No prior exposure
to chaos theory is assumed. Lectures 1 and 2 define classical chaos geometrically and build the
instruments that measure it, every one of which reads a phase-space trajectory. Quantization removes
that trajectory, and Lectures 3 and 4 rebuild the instruments from spectra and eigenstates instead:
level statistics, random matrix theory, the spectral form factor and eigenstate thermalization, as
much the working toolkit for thermalization in many-body physics as for holography. What they return
is a classification and a count rather than a rate. Lecture 5 recovers the rate from the squared
commutator, treats scrambling as operator growth, follows the commutator to the out-of-time-order
correlator, marks where that diagnostic stops being reliable, and reaches the butterfly velocity and
the chaos bound. Lecture 6 constructs the eternal black hole, the thermofield double and the
shock-wave geometry behind that bound, then follows two probes whose exponents discriminate between
backgrounds where the horizon result cannot, one tracking a thermodynamic phase transition, the other
resolving an anisotropy induced by a background field. That last part is research rather than
review, and every figure throughout is computed from the scripts supplied with the source rather
than redrawn from the literature. Every lecture closes with exercises. Claude
Opus 5, an AI assistant developed by Anthropic, worked as a research assistant under the author's
direction.
\end{abstract}

\clearpage

\tableofcontents

\clearpage

\section*{Foreword}
\addcontentsline{toc}{section}{Foreword}

These notes are aimed primarily at researchers in string theory, especially those working on holography, who want to follow the current literature on scrambling, the chaos bound, and chaotic dynamics in black-hole and holographic QCD backgrounds.
No prior exposure to chaos theory is assumed: Lectures 1--4 build the classical and quantum diagnostics from scratch, and Lectures 5 and 6 apply them to holography.
Readers who already know their Lyapunov exponents from their OTOCs can skip ahead to Lecture 5 without losing the thread.

Lectures 1--4 draw throughout on Arul Lakshminarayan's lecture notes for the 2018 ICTS summer school on statistical physics~\cite{Lakshminarayan2018}, whose treatment of classical maps, quantum billiards and random matrix theory shaped both the material selected here and the order in which it is presented.

\section*{Prerequisites}
\addcontentsline{toc}{section}{Prerequisites}

We assume basic Hamiltonian mechanics, phase space, Poisson brackets, quantum time evolution, and elementary statistical mechanics.
Lectures 5 and 6 also assume the AdS/CFT dictionary and black-hole thermodynamics, both standard background for the intended audience.

\section*{Notation and conventions}
\addcontentsline{toc}{section}{Notation and conventions}

Throughout these notes, phase-space coordinates are denoted by $(q_i,p_i)$ and Poisson brackets by $\{\cdot,\cdot\}$. The largest classical Lyapunov exponent is written $\lambda$, or $\lambda_{\max}$ where the rest of the spectrum is in play, and the finite-time exponent from which it is computed is $\lambda_T$. The quantum Lyapunov exponent extracted from an out-of-time-order correlator is written $\lambda_L$; it is a different quantity, exceeding twice the classical one whenever the finite-time exponent fluctuates, for reasons given in Sec.~\ref{sec:L5_bracket}. 
For quantum systems, the Hamiltonian is denoted by $H$, the time-evolution operator by $U(t)$, and thermal expectation values by $\langle \cdot \rangle_\beta$, with $\beta$ the inverse temperature. Two local exceptions carry the same symbol by long convention: $\beta=8/3$ is a parameter of the Lorenz system in Lecture 1, and $\beta=1,2,4$ is the Dyson index of a random-matrix symmetry class in Lecture 4, written
$\beta_{\rm D}$ from Sec.~\ref{sec:L4_sff} onward, where a physical temperature enters the same
subsection.

Three further symbols carry more than one meaning across the six lectures, in each case
because the subfield in question has long used them that way.  $\Omega$ is the symplectic form
in Lecture 2, the phase-space volume $\Omega(E)$ in Lecture 3, the generator $\Omega_{ij}$ of
rotations among eigenvectors in Lecture 4, and in Lecture 6 both the solid angle
$d\Omega^{2}_{d-1}$ and the orbital angular velocity $\Omega_c$ at a photon sphere.  $\sigma$ is
a Lorenz parameter in Lecture 1, a Gaussian width in Lectures 3 and 4, a Pauli matrix in
Lectures 4 and 5, and the proper length along a string in Lecture 6.  $\Phi$ is the Hamiltonian
flow $\Phi^t$ in Lectures 1 to 3 and the electric potential $\Phi_e$ of a charged black hole in
Lecture 6, the superscript and the subscript keeping the two apart wherever both could be
meant.

Several further letters carry more than one meaning, again because the subfields do. They are
collected here rather than left to be inferred, since the Foreword invites entry at Lecture 5.
Where two meanings would otherwise have sat close together they were separated instead: the
Kruskal metric functions of Sec.~\ref{sec:L6_shock} are written $\mathcal{A}$ and $\mathcal{B}$
because $A$ and $B$ are the warp factor and the magnetic field a few pages later, and the
Hilbert-space dimension is $\mathcal{D}$ throughout because $D$ is the discriminant of Lecture 2.

\begin{center}
\begin{tabular}{@{}ll@{}}
\toprule
$A$ & a driving amplitude (L2); billiard area (L3); trace-formula amplitude $A_p$,\\
    & a tensor factor, and a fit coefficient in $\Delta_3$ (L4); either argument of a\\
    & Poisson bracket (L5); the gauge one-form and the warp factor $A(z)$ (L6)\\
$B$ & the baker's map (L1); a measurable set (L1); the second tensor factor and\\
    & the other fit coefficient in $\Delta_3$ (L4); the second argument of a bracket\\
    & (L5); the magnetic field (L6)\\
$D$ & the discriminant $\tau^2-4\Delta$ (L2); Hilbert-space dimension, as $\mathcal{D}$ (L4, L5);\\
    & the diffusion constant of a large-deviation variance (L5)\\
$\Delta$ & the determinant $\det A$ (L2); the mean level spacing (L3); the Dyson--Mehta\\
    & statistic $\Delta_3(L)$ (L4). A wave-packet width is $\sigma$ and the\\
    & predictability scales are $\Delta_{\rm mac}$, $\Delta_{\rm init}$, kept subscripted throughout\\
$F$, $f$ & a kinetic function $f(p)$ and an observable $f$ (L1); the vector field of the\\
    & flow and its Jacobian $DF$ (L2); the ETH envelope $f_O$ (L4); the\\
    & out-of-time-order correlator $F(t)$ (L5, L6); a gauge field strength, a gauge\\
    & kinetic function $f_i(\phi)$, and the metric function $f(r)$ (L6). The two metric\\
    & functions of Sec.~\ref{sec:L6_hqcd} are written $\mathcal{F}$ and $\mathcal{G}$ for this reason\\
$K$ & the spectral form factor (L4); the stability matrix of a circular orbit, and\\
    & the cubic coefficients $K_1,\ldots,K_5$ of the truncated string action (L6)\\
$L$ & billiard perimeter (L3); length of an interval of unfolded levels (L4);\\
    & number of sites, and the generalized exponent $L_{2q}$ (L5); the angular\\
    & momentum of a geodesic, and the quark separation (L6). The classical\\
    & scale in the Ehrenfest time is written $L_{\rm cl}$ (L3)\\
$\lambda$ & the Lyapunov exponent and its spectrum, throughout; the typical generalized\\
    & exponent $\lambda_1$, which is the smallest of a family and not the largest of a\\
    & spectrum (L5); a linearized rate at an unstable orbit (L6). A multiplier,\\
    & being a factor per iteration rather than a rate, is written $\Lambda$\\
$\mu$ & the invariant measure (L1); the EBK indices $\mu_i$ (L3); the Maslov index $\mu_p$,\\
    & the Haar measure, and the multipliers of a generalized Gibbs ensemble (L4);\\
    & operator weight and the Ruelle rate (L5); the screening mass, in the source's\\
    & own notation, and the chemical potential (L6)\\
$N$ & the counting function $N(k)$, $N(\varepsilon)$ (L3, L4); rank of the gauge group (L3, L6)\\
$q$ & position, throughout; the period of an orbit (L2); the local Hilbert-space\\
    & dimension, and the order of a generalized exponent (L5); the charges\\
    & $q_e$, $q_m$ (L6)\\
$\rho$ & a Lorenz parameter (L1); the boundary radius of a lima\c{c}on (L3); the level\\
    & density $\bar\rho$ (L3, L4); a density matrix (L3, L4, L6)\\
$S$ & the classical action $S_p$ of an orbit (L4); thermodynamic and entanglement\\
    & entropy (L4, L5, L6); the scattering phase $S=e^{i\chi}$ (L6); the Nambu--Goto\\
    & and Einstein--Maxwell--dilaton actions (L6)\\
$T$ & forcing period and averaging window (L1); integration time in $\lambda_T$ (L1, L2);\\
    & period of a periodic orbit $T_p$ (L2, L4); an antiunitary symmetry, a transpose,\\
    & a drive period $T_{\rm d}$ and the time ordering $\mathcal{T}$, all four in one\\
    & subsection (L4); temperature, and the transition temperature $T_p$ (L5, L6)\\
$V$ & a potential (L1, L3, L6); one of the two operators in an out-of-time-order\\
    & correlator (L3, L5). A phase-space volume is $\Omega$ and a box potential\\
    & is $U_{\rm box}$, both to keep clear of it\\
\bottomrule
\end{tabular}
\end{center}

\clearpage

\part{Classical Chaos}

\section{Lecture 1: Historical emergence of classical chaos}

\subsection{The determinism-predictability puzzle}

The aim of this first lecture is to understand how chaos entered physics through classical mechanics itself. We shall not begin with quantum systems, black holes, or holography. Instead, we begin with the older and apparently simpler question of classical determinism. The central puzzle is the following:

\begin{quote}
    How can equations that are completely deterministic produce motion that is effectively unpredictable?
\end{quote}

This question is subtle because it asks us to separate two ideas that are often confused. The first is \emph{determinism}: the statement that the present state of a system fixes its future. The second is \emph{predictability}: the practical ability to forecast that future using finite-precision information about the present state. Classical chaos does not contradict determinism. It shows that determinism need not imply long-term predictability~\cite{TelGruiz2006,Strogatz2015,Ott2002}.

\begin{figure}[!htbp]
\centering
\definecolor{bandI}{HTML}{2D5F99}
\definecolor{bandIfill}{HTML}{BCD2E9}
\definecolor{bandII}{HTML}{5485BC}
\definecolor{bandIII}{HTML}{11365E}
\definecolor{inkMuted}{HTML}{6B6B6B}
\def\lorenzIcon{%
  (0.166,0.658) (0.198,0.731) (0.290,0.656) (0.360,0.544) (0.385,0.448) (0.373,0.381) (0.330,0.358) (0.254,0.418)
  (0.171,0.592) (0.163,0.744) (0.254,0.708) (0.349,0.587) (0.395,0.476) (0.399,0.389) (0.372,0.334) (0.312,0.336)
  (0.218,0.450) (0.140,0.676) (0.180,0.776) (0.301,0.674) (0.388,0.541) (0.421,0.430) (0.419,0.345) (0.391,0.289)
  (0.329,0.285) (0.225,0.397) (0.123,0.668) (0.156,0.816) (0.301,0.699) (0.408,0.550) (0.452,0.431) (0.463,0.336)
  (0.458,0.261) (0.439,0.203) (0.399,0.170) (0.317,0.199) (0.174,0.406) (0.065,0.810) (0.189,0.857) (0.383,0.660)
  (0.487,0.511) (0.534,0.407) (0.569,0.335) (0.621,0.308) (0.707,0.369) (0.811,0.577) (0.830,0.781) (0.717,0.737)
  (0.601,0.594) (0.544,0.470) (0.531,0.371) (0.542,0.296) (0.578,0.250) (0.650,0.262) (0.769,0.417) (0.868,0.734)
  (0.794,0.829) (0.631,0.674) (0.530,0.524) (0.490,0.409) (0.474,0.318) (0.462,0.245) (0.444,0.190) (0.405,0.157)
  (0.326,0.180) (0.183,0.376) (0.060,0.796) (0.175,0.875) (0.378,0.672) (0.491,0.520) (0.541,0.416) (0.578,0.348)
  (0.633,0.327) (0.720,0.401) (0.814,0.608) (0.814,0.774) (0.703,0.714) (0.599,0.578) (0.552,0.461) (0.545,0.368)
  (0.566,0.302) (0.616,0.277) (0.708,0.344) (0.823,0.578) (0.841,0.807) (0.712,0.747) (0.585,0.592) (0.525,0.464)
  (0.506,0.362) (0.507,0.281) (0.521,0.218) (0.552,0.176) (0.617,0.176) (0.740,0.304) (0.887,0.681) (0.845,0.898)
  (0.642,0.725) (0.506,0.555) (0.446,0.439) (0.411,0.356) (0.366,0.312) (0.292,0.338) (0.188,0.499) (0.129,0.738)
  (0.209,0.771) (0.336,0.638) (0.411,0.506) (0.435,0.400) (0.430,0.318) (0.401,0.262) (0.340,0.254) (0.233,0.361)
  (0.116,0.654) (0.140,0.839) (0.298,0.717) (0.417,0.558) (0.469,0.436) (0.487,0.339) (0.497,0.262) (0.508,0.200)
  (0.531,0.155) (0.578,0.136) (0.677,0.195) (0.839,0.484) (0.910,0.896) (0.725,0.823) (0.537,0.621) (0.447,0.489)
  (0.403,0.400) (0.359,0.353) (0.291,0.372) (0.200,0.506) (0.148,0.706) (0.209,0.748) (0.318,0.639) (0.386,0.517)
  (0.407,0.418) (0.394,0.345) (0.351,0.313) (0.270,0.363) (0.168,0.553) (0.137,0.767) (0.238,0.746) (0.356,0.607)
  (0.417,0.482) (0.433,0.382) (0.421,0.306) (0.385,0.260) (0.311,0.277) (0.193,0.439) (0.104,0.743) (0.185,0.816)
  (0.341,0.663) (0.435,0.517) (0.471,0.404) (0.483,0.313) (0.488,0.240) (0.492,0.182) (0.500,0.135) (0.516,0.099)
  (0.554,0.080) (0.637,0.113) (0.798,0.342) (0.940,0.852) (0.787,0.902) (0.554,0.670) (0.434,0.529) (0.379,0.446)
  (0.332,0.409) (0.270,0.440) (0.203,0.557) (0.180,0.687) (0.234,0.696) (0.312,0.610) (0.359,0.512) (0.366,0.434)
  (0.340,0.392) (0.283,0.411) (0.209,0.524) (0.169,0.680) (0.217,0.717) (0.305,0.631) (0.363,0.525) (0.378,0.437)
  (0.359,0.380) (0.308,0.376) (0.229,0.466) (0.163,0.645) (0.187,0.741) (0.283,0.671) (0.361,0.554) (0.391,0.453)
  (0.384,0.378) (0.346,0.343) (0.273,0.383) (0.181,0.545) (0.148,0.737) (0.231,0.736) (0.338,0.614) (0.398,0.494)
  (0.411,0.398) (0.394,0.330) (0.346,0.305) (0.259,0.369) (0.155,0.581) (0.139,0.785) (0.254,0.736) (0.371,0.592)
  (0.427,0.467) (0.440,0.369) (0.429,0.293) (0.395,0.245) (0.325,0.252) (0.207,0.397) (0.100,0.719) (0.164,0.838)
  (0.331,0.685) (0.437,0.532) (0.481,0.415) (0.499,0.323) (0.514,0.251) (0.539,0.198) (0.590,0.177) (0.690,0.242)
  (0.841,0.529) (0.888,0.877) (0.715,0.795) (0.547,0.607) (0.467,0.474) (0.430,0.379) (0.394,0.315) (0.336,0.300)
  (0.240,0.389) (0.138,0.630) (0.150,0.800) (0.280,0.713) (0.389,0.567) (0.436,0.447) (0.446,0.351) (0.434,0.278)
  (0.401,0.230) (0.332,0.235) (0.212,0.375) (0.095,0.710) (0.154,0.853) (0.329,0.695) (0.443,0.537) (0.491,0.420)
  (0.513,0.329) (0.536,0.260) (0.577,0.220) (0.656,0.244) (0.786,0.430) (0.882,0.778) (0.779,0.834) (0.607,0.659)
  (0.509,0.510) (0.469,0.399) (0.447,0.314) (0.418,0.252) (0.366,0.228) (0.269,0.294) (0.136,0.556) (0.104,0.842)
  (0.255,0.772) (0.402,0.599) (0.472,0.466) (0.500,0.365) (0.519,0.285) (0.546,0.228) (0.598,0.206) (0.698,0.275)
  (0.840,0.554) (0.873,0.860) (0.709,0.778) (0.555,0.599) (0.483,0.466) (0.450,0.367) (0.424,0.294) (0.380,0.253)
  (0.300,0.282) (0.177,0.469) (0.102,0.775) (0.203,0.806) (0.358,0.644) (0.444,0.502) (0.477,0.392) (0.489,0.304)
  (0.496,0.233) (0.508,0.177) (0.531,0.136) (0.581,0.122) (0.686,0.193) (0.854,0.515) (0.909,0.920) (0.706,0.809)
  (0.521,0.609) (0.435,0.484) (0.390,0.403) (0.343,0.366) (0.271,0.404) (0.187,0.553) (0.159,0.722) (0.233,0.722)
  (0.330,0.611) (0.385,0.498) (0.395,0.408) (0.373,0.349) (0.318,0.342) (0.230,0.437) (0.149,0.649) (0.173,0.769)
  (0.286,0.686) (0.376,0.555) (0.413,0.445) (0.412,0.359) (0.383,0.304) (0.320,0.304) (0.219,0.423) (0.127,0.681)
  (0.169,0.801) (0.306,0.685) (0.403,0.542) (0.442,0.426) (0.447,0.335)
}

\def\wignerIcon{%
  (0.105,0.163) (0.121,0.228) (0.136,0.291) (0.152,0.352) (0.167,0.410) (0.182,0.464) (0.198,0.513) (0.213,0.557)
  (0.228,0.596) (0.244,0.628) (0.259,0.655) (0.275,0.675) (0.290,0.689) (0.305,0.696) (0.321,0.698) (0.336,0.695)
  (0.351,0.686) (0.367,0.673) (0.382,0.656) (0.397,0.635) (0.413,0.612) (0.428,0.586) (0.444,0.559) (0.459,0.531)
  (0.474,0.502) (0.490,0.473) (0.505,0.445) (0.520,0.417) (0.536,0.390) (0.551,0.364) (0.567,0.340) (0.582,0.317)
  (0.597,0.296) (0.613,0.277) (0.628,0.259) (0.643,0.243) (0.659,0.229) (0.674,0.216) (0.689,0.204) (0.705,0.195)
  (0.720,0.186) (0.736,0.178) (0.751,0.172) (0.766,0.166) (0.782,0.162) (0.797,0.158) (0.812,0.154) (0.828,0.152)
  (0.843,0.149) (0.859,0.148) (0.874,0.146) (0.889,0.145) (0.905,0.144) (0.920,0.143)
}

\def\poissonIcon{%
  (0.105,0.860) (0.121,0.821) (0.136,0.783) (0.152,0.748) (0.167,0.715) (0.182,0.684) (0.198,0.654) (0.213,0.626)
  (0.228,0.599) (0.244,0.574) (0.259,0.550) (0.275,0.528) (0.290,0.507) (0.305,0.487) (0.321,0.468) (0.336,0.450)
  (0.351,0.433) (0.367,0.417) (0.382,0.402) (0.397,0.387) (0.413,0.374) (0.428,0.361) (0.444,0.349) (0.459,0.338)
  (0.474,0.327) (0.490,0.317) (0.505,0.307) (0.520,0.298) (0.536,0.289) (0.551,0.281) (0.567,0.273) (0.582,0.266)
  (0.597,0.259) (0.613,0.253) (0.628,0.246) (0.643,0.241) (0.659,0.235) (0.674,0.230) (0.689,0.225) (0.705,0.220)
  (0.720,0.216) (0.736,0.212) (0.751,0.208) (0.766,0.204) (0.782,0.201) (0.797,0.197) (0.812,0.194) (0.828,0.191)
  (0.843,0.188) (0.859,0.186) (0.874,0.183) (0.889,0.181) (0.905,0.179) (0.920,0.177)
}

\newcommand{\icoKepler}[1]{%
  \draw[#1,line width=0.6pt] (0.53,0.5) ellipse [x radius=0.36, y radius=0.235];
  \fill[#1] (0.257,0.5) circle (0.062);
  \fill[#1] (0.89,0.5) circle (0.040);
  \draw[#1!50,line width=0.4pt,-{Latex[length=2.4pt]}] (0.845,0.62) arc (60:110:0.36);}
\newcommand{\icoThreeBody}[1]{%
  \draw[#1!70,line width=0.5pt,dash pattern=on 1.6pt off 1.0pt]
    (0.26,0.30) .. controls (0.10,0.62) and (0.36,0.88) .. (0.50,0.74);
  \draw[#1!70,line width=0.5pt,dash pattern=on 1.6pt off 1.0pt]
    (0.50,0.74) .. controls (0.68,0.60) and (0.86,0.80) .. (0.74,0.32);
  \draw[#1!70,line width=0.5pt,dash pattern=on 1.6pt off 1.0pt]
    (0.74,0.32) .. controls (0.58,0.12) and (0.34,0.14) .. (0.26,0.30);
  \fill[#1] (0.26,0.30) circle (0.085);
  \fill[#1] (0.50,0.74) circle (0.068);
  \fill[#1] (0.74,0.32) circle (0.055);}
\newcommand{\icoKAM}[1]{%
  \fill[#1] (0.5,0.5) circle (0.035);
  \draw[#1,line width=0.55pt] (0.5,0.5) ellipse [x radius=0.115, y radius=0.085];
  \draw[#1,line width=0.55pt] (0.5,0.5) ellipse [x radius=0.235, y radius=0.170];
  \draw[#1!60,line width=0.5pt,dash pattern=on 1.1pt off 1.2pt]
    (0.5,0.5) ellipse [x radius=0.355, y radius=0.255];}
\newcommand{\icoLorenz}[1]{%
  \begin{scope}[shift={(0.10,0.10)},scale=0.80]
    \draw[#1,line width=0.20pt] plot[smooth] coordinates {\lorenzIcon};
  \end{scope}}
\newcommand{\icoStadium}[1]{%
  \draw[#1,line width=0.6pt]
    (0.42,0.20) -- (0.58,0.20) arc (-90:90:0.30) -- (0.42,0.80) arc (90:270:0.30);
  \draw[#1!65,line width=0.5pt] (0.30,0.66) -- (0.70,0.36) -- (0.30,0.36) -- (0.70,0.66) -- cycle;
  \fill[#1!65] (0.30,0.66) circle (0.030); \fill[#1!65] (0.70,0.36) circle (0.030);
  \fill[#1!65] (0.30,0.36) circle (0.030); \fill[#1!65] (0.70,0.66) circle (0.030);}
\newcommand{\icoSpacing}[1]{%
  \begin{scope}[shift={(0.13,0.19)},xscale=0.76,yscale=0.74]
    \draw[#1!60,line width=0.55pt,dash pattern=on 1.2pt off 1.1pt]
      plot[smooth] coordinates {\poissonIcon};
    \draw[#1,line width=0.7pt] plot[smooth] coordinates {\wignerIcon};
    \draw[#1!50,line width=0.4pt] (0.08,0.11) -- (0.98,0.11);
  \end{scope}}
\newcommand{\icoScramble}[1]{%
  \fill[#1] (0.5,0.5) circle (0.115);
  \draw[#1!45,line width=0.45pt,dash pattern=on 1pt off 1.1pt] (0.5,0.5) circle (0.235);
  \draw[#1!25,line width=0.45pt,dash pattern=on 1pt off 1.1pt] (0.5,0.5) circle (0.355);
  \foreach \a in {35,107,179,251,323}{%
    \draw[#1!60,line width=0.45pt,-{Latex[length=2.6pt]}] (0.5,0.5) ++(\a:0.145) -- +(\a:0.205);}}
\newcommand{\icoBound}[1]{%
  \draw[#1!55,line width=0.5pt,dash pattern=on 1.2pt off 1.2pt] (0.14,0.74) -- (0.88,0.74);
  \draw[#1!50,line width=0.4pt] (0.14,0.16) -- (0.14,0.88);
  \draw[#1!50,line width=0.4pt] (0.14,0.16) -- (0.88,0.16);
  \draw[#1,line width=0.7pt] (0.16,0.21) .. controls (0.46,0.26) and (0.46,0.71) .. (0.88,0.733);}

\begin{tikzpicture}[x=1cm,y=1cm,
    stem/.style={inkMuted!45,line width=0.45pt},
    dot/.style={circle,draw=white,line width=0.8pt,inner sep=0pt,minimum size=4.6pt},
    card/.style={align=center,text width=2.85cm,inner sep=0pt},
    part/.style={font=\scriptsize\scshape,align=center,inner sep=0pt,anchor=north}]

  \fill[bandIfill] (-0.05,-0.105) rectangle (7.30,0.105);
  \fill[bandII]    (7.30,-0.105)  rectangle (10.95,0.105);
  \fill[bandIII]   (10.95,-0.105) rectangle (14.60,0.105);
  \draw[bandIII,line width=0.8pt,-{Latex[length=4pt]}] (14.60,0) -- (15.00,0);

  \node[part,bandI] at (3.63,-0.19) {Part I\\Classical Chaos};
  \node[part,bandII]         at (9.13,-0.19) {Part II\\Quantum Chaos};
  \node[part,bandIII]        at (12.78,-0.19) {Part III\\Scrambling and Holography};

  \newcommand{\ms}[7]{%
    \draw[stem] (#1,0.12) -- (#1,#2-0.03);
    \node[dot,fill=#3] at (#1,0) {};
    \begin{scope}[shift={(#1-0.58,#2)},x=1.16cm,y=1.16cm]
      \fill[white] (0.5,0.5) circle (0.485);
      \draw[#3!45,line width=0.55pt] (0.5,0.5) circle (0.485);
      #4{#3}
    \end{scope}
    \node[card,anchor=south] at (#1,#2+1.26) {%
      {\small\bfseries\color{#3}#5}\\[1.5pt]
      {\small #6}\\[1.5pt]
      {\scriptsize\color{inkMuted}Sec.~#7}};
  }

  \ms{2.48}{0.34}{bandI}{\icoThreeBody}{1890}{The three-body problem}{\ref{sec:L1_threebody}}
  \ms{6.13}{0.34}{bandI}{\icoLorenz}{1963}{The Lorenz attractor}{\ref{sec:L1_lorenz}}
  \ms{9.83}{0.34}{bandII}{\icoSpacing}{1984}{Level repulsion}{\ref{sec:L4_poisson_wd}}
  \ms{13.48}{0.34}{bandIII}{\icoBound}{2016}{The chaos bound}{\ref{sec:L5_bound}}

  \ms{0.82}{2.70}{bandI}{\icoKepler}{1687}{Newtonian determinism}{\ref{sec:L1_determinism}}
  \ms{4.30}{2.70}{bandI}{\icoKAM}{1954--63}{KAM theory}{\ref{sec:L2_kam}}
  \ms{8.00}{2.70}{bandII}{\icoStadium}{1971}{The trace formula}{\ref{sec:L4_trace_formula}}
  \ms{11.66}{2.70}{bandIII}{\icoScramble}{2008}{Fast scrambling}{\ref{sec:L5_growth}}
\end{tikzpicture}
\caption{Eight results that mark the route these notes follow, each labelled with the
section that treats it and coloured by the Part it belongs to.  The spine orders the entries and
is not a time scale: the first gap is two centuries and the last is eight years.  Two of the
glyphs are computed rather than drawn, the Lorenz attractor and the Wigner surmise against the
Poisson law, and the other six are schematic.}
\label{fig:lecture1_timeline}
\end{figure}
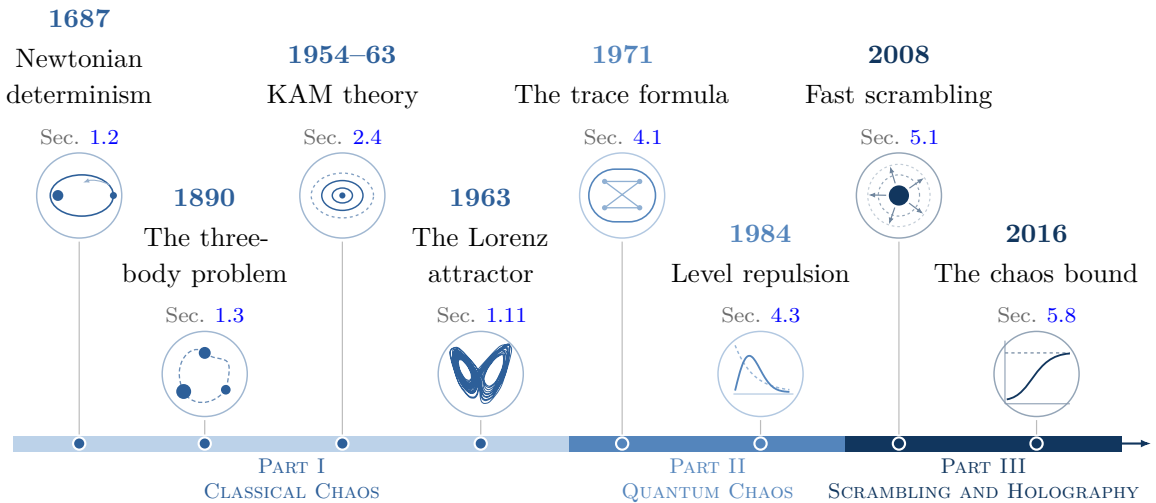

This lecture is mainly historical and geometric. The diagnostic tools of chaos, such as Poincar\'e sections, Lyapunov exponents, bifurcations and the KAM picture~\cite{LichtenbergLieberman1992,Ott2002,Strogatz2015}, will be developed in the next lecture. Here, our task is to understand why those tools became necessary in the first place. The historical arc we shall follow is summarized in Fig.~\ref{fig:lecture1_timeline}.

\subsection{Determinism is not predictability}
\label{sec:L1_determinism}
\label{sec:L1_predictability}

Classical mechanics is often presented as the paradigm of deterministic physics. Given the initial positions and velocities of all particles, and given the forces acting on them, the future motion is fixed. In Hamiltonian language, the state of a system with $d$ degrees of freedom is specified by the canonical coordinates $(q_1,\ldots,q_d,p_1,\ldots,p_d)$, which form a point in a $2d$-dimensional phase space~\cite{Arnold1989,Goldstein2002}. The Hamiltonian $H=H(q_1,\ldots,q_d,p_1,\ldots,p_d,t)$ generates time evolution through Hamilton's equations
\begin{equation}
    \dot q_i=\frac{\partial H}{\partial p_i},
    \qquad
    \dot p_i=-\frac{\partial H}{\partial q_i},
    \qquad i=1,\ldots,d.
    \label{eq:L1_hamilton_equations}
\end{equation}
Being of first order in time, these $2d$ equations generate a flow on phase space: one trajectory through every point, and never two through the same one. Writing $X$ for the phase-space point $(q_1,\ldots,q_d,p_1,\ldots,p_d)$, the state reached at time $t$ from the initial state $X(0)$ is $X(t)=\Phi^t(X(0))$, with $\Phi^t$ the Hamiltonian flow.

For an autonomous Hamiltonian, one with no explicit time dependence, $H=H(q,p)$, the energy is conserved:
\begin{equation}
    \frac{dH}{dt}
    =
    \sum_i\left(
    \frac{\partial H}{\partial q_i}\dot q_i
    +
    \frac{\partial H}{\partial p_i}\dot p_i
    \right)
    =
    \sum_i\left(
    \frac{\partial H}{\partial q_i}
    \frac{\partial H}{\partial p_i}
    -
    \frac{\partial H}{\partial p_i}
    \frac{\partial H}{\partial q_i}
    \right)
    =0,
\end{equation}
or equivalently $dH/dt=\partial H/\partial t+\{H,H\}=\partial H/\partial t$, which vanishes because $H$ carries no explicit time. The bracket is not what does the work: $\{H,H\}=0$ by antisymmetry alone, for a time-dependent $H$ just as much as for an autonomous one. Thus, autonomous Hamiltonian motion is restricted to constant-energy hypersurfaces.

The deterministic structure of \eqref{eq:L1_hamilton_equations} says that the future is fixed by the present. It does not say that the future is practically predictable from approximate knowledge of the present.

In any real experiment or numerical calculation, the initial state is known only with finite precision. Thus, instead of knowing an exact point $X(0)$ in phase space, one knows a small neighbourhood of possible initial points. If nearby initial points remain close under time evolution, the system is predictable. If nearby initial points separate rapidly, long-time prediction becomes impossible.

A basic signature of chaos is sensitive dependence on initial conditions. In its simplest form, this means that two nearby trajectories separate approximately as
\begin{equation}
    |\delta X(t)|\sim |\delta X(0)|e^{\lambda t},
    \label{eq:L1_sensitive_dependence}
\end{equation}
where $\lambda>0$ is a characteristic exponential growth rate. Enough of what $\lambda$ means is wanted later in this lecture to be worth setting down at once. Carry a small deviation $\delta X(t)$ along the flow linearized about a reference trajectory, and put
\begin{equation}
    \lambda=\lim_{t\to\infty}\frac{1}{t}\log\frac{|\delta X(t)|}{|\delta X(0)|}.
    \label{eq:L1_lyapunov_working}
\end{equation}
A system with $d$ degrees of freedom has $2d$ of these numbers, one for each independent direction the deviation can point along, and a deviation chosen at random picks up the largest. For a Hamiltonian system they sum to zero, which is Liouville's theorem read in the tangent space. Lecture 2 restates this as \eqref{eq:L2_lyapunov_def} and supplies what it leaves out, namely the conditions under which the limit exists and the invariant measure with respect to which almost every starting point returns the same value. Until then, \eqref{eq:L1_sensitive_dependence} is the statement that the largest of the $2d$ numbers is positive.

Suppose prediction becomes useless when the uncertainty reaches a macroscopic scale $\Delta$, so that $\Delta\sim|\delta X(0)|e^{\lambda t_{\rm p}}$. The predictability time is then
\begin{equation}
    t_{\rm p}
    \sim
    \frac{1}{\lambda}
    \log\left(\frac{\Delta}{|\delta X(0)|}\right).
    \label{eq:L1_predictability_time}
\end{equation}
It says that improving the initial precision by many orders of magnitude increases the prediction time only logarithmically. In a chaotic system, arbitrarily accurate initial data would be needed for arbitrarily long prediction~\cite{Lorenz1963,Ott2002}.

This is the first conceptual lesson:

\begin{quote}
    Classical chaos is deterministic short-time evolution together with limited long-time predictability.
\end{quote}

\subsection{Poincar\'e and the three-body problem}
\label{sec:L1_threebody}

The historical root of chaos is the gravitational three-body problem~\cite{Poincare1890,Poincare1892,BarrowGreen1997}. The two-body problem is one of the great successes of Newtonian mechanics. It is exactly solvable: after reducing to the centre-of-mass frame, the relative motion is equivalent to that of one particle in a central potential. Conservation of energy and angular momentum leads to Keplerian orbits.

The three-body problem is qualitatively different. Adding one more gravitating body changes the nature of the problem, not just its length. The motion remains deterministic, but the general problem is no longer solvable in the same sense. The ten classical integrals of the general problem, energy, momentum, angular momentum and the centre of mass, are still there and are not enough. Bruns~\cite{Bruns1887} and then Poincar\'e~\cite{Poincare1890} showed that no further independent integrals of the relevant kinds exist, so there is nothing left with which to reduce the dynamics to a set of elementary motions.

\begin{figure}[!htbp]
    \centering
    \includegraphics[width=0.95\textwidth]{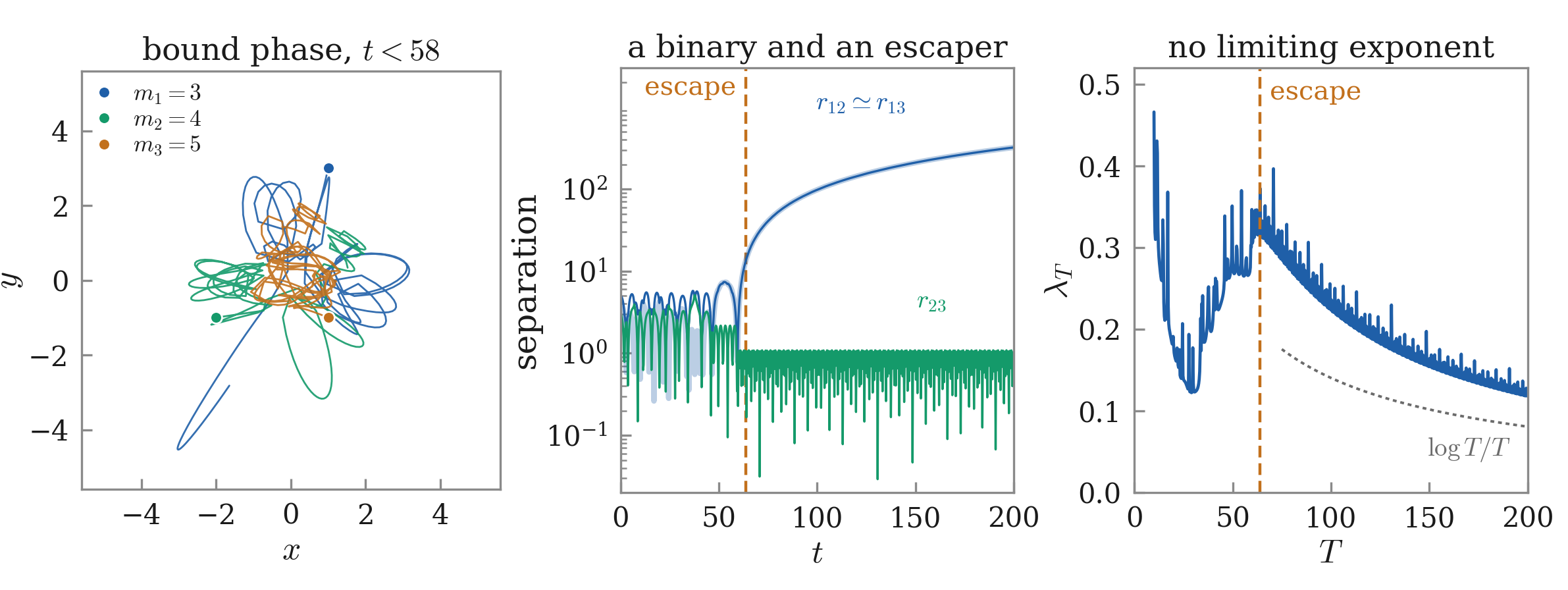}
    \caption{The Pythagorean three-body problem: masses $m_1=3$, $m_2=4$ and $m_3=5$ in units with $G=1$, released from rest at the vertices of a $3$--$4$--$5$ right triangle, each mass opposite the side of its own length, at $(1,3)$, $(-2,-1)$ and $(1,-1)$~\cite{Burrau1913,SzebehelyPeters1967}. The centre of mass is at the origin and stays there. Left: the motion up to $t=58$, with repeated close encounters and no visible periodicity. Middle: the three separations $r_{ij}$ on a logarithmic scale. Right: the finite-time exponent $\lambda_T$, obtained by carrying a tangent vector along the linearized flow and renormalizing it every $\Delta t=0.25$~\cite{Benettin1976}. The first ten time units are omitted, since $\lambda_T$ there is still recording the alignment of the initial vector rather than the dynamics. It sits near $0.3$ through the bound phase and peaks at $0.4$ as the ejection happens, after which it falls away. That fall is physics rather than a numerical failure. The Pythagorean problem disintegrates~\cite{SzebehelyPeters1967}: the two heavier masses settle into a binary and the lightest escapes on a hyperbolic orbit, at $t\approx64$ here. The middle panel is where one sees it, with $r_{12}$ and $r_{13}$ lying on top of each other and growing linearly while $r_{23}$ goes on oscillating at order unity. What is left is a Kepler pair and a free particle, an integrable system, so the tangent vector grows linearly instead of exponentially and $\lambda_T$ decays like $\log T/T$, down to $0.13$ at $t=200$. The quantity worth quoting is therefore a finite-time exponent over the bound phase, $\lambda_T\approx0.3$ for $t\lesssim64$, and over that window it makes the sensitivity in \eqref{eq:L1_sensitive_dependence} quantitative.}
    \label{fig:L1_three_body_problem}
\end{figure}

Poincar\'e's work revealed that this failure was structural, not merely technical. The geometry of the solutions themselves could be extraordinarily complicated. Poincar\'e found that trajectories in the restricted three-body problem could depend sensitively on their initial conditions, and that their phase-space structure could not be understood using only the traditional language of explicit closed-form solutions. Fig.~\ref{fig:L1_three_body_problem} illustrates both features in a specific case.

This marked a change in the philosophy of mechanics. Instead of asking only \emph{can we solve the equations explicitly?}, one began to ask \emph{what is the qualitative structure of the space of solutions?} This is the beginning of the modern theory of dynamical systems.

\subsection{Phase space and no-crossing of trajectories}

Regular and chaotic motion are most naturally told apart by the geometry of trajectories in phase space, which is why the rest of this lecture works there rather than with solutions written out as functions of time. For $d$ degrees of freedom that space is $2d$-dimensional, a point in it fixes the complete instantaneous state of the system, and the evolution of that state is a curve.

\begin{figure}[!htbp]
    \centering
    \includegraphics[width=0.86\textwidth]{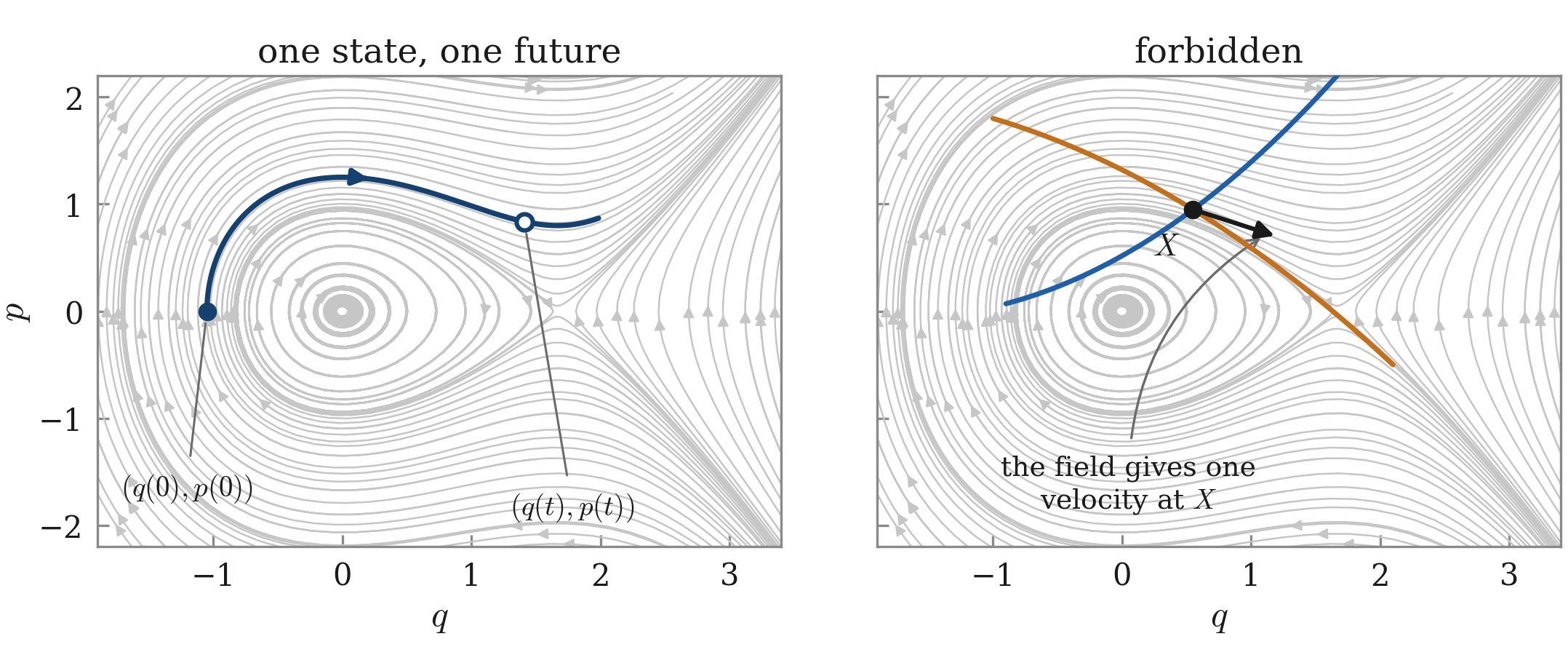}
    \caption{Left: the streamlines of a one-degree-of-freedom Hamiltonian flow, drawn from the
field itself rather than sketched, with one integral curve picked out from \((q(0),p(0))\) to
\((q(t),p(t))\).  Right: what cannot happen.  The field assigns exactly one velocity vector to
the point \(X\), drawn there, and neither of the two curves through \(X\) is tangent to it, so
neither can solve Hamilton's equations at that point.  At most one trajectory leaves any
phase-space point, so trajectories neither cross each other nor cross themselves.  The
restriction bites hardest in low dimensions: with one degree of freedom, conservation of energy
already confines the motion to a curve, and a curve that may not cross itself has nowhere
complicated to go.}
    \label{fig:L1_one_dof_phase_space}
\end{figure}

No two trajectories of an autonomous system can meet in phase space. The reason is simple. Were they to meet, the point at which they did so would be a single initial condition with two different futures. This would violate uniqueness of solutions of ordinary differential equations, as sketched in Fig.~\ref{fig:L1_one_dof_phase_space}.

The word \emph{autonomous} is doing real work here. If the Hamiltonian depends explicitly on time, the vector field at a given $(q,p)$ changes from moment to moment, and a curve drawn in the $(q,p)$ plane may cross itself and other curves as often as it likes. Sec.~\ref{sec:L1_onehalf} plots trajectories that do exactly that. Uniqueness is not lost, only projected away: in the extended phase space $(q,p,t)$, where the flow is autonomous again by construction, the trajectories still do not meet. That extra dimension is the half in one and a half.

This no-crossing property is extremely restrictive in low dimensions. For a one-degree-of-freedom autonomous Hamiltonian system, phase space is two-dimensional. Conservation of energy restricts the motion to one-dimensional curves,
\begin{equation}
    H(q,p)=E.
\end{equation}
Since the trajectory is confined to such a curve and cannot cross itself in phase space, the motion is geometrically too simple to be chaotic~\cite{Arnold1989,LichtenbergLieberman1992}. The sharp statement behind this is the Poincar\'e--Bendixson theorem, which leaves a planar flow nothing to do at long times but approach a fixed point or a closed orbit.

For higher-dimensional systems, however, the situation changes. A two-degree-of-freedom autonomous Hamiltonian system has a four-dimensional phase space, so the energy surface is three-dimensional, giving a trajectory enough room to wind, stretch, fold, and explore complicated regions~\cite{Arnold1989,LichtenbergLieberman1992}.

\subsection{Liouville's theorem and Poincar\'e recurrence}
\label{sec:L1_liouville}

Phase space carries a second structural constraint besides no-crossing, and it is the one that fixes what a Hamiltonian system can do at long times. Regard \eqref{eq:L1_hamilton_equations} as a velocity field on phase space, attaching the components $\dot q_i$ and $\dot p_i$ to the point $(q,p)$. Its divergence is
\begin{equation}
    \sum_{i=1}^{d}\left(\frac{\partial\dot q_i}{\partial q_i}+\frac{\partial\dot p_i}{\partial p_i}\right)
    =\sum_{i=1}^{d}\left(\frac{\partial^2H}{\partial q_i\,\partial p_i}-\frac{\partial^2H}{\partial p_i\,\partial q_i}\right)=0,
    \label{eq:L1_liouville}
\end{equation}
since mixed partial derivatives commute. The flow is incompressible: a region of phase space carried along by the dynamics may be deformed without limit, but its volume never changes. This is Liouville's theorem~\cite{Arnold1989,Goldstein2002}, which is a different statement from the Liouville--Arnold theorem of Sec.~\ref{sec:L1_integrability} despite the shared name. One consequence is immediate. A Hamiltonian system can have no attractor, since an attractor is a set that neighbouring volumes shrink onto, and nothing here shrinks.

Volume conservation has a second consequence, drawn by Poincar\'e, that is stronger than it looks. Take a region $A$ of positive volume inside a bounded region that the motion never leaves, and follow the preimages $A$, $\Phi^{-T}A$, $\Phi^{-2T}A$ and so on under the flow sampled at intervals $T$. Every one of them has the volume of $A$. Were they all disjoint, the volume they occupy together would grow past any bound, which boundedness forbids. Two of them therefore overlap, and carrying that overlap forward shows that some points of $A$ come back to $A$. The sharp statement is that almost every point of $A$ returns, and returns infinitely often~\cite{Poincare1890,Arnold1989}.

It is worth being exact about what that argument used, because the hypotheses are weaker than the conclusion suggests. It used volume conservation and a bounded phase space, and nothing else. Nonlinearity played no part, nor did sensitivity to initial conditions, nor anything else separating a chaotic system from an integrable one. Recurrence holds for the pendulum as surely as for the three-body problem, which is why it appears among the ingredients assembled in Sec.~\ref{sec:L1_ingredients} rather than as a symptom of chaos. What the argument does not supply is a time. Poincar\'e's theorem promises a return without saying when, and for a gas of $10^{23}$ particles the wait exceeds any interval that means anything physically, which is how Boltzmann answered Zermelo's objection~\cite{Zermelo1896}. Sec.~\ref{sec:L1_ergodic} recovers the missing number once one further property of the dynamics is in hand.

\subsection{Integrability}
\label{sec:L1_integrability}

It is worth having two solvable systems in front of us before saying what solvability means. The first is the harmonic oscillator,
\begin{equation}
    H(q,p)=\frac{p^2}{2m}+\frac{1}{2}m\omega^2q^2,
    \label{eq:L1_harmonic_oscillator}
\end{equation}
with $\dot q=p/m$ and $\dot p=-m\omega^2q$. Its trajectories are the level sets $H=E$, which are ellipses, and the motion is periodic, stable and exactly solvable.

\begin{figure}[!htbp]
    \centering
    \includegraphics[width=0.88\textwidth]{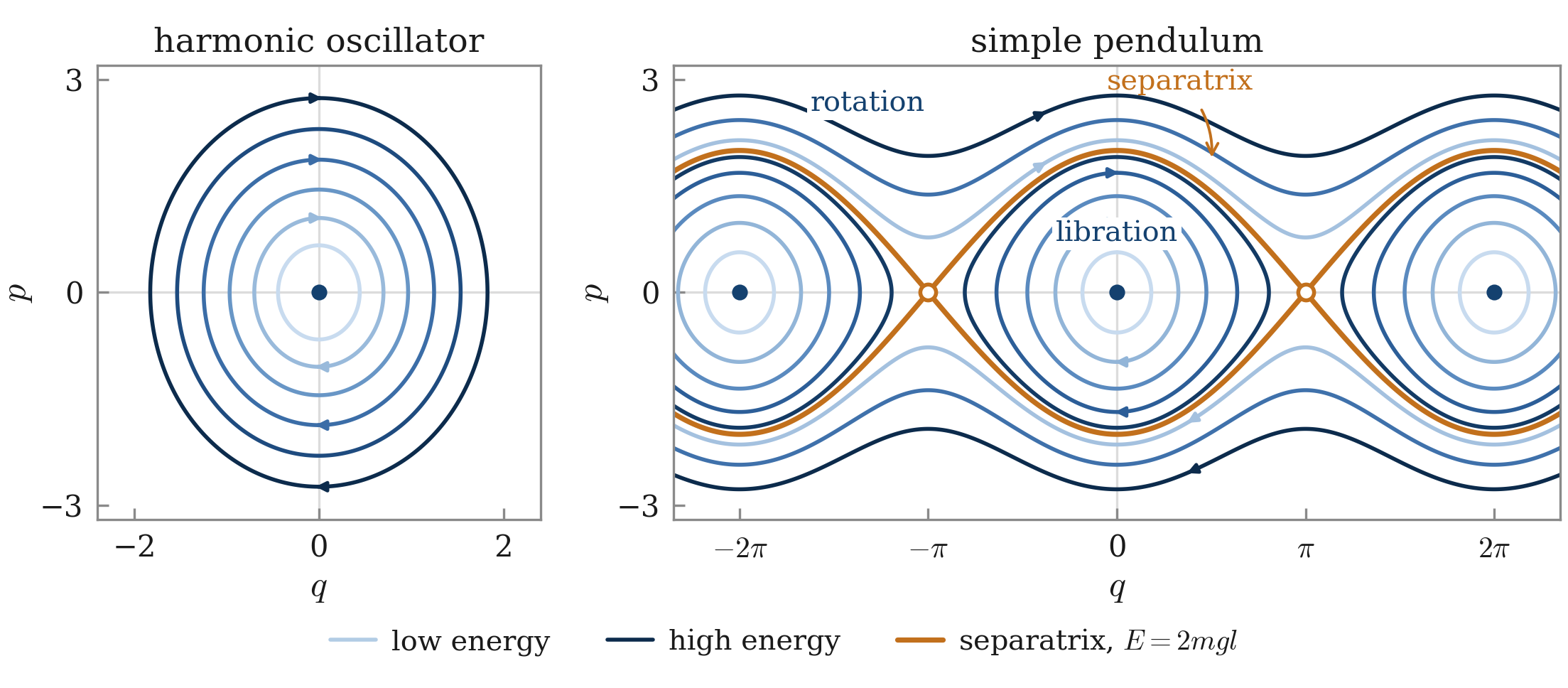}
    \caption{Level sets of \eqref{eq:L1_harmonic_oscillator} and \eqref{eq:L1_pendulum_hamiltonian}, contoured exactly rather than sketched and shaded light to dark with increasing energy. Left: the oscillator at $m=1$ and $\omega=1.5$, whose level sets are ellipses of semi-axis ratio $m\omega$, measured at $1.5000$, so the portrait has a single centre and nothing else at any energy. Right: the pendulum at $m=l=g=1$. Filled dots are the stable equilibria at $q=2\pi n$ and open circles the unstable ones at $q=(2n+1)\pi$; the orange curve through the latter is the exact level set $E=2mgl=2$, with $H=2.0000$ at $(\pi,0)$ against $0$ at the origin. Below it the motion librates, above it $q$ runs through all values, and the lowest rotation branch drawn crosses $q=0$ at $p=2.14$. Arrowheads sit where the direction of the flow was computed from Hamilton's equations rather than assumed: at $(q,p)=(0,1)$ they give $\dot q=1$ and $\dot p=0$, so the circulation is clockwise. Every trajectory in either panel is a level set, which is what one degree of freedom and a conserved energy buy, and is the reason neither system can be chaotic however nonlinear it is.}
    \label{fig:L1_oscillator_pendulum}
\end{figure}

The second is the simple pendulum,
\begin{equation}
    H(q,p)=\frac{p^2}{2ml^2}+mgl(1-\cos q),
    \label{eq:L1_pendulum_hamiltonian}
\end{equation}
with $\dot q=p/ml^2$ and $\dot p=-mgl\sin q$, where the $\sin q$ is what makes the system nonlinear. Its stable equilibria sit at $(q,p)=(2\pi n,0)$ and its unstable ones at $(q,p)=((2n+1)\pi,0)$, with $n\in\mathbb{Z}$. The separatrix passing through the unstable points separates oscillatory motion from rotational motion~\cite{Arnold1989,Strogatz2015}. Both phase portraits are compared in Fig.~\ref{fig:L1_oscillator_pendulum}.

The pendulum already has richer geometry than the harmonic oscillator. It has stable points, unstable points, rotations, oscillations, and a separatrix. Nevertheless, it is still not chaotic as an autonomous one-degree-of-freedom Hamiltonian system: nonlinearity is necessary for chaos, but not sufficient on its own~\cite{Strogatz2015,LichtenbergLieberman1992}.

What the two systems share is not linearity, which only the oscillator has, but a conserved quantity that pins every trajectory to a curve. Integrability is that observation made general: a system is solvable because it carries enough conserved quantities.

An autonomous Hamiltonian system with $d$ degrees of freedom is called Liouville integrable if there exist $d$ functionally independent conserved quantities $F_1,\ldots,F_d$, one of which may be taken to be $H$ itself, that are in involution,
\begin{equation}
    \{F_i,F_j\}=0,
    \qquad i,j=1,\ldots,d.
    \label{eq:L1_involution}
\end{equation}
The Liouville--Arnold theorem~\cite{Arnold1989} then guarantees, on any compact connected component of a joint level set $F_i=c_i$, action-angle variables $(I_i,\theta_i)$ in which $H=H(I_1,\ldots,I_d)$, so that the equations of motion become
\begin{equation}
    \dot I_i=0,
    \qquad
    \dot\theta_i=\omega_i(I).
\end{equation}
The actions are conserved and the angles increase linearly in time. The motion lies on invariant tori, and depending on the frequency ratios it is periodic or quasiperiodic on them. Compactness is what makes the leaves tori and is a real hypothesis rather than a formality: without it a component is $\mathbb{T}^k\times\mathbb{R}^{d-k}$, some directions run off to infinity instead of closing, and the KAM story of Lecture 2, which is about deforming tori, has nothing to act on.

In this sense, integrable systems are geometrically ordered. Their phase spaces are foliated by invariant tori. Chaos appears when this ordered structure is partially or completely destroyed~\cite{Kolmogorov1954,Arnold1963,Moser1962}.

It is useful to remember the following principle:

\begin{quote}
    Integrable systems are special; nonintegrable systems are generic.
\end{quote}

This statement should not be overinterpreted. Nonintegrability does not automatically mean that every trajectory is chaotic. Many systems have both regular and chaotic regions. But the loss of integrability is what allows chaos to enter Hamiltonian mechanics.

\subsection{From one degree of freedom to one-and-a-half}
\label{sec:L1_onehalf}

A one-degree-of-freedom autonomous Hamiltonian system is automatically Liouville integrable, since $H$ itself supplies the single conserved quantity required and is trivially in involution with itself. This is the general reason behind the two phase portraits of the previous subsection: such systems can never be chaotic. There are two natural ways out:

\begin{enumerate}
    \item increase the number of degrees of freedom;
    \item introduce explicit time dependence.
\end{enumerate}

The first route leads to autonomous Hamiltonian systems with $d\geq 2$. The second route is especially useful pedagogically. A one-degree-of-freedom Hamiltonian with any explicit time dependence is called a one-and-a-half-degree-of-freedom system~\cite{LichtenbergLieberman1992,Ott2002}, the extended phase space $(q,p,t)$ being three-dimensional whether or not the time dependence repeats. Its phase space is still described by $(q,p)$ at a fixed time. What periodicity adds is not the name but the tool of the next paragraph.

Consider a Hamiltonian of the form
\begin{equation}
    H(q,p,t)=f(p)+g(t)V(q),
    \qquad
    g(t+T)=g(t).
    \label{eq:L1_periodic_hamiltonian}
\end{equation}
Because the Hamiltonian is periodic in time, we may observe the system once per period, at $t_n=t_0+nT$. This produces a sequence of points $(q_n,p_n)=\big(q(t_n),p(t_n)\big)$, and the evolution from one to the next defines the stroboscopic map $(q_n,p_n)\mapsto(q_{n+1},p_{n+1})$.

The stroboscopic map is an important bridge between continuous-time dynamics and discrete-time dynamics. Periodic motion appears as a fixed point or a finite cycle of the map, quasiperiodic motion as a smooth invariant curve, and chaotic motion as a scattered filamentary spray. Figure~\ref{fig:L1_stroboscopic_map} shows all three at once, for the two systems compared below.

The distinction between linear and nonlinear dynamics can be seen clearly by comparing a driven harmonic oscillator and a driven pendulum.

\begin{figure}[!htbp]
    \centering
    \includegraphics[width=0.94\textwidth]{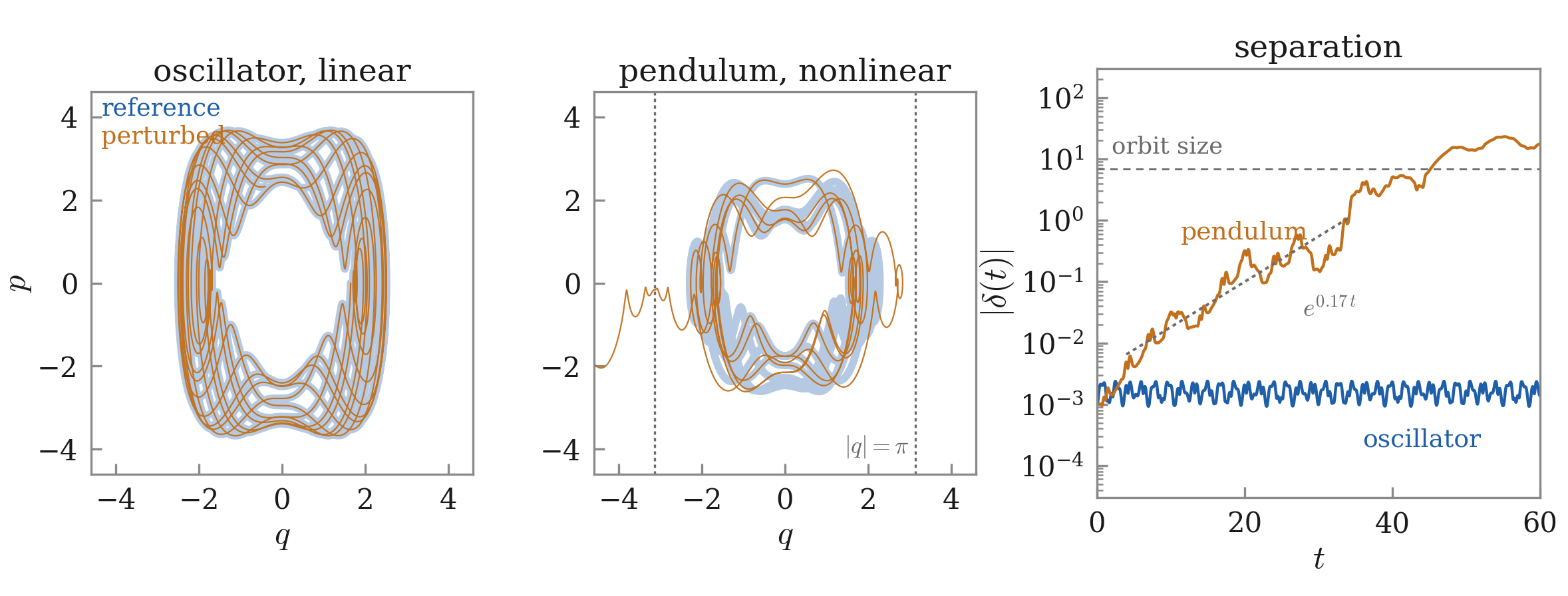}
    \caption{Two nearby trajectories of the driven harmonic oscillator (left) and the driven pendulum (middle), from \eqref{eq:L1_driven_ho} and \eqref{eq:L1_driven_pendulum} with $g_0=1.4$ and $\epsilon=4$, released from rest at $q(0)=1.65$ for the reference orbit and $q(0)=1.65+10^{-3}$ for the perturbed one, shown for $t\le44$. Right: the phase-space separation $|\delta(t)|$ for both, out to $t=60$. In the oscillator the perturbation neither grows nor decays, staying between $9.2\times10^{-4}$ and $2.4\times10^{-3}$ over the whole run. Linearity is not the reason, since a linear system can perfectly well be unstable. The reason is that the one-period map is elliptic here, with $\operatorname{tr}M=+0.394$ against the threshold $|\operatorname{tr}M|=2$; raising $\epsilon$ at fixed $g_0$ carries the trace through $-2$ at $\epsilon=11.11$, and past that the same linear system separates exponentially. In the pendulum the separation grows like $e^{0.17\,t}$ until it reaches the size of the orbit, and by $t=41.0$ the perturbed trajectory has passed $|q|=\pi$ and gone over the top while the reference one is still librating, its angle staying inside $[-2.29,+2.29]$ throughout. A difference of $10^{-3}$ in the initial angle has turned libration into rotation, which is \eqref{eq:L1_sensitive_dependence} made concrete. The fitted rate depends on the window and on the direction of the nudge, running from $0.14$ to $0.22$ across the variants checked, so the exponent here is a scale and not a measurement.}
    \label{fig:L1_driven_oscillator_pendulum}
\end{figure}

\begin{figure}[!htbp]
    \centering
    \includegraphics[width=\textwidth]{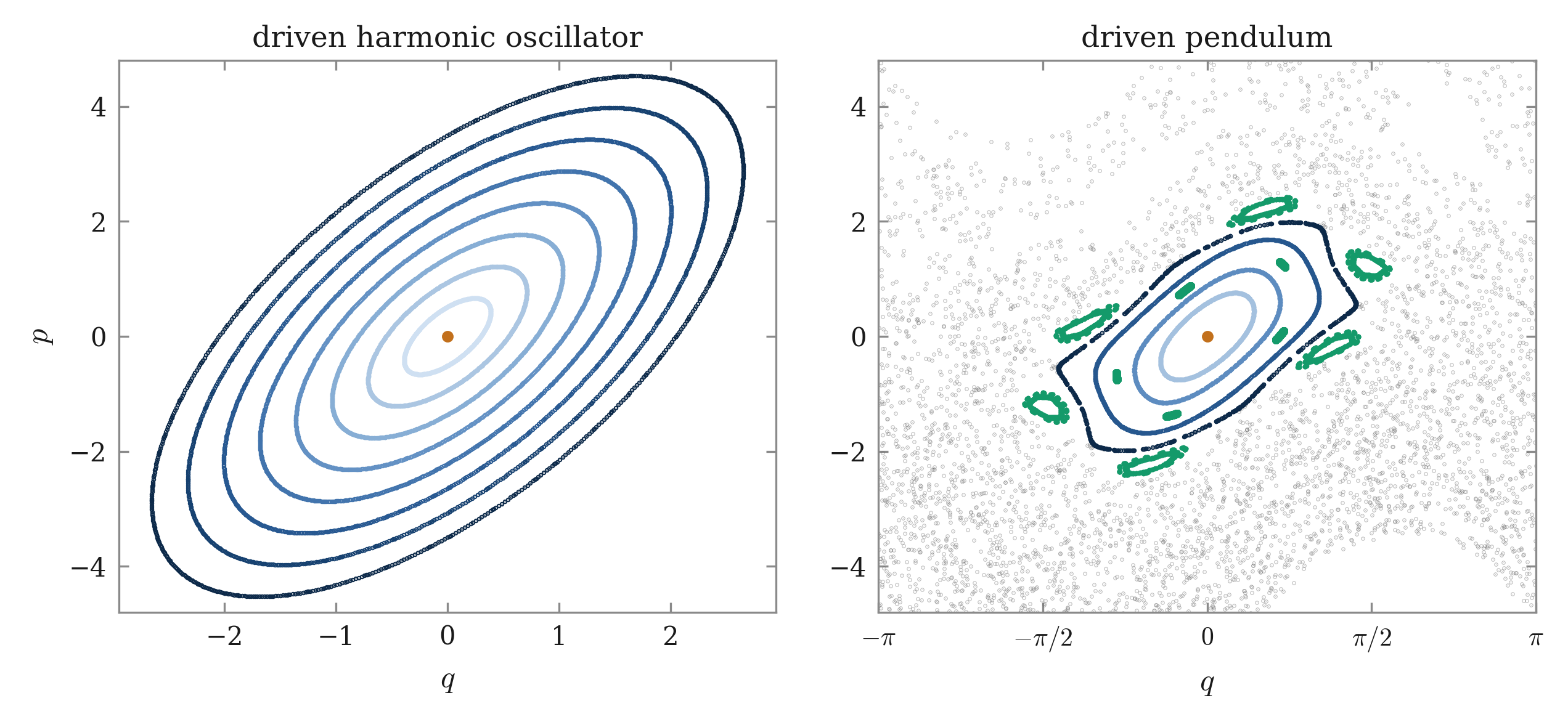}
    \caption{Stroboscopic maps of the two systems of \eqref{eq:L1_driven_ho} and
    \eqref{eq:L1_driven_pendulum} at $g_0=1.4$ and $\epsilon=4$, sampled once per drive period
    from initial conditions released at rest. The orange dot is the fixed point at the origin,
    which both systems carry. Left: the driven oscillator, where the map is the single matrix
    $M$ of the paragraph above. Integrating the two basis solutions over one period gives
    $\det M=1.000$ to eight places, and the elliptic trace quoted in
    Fig.~\ref{fig:L1_driven_oscillator_pendulum} then closes every orbit onto an ellipse
    whatever its amplitude, so one family of nested curves is the whole picture. Right: the
    driven pendulum, with $q$ wrapped to $(-\pi,\pi]$. The same construction returns the three
    behaviours named above, sorted by nothing but the initial amplitude. Blue: smooth invariant
    curves. Teal: two periodic orbits, a five-cycle whose islands measure $0.03$ across and so
    plot as five points, and a six-cycle whose islands are wide enough to see, each period
    measured by splitting the orbit into $m$ interleaved subsequences and taking the smallest
    $m$ at which their spread collapses. Grey: orbits that pass $|q|=\pi$, reach
    $|p|\approx9.4$ and fill the frame without settling. Two remarks. The cycles are
    Poincar\'e--Birkhoff twice over: the rotation number is $0.218$ at the origin and falls as
    the amplitude grows, so it crosses $1/5$ and then $1/6$, and each crossing leaves a chain
    behind. And that a $400\%$ modulation leaves so much regular structure stops being
    surprising once the drive is compared with the pendulum rather than with unity, since
    $\Omega/\omega_0=5.3$ here and a fast drive averages rather than resonates.}
    \label{fig:L1_stroboscopic_map}
\end{figure}

A parametrically driven harmonic oscillator may be written schematically as
\begin{equation}
    H_{\rm ho}(q,p,t)
    =
    \frac{1}{2}p^2
    +
    \frac{1}{2}g_0\left(1+\epsilon\sin(2\pi t)\right)q^2.
    \label{eq:L1_driven_ho}
\end{equation}
Here $g_0$ is the squared frequency of the undriven oscillator and $\epsilon$ the fractional depth of the modulation, with masses and lengths set to unity; the drive has period $T=1$ in the notation of \eqref{eq:L1_periodic_hamiltonian}. At the $\epsilon=4$ of Fig.~\ref{fig:L1_driven_oscillator_pendulum} the bracket $1+\epsilon\sin(2\pi t)$ is negative over $42\%$ of every period, so the restoring force reverses and the oscillator spends close to half of each cycle inverted, which is what makes the stability of that figure's left panel worth remarking on. The pairing of these two examples, and the idea of separating them by linearity rather than by anything else, follows~\cite{Lakshminarayan2018}; the parameters and the perturbed variable used here are not those of that source.
The equations of motion are linear in $q$ and $p$, and the stroboscopic map inherits it. Solutions of a linear equation superpose, so if $(q(t),p(t))$ solves them then so does any linear combination of solutions, and advancing every one of them by a period is therefore a linear operation. The map is multiplication by a fixed $2\times2$ matrix $M$, the same matrix acting on every initial condition, with $\det M=1$ because the flow preserves phase-space area. Its trace alone then fixes the behaviour, by the classification Sec.~\ref{sec:L2_fixed_points} works out. Such a system can be unstable, if the trace leaves $[-2,2]$, but it cannot produce the bounded phase-space complexity a nonlinear one can: one matrix does the same thing to every orbit at once, so amplitude only rescales the picture.

A parametrically driven pendulum may be written as
\begin{equation}
    H_{\rm pend}(q,p,t)
    =
    \frac{1}{2}p^2
    +
    g_0\left(1+\epsilon\sin(2\pi t)\right)(1-\cos q).
    \label{eq:L1_driven_pendulum}
\end{equation}
with $\dot q=p$ and $\dot p=-g_0\left(1+\epsilon\sin(2\pi t)\right)\sin q$. Here the nonlinearity of $\sin q$ is crucial. For sufficiently strong driving, nearby initial conditions may separate rapidly, while the angular nature of the coordinate keeps the motion effectively bounded. This combination leads naturally to chaos.

Thus, the driven pendulum is one of the simplest systems in which deterministic chaos appears. Fig.~\ref{fig:L1_driven_oscillator_pendulum} contrasts the linear and the nonlinear case directly.

The name covers two different systems, and they are worth keeping apart. The parametrically driven Hamiltonian pendulum written here belongs with the periodically driven nonlinear oscillators whose phase space Lecture 2 takes apart~\cite{Chirikov1979,LichtenbergLieberman1992}. Its damped, additively forced cousin is the one that has been built and measured, and there the route to chaos runs through a period-doubling cascade onto attractors~\cite{dHumieresEtAl1982}, which is a feature no Hamiltonian version can have, for the reason Sec.~\ref{sec:L1_lorenz} gives.

\subsection{Instability is not chaos}
\label{sec:L1_ingredients}

Instability alone is not chaos. Consider the inverted harmonic oscillator,
\begin{equation}
    H(q,p)=\frac{p^2}{2m}-\frac{1}{2}m\omega^2q^2.
\end{equation}
The equilibrium point at the origin is unstable. Nearby trajectories separate exponentially. However, the particle simply runs away to infinity. There is no repeated return, no folding, and no complicated bounded exploration of phase space.

Chaos requires more than local instability~\cite{Ott2002}. Devaney's definition packages the requirements differently, as transitivity together with dense periodic orbits, from which sensitive dependence already follows~\cite{Banks1992}. The version used in this lecture requires the combination of:

\begin{enumerate}
    \item local separation of nearby trajectories;
    \item boundedness or recurrence, in the sense of Sec.~\ref{sec:L1_liouville};
    \item repeated stretching and folding of phase-space regions.
\end{enumerate}

The first ingredient gives sensitivity to initial conditions. The second prevents the system from escaping trivially. The third produces complicated geometry.

An unstable system need not be chaotic. A chaotic system, however, typically contains instability in the form of exponential sensitivity to initial conditions.

\subsection{Stretching and folding}
\label{sec:L1_stretchfold}

The third ingredient is the one with a geometry of its own~\cite{Smale1967,Ott2002}. Take a small blob of initial conditions. Regular dynamics may rotate or shear it, but the blob stays a blob. Chaotic dynamics stretches it along unstable directions and squeezes it along stable ones, and because the motion is bounded the stretched region has to be folded back into the space available. Iterating produces filaments that are finer at every step.

One system does this and nothing else, which is why it is worth meeting before the general machinery of Lecture 2. The baker's map acts on the unit square by
\begin{equation}
    B(q,p)=
    \begin{cases}
        \left(2q,\ \tfrac{1}{2}p\right), & 0\leq q<\tfrac{1}{2},\\[4pt]
        \left(2q-1,\ \tfrac{1}{2}(p+1)\right), & \tfrac{1}{2}\leq q\leq 1.
    \end{cases}
    \label{eq:L1_baker}
\end{equation}
The two lines are the two instructions a baker follows, which is where the map gets its
name~\cite{Hopf1937}. Stretch the dough to twice its length and half its thickness, then sever it at the midpoint and set the far piece down on the near one. The first operation is what separates neighbours; the second is what keeps them inside the square. Nothing else in the map does anything else. The treatment here follows that of~\cite{Lakshminarayan2018}.

\begin{figure}[!htbp]
    \centering
    \includegraphics[width=0.94\textwidth]{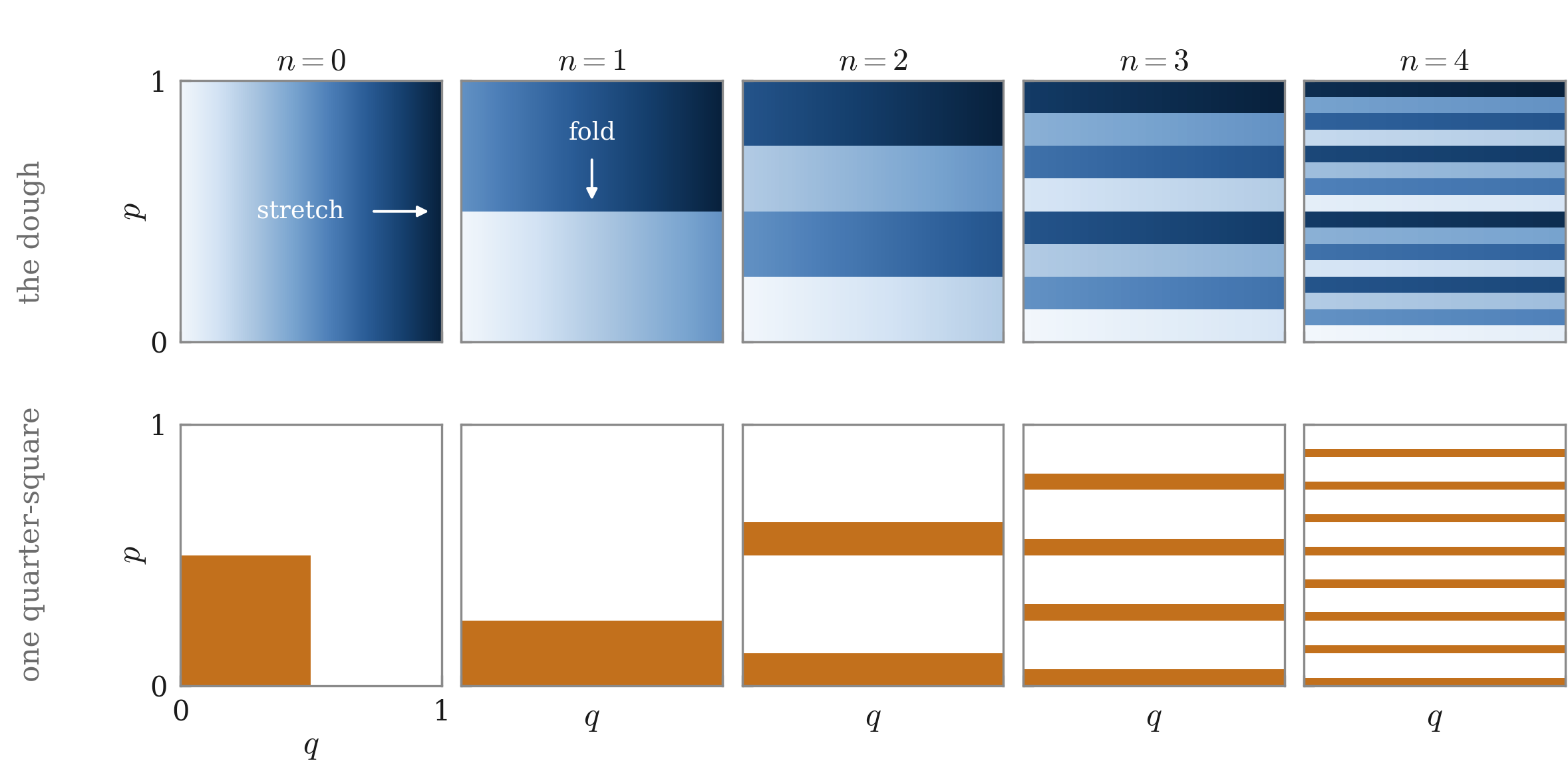}
    \caption{The baker's map, \eqref{eq:L1_baker}, iterated on the unit square. Top: the dough, shaded by the value of $q$ each point started from, light at $q=0$ and dark at $q=1$. After $n$ steps the square is $2^n$ horizontal bands. A band is not a copy of the original gradient. Inside one the shade still runs left to right, but over a range of only $2^{-n}$, and which slice of the gradient a band carries is fixed by reversing the first $n$ bits of $p$. That is the shift of \eqref{eq:L1_baker_symbol} read off the picture: cut and stack, made visible. Bottom: one quarter-square of initial conditions carried through the same steps. Its area is exactly $1/4$ at every $n$ while its shape becomes $2^{n-1}$ strips of thickness $2^{-n-1}$ spanning the full width. Both rows are computed by carrying each output pixel backwards through the inverse map rather than by sampling forwards, so the images are exact at every $n$ and the strips do not alias. The shading ramp is monotonic in luminance, so the ordering survives a greyscale print.}
    \label{fig:L1_stretching_folding}
\end{figure}

Its Jacobian is $\mathrm{diag}(2,\tfrac{1}{2})$ at every point away from the cut, so the map preserves area, and its expansion rate is the same at every point rather than only in the long-run average. Uniform hyperbolicity in that strict sense is what makes the next two statements exact rather than asymptotic. The two Lyapunov exponents are therefore exactly
\begin{equation}
    \lambda_\pm=\pm\log 2,
    \label{eq:L1_baker_lyapunov}
\end{equation}
with no limit to take and no numerical fit, which is a luxury no system in these notes will offer again. Their sum vanishes, as it must for an area-preserving map. Fig.~\ref{fig:L1_stretching_folding} shows the map acting on the square.

The map is also solvable, and that sharpens the lesson of Sec.~\ref{sec:L1_predictability} more than any estimate could. Write both coordinates in binary, $q=0.a_0a_1a_2\ldots$ and $p=0.a_{-1}a_{-2}a_{-3}\ldots$, and record the state as one doubly infinite string of bits with a marker separating the halves,
\begin{equation}
    (p\,|\,q)=\ldots a_{-3}a_{-2}a_{-1}\circ a_0a_1a_2\ldots
    \label{eq:L1_baker_symbol}
\end{equation}
Doubling $q$ discards $a_0$ and promotes everything behind it; halving $p$ and adding $a_0/2$ pushes that same bit onto the front of $p$. The bits themselves are never touched. All that one step of \eqref{eq:L1_baker} does is advance the marker $\circ$ by a single place, and there is nothing else to the dynamics. In the same notation $q_n=2^nq_0 \bmod 1$ in closed form. Chaos here is not a failure to solve the equations of motion. The equations are solved, and the solution is chaotic.

The shift also says precisely what becomes of an imprecisely known state. Suppose $q$ and $p$ are held in double precision, fifty-three bits of mantissa, a relative accuracy near $10^{-16}$. Each step moves one unknown bit into the position that matters, so $q$ loses one bit per iteration and the last of it is gone after fifty-three steps. Equation~\eqref{eq:L1_predictability_time} said the predictability time grows only logarithmically with the initial precision; here that statement is exact and the constant is fixed, one bit per step, so a further factor of ten in precision buys $\log_2 10\approx3.3$ more iterations however good the equipment.

What becomes of the uncertain region is the other half of the story, drawn in the lower row of Fig.~\ref{fig:L1_stretching_folding}. The quarter square $[0,\tfrac12]^2$ has area one quarter and keeps it forever. Its shape does not survive: after $n$ steps it is $2^{n-1}$ strips spanning the whole square, each of thickness $2^{-n-1}$. The dyadic corner is what makes the count that clean rather than what makes it happen; a square of the same size sitting at $[0.1,0.6]\times[0,\tfrac12]$ is cut by the map at $q=\tfrac12$ and lands as two partial strips at different heights, and shreds from there in the same way. The uncertainty has been redistributed rather than amplified. Long before the region is large it has stopped being a region in any useful sense, and quoting an error bar on the state has stopped meaning anything, which is a sharper statement than saying the error grows.

The same string picture counts the periodic orbits, and Lecture 4 will need the count. A state returns to itself after $k$ steps exactly when its bit string repeats with period $k$, so the points fixed by $B^k$ are those built by repeating a block of $k$ bits, and there are exactly $2^k$ of them, a count that includes every period dividing $k$. They are dense: any point of the square has a periodic point within $2^{-K}$ of it, of period $2K$, got by repeating the block formed from its own leading $K$ bits on either side of the marker. All of them are unstable and all share the exponent $\log 2$. Their number grows as $e^{k\log 2}$, so for this map the rate at which periodic orbits proliferate is equal to the Lyapunov exponent. That equality of topological entropy with the exponent is special to uniform hyperbolicity; in general the two are related by Pesin's formula and by the Ruelle inequality~\cite{Pesin1977}. Section~\ref{sec:L4_trace_formula} needs both numbers at once, and the competition between an exponentially proliferating sum and exponentially damped terms is exactly what makes the periodic-orbit expansion there difficult to control.

Real systems are less tidy in three ways. There is no clean cut, so the folding is smooth rather than a stacking; the unstable direction turns as the orbit moves, instead of lying along a coordinate axis; and the stretching rate varies from place to place, so only its long-time average is well defined. Lecture 2 builds the machinery that copes with all three. The mechanism that machinery is measuring is the one written in \eqref{eq:L1_baker}.

\subsection{Ergodicity and mixing}
\label{sec:L1_ergodic}

Recurrence says that a trajectory comes back. It does not say that a trajectory goes everywhere, and the gap between those two statements is where statistical mechanics lives.

A flow is \emph{ergodic} on an energy shell when the shell holds no invariant subset of intermediate size: any set the dynamics carries into itself has either zero measure or the measure of the whole shell. Birkhoff's ergodic theorem turns that into the statement one actually uses~\cite{Birkhoff1931}. For an ergodic flow and any integrable observable $f$,
\begin{equation}
    \lim_{T\to\infty}\frac{1}{T}\int_0^{T}\!f\big(\Phi^t X_0\big)\,dt
    \;=\;\frac{\int f\,d\mu}{\int d\mu}
    \label{eq:L1_birkhoff}
\end{equation}
for almost every starting point $X_0$, with $\mu$ the invariant measure on the shell. Which measure that is matters here, since the claim below is that statistical mechanics rests on this equality. It is not the surface element $dA$ induced on $H=E$ by the ambient metric, which the flow does not preserve, but $d\mu=dA/\|\nabla H\|$, the Liouville measure disintegrated onto the shell. That weight is the microcanonical ensemble. One long trajectory averages to the same thing as the shell. That equality is what licenses the exchange of a time average for a phase-space average, and equilibrium statistical mechanics rests on it.

It is tempting to read \eqref{eq:L1_birkhoff} as a statement about chaos, and it is not. Take the rigid rotation of the two-torus, $(q,p)\mapsto(q+\alpha,\,p+\beta)$ modulo one, with $1$, $\alpha$ and $\beta$ rationally independent; the figures below take $\alpha=(\sqrt5-1)/12$ and $\beta=(\sqrt2-1)/6$. No orbit ever closes, every orbit is dense, and the time average of an observable converges to its average over the torus. Counting the fraction of one orbit that lands in the test rectangle $A=[0.13,0.47)\times[0.09,0.53)$, of area $0.1496$, gives $0.1540$ after $10^{3}$ iterates, $0.14949$ after $10^{5}$, and $0.1496001$ after $10^{7}$. The rotation is ergodic. It is also about as far from chaotic as a system gets: it is linear, its Lyapunov exponents all vanish, and two nearby points stay exactly as far apart as they began.

\begin{figure}[!htbp]
    \centering
    \includegraphics[width=0.96\textwidth]{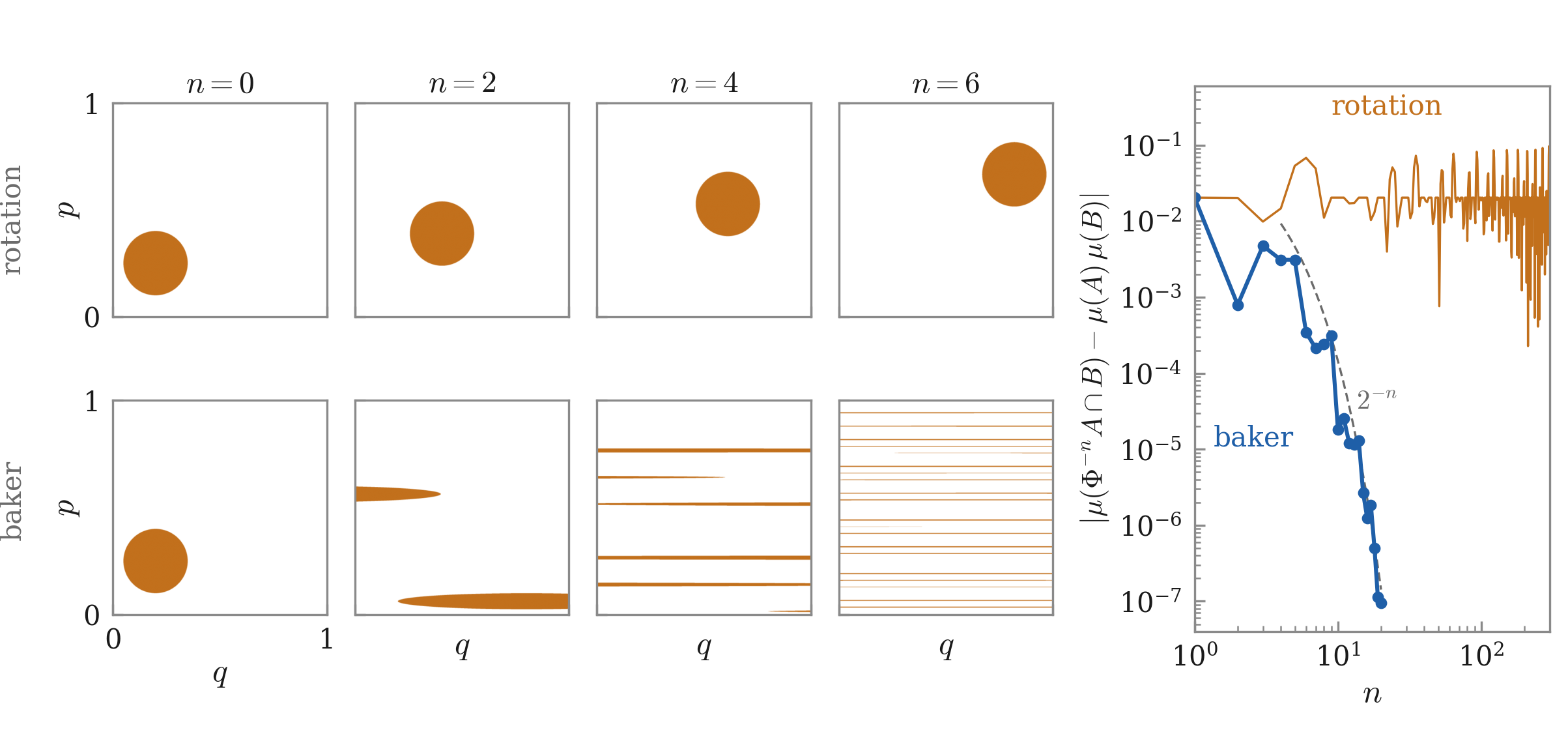}
    \caption{Ergodic is not the same as mixing. One disc of initial conditions, radius $0.15$, under two area-preserving maps of the unit square. Top: a rigid rotation $(q,p)\mapsto(q+\alpha,p+\beta)$ on the torus, that is with the edges identified, with $1$, $\alpha$ and $\beta$ rationally independent. Every orbit is dense and time averages converge to phase-space averages, the relative gap falling from $2.9\times10^{-2}$ after $10^{3}$ steps to $6.7\times10^{-7}$ after $10^{7}$, so the map is ergodic, yet the disc is carried about without ever changing shape. Bottom: the baker's map \eqref{eq:L1_baker} on the same disc, drawn out into a band inside each of the $2^{n}$ strips of thickness $2^{-n}$ that it reaches, thinner than the strip itself, while the area stays fixed at $\pi(0.15)^2$. Right: the quantity that tells the two apart, the discrepancy in \eqref{eq:L1_mixing} for one fixed pair of test rectangles, with $\Phi$ the map of the row in question and $n$ the number of steps. For the baker's map it falls from $2.1\times10^{-2}$ at $n=1$ to $9.5\times10^{-8}$ at $n=20$, at a fitted rate of $0.62$ per step against $\log2=0.69$; for the rotation it stays of order $10^{-2}$ out to three hundred steps with no trend. The images are computed by carrying each pixel backwards through the inverse map, and block-averaging a threefold supersampled grid, which removes the gaps a forward scatter would leave. At the largest $n$ drawn a band is still about three display pixels across, so what is shown is resolved rather than aliased; several steps further on it would not be. The curves are evaluated in closed form, which is what a pixel estimate cannot do once $\Phi^{-n}A$ has become $2^{16}$ slivers, and they agree with a four-million-point Monte Carlo check to $2.5\times10^{-4}$, which is that check's own noise level.}
    \label{fig:L1_ergodic_mixing}
\end{figure}

What the rotation never does is spread anything out. A disc of initial conditions is carried around the torus and stays the same disc, as the top row of Fig.~\ref{fig:L1_ergodic_mixing} shows. The property that separates this from the bottom row is \emph{mixing}: for any two measurable sets $A$ and $B$,
\begin{equation}
    \mu\big(\Phi^{-t}A\cap B\big)\;\longrightarrow\;\mu(A)\,\mu(B)
    \qquad\text{as }t\to\infty,
    \label{eq:L1_mixing}
\end{equation}
with $\Phi^{-t}$ read as the $n$-th inverse iterate when the dynamics is a map~\cite{ArnoldAvez1968}. The left side is the part of $B$ that will be sitting in $A$ at time $t$. Mixing says that this fraction eventually forgets where $B$ is, so that the image of $A$ is spread evenly through the shell on whatever scale one chooses to resolve. For the baker's map of \eqref{eq:L1_baker}, the discrepancy in \eqref{eq:L1_mixing} for a fixed pair of test rectangles falls from $2.0\times10^{-2}$ at one step to $1.3\times10^{-6}$ at sixteen, with an envelope that tracks $2^{-n}$. For the rotation the same discrepancy is still of order $10^{-2}$ after three hundred steps and shows no downward trend at all.

The coding of Sec.~\ref{sec:L1_stretchfold} settles the baker's case without any of this arithmetic. In the symbol string $(p\,|\,q)$ the map is the left shift, and under the uniform measure the bits are independent fair coin flips, so the baker's map is a Bernoulli shift wearing a disguise. Independence of the symbols is mixing written in a different alphabet.

Mixing implies ergodicity, and the rotation shows that the implication runs only one way. The distinction earns its keep because it is mixing, not ergodicity, that makes probabilistic language honest. In a mixing system the chance of finding the trajectory in a given region approaches that region's share of the volume no matter where the trajectory was released, so the initial condition is forgotten and the equilibrium ensemble stops being a convenience and becomes a description of what there is to know. This is the classical ancestor of thermalization, and Sec.~\ref{sec:L4_eth} takes up its quantum form.

Ergodicity also supplies the number Poincar\'e's theorem withheld. Kac's lemma says that in an ergodic system the mean time to first return to a region, averaged over starting points inside it, is exactly the reciprocal of the measure of that region~\cite{Kac1947}. The reason is short: an ergodic orbit spends a fraction $\mu(A)$ of its time inside $A$, so if it is inside on average once every $k$ steps then $k\,\mu(A)$ must be one. On the baker's map a quarter of the unit square gives a measured mean return time of $3.998$ against $4$, and a half gives $2.004$ against $2$. Recurrence is generic, and how long it takes is fixed by nothing more than how much room the region takes up.

\subsection{Lorenz and the butterfly effect}
\label{sec:L1_lorenz}

Although Poincar\'e had already uncovered essential aspects of chaotic dynamics, the modern revival of chaos theory was strongly shaped by computation and meteorology. Edward Lorenz studied simplified models of atmospheric convection and observed that extremely small changes in initial numerical data could lead to drastically different long-time outcomes~\cite{Lorenz1963}. Nine years later a talk of his went out under the title ``Does the Flap of a Butterfly's Wings in Brazil Set Off a Tornado in Texas?''~\cite{Lorenz1972}, and that title is where the phenomenon got the name it now carries. The title was not his. It was supplied by Philip Merilees, who convened the session, while Lorenz was out of the country; Lorenz's own metaphor had been a sea gull, and he never used the phrase \emph{butterfly effect} himself. It reached general circulation through the popular literature that followed~\cite{Gleick1987,Hilborn2004}.

Truncating Saltzman's Fourier expansion for two-dimensional convection~\cite{Saltzman1962} to three modes, which is the debt the 1963 paper acknowledges in its opening, Lorenz obtained
\begin{equation}
    \dot x=\sigma(y-x),
    \qquad
    \dot y=x(\rho-z)-y,
    \qquad
    \dot z=xy-\beta z,
    \label{eq:L1_lorenz_system}
\end{equation}
with the standard parameter values $\sigma=10$, $\rho=28$, $\beta=8/3$. There are only three variables, and only two nonlinear terms, $xz$ and $xy$. That so little structure already destroys long-time predictability is precisely the point.

\begin{figure}[!htbp]
    \centering
    \includegraphics[width=0.92\textwidth]{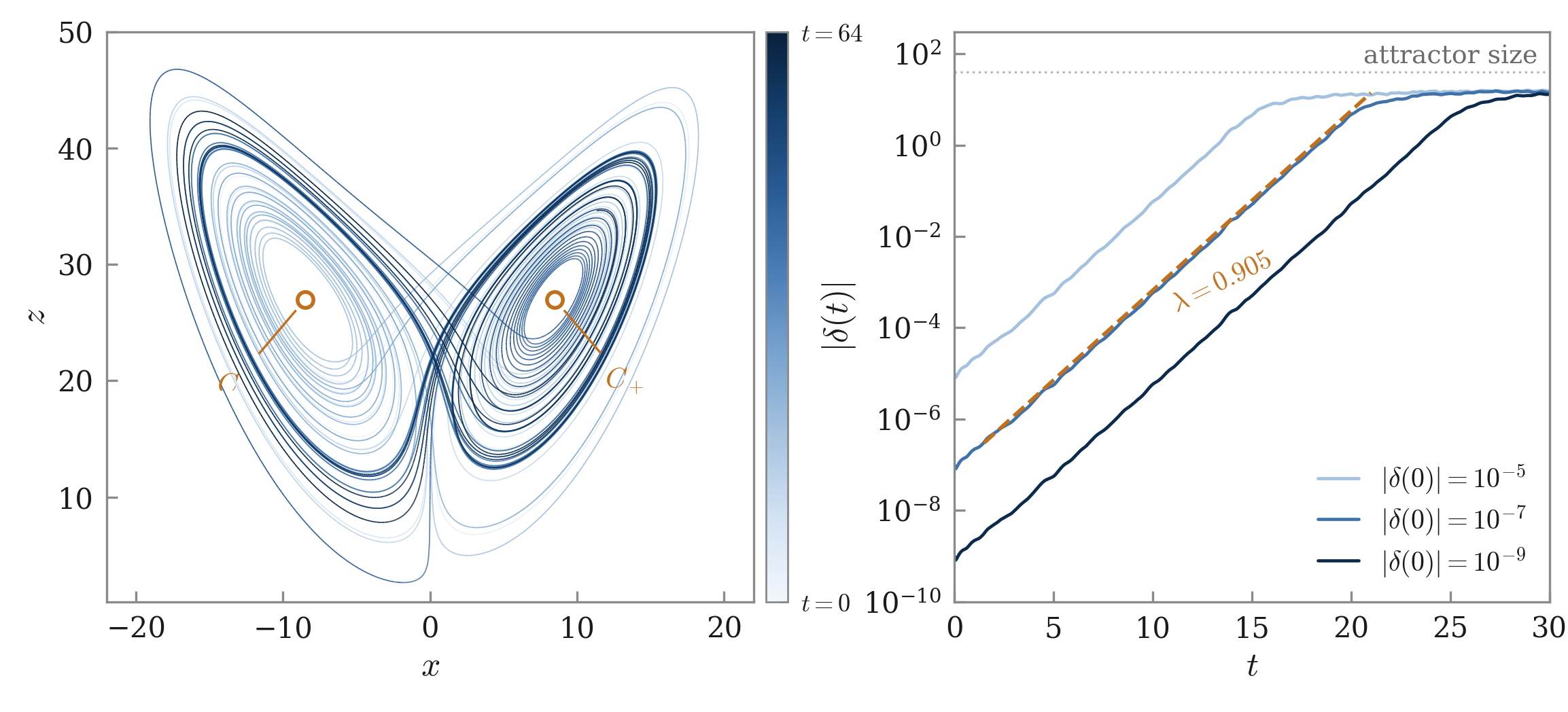}
    \caption{Left: the Lorenz attractor, obtained by integrating \eqref{eq:L1_lorenz_system}
at the standard parameter values \(\sigma=10\), \(\rho=28\), \(\beta=8/3\), shaded by time along
the trajectory.  It circulates around the two unstable fixed points
\(C_\pm=\big(\pm\sqrt{\beta(\rho-1)},\pm\sqrt{\beta(\rho-1)},\rho-1\big)=(\pm8.485,\pm8.485,27)\)
and switches
between the lobes in an aperiodic sequence, never closing on itself.  Right: the sensitivity that
produces it, measured.  Pairs of trajectories started \(10^{-5}\), \(10^{-7}\) and \(10^{-9}\)
apart separate exponentially at the rate \(\lambda\) marked, then saturate at the diameter of the
attractor; the saturation time moves by only \(\log 10/\lambda\) per decade of initial precision,
which is \eqref{eq:L1_predictability_time} in a picture.  The curves are ensemble averages of
\(\log|\delta(t)|\) over 600 decorrelated starting points, since a single pair fluctuates far too
strongly to fit.  The dashed line is not a fit to them: it carries the exponent measured
independently by the Benettin method, which the ensembles reproduce to better than one per cent.
Although the flow contracts phase-space volume at the constant rate
\eqref{eq:L1_lorenz_contraction}, nearby trajectories on the attracting set still separate, so
the set acquires fine transverse structure rather than the smooth geometry of a curve or surface.}
    \label{fig:L1_lorenz}
\end{figure}

The butterfly effect is often misunderstood. The point is not that a butterfly literally causes a storm. The point is that nonlinear deterministic systems can amplify small perturbations so strongly that long-term prediction becomes impossible in practice.

Lorenz's work was important for several reasons. First, it showed that chaos was not confined to celestial mechanics. Second, it made sensitive dependence visible through numerical experiments. Third, it connected chaos to a practical problem: weather prediction. Fourth, it helped establish the modern role of computation in nonlinear dynamics.

One structural point deserves emphasis, since \eqref{eq:L1_lorenz_system} differs in kind from every system considered so far in this lecture. The Lorenz system is not Hamiltonian. It is dissipative, and the phase-space volume element contracts at the constant rate
\begin{equation}
    \frac{\partial\dot x}{\partial x}
    +\frac{\partial\dot y}{\partial y}
    +\frac{\partial\dot z}{\partial z}
    =-(\sigma+1+\beta)<0,
    \label{eq:L1_lorenz_contraction}
\end{equation}
so any bundle of initial conditions shrinks in volume as it evolves. This is what permits an attractor at all. Hamiltonian flow preserves phase-space volume, so no Hamiltonian system can possess one, and nothing resembling Fig.~\ref{fig:L1_lorenz} could have appeared in the earlier sections of this lecture.

The Lorenz system thus introduced one of the iconic images of chaos: a strange attractor. Unlike a fixed point or a limit cycle, a strange attractor supports aperiodic motion while attracting nearby trajectories. It has structure, but not the simple geometry of a smooth curve or surface. Volume contraction and exponential separation are not in conflict here. The volume of a bundle shrinks overall, while distances within it still stretch along the unstable directions and contract more strongly along the stable ones. This is the stretching and folding of Fig.~\ref{fig:L1_stretching_folding} acting on a set that is itself collapsing, and it is what gives the attractor its fine transverse structure.

The central lesson of this lecture is that determinism and predictability are not the same thing. Hamilton's equations fix the future exactly once the present is known, yet nearby initial conditions can separate exponentially fast, so a deterministic system becomes practically unpredictable beyond a finite horizon set by the available precision and the system's Lyapunov exponent. This is a genuine revision of what it means to ``solve'' a physical system: exact equations no longer guarantee exact forecasts.

This revision rests on precise geometric structure, not on vague intuition. Phase-space trajectories cannot cross, which is why a single autonomous degree of freedom is too confined to be chaotic: the trajectory is trapped on the one-dimensional curve $H(q,p)=E$. Chaos requires enough room to move, through extra degrees of freedom or explicit time dependence, and even then it requires more than local instability. Sensitivity to nearby initial conditions, boundedness of the motion, and repeated stretching and folding of phase-space regions~\cite{Smale1967,Ott2002} all have to act together. Each of the two examples above fails in a way one can point at. The inverted oscillator has the first ingredient and lacks the second: its trajectories separate exponentially and then leave, so nothing is ever folded back onto itself. The pendulum has the first two and lacks the third: there is no room in which to fold. That the criterion can be failed in an identifiable way, and differently in each case, is what makes it useful rather than circular.

The three-body problem and the Lorenz system show that this is not a special pathology of exotic equations. Poincar\'e found that the loss of integrability in the three-body problem could generate solution spaces too complicated for any explicit closed form~\cite{Poincare1890,Poincare1892}. Lorenz found the same phenomenon numerically, in a system built for an entirely different purpose~\cite{Lorenz1963}. Simple, deterministic equations turned out to encode far more structure than anyone had expected. This is also the route by which deterministic mechanics connects to statistical descriptions, made precise by the mixing condition of Sec.~\ref{sec:L1_ergodic}: once the dynamics forgets where a trajectory started, probabilistic language becomes honest even though nothing in the underlying equations is random.

What this lecture has not done is measure anything. Trajectories separate, regions stretch and fold, integrable structure breaks, and every one of those statements has been geometric. None of them returns a number that could be computed for a given Hamiltonian and compared against another. Lecture 2 supplies the instruments that do, beginning with the exponent that \eqref{eq:L1_lyapunov_working} so far only sketches.

\subsection{Exercises}

\begin{exercise}
Assume that two nearby trajectories separate as
\begin{equation}
    \delta(t)=\delta(0)e^{\lambda t}.
\end{equation}
Let $\delta(0)=10^{-8}$, $\lambda=0.5$, and suppose prediction fails when $\delta(t)\sim 1$. Estimate the predictability time, and repeat for $\delta(0)=10^{-12}$. Then account for the difference between the two answers without recomputing either: show that the number of correct decimal digits falls linearly in time, at $\lambda/\ln 10$ per unit time, so that four extra digits of initial precision buy $4\ln10/\lambda$ more time whatever $\lambda$ is. Evaluate that rate here, and say how many digits per unit time the Lorenz system loses at the exponent measured in Fig.~\ref{fig:L1_lorenz}.
\end{exercise}

\begin{exercise}
Consider the one-degree-of-freedom autonomous Harper Hamiltonian~\cite{Lakshminarayan2018,Harper1955}
\begin{equation}
    H(q,p)=\cos(2\pi p)+g\cos(2\pi q),
\end{equation}
where $g>0$; for $g<0$ the portrait is the $|g|$ portrait shifted by half a period in $q$, so the critical value is $|g|=1$. Since both $q$ and $p$ appear periodically, one may regard the phase space as a torus with
\begin{equation}
    q\sim q+1,
    \qquad
    p\sim p+1.
\end{equation}
Sketch the phase-space trajectories, i.e. the constant-energy curves
\begin{equation}
    H(q,p)=E,
\end{equation}
for representative values of $g$. In particular, discuss how the qualitative structure of the phase portrait changes as $g$ is varied through the critical value $g=1$.

\textbf{Hint:} The fixed points, obtained by setting $\dot q=\dot p=0$, are the same four for every $g$, and so are their stability types, so they are not where the transition lives. Look instead at the \emph{energies} of the two saddles, $E=1-g$ and $E=g-1$, and at what happens to the level set through them when those two energies coincide.
\end{exercise}

\begin{exercise}
\label{ex:L1_baker_coding}
The coding of Sec.~\ref{sec:L1_stretchfold} does more work than Sec.~\ref{sec:L1_ergodic} asks
of it.
\begin{enumerate}
    \item[(a)] The binary expansion is ambiguous on the dyadic rationals, where
    \(0.0111\ldots=0.1000\ldots\). Show that the ambiguous set has Lebesgue measure zero, and
    that off it the coding is a bijection conjugating the baker's map to the left shift.
    \item[(b)] Under Lebesgue measure the bits are independent fair coin flips. Use that to
    prove mixing directly, and to prove something stronger than
    \eqref{eq:L1_mixing}: if \(A\) and \(B\) each depend on finitely many bits, then
    \(\mu(B^{-n}A\cap B)=\mu(A)\mu(B)\) \emph{exactly} once \(n\) exceeds the span of the two
    sets of bit positions. The correlation does not decay; it stops.
    \item[(c)] Rectangles with dyadic corners are exactly the sets of part (b). Check
    numerically that a pair with non-dyadic corners, such as the pair used for
    Fig.~\ref{fig:L1_ergodic_mixing}, gives a discrepancy that falls like \(2^{-n}\) instead of
    terminating, and explain the difference in one sentence.
    \item[(d)] Kac's lemma was checked in the text on two dyadic regions. Check it on a region
    whose measure is not a power of one half, and say why the lemma needs ergodicity but not
    mixing.
\end{enumerate}
\end{exercise}

\clearpage

\section{Lecture 2: Diagnostic tools of classical chaos}

Lecture 1 placed classical chaos in the gap between determinism and predictability. The
equations of motion fix the future uniquely, yet finite precision on the initial data buys
only a finite horizon of useful prediction. The geometric side of that discussion is what we
build on here: an autonomous Hamiltonian system defines a flow on phase space, energy
conservation confines the flow to a level set, trajectories cannot cross, and one autonomous
degree of freedom leaves too little room for sustained chaos.

That leaves a practical question. Given a Hamiltonian, how does one decide whether its
motion is regular or chaotic, and in which part of phase space? Five instruments do most of
the work~\cite{Ott2002,LichtenbergLieberman1992}: linear stability near fixed points and
periodic orbits, bifurcations, KAM theory, Poincar\'e sections, and Lyapunov exponents.
They are taken in that order for a reason: local structure first, then how that structure
changes as a parameter is moved, then what survives when an integrable system is perturbed,
then a way to see the result, and finally a way to put a number on it.

\subsection{Hamiltonian flow on phase space}

Lecture 1 wrote Hamilton's equations as \eqref{eq:L1_hamilton_equations}, \(2d\) coupled
first-order equations for the components of \(X=(q_1,\ldots,q_d,p_1,\ldots,p_d)\). Everything
here is easier if they are read instead as a single vector field on phase space,
\(\dot X=F(X)\), of which the Hamiltonian case is the one where
\begin{equation}
F(X)=J\nabla H(X),
\qquad
J=
\begin{pmatrix}
0 & \mathbb{I}_d \\
-\mathbb{I}_d & 0
\end{pmatrix},
\label{eq:L2_symplectic_flow}
\end{equation}
with \(\nabla H\) the gradient of \(H\) with respect to all \(2d\) phase-space coordinates
and \(J\) the standard symplectic matrix. Antisymmetry of \(J\) is what separates this from
a gradient flow: since \(\nabla H\cdot J\nabla H=0\), the field \(F\) is everywhere tangent
to the level sets of \(H\), which is why \(H\) is conserved.

Writing the solution as \(X(t)=\Phi^t(X_0)\) lets us treat the dynamics as a flow on phase
space~\cite{Arnold1989}. The same
deterministic flow that organizes an integrable system into invariant tori can stretch, fold
and mix nearby regions in a chaotic one. The diagnostic problem is to decide which is
happening.

\subsection{Fixed points, linear stability, and the trace--determinant diagram}
\label{sec:L2_fixed_points}

Most diagnostic tools begin with the local analysis of fixed points and periodic orbits,
because chaotic motion is organized around them. Islands, separatrices and chaotic layers
all take their shape from the local stability of the fixed points and periodic orbits they
surround~\cite{LichtenbergLieberman1992,Meiss1992}. The selection and ordering of the material
in this subsection and in Sec.~\ref{sec:L2_lyapunov} follow~\cite{Lakshminarayan2018}.

A fixed point of the flow \(\dot X=F(X)\) is a point \(X_\ast\) with \(F(X_\ast)=0\):
start there and the system stays there for all time. To decide whether nearby trajectories
stay nearby, perturb it, \(X(t)=X_\ast+\xi(t)\) with \(|\xi|\ll 1\), substitute into the
equation of motion and keep only the terms linear in \(\xi\),
\begin{equation}
\dot \xi = DF(X_\ast)\,\xi,
\end{equation}
where \(DF(X_\ast)\) is the Jacobian matrix of the vector field evaluated at the fixed
point. The local stability of \(X_\ast\) is therefore fixed by the eigenvalues of
\(DF(X_\ast)\): perturbations that decay make it stable, perturbations that grow make it
unstable, so stability is \(\operatorname{Re}\lambda_i<0\) for every \(i\). No Hamiltonian
equilibrium meets that condition. For a Hamiltonian flow \(DF=J\,\mathrm{Hess}\,H\), whose
eigenvalues come in quadruples \(\pm\lambda,\pm\bar\lambda\), so a decaying direction always
has a growing partner and asymptotic stability is unavailable.

Fixed points of maps matter as much as fixed points of flows here, since such maps arise
naturally from Poincar\'e sections and stroboscopic sampling. For a two-dimensional map
\(X_{n+1}=M(X_n)\) with \(X_n=(q_n,p_n)\), a fixed point satisfies \(X_\ast=M(X_\ast)\), and
the same perturbation \(X_n=X_\ast+\xi_n\) with \(|\xi_n|\ll 1\), kept to linear order, gives
\begin{equation}
\xi_{n+1}=A\,\xi_n,
\qquad
A=DM(X_\ast),
\end{equation}
where \(A\) is the Jacobian, or tangent matrix, of the map at the fixed point. If the
eigenvalues of \(A\) have modulus smaller than one, nearby points approach the fixed
point. If at least one eigenvalue has modulus greater than one, nearby points move away.

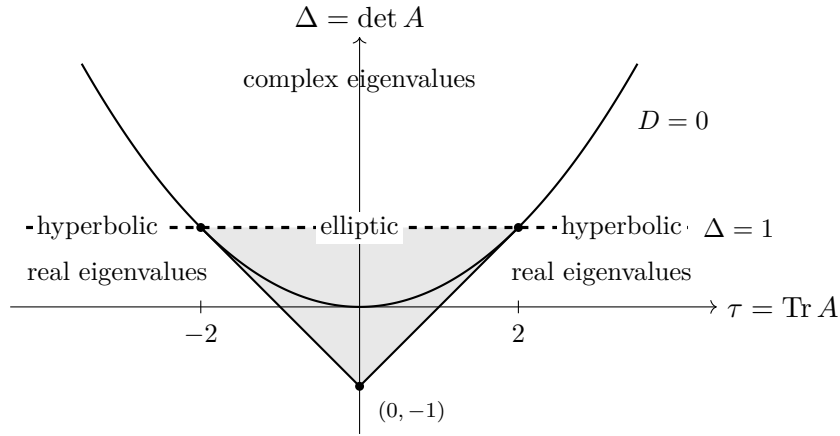
\begin{figure}[!htbp]
\centering
\begin{tikzpicture}[scale=1.05]
    \fill[black!9] (-2,1) -- (2,1) -- (0,-1) -- cycle;

    \draw[->] (-4.4,0) -- (4.5,0)
        node[right] {\(\tau=\operatorname{Tr}A\)};
    \draw[->] (0,-1.6) -- (0,3.4)
        node[above] {\(\Delta=\det A\)};

    \draw[thick, samples=120, domain=-3.5:3.5, smooth]
        plot (\x,{0.25*\x*\x});
    \node[font=\small] at (3.95,2.35) {\(D=0\)};

    \draw[thick] (0,-1) -- (2,1);
    \draw[thick] (0,-1) -- (-2,1);

    \draw[dashed, very thick] (-4.2,1) -- (4.2,1);
    \node[right,font=\small] at (4.2,1) {\(\Delta=1\)};

    \fill (-2,1) circle (1.6pt);
    \fill (2,1) circle (1.6pt);
    \fill (0,-1) circle (1.6pt);
    \node[below right,font=\scriptsize] at (0.10,-1.06) {\((0,-1)\)};

    \node[font=\small, fill=white, inner sep=1.6pt] at (0,1) {elliptic};
    \node[font=\small, fill=white, inner sep=1.6pt] at (-3.3,1) {hyperbolic};
    \node[font=\small, fill=white, inner sep=1.6pt] at (3.3,1) {hyperbolic};

    \node[font=\small] at (0,2.85) {complex eigenvalues};
    \node[font=\small] at (-3.05,0.42) {real eigenvalues};
    \node[font=\small] at (3.05,0.42) {real eigenvalues};

    \draw (-2,0.08) -- (-2,-0.08);
    \draw (2,0.08) -- (2,-0.08);
    \node[below,font=\small] at (-2,-0.10) {\(-2\)};
    \node[below,font=\small] at (2,-0.10) {\(2\)};
\end{tikzpicture}
\caption{The \((\tau,\Delta)\) plane for the tangent matrix \(A\) at a fixed
point of a two-dimensional map. The parabola \(\Delta=\tau^2/4\) is the discriminant
\(D=0\), separating real eigenvalues below from complex-conjugate pairs above. The shaded
triangle with vertices \((\pm2,1)\) and \((0,-1)\) is the region \(|\Lambda_{1,2}|<1\) of
asymptotic stability; its three edges are the three ways a fixed point of a map loses
stability, saddle-node, period doubling and Neimark--Sacker, in the order the text works
through them.
Area-preserving maps are confined to the dashed line \(\Delta=1\), the top edge, for every
value of any parameter: they never enter the interior. Between the corners the eigenvalues
sit on the unit circle (elliptic), outside them they are real and reciprocal (hyperbolic).}
\label{fig:trace_det_diagram}
\end{figure}

For a \(2\times 2\) matrix the eigenvalues are fixed entirely by the trace
\(\tau=\operatorname{Tr}A\) and the determinant \(\Delta=\det A\), through the characteristic
equation
\begin{equation}
\Lambda^2-\tau\Lambda+\Delta=0,
\end{equation}
so the pair \((\tau,\Delta)\) determines the local type of the fixed point. The capital is
deliberate: an eigenvalue of \(A\) is a multiplier, a factor per iteration of the map, while
the rate per unit time this lecture writes \(\lambda\) is its logarithm divided by a time, and
Sec.~\ref{sec:L2_poincare} does that conversion. The discriminant of the quadratic,
\(D=\tau^2-4\Delta\), separates the two cases: \(D>0\) gives real eigenvalues,
\(D<0\) a complex-conjugate pair, and the boundary between them is the parabola
\(\Delta=\tau^2/4\). This gives the trace--determinant diagram of Fig.~\ref{fig:trace_det_diagram}.

For a map, stability means \(|\Lambda_{1,2}|<1\), which confines the eigenvalues to the open
triangle bounded by
\begin{equation}
\Delta=1,
\qquad
\Delta=\tau-1,
\qquad
\Delta=-\tau-1 ,
\label{eq:L2_stability_lines}
\end{equation}
with vertices at \((\pm 2,1)\) and \((0,-1)\). Its three edges are the three ways a fixed
point of a map loses stability: an eigenvalue leaving through \(+1\) on \(\Delta=\tau-1\)
(saddle-node), through \(-1\) on \(\Delta=-\tau-1\) (period doubling, which we return to in
the next subsection), and a complex-conjugate pair crossing the unit circle on \(\Delta=1\)
(Neimark--Sacker)~\cite{Neimark1959,Sacker1964}.

Poincar\'e and stroboscopic maps obtained from Hamiltonian systems are area-preserving, so
\(\Delta=\det A=1\), and the Hamiltonian case is the whole line \(\Delta=1\), the top edge of that triangle together
with its extension past the two corners, rather than any part of the interior. Two consequences follow, and both matter later. First, an area-preserving fixed
point is never asymptotically stable: an elliptic point is marginally stable, surrounded by
invariant curves that neither approach it nor recede from it.  Only the first half of that
follows from \(\Delta=1\).  The invariant curves are Moser's twist theorem, the fixed-point
analogue of the KAM statement of Sec.~\ref{sec:L2_kam}, and it comes with hypotheses: the
rotation number at the point has to avoid the low-order resonances, \(\tau=0\) and
\(\tau=-1\) among them, and a twist coefficient has to be nonzero.  At the \(1{:}3\)
resonance an elliptic point is generically unstable rather than surrounded by curves. Second, since \(\Delta\)
equals one for every value of any parameter, such a map never crosses the \(\Delta=1\) edge
and so never undergoes a Neimark--Sacker bifurcation. That last is a two-dimensional statement.
In a \(2n\)-dimensional symplectic map with \(n\geq2\) the eigenvalues come in quadruples
\(\Lambda,\Lambda^{-1},\bar\Lambda,\bar\Lambda^{-1}\) and can leave the unit circle when two
pairs collide on it, so the higher-dimensional sections of Sec.~\ref{sec:L2_poincare} are not
protected in the same way. The corners \((\pm2,1)\) are
the only places where the fixed point itself can change stability type, an eigenvalue reaching
\(+1\) or \(-1\) there.  They are not the only bifurcations available: whenever
\(\tau=2\cos(2\pi p/q)\), which happens at every rational rotation number strictly inside
\(|\tau|<2\), a satellite orbit of period \(q\) is born around the elliptic point.  That is
the map-level origin of the island chains of the next subsection.

Along this line the eigenvalue equation becomes \(\Lambda^2-\tau\Lambda+1=0\), so
\(\Lambda_1\Lambda_2=1\): if one eigenvalue expands a perturbation, the other contracts it by
the reciprocal amount. This is the local linear expression of area preservation, and it
gives the classification
\begin{equation}
|\tau|<2:\ \text{elliptic},
\qquad
|\tau|>2:\ \text{hyperbolic},
\qquad
|\tau|=2:\ \text{parabolic or marginal}.
\end{equation}

When \(|\tau|<2\), the eigenvalues are complex conjugates on the unit circle,
\(\Lambda_{1,2}=e^{\pm i\alpha}\), and the linearized motion is locally rotational. Such a
fixed point is called elliptic. In
Hamiltonian phase space, elliptic fixed points are typically surrounded by regular
quasi-periodic curves. These are the centers of stability islands.

When \(|\tau|>2\) the eigenvalues are real, so one direction expands and the other
contracts. Such a fixed point is called hyperbolic. The sign matters for what the iterates look
like: at \(\tau<-2\) both eigenvalues are negative and the point is inverse hyperbolic, each
iterate exchanging the two branches of both manifolds, which is the regime an orbit enters
immediately past the period doubling at \(\tau=-2\).
Hyperbolic fixed points possess stable and unstable manifolds. The stable manifold is
the set of initial conditions that approach the fixed point under forward iteration; the
unstable manifold is the set of initial conditions that approach it under backward
iteration.

These manifolds are where local stability stops being merely local. If the stable and
unstable manifolds of a hyperbolic point intersect transversally at a point other than the
fixed point itself, the intersection cannot be isolated: every image and preimage of that
homoclinic point is another transverse intersection, and they accumulate on the fixed point
itself. Both manifolds therefore have to fold back through each other infinitely often while
the area they sweep stays bounded, which is what forces the excursions to grow ever longer and
thinner. The Smale--Birkhoff theorem turns this geometry into a
statement about dynamics. Some iterate of the map contains a Smale
horseshoe, and therefore an invariant Cantor set on which the dynamics is conjugate to a
shift on two symbols, with positive topological entropy~\cite{Smale1967,GuckenheimerHolmes1983}.
This is the sense in which a hyperbolic point organizes chaos. The eigenvalues of \(A\)
certify none of it: they describe the linearization, and a positive \(\log|\Lambda|\) at one
fixed point says nothing about whether orbits keep returning to its neighbourhood.

When \(|\tau|=2\), the eigenvalues collide at \(\Lambda=1\) or \(\Lambda=-1\). This is the
marginal case, sitting at the corners \((\pm 2,1)\) of the stability triangle.
For an area-preserving map the two slanted edges degenerate to these corners, which is where
the fixed point changes type: at \(\tau=2\) an eigenvalue reaches \(+1\), at \(\tau=-2\) it
reaches \(-1\) and the orbit period-doubles.  Bifurcations of the satellite orbits happen
throughout the interval, as noted above. In parameter-dependent
systems this is where an elliptic point turns hyperbolic, or the reverse.

The trace--determinant diagram thus gives a compact geometric summary of local stability,
and for Hamiltonian maps it explains why elliptic and hyperbolic points are the two local
building blocks of mixed phase space~\cite{Meiss1992,LichtenbergLieberman1992}. Elliptic
points organize regular islands, hyperbolic points organize unstable manifolds and the
chaotic layers that grow out of them.

\subsection{Bifurcations}
\label{sec:L2_bifurcations}

The three edges of the stability triangle are each a bifurcation: a qualitative change in
the phase-space structure as a control parameter is varied. Writing the parameter dependence explicitly as \(\dot X=F(X;\mu)\), or \(X_{n+1}=M_\mu(X_n)\)
for a map, the question is what becomes of fixed points and periodic orbits as \(\mu\) is
tuned across one of those edges.

The simplest instance is the saddle-node bifurcation, in which two fixed points are created
or destroyed together. For a one-dimensional flow the normal form is
\begin{equation}
\dot x=\mu-x^2 ,
\end{equation}
with fixed points \(x_\ast=\pm\sqrt{\mu}\) for \(\mu>0\), none for \(\mu<0\), and a collision
of the two at \(\mu=0\). The corresponding map, \(x_{n+1}=x_n+\mu-x_n^{2}\), says the same
thing in the language the triangle above is drawn in: its fixed points sit at the same places,
their multipliers are \(1\mp2\sqrt{\mu}\), and both reach \(+1\) as the pair collides, which
is the edge \(\Delta=\tau-1\).

Period doubling is the case that matters most for maps. A fixed point of a one-dimensional
map loses stability when \(f'(x_\ast)=-1\), and a stable period-two orbit typically appears in
its place. Iterating the process gives the
period-doubling cascade, whose parameter intervals shrink geometrically with the Feigenbaum
ratio \(\delta\simeq 4.6692\)~\cite{Feigenbaum1978,Feigenbaum1979}. Area preservation changes
the number without changing the phenomenon: families of area-preserving maps have their own
cascade, with
\(\delta\simeq 8.7211\)~\cite{GreeneMacKayVivaldiFeigenbaum1981,Bountis1981,LichtenbergLieberman1992}. The universality comes from the
renormalization structure, which survives area preservation.

Area preservation also constrains what happens to a whole family of periodic points at once,
not merely to one fixed point, and there it takes a sharper form still. Consider a resonant
torus, one whose frequencies stand in a rational ratio
\(p/q\); it meets a surface of section in a circle every point of which is a period-\(q\)
point of the section map. Perturb it, keeping the twist condition that the rotation number
varies monotonically across the annulus. The Poincar\'e--Birkhoff theorem~\cite{Birkhoff1913} says that the
circle does not survive intact: an even number \(2kq\) of period-\(q\) points of the section
map remains, for some positive integer \(k\) that is one in the generic case,
alternating around the circle between elliptic and
hyperbolic~\cite{LichtenbergLieberman1992,Meiss1992}. The tori and sections this statement
refers to, and the rotation number and twist condition it uses, are set up properly in the next
two subsections. That alternation is the origin of the
island chains on Poincar\'e sections, with elliptic points at the island centres and
hyperbolic points in the necks between them. The chains return in the KAM picture below.

A bifurcation is a change in the structure of the dynamics, and chaos may follow a sequence
of them, but one bifurcation on its own does not make a system chaotic. What settles how much
of phase space chaos can eventually occupy is not any single bifurcation but what a
perturbation leaves standing, and that is the subject of the next subsection.

\subsection{Integrable tori and the KAM picture}
\label{sec:L2_kam}

Lecture 1 introduced Liouville integrability: an autonomous Hamiltonian system with \(d\)
degrees of freedom is integrable when it carries \(d\) independent conserved quantities in
involution, and the Liouville--Arnold theorem then supplies action-angle variables
\((I_i,\theta_i)\) in which \(H=H_0(I)\), so that \(\dot I_i=0\) and
\(\dot\theta_i=\omega_i(I)=\partial H_0/\partial I_i\). The actions are frozen and the angles
wind at fixed rates, \(\theta_i(t)=\theta_i(0)+\omega_i(I)\,t\). What matters for the
diagnostics is the geometry this implies~\cite{Arnold1989}.

Each angle is periodic, so a level set of the actions is a \(d\)-torus, and phase space is
foliated by these invariant tori. Take the case a Poincar\'e section can display, \(d=2\).
The tori are two-dimensional surfaces in a four-dimensional phase space; fixing the energy
confines the motion to a three-dimensional shell, and a section through that shell meets each
torus in a closed curve. A smooth closed curve on a Poincar\'e section is therefore the
signature of an intact torus.

The two frequencies define the rotation number \(W=\omega_1/\omega_2\), the average advance
in one angle per turn of the other, equivalently the average rotation per crossing of the
section. Its arithmetic decides what happens under perturbation. If \(W=p/q\)
is rational, every orbit on the torus closes after \(q\) crossings and the entire torus is
periodic; this is the resonant case to which the Poincar\'e--Birkhoff theorem applies. If
\(W\) is irrational, a single orbit covers the torus densely and never repeats.

Now perturb. Write a nearly integrable Hamiltonian as
\begin{equation}
H(I,\theta)=H_0(I)+\epsilon H_1(I,\theta),
\end{equation}
with \(\epsilon\) controlling the strength of the perturbation. For general \(d\) the
resonance condition generalizes the rational-\(W\) case: a torus is resonant when its
frequency vector \(\omega(I)=(\omega_1(I),\ldots,\omega_d(I))\) admits a nonzero integer
vector \(m\in\mathbb{Z}^d\) with \(m\cdot\omega(I)=0\)~\cite{LichtenbergLieberman1992,Chirikov1979}.
Resonant tori are the fragile ones, and Sec.~\ref{sec:L2_bifurcations} has already said what
becomes of them. At the hyperbolic points of the chain that replaces such a torus the stable
and unstable manifolds cross transversally and then go on crossing infinitely often, the
homoclinic tangle that supplies the horseshoe of Sec.~\ref{sec:L2_fixed_points}.

The Kolmogorov--Arnold--Moser theorem says what happens to the rest. If \(H_0\) is
nondegenerate, \(\det(\partial^2H_0/\partial I^2)\neq 0\), so that the frequencies vary with
the actions, and if the perturbation is small enough and the frequency vector is sufficiently
far from resonance, the torus survives, smoothly
deformed~\cite{Kolmogorov1954,Arnold1963,Moser1962}. Smoothness is a hypothesis and not a
technicality: \(H\) is taken analytic, or \(C^{r}\) with \(r\) large enough in \(d\), and
the theorem is false without it. The precise
non-resonance requirement is the Diophantine condition
\begin{equation}
|m\cdot \omega(I)|
\geq
\frac{\gamma}{|m|^{\nu}},
\qquad
m\in\mathbb{Z}^d\setminus\{0\},
\label{eq:L2_diophantine}
\end{equation}
for positive constants \(\gamma\) and \(\nu>d-1\), with \(|m|\) any fixed norm on
\(\mathbb{Z}^d\), since whether the condition can be met at some \(\nu>d-1\) does not depend
on which norm is chosen. It says that \(\omega\) cannot be
approximated too well by any rational relation. The exponent is written \(\nu\) here to avoid
collision with the trace \(\tau\) of Sec.~\ref{sec:L2_fixed_points}.

What makes the theorem striking is the measure statement that accompanies it. The resonant
tori, which are destroyed, form a dense set: arbitrarily close to any torus there is a
resonant one. Yet the surviving Diophantine tori occupy positive measure, and that measure
tends to the full measure of the region as
\(\epsilon\to 0\)~\cite{Arnold1963,Moser1962,LichtenbergLieberman1992}. The set removed has
measure of order \(\sqrt\epsilon\), which is what makes destruction dense but thin a
quantitative statement rather than a rhetorical one.
This is why a Poincar\'e section of a weakly perturbed system looks the way it does, with
smooth curves filling most of the picture and island chains and thin chaotic layers threaded
between them. As \(\epsilon\) grows the surviving set shrinks, the layers widen and merge,
and the regular curves give way to a connected chaotic sea.

For H\'enon--Heiles there is no separate \(\epsilon\) to tune, and it is worth saying what
plays its part. Rescaling \((x,y,p_x,p_y)\to\sqrt{E}\,(x,y,p_x,p_y)\) takes the Hamiltonian
into the harmonic part plus \(\sqrt{E}\) times the cubic, at fixed energy one, so the
perturbation parameter is \(\epsilon=\sqrt{E}\) exactly. Between the two panels of
Fig.~\ref{fig:henon_heiles_sections} it therefore rises only from \(0.289\) to \(0.354\), a
change of twenty-two per cent, while the section goes from smooth curves nearly everywhere to a
connected sea. Small in \(\epsilon\) is not small in what the picture looks like.

For \(d=2\) one can ask which torus is the most robust. Greene's residue criterion tracks the
stability of the periodic orbits whose rotation numbers converge to a given \(W\), through the
residue
\begin{equation}
    R=\frac{2-\operatorname{Tr}M_q}{4}
\end{equation}
of the period-\(q\) orbit, which is the trace classification of Sec.~\ref{sec:L2_fixed_points}
rewritten: \(0<R<1\) is elliptic, and \(R<0\) or \(R>1\) is hyperbolic. Watching \(R\) along
the sequence
led to the conjecture that the last curves to break are those with noble rotation number, a
continued-fraction expansion ending in an infinite tail of ones, and that in the standard map
the very last is the golden-mean circle \(W=(\sqrt{5}-1)/2\)~\cite{Greene1979}. Which noble
circle survives longest is map-dependent. The criterion remains a
conjecture, with parts of it proved for Aubry--Mather sets that are smooth invariant circles or
uniformly hyperbolic~\cite{FalcoliniDeLaLlave1992}, but it is the standard numerical route
to the critical perturbation strength.

\subsection{Poincar\'e sections}
\label{sec:L2_poincare}

\begin{figure}[!htbp]
    \centering
    \includegraphics[width=0.94\textwidth]{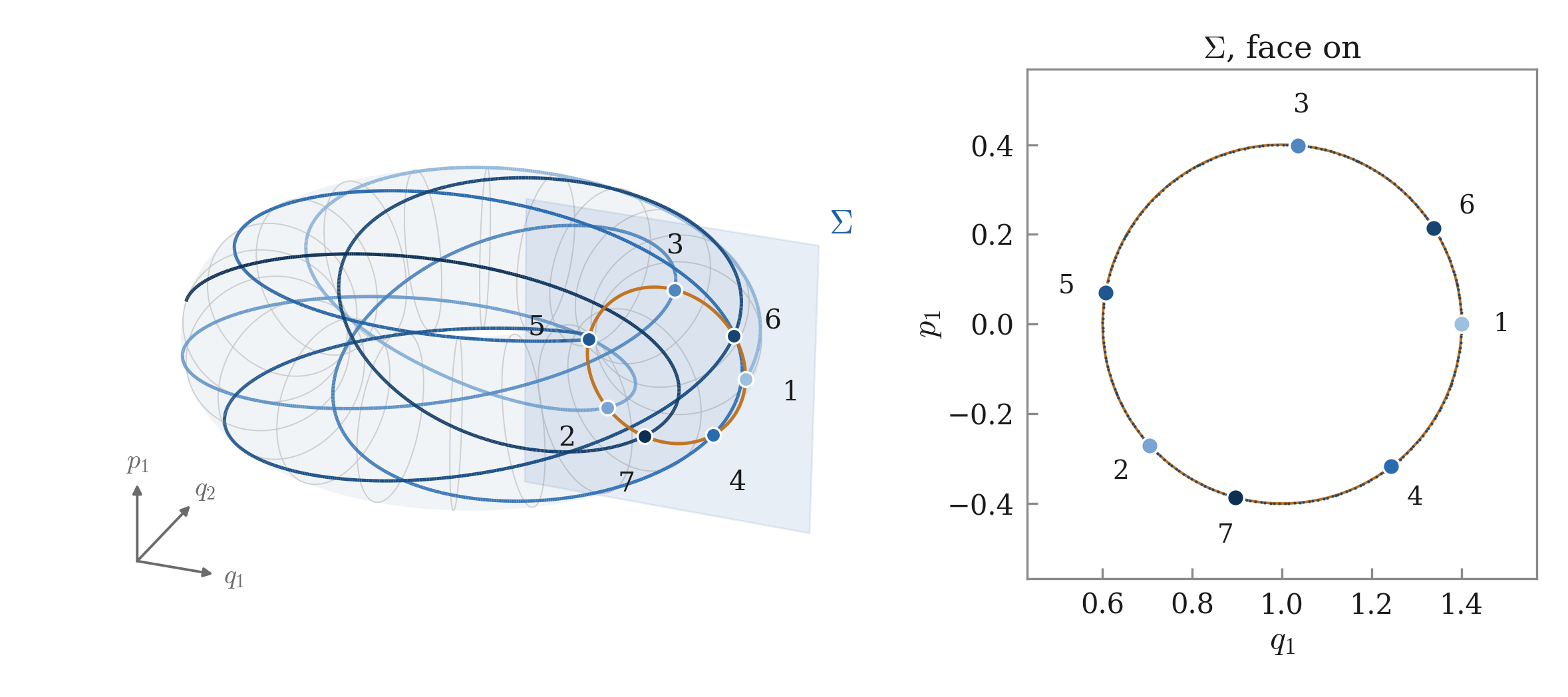}
    \caption{Left: a single quasi-periodic trajectory winding on an invariant torus, drawn in
    the \((q_1,q_2,p_1)\) projection with tube radius \(r=0.4\) about a circle of radius
    \(R=1\), and a surface of section \(\Sigma\) placed transverse to the flow.  Colour runs
    along the trajectory in the direction of increasing time.  Here \(\Sigma\) is the
    half-plane \(q_2=0\), \(q_1>0\), taken in the direction of increasing \(q_2\), and the
    flow is nowhere tangent to it: \(\dot q_2\) on \(\Sigma\) is \(R+r\cos\theta_1\) times the
    angular frequency, which never drops below \(0.6\).  The seven marked points are the first
    seven crossings, numbered in the order they occur and carrying the colour of the time at
    which they happen.  Right: the same seven points seen in \(\Sigma\) itself.  They lie on
    the circle of radius \(r\) about \((R,0)\) in which \(\Sigma\) cuts the torus, and
    successive crossings advance around it by \(2\pi W\) with \(W=(\sqrt5-1)/2\) the rotation
    number.  Since \(W\) is irrational no crossing repeats and the orbit fills the circle: the
    dark points strung along the orange curve are \(240\) later crossings, and the largest gap
    among the first \(N\) of them shrinks like \(1/N\), with \(N\) times that gap staying
    between \(7.3\) and \(8.3\) out to \(N=247\).  By the three-gap
    theorem~\cite{Sos1958,Swierczkowski1959} those
    gaps take at most three distinct values at any \(N\), and exactly two when \(N\) is a
    Fibonacci number, which holds here at \(N=5,8,13,21,34\).  The layout of this figure
    follows the interactive visualization of Ref.~\cite{Stein2018}.}
    \label{fig:poincare_section_schematic}
\end{figure}

A Poincar\'e section converts a continuous-time flow into a discrete-time map by recording
successive intersections of a trajectory with a chosen lower-dimensional surface. The
construction is shown in Fig.~\ref{fig:poincare_section_schematic}.

Choose a hypersurface \(\Sigma\) in phase space. For example, in a two-degree-of-freedom
Hamiltonian system with coordinates \((q_1,q_2,p_1,p_2)\), one may choose
\begin{equation}
\Sigma:\quad q_2=0,\qquad p_2>0.
\end{equation}
The condition \(p_2>0\) fixes the direction of crossing and prevents counting the same
oscillation twice. Two requirements make the construction work. The surface must be
transverse to the flow, so that no trajectory of interest grazes it tangentially, and orbits
must actually come back to it. Where either fails the return map is not defined, which in
practice is what limits how far a section can be pushed before it starts to mislead.

The Poincar\'e map is then \(X_{n+1}=P(X_n)\), with \(X_n\) the \(n\)-th intersection of the
trajectory with \(\Sigma\)~\cite{Poincare1892,Meiss1992}.

This map is area-preserving, which is the fact Sec.~\ref{sec:L2_fixed_points} assumed when it
set \(\Delta=\det A=1\). The reason is that the symplectic form \(\Omega=\sum_i dp_i\wedge
dq_i\) is conserved by the Hamiltonian flow, so its restriction to the section is preserved by
the return map.  Its sign is opposite to the one carried by the matrix \(J\) of
\eqref{eq:L2_symplectic_flow}, whose two-form is \(\sum_i dq_i\wedge dp_i\); that is a choice
of convention and not a disagreement, and nothing here or later depends on which is made. For \(d=2\) that restriction is easy to see explicitly: on
\(\Sigma=\{q_2=0\}\) one has \(dq_2=0\), so
\begin{equation}
\Omega\big|_\Sigma = dp_1\wedge dq_1 ,
\end{equation}
which is precisely the area element of the \((q_1,p_1)\) plane the section is plotted in.
Hence \(P\) preserves area, the eigenvalues of its linearization multiply to one, and
everything said earlier about elliptic and hyperbolic fixed points rests on this. For
\(d>2\) the same argument gives a symplectic form on the \((2d-2)\)-dimensional section
rather than an area.

A periodic orbit of the flow shows up on the section as a finite set of points, and if it
pierces the surface \(q\) times per period it is a fixed point of \(P^{q}\). Its stability is
therefore read exactly as in Sec.~\ref{sec:L2_fixed_points}, from the linearisation of that
return map. That linearisation has a name of its own, the \emph{monodromy matrix} \(M_p\) of
the orbit, and one feature of it is worth stating because Lecture 4 relies on it. \(M_p\) is
the \((2d-2)\times(2d-2)\) map on the section, not the full \(2d\times2d\) linearisation of
the flow. The full object always carries the eigenvalue \(+1\) twice: once along the orbit,
since displacing a point along its own trajectory only slides it to a later time on the same
trajectory, and once again because the multipliers of a symplectic matrix come in reciprocal
pairs and \(1\) is its own reciprocal. Anything built from \(\det(1-M)\) would vanish
identically for the full matrix, which is why the reduced block is the object that appears in
a trace formula.

Being symplectic, \(M_p\) has multipliers in reciprocal pairs \(\Lambda,1/\Lambda\), which is
the fixed-point reciprocity of Sec.~\ref{sec:L2_fixed_points} now attached to an orbit. For
\(d=2\) an unstable orbit has \(\Lambda_p=e^{\lambda_pT_p}\) with \(T_p\) the period, which is
what converts a multiplier per traversal into a rate per unit time. Moving the base point
around the orbit conjugates \(M_p\) and leaves its multipliers alone, so they belong to the
orbit rather than to the place one chose to cut it, and the \(r\)-th traversal has monodromy
\(M_p^{\,r}\).

The dimension count explains when sections are useful. For \(d\) degrees of freedom the
phase space is \(2d\)-dimensional, energy conservation restricts the motion to a
\((2d-1)\)-dimensional shell, and taking a section removes one dimension more, leaving
\(2d-2\). For \(d=2\) that is a two-dimensional picture one can simply plot, which is why
sections are the tool of choice for two-degree-of-freedom systems. The dimension does more than
decide what can be drawn. A surviving invariant curve on a two-dimensional section is a closed
curve, a closed curve has an inside, and the inside is mapped to itself, so an orbit starting
within it can never get out. That is why the chaotic layers of the previous paragraph are thin
and threaded between the curves rather than joined up: each surviving torus is a wall. For
\(d=3\) the section is already four-dimensional and no longer something to look at, and the
tori in it have codimension two rather than one. A codimension-two set does not separate the
space it sits in, so the chaotic layers are connected to one another however small
\(\epsilon\) is, and an orbit can in principle wander along them across the whole shell. That
slow leakage is Arnold diffusion~\cite{Arnold1964,LichtenbergLieberman1992}, and nothing in
these lectures rules it out. This is the practical reason we need a
diagnostic that returns a number rather than a picture, which is the subject of the next
subsection.

\begin{table}[!htbp]
\centering
\begin{tabular}{@{}lp{0.55\textwidth}@{}}
\toprule
Object on the section & Dynamical interpretation \\
\midrule
single point & period-one orbit of the section map \\
finite set of points & periodic orbit \\
smooth closed curve & quasi-periodic motion on an invariant torus \\
island chain & resonant torus broken into elliptic and hyperbolic orbits \\
irregular scattered region & chaotic motion \\
\bottomrule
\end{tabular}
\caption{How to read a Poincar\'e section. Each row is a feature of the plotted point set and
what it says about the flow underneath. Fig.~\ref{fig:henon_heiles_sections} shows the last
three of them in one system and at one energy.}
\label{tab:L2_section_dictionary}
\end{table}

Thus Poincar\'e sections provide a geometric diagnostic of regularity and chaos, and
Table~\ref{tab:L2_section_dictionary} is the dictionary. What a nearly integrable system
usually shows is the mixed phase space of Sec.~\ref{sec:L2_fixed_points}: smooth invariant
curves, island chains and chaotic layers in one picture.

\begin{figure}[!htbp]
    \centering
    \includegraphics[width=0.94\textwidth]{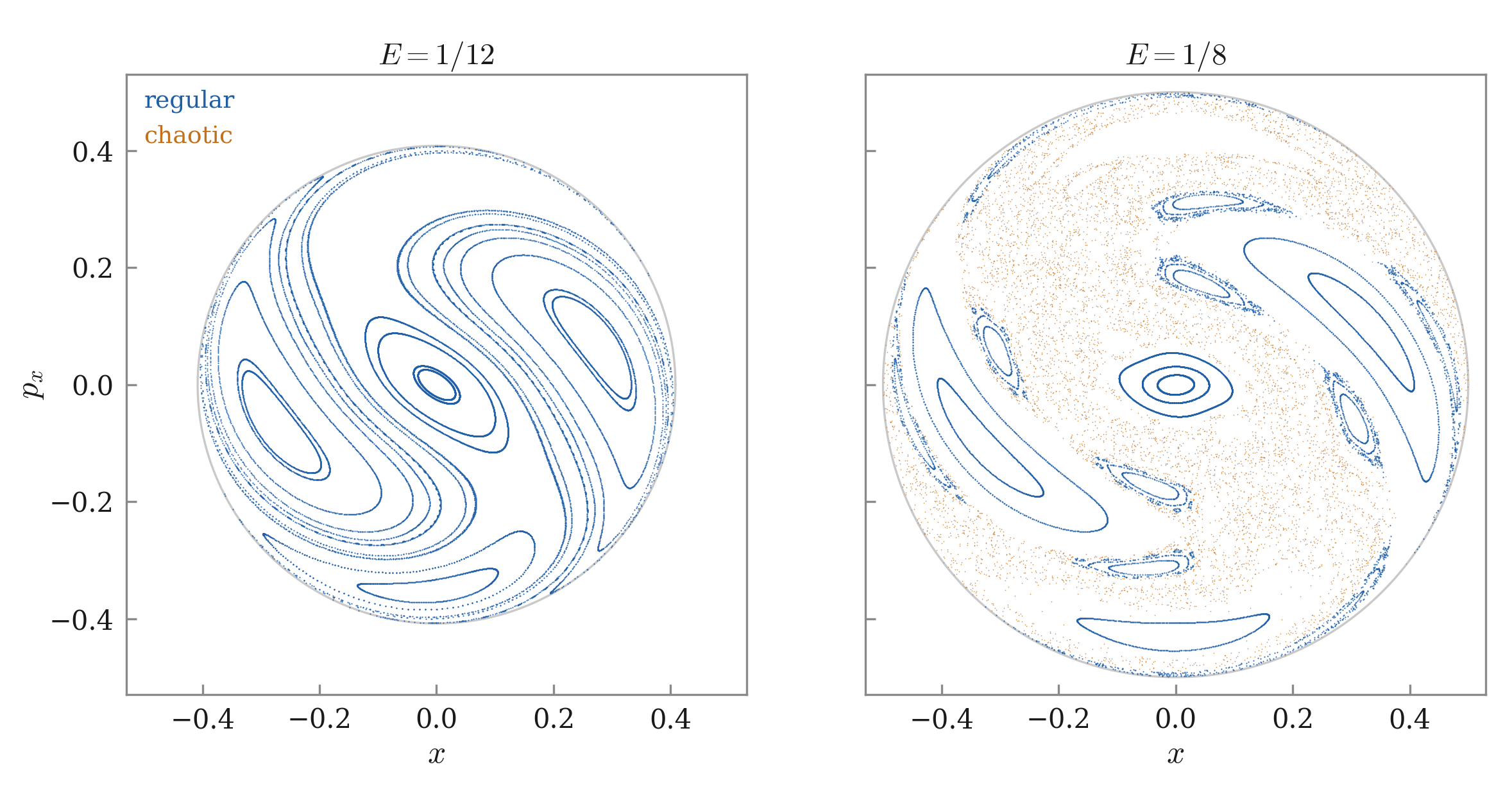}
    \caption{Poincar\'e sections of the H\'enon--Heiles Hamiltonian, taken on \(y=0\) with
    \(p_y>0\) and plotted in the \((x,p_x)\) plane.  Each panel follows \(30\) trajectories for
    \(700\) crossings, located by event detection rather than by interpolation between stored
    steps, with energy conserved to \(3\times10^{-11}\).  The grey circle is the bound
    \(x^2+p_x^2=2E\) that energy conservation puts on the section, and the points fill it to the
    last figure, \(0.4082\) and \(0.4999\) against \(\sqrt{1/6}\) and \(\sqrt{1/4}\).  Every
    orbit carries a tangent vector, so regular and chaotic are told apart by measurement rather
    than by eye: blue below \(\lambda_T=0.01\), orange above, the threshold falling in the gap
    between the two populations.  Left, \(E=1/12\): nested invariant curves and island chains,
    and not one of the thirty orbits is chaotic, the largest exponent being \(0.0015\).  Thin
    stochastic layers do exist at this energy, but they are too narrow for this sampling to
    catch, and a scan of \(230\) further starting points along the two axes did not find one
    either.  Right, \(E=1/8\): twelve of the thirty orbits have \(\lambda_T\ge0.017\) against
    \(0.008\) for the largest regular one, and between them they contribute forty per cent of
    the plotted points, in one connected sea with islands of regular motion surviving inside
    it.}
    \label{fig:henon_heiles_sections}
\end{figure}

Figure~\ref{fig:henon_heiles_sections} shows this happening in the H\'enon--Heiles
system~\cite{HenonHeiles1964}, the model of Exercise~\ref{ex:L2_henon_heiles} at the end of the lecture. The
potential confines the motion only below the escape energy \(E=1/6\). The pair worth comparing,
and the pair H\'enon and Heiles themselves compared, is \(E=1/12\) and \(E=1/8\), one half and
three quarters of that ceiling, because the transition from regular to largely chaotic falls
between them. At \(E=1/12\) the section is dominated by smooth curves and the island chains around them, the
signature of surviving tori and of resonances broken according to Poincar\'e--Birkhoff. At
\(E=1/8\) the chaotic layers have widened and merged into a connected sea, with islands of
regular motion surviving inside it. Calling those islands a remnant would misread the panel.
Sampled uniformly over the section rather than along the two axes, the regular set still holds
about half the area, \(48\%\) against \(52\%\) for the sea, so what the right-hand panel
shows is a mixed phase space and not one that has gone over to chaos. The two panels are the same system at two
energies, which is what makes the KAM statement concrete: as the effective perturbation
grows, the surviving set shrinks.

\subsection{Lyapunov exponents}
\label{sec:L2_lyapunov}

A Poincar\'e section shows the geometry but returns a picture; the Lyapunov exponent
returns a number, and it keeps working when the picture cannot be drawn. It measures the
exponential rate at which neighbouring trajectories separate.

Let \(\dot X=F(X)\) and let \(X(t)\) be a reference trajectory. A small deviation \(\xi(t)\)
obeys the variational equation Sec.~\ref{sec:L2_fixed_points} wrote down at a fixed point, with
the base point now carried along the orbit rather than standing still, which is the whole of
what generalizing linear stability from a point to an orbit amounts to,
\begin{equation}
\dot \xi = DF(X(t))\,\xi ,
\label{eq:L2_variational}
\end{equation}
and one sets
\begin{equation}
\lambda(X_0,\xi_0)
=
\lim_{t\to\infty}
\frac{1}{t}
\log
\frac{\|\xi(t)\|}{\|\xi(0)\|} .
\label{eq:L2_lyapunov_def}
\end{equation}
That this limit exists at all is the content of the Oseledec multiplicative ergodic
theorem~\cite{Oseledec1968,EckmannRuelle1985}, which guarantees it, whenever
\(\log^{+}\|D\Phi^{1}\|\) is integrable against the measure, for almost every \(X_0\)
with respect to an invariant measure \(\mu\), and which shows that \(\lambda\) takes only finitely
many values as \(\xi_0\) is varied. Almost every with respect to which measure matters: the
values are constant almost everywhere only on an ergodic component. That is not a formality for
the example this lecture uses, since the H\'enon--Heiles section at \(E=1/8\) is about half
regular and half chaotic and the two halves carry quite different exponents. These are the Lyapunov spectrum \(\lambda_1\geq\lambda_2\geq\cdots\geq\lambda_{2d}\), with
\(\lambda_1+\cdots+\lambda_k\) giving the growth rate of a \(k\)-dimensional volume element in
the tangent space. The deviation vectors returning an exponent smaller
than \(\lambda_1\) lie in a proper subspace of the tangent space, hence in a set of measure
zero, so the limit above returns \(\lambda_1\equiv\lambda_{\max}\) for almost any choice of
deviation vector. This is what licenses the numerical recipe below, where one
simply picks a random \(\xi_0\).

When \(\lambda_{\max}>0\) the separation grows as
\(\|\xi(t)\|\sim\|\xi(0)\|e^{\lambda_{\max}t}\), and the horizon over which prediction
survives is the logarithmic estimate~\eqref{eq:L1_predictability_time} of Lecture 1, with
\(\lambda_{\max}\) now defined precisely rather than as a generic growth rate.

For a map \(X_{n+1}=M(X_n)\) the tangent evolution is \(\xi_{n+1}=DM(X_n)\xi_n\) and the
same definition applies with \(t\) replaced by the iteration number \(n\).

Symplectic structure constrains the spectrum sharply, and in two steps rather than one.
Liouville's theorem, \eqref{eq:L1_liouville}, preserves phase-space volume, which gives
\begin{equation}
\sum_{i=1}^{2d}\lambda_i=0,
\end{equation}
and no more than that. The stronger statement, that the exponents come in pairs
\((\lambda,-\lambda)\) so that whatever a direction expands its partner contracts at the same
rate, needs the symplectic form itself, and for \(d\geq2\) it says more than the sum rule
does. This is the same reciprocity we met for the
eigenvalues of an area-preserving map in Sec.~\ref{sec:L2_fixed_points}, now for the flow.

At least two of the exponents vanish for an autonomous Hamiltonian flow. Displacement along the
trajectory neither grows nor decays, giving one zero, and the pairing then forces its partner
to vanish as well; that second neutral direction is the one transverse to the energy shell.
Hence for \(d=2\) the spectrum reads \(\{\lambda,\,0,\,0,\,-\lambda\}\), so a single number
characterizes each orbit~\cite{LichtenbergLieberman1992}. For the
H\'enon--Heiles sections of Fig.~\ref{fig:henon_heiles_sections}, \(\lambda\) vanishes on the
invariant curves and is positive throughout the chaotic sea.

The infinite-time limit is never taken numerically. One computes instead the finite-time
exponent
\begin{equation}
\lambda_T(X_0,\xi_0)
=
\frac{1}{T}
\log
\frac{\|\xi(T)\|}{\|\xi(0)\|},
\label{eq:L2_finite_time}
\end{equation}
which depends on the initial condition, the initial direction and the integration time. How
\(\lambda_T\) approaches its limit is itself the diagnostic. On an invariant torus the
frequencies depend on the actions, so neighbouring trajectories shear apart linearly rather
than exponentially, \(\|\xi(T)\|\sim T\), and therefore \(\lambda_T\sim(\log T)/T\to 0\).
In a chaotic region \(\lambda_T\) instead settles onto a positive plateau. Plotting
\(\lambda_T\) against \(T\) on logarithmic axes separates the two cleanly: regular orbits fall
off with asymptotic slope \(-1\), approached slowly because of the logarithm, while chaotic
ones flatten. This distinguishes genuine decay from slow
convergence, which a single value of \(\lambda_T\) cannot do.

Direct integration of \(\xi\) overflows once the separation has grown by many orders of
magnitude, and it also leaves the linear regime in which the variational equation is valid.
The standard remedy is periodic renormalization~\cite{Benettin1976,Benettin1980a,Benettin1980b}. Evolve
the trajectory and the deviation together over a short interval \(\Delta t\), short enough
that \(\xi\) stays in the linear regime and long enough that \(\log a_k\) is not dominated by
round-off, which in practice means rescaling once the deviation has grown by about an order of
magnitude, record the growth
factor \(a_k=\|\xi(k\Delta t)\|/\|\xi((k-1)\Delta t)\|\), rescale \(\xi\) back to its original
length, and repeat. After \(N\) steps
\begin{equation}
\lambda_T
\simeq
\frac{1}{N\Delta t}
\sum_{k=1}^{N}\log a_k .
\label{eq:L2_renormalization}
\end{equation}
This returns \(\lambda_{\max}\) alone. The full spectrum requires \(2d\) deviation vectors
evolved simultaneously, with Gram--Schmidt re-orthonormalization at each step; the growth
rates of the successive Gram--Schmidt norms then converge to
\(\lambda_1,\ldots,\lambda_{2d}\)~\cite{Benettin1980b}.  A system known only through a measured
time series, where the tangent space has to be reconstructed rather than integrated, needs a
different recipe~\cite{Wolf1985}.

In practice one checks that energy is conserved to the required accuracy, that \(\lambda_T\)
has converged in \(T\) rather than merely stopped changing over the window examined, and that
the answer is insensitive to the initial separation and direction.

No single diagnostic settles the question on its own, either, which is why this lecture built
five instruments rather than one. A Poincar\'e section may show scattered points suggestive of
chaos, and the impression still needs checking against an exponent. A positive finite-time
exponent may be recording a transient, or an integration step too coarse for the orbit being
followed. What one wants is agreement between the geometry, the local stability, the parameter
dependence and the measured growth rate.

The diagnostics of this lecture all rest on the classical notion of a trajectory. Linear
stability asks what the flow does to trajectories beside a fixed one, a bifurcation is a change
in that answer as a parameter moves, a Poincar\'e section follows one trajectory through phase
space, a Lyapunov exponent measures the separation of two, and KAM theory is a statement about
the invariant tori they lie on.

Quantum mechanics supplies none of them. The question that opens Lecture 3 is
therefore forced on us: if quantum mechanics has no classical trajectories, what replaces
classical chaos?

\subsection{Exercises}

\begin{exercise}
\label{ex:L2_henon_heiles}
Consider the H\'enon--Heiles Hamiltonian~\cite{HenonHeiles1964}
\begin{equation}
H=
\frac{1}{2}(p_x^2+p_y^2)
+
\frac{1}{2}(x^2+y^2)
+
x^2y-\frac{1}{3}y^3.
\end{equation}
Construct a Poincar\'e section using
\begin{equation}
y=0,
\qquad
p_y>0.
\end{equation}
For fixed energy \(E\), determine \(p_y\) from the energy constraint and plot the points
\begin{equation}
(x,p_x)
\end{equation}
at successive crossings. On the section the constraint reads
\(p_y=+\sqrt{2E-x^{2}-p_x^{2}}\), so the admissible starting points are exactly the open disc
\(x^{2}+p_x^{2}<2E\), of radius \(0.408\) at \(E=1/12\) and \(0.500\) at \(E=1/8\). Use a
symplectic integrator, or check that the energy is conserved to better than \(10^{-9}\) over
the run and locate the crossings by root-finding rather than by interpolating between steps: a
non-symplectic method at loose tolerance manufactures drift across the invariant curves and
turns the low-energy panel into a mess, which is the hazard Sec.~\ref{sec:L2_lyapunov} warns
about in its own setting.

Compare a low energy with one closer to the escape energy \(E=1/6\); your sections should
reproduce Fig.~\ref{fig:henon_heiles_sections}. Then measure what the figure only shows.
Count the cells of a coarse grid on the section that a single long chaotic orbit visits, take
that as the chaotic fraction, and locate the energy at which it first passes one half.
\end{exercise}

\begin{exercise}
\label{ex:L2_driven_pendulum}
A driven damped pendulum is a standard example of a nonlinear system in which
regular and chaotic motion can both occur, depending on the driving parameters~\cite{BakerGollub1996}. Its
equation of motion is
\begin{equation}
\frac{d^2\theta}{dt^2}
=
-\frac{g}{l}\sin\theta
-\gamma\frac{d\theta}{dt}
+
A\sin(\omega_d t),
\end{equation}
where \(\theta\) is the angular displacement, \(g\) is the acceleration due to gravity,
\(l\) is the length of the pendulum, \(\gamma\) is the damping coefficient, \(A\) is the
driving amplitude, and \(\omega_d\) is the driving frequency.

Work in dimensionless units with
\begin{equation}
\frac{g}{l}=1.
\end{equation}
For fixed
\begin{equation}
\gamma=0.5,
\qquad
\omega_d=\frac{2}{3},
\end{equation}
numerically estimate the largest Lyapunov exponent for two different driving
amplitudes, for example
\begin{equation}
A=0.5
\qquad
\text{and}
\qquad
A=1.2.
\end{equation}
Use the renormalization procedure of Sec.~\ref{sec:L2_lyapunov} rather than simply following
two nearby trajectories, which leaves the linear regime once the separation has grown. That
recipe needs the variational equation, which here is
\begin{equation}
    \ddot{\delta\theta}=-\cos\theta\,\delta\theta-\gamma\,\dot{\delta\theta} ,
\end{equation}
evaluated along the trajectory. Two further instructions matter more than they look. Discard a
transient of several hundred drive periods before accumulating anything, since at \(A=0.5\) the
approach to the attractor is slow and a student who starts the average at \(t=0\) gets a number
that means nothing. And work in \((\theta,\dot\theta)\), where there are two exponents summing
to \(-\gamma\); autonomising the drive by adding a phase variable gives three, one of them zero
by construction, and reporting that zero as \(\lambda_{\max}\) destroys the comparison the
exercise is for.

Discuss how the sign and magnitude of the largest Lyapunov exponent distinguish regular from
chaotic motion. Note that this system is dissipative, so the constraints derived in
Sec.~\ref{sec:L2_lyapunov} do not apply to it: the exponents no longer sum to zero but to
\(-\gamma\), and they do not pair as \((\lambda,-\lambda)\). Volume in phase space contracts
here rather than being preserved.
\end{exercise}

\begin{exercise}
\label{ex:L2_standard_map}
The standard map
\begin{equation}
    p_{n+1}=p_n+K\sin q_n,
    \qquad
    q_{n+1}=q_n+p_{n+1}
\end{equation}
on the torus is the canonical area-preserving map, and its fixed points sit exactly on the
corners of the diagram of Sec.~\ref{sec:L2_fixed_points}.
\begin{enumerate}
    \item[(a)] Show that the tangent matrix is
    \(M=\bigl(\begin{smallmatrix}1+K\cos q&1\\K\cos q&1\end{smallmatrix}\bigr)\), that
    \(\det M=1\) for every \(K\) and every \(q\), and hence that the map is confined to the top
    edge \(\Delta=1\) of the diagram, as Sec.~\ref{sec:L2_fixed_points} says every
    area-preserving map must be.
    \item[(b)] The fixed points are \(p=0\) with \(q=0\) and \(q=\pi\). Compute \(\tau\) at each.
    Show that \(q=0\) is hyperbolic for every \(K>0\), that \(q=\pi\) is elliptic for
    \(0<K<4\), and that at \(K=4\) it reaches the corner \(\tau=-2\).
    \item[(c)] Identify the bifurcation at \(K=4\) from Sec.~\ref{sec:L2_bifurcations}, and name
    the stability type of \(q=\pi\) for \(K>4\). Confirm numerically by iterating a small
    neighbourhood of \(q=\pi\) at \(K=3.9\) and at \(K=4.5\) and watching what the two branches
    of each manifold do from one iterate to the next.
    \item[(d)] Greene's criterion of Sec.~\ref{sec:L2_kam} concerns the golden circle
    \(W=(\sqrt5-1)/2\). Compute the residue of the periodic orbits whose rotation numbers are
    the continued-fraction convergents \(1/1,1/2,2/3,3/5,5/8,\ldots\) at \(K\) below and above
    \(K_c\simeq0.9716\), and see which way the residues run in each case.
\end{enumerate}
\end{exercise}
\clearpage

\part{Quantum Chaos}

\section{Lecture 3: Why quantum chaos is subtle}

Lectures 1 and 2 built the classical theory of chaos on the geometry of phase
space. A Hamiltonian system with \(d\) degrees of freedom is a point
\(X=(q_1,\ldots,q_d,p_1,\ldots,p_d)\) moving under the flow \eqref{eq:L2_symplectic_flow}, \(\dot X=J\nabla H(X)\) with \(J\) the
antisymmetric matrix that pairs each coordinate with its conjugate momentum,
and every diagnostic we developed reads that flow directly. A Poincar\'e section
records where a trajectory pierces a surface. A Lyapunov exponent measures how
fast two trajectories separate. KAM theory tracks which invariant tori survive a
perturbation. Even the statistical language of ergodicity and mixing, Sec.~\ref{sec:L1_ergodic},
describes what a single trajectory does over long times.

Quantization removes the object all of these diagnostics act on. A quantum system
carries no sharply defined trajectory in phase space, so there is nothing to
pierce a surface, nothing to separate from a neighbour, and no torus to destroy.
This lecture asks what becomes of the classical picture once that object is gone:
which classical notions survive quantization, which fail outright, and where the
signatures of classical chaos are hidden in a theory built from wavefunctions,
operators and spectra. The diagnostics that answer the last question, level
spacing statistics, random matrix theory, the spectral form factor and eigenstate
thermalization, are developed in Lecture 4.
Here we work out why they are needed at all~\cite{Haake2010,Lakshminarayan2018}.

\subsection{Why the classical definition cannot be copied}
\label{sec:L3_obstruction}

Start from the classical diagnostic and locate exactly where it breaks. Two nearby
initial conditions \(X_A(0)\) and \(X_B(0)\) define two trajectories whose
separation grows as
\begin{equation}
    |\delta X(t)| \sim |\delta X(0)|\,e^{\lambda t},
\end{equation}
and a positive \(\lambda\), sustained over long times on an orbit that stays bounded and
keeps returning, is the classical diagnostic. Sec.~\ref{sec:L2_lyapunov} turned the exponent into a number,
and Sec.~\ref{sec:L1_ingredients} said why the number by itself will not do: the inverted
oscillator carries a positive \(\lambda\) for all time and is not chaotic. The exponent is
nonetheless the ingredient quantization is about to take away, so it is the one to follow. Writing it down at all takes two things: a pair of sharply specified points, and a
separation between them that we are free to send to zero.

Quantum mechanics supplies neither. A state is a vector \(|\psi\rangle\) in a
Hilbert space \(\mathcal{H}\), or a density matrix \(\rho\), evolving under
\begin{equation}
    i\hbar\frac{d}{dt}|\psi(t)\rangle = H|\psi(t)\rangle,
    \qquad
    |\psi(t)\rangle = U(t)|\psi(0)\rangle,
    \qquad
    U(t)=e^{-iHt/\hbar},
\end{equation}
for a time-independent \(H\). Unitarity then fixes the overlap of any two states
for all time,
\begin{equation}
    \langle \psi_A(t)|\psi_B(t)\rangle
    = \langle \psi_A(0)|U^\dagger(t)U(t)|\psi_B(0)\rangle
    = \langle \psi_A(0)|\psi_B(0)\rangle,
\end{equation}
so two nearby quantum states never separate in the Hilbert-space norm, however
chaotic the underlying classical system. Taken at face value this reads as a proof
that quantum chaos cannot exist~\cite{Haake2010}.

The argument compares the wrong pair of objects, and the resolution below
follows the presentation of~\cite{Lakshminarayan2018}. A classical phase-space point has
no quantum counterpart, but a classical phase-space density does, and densities
behave the same way. Liouville evolution is \(\rho(X,t)=\rho(\Phi^{-t}X,0)\), and
since the flow \(\Phi^t\) preserves phase-space volume, the overlap of any two
densities is conserved,
\begin{equation}
    \int d\Gamma\, \rho_A(X,t)\,\rho_B(X,t)
    =
    \int d\Gamma\, \rho_A(X,0)\,\rho_B(X,0),
\end{equation}
which is precisely the classical counterpart of the constancy of
\(\langle \psi_A(t)|\psi_B(t)\rangle\). It holds in the most strongly chaotic
classical systems we know. The same remark disposes of the related worry that chaos
requires nonlinear equations of motion while the Schr\"odinger equation is linear.
Trajectories obey nonlinear equations, but the Perron--Frobenius operator that
transports classical densities is linear~\cite{Ott2002,Lakshminarayan2018}, and densities are what
quantum mechanics generalizes. Linearity of the evolution law is a feature shared
by both theories and settles nothing.

The genuine obstruction is kinematic. For a canonical pair,
\begin{equation}
    [\hat q,\hat p]=i\hbar
    \qquad\Longrightarrow\qquad
    \Delta q\,\Delta p\geq \frac{\hbar}{2},
\end{equation}
a property of the operator algebra rather than of any measuring apparatus. The
classical construction requires an exact initial condition \((q(0),p(0))\) and then
requires perturbing it by an arbitrarily small \(\delta X(0)\), and quantum
mechanics puts a floor under both. The definition
\begin{equation}
    \lambda=\lim_{t\to\infty}\;\lim_{|\delta X(0)|\to 0}
    \frac{1}{t}\log\frac{|\delta X(t)|}{|\delta X(0)|}
\end{equation}
therefore has no quantum analogue, because the inner limit cannot be taken.

Counting states makes the same obstruction concrete. Quantum mechanics partitions
phase space into cells of volume \((2\pi\hbar)^d\), so a bounded classical region of
phase-space volume \(\Omega\) supports only
\begin{equation}
    N \simeq \frac{\Omega}{(2\pi\hbar)^d}
\end{equation}
states. A bounded system thus has only finitely many states below any given energy and a
discrete spectrum, which is all the argument needs, and by
\(|\psi(t)\rangle=\sum_n c_n e^{-iE_n t/\hbar}|n\rangle\) its evolution is almost
periodic: it returns arbitrarily close to its initial state, which is a
theorem~\cite{BocchieriLoinger1957} rather than a heuristic. Chaos in the
long-time classical sense is unavailable to it. Whatever survives of classical chaos
must live at times short enough that the discreteness of the spectrum has yet to be
resolved, a restriction made quantitative in Sec.~\ref{sec:L3_ehrenfest}.

None of this leaves the subject empty. Unitarity forbids exponential separation of
states and forbids nothing else on the list: spectra can be rigid, eigenstates can
be complicated, wave packets can spread, operators can grow, and many-body systems
can thermalize. Locating those signatures is the business of the rest of this
lecture.

\subsection{Hilbert space and the semiclassical limit}
\label{sec:L3_semiclassical}

Trading phase space for Hilbert space changes which questions are natural to ask.
Classically the central object is a flow
\begin{equation}
    \Phi^t:\Gamma\to\Gamma,
\end{equation}
and the questions asked of it are geometric: which orbits close, which manifolds are
invariant, how a small volume element is stretched. Quantum mechanically the central
object is a one-parameter family of unitaries
\begin{equation}
    U(t):\mathcal{H}\to\mathcal{H},
\end{equation}
whose entire content, for a time-independent Hamiltonian, sits in the spectral problem
\(H|n\rangle=E_n|n\rangle\). The replacements for the geometric questions are questions
about \(\{E_n\}\) and \(\{|n\rangle\}\): how the levels are spaced, how the
eigenfunctions distribute themselves over the accessible region, how correlation
functions built from them decay. Quantum chaos is the study of how the regular and
chaotic alternatives for \(\Phi^t\) show up in those data. The phrase is a mild abuse
of language, since the chaos belongs to the classical limit and what the quantum
system carries are its signatures.

The correspondence principle links the two descriptions, with the ratio of \(\hbar\) to
a characteristic classical action as control parameter,
\begin{equation}
    \hbar_{\rm eff}\equiv \frac{\hbar}{S_{\rm cl}}\ll 1,
\end{equation}
where \(S_{\rm cl}\) may be an action variable in an integrable system or a phase-space
area scale in a bounded one. The semiclassical limit is \(\hbar_{\rm eff}\to 0\). The counting
estimate of the previous subsection then reads \(N\sim\hbar_{\rm eff}^{-d}\) up to numerical
factors, so this is the limit in which the number of accessible states grows without bound and
level spacings become dense.

For integrable systems the correspondence is well organized. Given action-angle
variables \((I_i,\theta_i)\), semiclassical quantization selects those tori whose
actions satisfy
\begin{equation}
    \oint_{\gamma_i} p\cdot dq = 2\pi\hbar\left(n_i+\frac{\mu_i}{4}\right),
    \qquad
    n_i\in\mathbb{Z}_{\geq 0},
\end{equation}
with \(\gamma_i\) the \(i\)th irreducible cycle on the torus and \(\mu_i\) the Maslov
index counting caustic crossings along it; equivalently
\(I_i=\hbar\left(n_i+\mu_i/4\right)\). This is the Einstein--Brillouin--Keller
construction~\cite{Keller1958}. Each quantized torus supplies one energy level, so the
classical invariant tori are imprinted directly on the spectrum. A \(d\)-dimensional
integrable system accordingly has its levels labelled by a \(d\)-dimensional lattice of
quantum numbers rather than by a single ordinal, which is the structural reason a
generic integrable system has uncorrelated levels~\cite{BerryTabor1977}, the Poisson
case taken up in Lecture 4. Systems of harmonic oscillators evade it, and the reason is worth naming, since it is the
hypothesis the argument turns on. Berry and Tabor need the energy to depend nonlinearly on the
actions, so that the level surface in action space is curved and lattice points adjacent on it
land at separated energies. For \(H=\sum_i\omega_iI_i\) that surface is a plane instead, the
degeneracies are systematic rather than accidental, and what the spacing distribution looks
like is decided by the arithmetic of the ratios \(\omega_i/\omega_j\) rather than by any
statistical argument~\cite{BerryTabor1977}. The Poisson case of Lecture 4 assumes the
curvature.

Chaotic systems admit no such construction. Global invariant tori are absent, and the
classical skeleton that survives is the set of unstable periodic orbits. These occupy
measure zero in the chaotic sea, yet they control the oscillatory part of the density of
states through Gutzwiller's trace formula~\cite{Gutzwiller1971,Gutzwiller1990}, written
down and put to work in Lecture 4. The structural lesson is worth stating on its own:
quantization reorganizes classical information rather than transcribing it. What passes
through are the special invariant structures, tori in the integrable case and unstable
periodic orbits in the chaotic one, and what is lost is the generic trajectory.

\subsection{Wave packets, the Ehrenfest time, and the Heisenberg time}
\label{sec:L3_ehrenfest}

Classical trajectories do reappear in quantum mechanics, for a while. Consider a
Gaussian packet centred at \((q_0,p_0)\),
\begin{equation}
    \psi(q,0)\sim
    \exp\left[-\frac{(q-q_0)^2}{4\sigma^2}
    +\frac{i}{\hbar}p_0(q-q_0)\right],
\end{equation}
which saturates the uncertainty bound and is therefore as close to a phase-space point
as quantum mechanics permits. Ehrenfest's theorem~\cite{Ehrenfest1927} governs the
motion of its centre. For \(H=\hat p^{2}/2m+V(\hat q)\),
\begin{equation}
    \frac{d}{dt}\langle \hat q\rangle = \frac{\langle \hat p\rangle}{m},
    \qquad
    \frac{d}{dt}\langle \hat p\rangle = -\langle V'(\hat q)\rangle,
\end{equation}
and both relations are exact. They reduce to the classical equations of motion when
\(\langle V'(\hat q)\rangle\) may be replaced by \(V'(\langle \hat q\rangle)\), which
requires the packet to be narrow on the scale over which \(V'\) varies. Expanding about
\(\langle\hat q\rangle\),
\begin{equation}
    \langle V'(\hat q)\rangle
    = V'(\langle\hat q\rangle)
    + \tfrac{1}{2}V'''(\langle\hat q\rangle)\,(\Delta q)^{2}
    + \cdots,
\end{equation}
so the leading correction is set by the width of the packet. For quadratic Hamiltonians
\(V'''=0\) and the correspondence is exact for all time: Gaussians stay Gaussian and
their centres run along classical trajectories. Anharmonicity spoils it at a rate fixed
by the growth of \(\Delta q\).

In a chaotic system that growth is exponential. The packet is stretched along unstable
directions and squeezed along stable ones, much as a small classical blob of density
would be, with the difference that a quantum packet has a floor on its phase-space area
and a classical blob does not. Write \(\sigma_0\) for the minimal width, which is the
\(\sigma\) of the packet above resolved along the unstable direction, and \(L_{\rm cl}\) for
the classical scale of the region the motion can reach. Then
\begin{equation}
    \sigma(t)\sim \sigma_0\,e^{\lambda t},
\end{equation}
and the packet ceases to be localized once \(\sigma(t)\) reaches \(L_{\rm cl}\). Setting
\(L_{\rm cl}\sim\sigma_0 e^{\lambda t_E}\) defines the Ehrenfest
time~\cite{BermanZaslavsky1978},
\begin{equation}
    t_E \sim \frac{1}{\lambda}\log\frac{L_{\rm cl}}{\sigma_0}
    \sim \frac{1}{\lambda}\log\frac{1}{\hbar_{\rm eff}},
\end{equation}
the second form following because \(\sigma_0\) cannot be made smaller than \(\hbar\) permits.
How much smaller depends on the packet. One spread across the whole classical momentum range has
\(\sigma_0/L_{\rm cl}\sim\hbar_{\rm eff}\) and gives the estimate as written; the balanced
packet used elsewhere in this lecture has
\(\sigma_0/L_{\rm cl}\sim\sqrt{\hbar_{\rm eff}}\) and gives
\(t_E\sim(2\lambda)^{-1}\log(1/\hbar_{\rm eff})\). The logarithm is the content and the
factor in front of it is a choice of packet, which is worth knowing before
Exercise~\ref{ex:L3_timescales} asks for a number.
It shares its logarithmic structure with the classical predictability time,
\begin{equation}
    t_{\rm pred}\sim \frac{1}{\lambda}\log\frac{\Delta_{\rm mac}}{\Delta_{\rm init}},
\end{equation}
and the two logarithms have different reasons behind them. Classically the limitation is our ignorance of the
initial condition, and a better measurement pushes it back without bound. Quantum
mechanically the floor is kinematic and cannot be lowered at all.

The logarithm is the whole reason quantum chaos is subtle, because it makes two limits
fail to commute. Consider the squared commutator
\begin{equation}
    C(t)=-\frac{1}{\hbar^{2}}\big\langle [\hat q(t),\hat p(0)]^{2}\big\rangle ,
\end{equation}
whose classical limit at fixed \(t\) is \(\{q(t),p(0)\}^{2}\sim e^{2\lambda t}\). Fix
\(\hbar>0\) instead and let \(t\to\infty\): the spectrum is discrete, \(C(t)\) is an
almost periodic and bounded function of time, and the growth rate extracted from it
vanishes. Hence
\begin{equation}
    \lim_{\hbar\to 0}\ \lim_{t\to\infty}\ \frac{1}{2t}\log C(t) = 0,
    \qquad
    \lim_{t\to\infty}\ \lim_{\hbar\to 0}\ \frac{1}{2t}\log C(t) = \lambda .
\end{equation}
The second of these is a statement about a state localized at a phase point, where the
classical limit runs along a single trajectory and the Oseledec theorem of
Sec.~\ref{sec:L2_lyapunov} supplies \(\lambda\) for almost every starting point. Averaging
the correlator over phase space before taking the logarithm returns a larger rate, strictly
larger whenever the finite-time exponent fluctuates at all,
for reasons taken up in
Sec.~\ref{sec:L5_bracket}~\cite{RozenbaumGaneshanGalitski2017,PappalardiKurchan2023}.

The classical limit is singular, in the way the short-wavelength limit of wave optics is
singular, and \(t_E\) is the scale at which the two orders part company. Every honest
statement about a ``quantum Lyapunov exponent'' is therefore a statement about the window
\(t\lesssim t_E\), or about a quantity defined without pushing either limit to its end.

A second time scale comes from the discreteness itself. After a time \(t\) the finest
energy difference that can have been resolved is \(\hbar/t\), so levels separated by the
mean spacing \(\Delta\) become resolved at the Heisenberg time
\begin{equation}
    t_H = \frac{2\pi\hbar}{\Delta} = 2\pi\hbar\,\bar\rho(E),
\end{equation}
with \(\bar\rho\) the smooth density of states. The counting estimate of
Sec.~\ref{sec:L3_obstruction} gives \(\bar\rho(E)=\Omega(E)/(2\pi\hbar)^{d}\), where
\(\Omega(E)\) is the classical energy-shell volume, hence
\begin{equation}
    t_H \sim \hbar^{-(d-1)} .
\end{equation}
Set the two scales side by side: \(t_E\) grows like \(\log(1/\hbar)\), while \(t_H\)
grows like a power of \(1/\hbar\). For \(d\ge 2\) the ratio \(t_E/t_H\) vanishes in the
semiclassical limit, so the window
\begin{equation}
    t_E \ll t \ll t_H
\end{equation}
opens and widens without bound. Inside it a wave packet has already lost its classical
trajectory, while the spectrum remains unresolved and the almost periodicity of
Sec.~\ref{sec:L3_obstruction} has yet to make itself felt. This window is where quantum
chaos lives, and the diagnostics of Lecture 4 are built to probe it. The spectral form
factor in particular is designed to separate the ramp that fills this window from the
plateau that sets in at \(t_H\).

Both scales have counterparts in the many-body and holographic settings of Lectures 5
and 6, and the correspondence is worth flagging now. In a gauge theory with a classical
gravity dual the role of \(\hbar_{\rm eff}\) is played by a quantity of order
\(1/N^{2}\), and the logarithmic scale becomes the scrambling time
\begin{equation}
    t_* \sim \frac{1}{\lambda_L}\log N^{2},
\end{equation}
with \(\lambda_L\) the exponent that Lecture 5 defines through the growth of a squared
commutator, conjectured in~\cite{SekinoSusskind2008} to be the fastest available to any system whose
interactions involve bounded clusters of degrees of freedom.
A system of entropy \(S\), meanwhile, has level spacing of order \(e^{-S}\) and hence a
Heisenberg time \(t_H\sim e^{S}\). The same hierarchy reappears, with the logarithm and
the exponential separated by an enormous margin, and for the same reason: exponential
growth runs until the finiteness of the Hilbert space stops it.

Nothing breaks down at \(t_E\) except the single-trajectory approximation itself. Beyond
it the signatures of classical chaos persist, relocated into interference patterns,
spectra, eigenstates, correlation functions and operator growth. The remaining
subsections take up where to look for them.

\subsection{Wigner and Husimi functions}
\label{sec:L3_wigner}

\begin{figure}[!htbp]
    \centering
    \includegraphics[width=\linewidth]{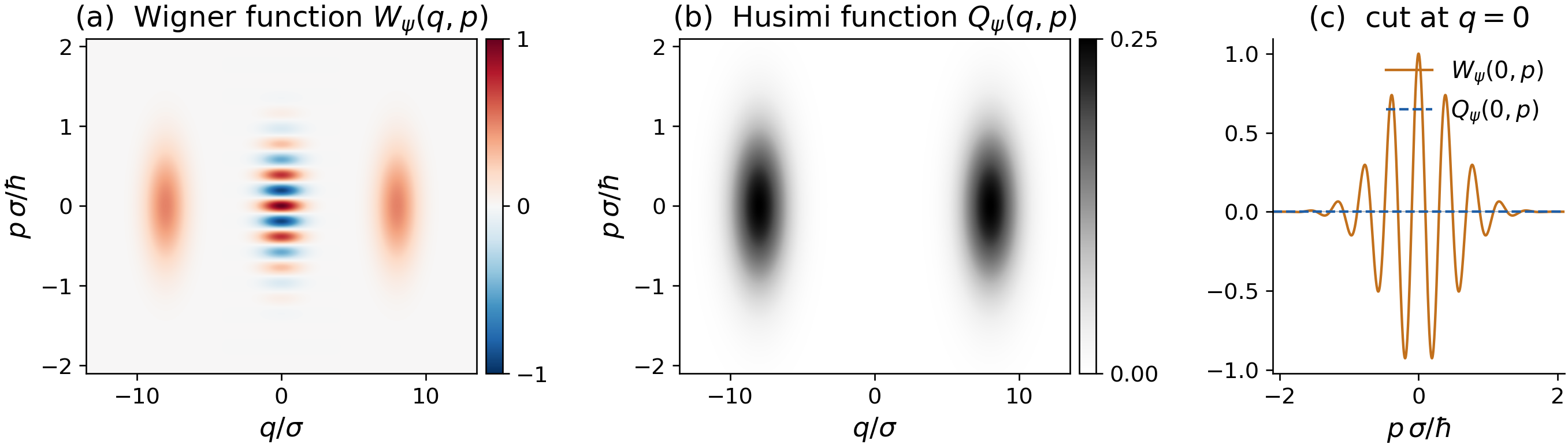}
    \caption{Wigner and Husimi functions of the cat state
    \(\psi(q)\propto e^{-(q-a)^{2}/4\sigma^{2}}+e^{-(q+a)^{2}/4\sigma^{2}}\) with
    \(a=8\sigma\), in units \(\hbar=\sigma=1\), all panels on a common scale with \(|W|\)
    normalised to unity at its largest. (a) \(W\) has two positive lobes at \(q=\pm a\) and,
    at \(q=0\), an interference term oscillating in \(p\) with period \(\pi\hbar/a\),
    alternating in sign and reaching twice the height of either lobe. (b) The Husimi
    function, \(W\) smoothed over a coherent state of width \(\sigma\), is non-negative and
    retains only the lobes. (c) Cut along \(p\) at \(q=0\). The smoothing suppresses the
    interference term by \(e^{-a^{2}/4\sigma^{2}}\), which at \(a=8\sigma\) leaves a
    residual of order \(10^{-7}\), too small to see on this scale.  Panel (b) drops it
    altogether.  What survives is not simply a fainter copy of the Wigner fringes: it
    oscillates at half their frequency, since the smoothing acts on the two lobes
    separately as well as on the cross term.}
    \label{fig:wigner_cat}
\end{figure}

Exact quantum trajectories in phase space are unavailable, but phase-space \emph{functions}
representing quantum states are not, and they retain a good deal of classical intuition.
The standard construction is the Wigner function~\cite{Wigner1932}. For a pure state
\(\psi(q)\) in one dimension,
\begin{equation}
    W_\psi(q,p)
    =
    \frac{1}{2\pi\hbar}
    \int_{-\infty}^{\infty}
    dX\,
    e^{-ipX/\hbar}
    \psi^*\!\left(q-\frac{X}{2}\right)
    \psi\!\left(q+\frac{X}{2}\right),
\end{equation}
which reproduces the correct marginals,
\begin{equation}
    \int dp\, W_\psi(q,p)=|\psi(q)|^2,
    \qquad
    \int dq\, W_\psi(q,p)=|\tilde\psi(p)|^2 .
\end{equation}
With both marginals genuine probability densities, \(W_\psi\) is a natural candidate for a
phase-space probability distribution, and it fails to be one only by taking negative
values. The negativity carries information rather than signalling a defect: it marks
quantum interference, and the sign-changing fringes it produces occupy phase-space cells
of area of order \(\hbar\). Hudson's theorem makes this exact~\cite{Hudson1974}. Among
pure states, \(W_\psi\ge 0\) everywhere if and only if \(\psi\) is Gaussian. The
Gaussian packets are therefore exactly the pure states admitting a literal classical
phase-space reading, and the minimal-uncertainty packet of Sec.~\ref{sec:L3_ehrenfest} is one
of them, which is why it was the right object with which to go looking for classical
trajectories.  What the theorem asks for is Gaussianity and not saturation of
\(\Delta q\,\Delta p=\hbar/2\); a chirped packet
\(\psi\propto e^{-(1+ic)q^{2}/4\sigma^{2}}\) has
\(\Delta q\,\Delta p=\sqrt{1+c^{2}}\,\hbar/2\), more than three times the minimum at
\(c=3\), and a Wigner function that stays positive everywhere. Fig.~\ref{fig:wigner_cat}(a) shows
what happens when two such packets are superposed. The two lobes stay positive, and between
them, in a region where the state has almost no probability at all, \(W\) develops an
oscillating pattern that alternates in sign and is taller than either lobe.

The evolution of \(W\) is governed by the Moyal bracket~\cite{Moyal1949}, the Wigner
transform of \(-i[H,\cdot]/\hbar\) and so the exact quantum counterpart of the Poisson
bracket rather than an approximation to it,
\begin{equation}
    \frac{\partial W}{\partial t}=\{H,W\}_{\rm MB},
\end{equation}
and for \(H=p^2/2m+V(q)\) the bracket admits an explicit expansion in \(\hbar\),
\begin{equation}
    \frac{\partial W}{\partial t}
    =
    \{H,W\}
    \;-\;\frac{\hbar^{2}}{24}\,V'''(q)\,\frac{\partial^{3}W}{\partial p^{3}}
    \;+\;O(\hbar^{4}),
\end{equation}
with \(\{H,W\}\) the classical Poisson bracket. At leading order the Wigner function obeys
the classical Liouville equation, the first correction entering at \(O(\hbar^{2})\). Two features of the correction repay attention.
It is proportional to \(V'''\), the same third derivative that controlled the Ehrenfest
correction \(\tfrac{1}{2}V'''(\langle q\rangle)(\Delta q)^{2}\) of
Sec.~\ref{sec:L3_ehrenfest}, so the exactness of the classical limit for quadratic
Hamiltonians is one fact surfacing in two places. And it is a third derivative in
momentum, carrying no definite sign and therefore unable to act as diffusion; this is the
term through which an initially positive \(W\) grows its negative regions.

The correction also fixes when the classical reading expires. Chaotic evolution folds
\(W\) onto ever finer scales, its width along the contracting direction shrinking as
\(\delta p(t)\sim\delta p_0 e^{-\lambda t}\). Each momentum derivative costs a factor
\(1/\delta p\), and writing \(\ell^{2}\equiv V'/V'''\) for the length on which the anharmonicity is felt,
the correction measured against the leading term is of order
\(\hbar^{2}/24\ell^{2}\delta p(t)^{2}\), reaching unity at
\begin{equation}
    t\sim\frac{1}{\lambda}\log\frac{\ell\,\delta p_0}{\hbar}\sim t_E .
\end{equation}
The semiclassical description of the Wigner function thus fails at the Ehrenfest time, by
the same mechanism and with the same logarithm that defeated the wave packet. The two
breakdowns are one breakdown.

Positivity can be bought back. The Husimi function~\cite{Husimi1940} is the overlap of the
state with a coherent state \(|\alpha_{q,p}\rangle\) centred at \((q,p)\),
\begin{equation}
    Q_\psi(q,p)=\frac{1}{2\pi\hbar}\big|\langle \alpha_{q,p}|\psi\rangle\big|^{2}\ \ge 0,
\end{equation}
with \(\sigma\) the width of the coherent state. That is the same object as a smoothed
Wigner function, and the equivalence is two lines. The overlap of two states is the phase-space
integral of their Wigner functions,
\begin{equation}
    \big|\langle\alpha_{q,p}|\psi\rangle\big|^{2}
    =2\pi\hbar\int dq'\,dp'\;W_{\alpha_{q,p}}(q',p')\,W_\psi(q',p') ,
\end{equation}
and the Wigner function of a coherent state is a Gaussian centred on \((q,p)\), of width
\(\sigma\) in position and \(\hbar/2\sigma\) in momentum. So \(Q_\psi\) is \(W_\psi\)
convolved with that Gaussian, whose area is of order \(\hbar\) however the two widths are
balanced. On the axes of Fig.~\ref{fig:wigner_cat}, which are \(q/\sigma\) and
\(p\sigma/\hbar\), the smoothing kernel is one unit wide in the first and half a unit in the
second. That smoothing coarse-grains over exactly the \(\hbar\)-scale cells in
which the fringes live, so positivity and the loss of sub-\(\hbar\) structure are a single
operation seen twice. Panels (b) and (c) of Fig.~\ref{fig:wigner_cat} make the trade
visible: the lobes survive the smoothing, the fringes disappear, and the region between the
lobes is returned to the near-zero density that the position distribution had there all
along. The trade is often worth making. Because \(Q_\psi\) may be plotted as an ordinary
density, it is the representation in which the phase-space portrait of an eigenstate can be
read by eye, including the enhancement along unstable periodic orbits taken up in the next
subsection.

\subsection{Where quantum signatures of chaos appear}
\label{sec:L3_signatures}

With trajectories unavailable, the question becomes where the regular and chaotic
alternatives of the classical limit deposit themselves in the quantum data. Three places
carry most of the answer, and each is developed further in a later lecture.

\paragraph{1. Spectra.}
For a bounded system the spectrum is discrete,
\begin{equation}
    E_0\leq E_1\leq E_2\leq \cdots,
\end{equation}
and its statistical arrangement remembers the classical dynamics. Integrable systems show
uncorrelated spacings, for the reason given in Sec.~\ref{sec:L3_semiclassical}.
Chaotic systems show level repulsion and the spectral correlations of random matrix
theory~\cite{BohigasGiannoniSchmit1984}. Reading either pattern requires resolving
individual levels, so this is a statement about times of order \(t_H\) and beyond.
Lecture 4 is devoted to it.

\paragraph{2. Eigenstates.}
The stationary states \(H|n\rangle=E_n|n\rangle\) carry dynamical information of their own.
For integrable systems the eigenstates organize themselves around quantized invariant tori.
For chaotic systems there is a theorem, which is worth pausing on: most statements
separating regular from chaotic quantum behaviour are conjectures supported by numerics, and
this one is not.

Let the classical flow be ergodic on the energy shell. The quantum ergodicity theorem of
Shnirelman, Zelditch and Colin de
Verdi\`ere~\cite{Shnirelman1974,Zelditch1987,ColindeVerdiere1985} states that a subsequence
of eigenstates of density one equidistributes: the densities \(|\psi_n(q)|^2\) converge
weakly to the uniform measure on the billiard domain, or on \(\{V<E\}\) for two degrees of
freedom in a potential; in \(d\) degrees of freedom the limit carries the density of states of
the momentum shell, \(\propto(E-V(q))^{d/2-1}\), and is uniform only for \(d=2\). In the
sharper phase-space form the Husimi functions \(Q_{\psi_n}\) of Sec.~\ref{sec:L3_wigner}
converge to the Liouville measure on the energy shell, and the position statement is that
measure's momentum marginal, which is where the factor above comes from. Ergodicity of the classical flow thus passes intact through
quantization, asserted now of almost every eigenfunction rather than of almost every
trajectory.

Density one is not all eigenstates. The theorem leaves room for an exceptional subsequence
of density zero that fails to equidistribute. Scarring, taken up next, is a different
phenomenon and does not need that room.
Berry described the typical chaotic eigenstate as behaving locally like a random
superposition of plane waves of fixed wavelength~\cite{Berry1977}, and Voros gave the
phase-space counterpart, semiclassical ergodicity of the Wigner function~\cite{Voros1979};
equidistribution is the weakest of that picture's consequences rather than its content: the
random-wave model also predicts Gaussian amplitude statistics and \(J_0(k|q-q'|)\) spatial
correlations, neither of which follows from weak convergence of \(|\psi_n|^2\). Some eigenstates
instead carry conspicuous extra density along individual unstable periodic orbits, the
scarring reported in~\cite{Heller1984}. The mechanism is a competition between recurrence and
instability. A wave packet launched on a periodic orbit of period \(T\) returns to its
starting neighbourhood once per period, so its autocorrelation
\(\langle\phi(0)|\phi(t)\rangle\) has peaks at \(t=nT\), while the instability of the orbit
drains those peaks at a rate set by \(\lambda\). A signal recurring with spacing \(T\) and
decaying over a time \(\lambda^{-1}\) Fourier transforms into a comb of spectral features
spaced by \(\hbar\omega=2\pi\hbar/T\) and broadened to width of order \(\hbar\lambda\). The
orbit therefore imprints itself across the band of eigenstates under each feature rather
than on any single one, and the number of them it reaches is the spacing divided by the width,
\begin{equation}
    \frac{2\pi\hbar/T}{\hbar\lambda}=\frac{2\pi}{T\lambda}=\frac{\omega}{\lambda} ,
\end{equation}
so their density along the orbit is enhanced by that factor. Scarring survives, in short, wherever an orbit repeats faster than it
destabilizes.

None of this contradicts quantum ergodicity, and the two are easy to run together. The
enhancement is a statement at the scale of \(\hbar\), about the weight an eigenfunction carries
in a phase-space tube around one orbit, and that tube shrinks as \(\hbar\to0\), so a scarred
eigenfunction generically sits inside the density-one equidistributing subsequence rather than
outside it. The same point is made from the other side in~\cite{Kaplan1999}: scarring is a deviation from
random matrix statistics, which is a far more demanding standard than the equidistribution the
theorem delivers. The exceptional set is occupied by other material. For almost every aspect
ratio the Bunimovich stadium is not quantum uniquely
ergodic~\cite{Hassell2010}, the obstruction being bouncing-ball modes that keep a finite
fraction of the mass on
the measure-zero family of vertical bounces, and quantized cat maps carry
strongly scarred sequences at arithmetically special values of
\(\hbar\)~\cite{FaureNonnenmacherDeBievre2003}. Whether any such sequence survives on a
negatively curved manifold is the subject of the quantum unique ergodicity conjecture of
Rudnick and Sarnak, which says the exceptional set is empty there~\cite{RudnickSarnak1994}.

\paragraph{3. Time-dependent observables.}
The third signature is the one Sec.~\ref{sec:L3_ehrenfest} has already used. Operators in
the Heisenberg picture,
\begin{equation}
    A(t)=U^\dagger(t)A(0)U(t),
\end{equation}
grow complicated in time even while unitarity holds every state overlap fixed, and the
semiclassical correspondence
\begin{equation}
    \frac{1}{i\hbar}[A,B]\quad\longrightarrow\quad \{A,B\}
\end{equation}
converts the classical stretching rate \(\{q(t),p(0)\}=\partial q(t)/\partial q(0)\) into
the growth of a commutator. Promoting \(\hat q\) and \(\hat p\) to general Hermitian
operators gives the out-of-time-order correlator
\begin{equation}
    C(t)=-\big\langle [W(t),V(0)]^{2}\big\rangle_\beta ,
\end{equation}
in which \(W\) and \(V\) denote arbitrary few-body operators, unrelated to the Wigner
function and the potential of Sec.~\ref{sec:L3_wigner}. Lecture 5 develops this in
detail~\cite{LarkinOvchinnikov1969,MaldacenaShenkerStanford2016}. The conceptual point is
that quantum sensitivity to initial conditions registers as the growth of
non-commutativity between an early observable and a late one.

\subsection{Quantum billiards as a clean laboratory}
\label{sec:L3_billiards}

Billiards strip the problem down to geometry. Classically a billiard is a point particle
moving freely inside a bounded domain \(D\) and reflecting elastically from the boundary,
with Hamiltonian
\begin{equation}
    H=\frac{p_x^{2}+p_y^{2}}{2m},
\end{equation}
so every nontrivial feature of the dynamics comes from the shape of \(\partial D\) alone. It
is also where the control parameter of Sec.~\ref{sec:L3_semiclassical} becomes concrete. A
billiard of area \(A\) offers one classical scale, \(\sqrt{A}\), and one quantum scale, the
de Broglie length \(\ell_{\rm dB}=\hbar/mv=1/k\), so
\(\hbar_{\rm eff}\sim\ell_{\rm dB}/\sqrt{A}=1/k\sqrt{A}\): the semiclassical limit here is
short wavelength against the size of the domain and nothing besides.
Exercise~\ref{ex:L3_timescales} uses it in that form.
Quantum mechanically the same system is the Dirichlet eigenvalue problem for the Laplacian,
\begin{equation}
    -\frac{\hbar^{2}}{2m}\nabla^{2}\psi = E\psi \quad\text{in } D,
    \qquad
    \psi\big|_{\partial D}=0,
\end{equation}
equivalently
\begin{equation}
    -\nabla^{2}\psi_n = k_n^{2}\psi_n,
    \qquad
    E_n=\frac{\hbar^{2}k_n^{2}}{2m}.
\end{equation}
The classical problem asks about rays and the quantum problem about standing waves, with
the same boundary controlling both. The stadium's levels and eigenfunctions were computed
numerically early~\cite{McDonaldKaufman1979}, and what those pictures showed is what made the
question concrete.

Which shapes are chaotic, and by what mechanism, has sharp answers. Rectangles and circles
are integrable, conserving \(|p_x|\) and \(|p_y|\) in the first case and angular momentum
in the second; ellipses are integrable less obviously, conserving the product of the
angular momenta about the two foci~\cite{Berry1981,Lakshminarayan2018}. Chaos arrives by one of two routes. Sinai's is
dispersing: a convex obstacle inside the domain scatters a parallel bundle of rays apart at
every encounter~\cite{Sinai1970}. Bunimovich's is the less intuitive one, since a boundary that focuses rays
can randomize them just as effectively~\cite{Bunimovich1979}. In the stadium, two
semicircular caps joined by straight segments, a bundle reflected from a cap converges to a
focus, overshoots it, and diverges thereafter. Defocusing on its own is not yet the mechanism,
since a circular boundary does the same at every bounce and is integrable: there the divergence
picked up between reflections is exactly undone by the next one, and the section map is a rigid
rotation at fixed angle of incidence. What the straight segments add is free flight that no cap
has to compensate, and once that balance is broken the overshoot accumulates over many bounces
into exponential separation. The stadium is the example we return to in Lectures 4 and 5.
Between the integrable and the fully chaotic extremes lies a continuum. The lima\c{c}on
family \(\rho(\phi)=1+\epsilon\cos\phi\) is the integrable circle at \(\epsilon=0\) and the
cardioid at the chaotic end, \(\epsilon=1\), while at intermediate \(\epsilon\) it shows the
mixed phase space of Lecture 2, surviving KAM tori interleaved with a chaotic
sea~\cite{Robnik1983,Lakshminarayan2018}; the polar form of the family used here is the
second of these, since Ref.~\cite{Robnik1983} parameterises it through a conformal map.

\begin{figure}[!htbp]
    \centering
    \includegraphics[width=0.86\linewidth]{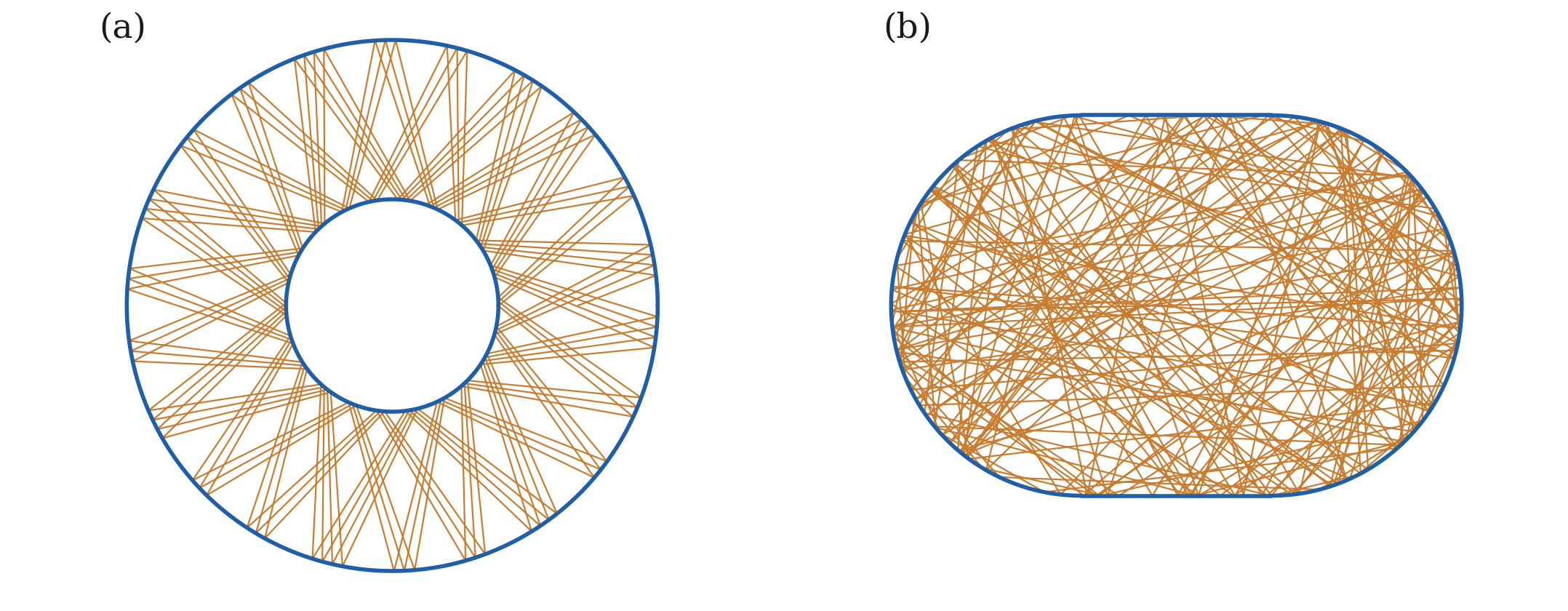}
    \caption{Single classical trajectories in (a) an integrable annular billiard, 150
    reflections, and (b) a Bunimovich stadium, 190 reflections.  The two domains have the same
    area, \(2+\pi\), and are drawn at the same scale, so what is being compared is shape
    rather than size.  Along the annulus orbit the angular momentum about the centre holds at
    \(0.3533\) to a spread of \(1.4\times10^{-15}\), and since that is below the inner
    radius \(0.558\) the orbit reaches the inner wall rather than missing it.  In the
    stadium a change of \(10^{-9}\) in the launch angle grows to \(0.66\) by the \(190\)th
    reflection, at about \(0.05\) per reflection while the separation is still small.}
    \label{fig:billiard_comparison}
\end{figure}

Fig.~\ref{fig:billiard_comparison} puts the two behaviours side by side. Reflection from
either concentric circle preserves angular momentum about the centre, so the annulus is
integrable and a single orbit fills the sub-annulus bounded by its caustic, tracing a rosette
without closing; the stadium conserves only the energy,
and one orbit covers the domain without pattern.

Weyl's law makes the state counting of Sec.~\ref{sec:L3_obstruction} concrete here. For a
planar domain of area \(A\) and perimeter \(L\) with Dirichlet conditions, the number of
modes with wavenumber below \(k\) is
\begin{equation}
    N(k)=\frac{A}{4\pi}k^{2}-\frac{L}{4\pi}k+o(k),
\end{equation}
the leading term of which is Weyl's~\cite{Weyl1911} and the perimeter term of which was proved
with the \(o(k)\) remainder much later~\cite{Ivrii1980}, on the hypothesis that the periodic
billiard trajectories have measure zero; the question of how much of the shape the spectrum
carries is the one made famous by Kac~\cite{Kac1966}. The leading term is exactly the classical
phase-space volume divided by
\((2\pi\hbar)^{2}\), while the subleading term records the suppression of states by a
Dirichlet wall. In energy variables \(N(E)=AmE/2\pi\hbar^{2}\), so the mean level density
\begin{equation}
    \bar\rho=\frac{Am}{2\pi\hbar^{2}}
\end{equation}
is independent of energy, and the Heisenberg time of Sec.~\ref{sec:L3_ehrenfest} evaluates
to
\begin{equation}
    t_H=2\pi\hbar\,\bar\rho=\frac{Am}{\hbar},
\end{equation}
growing as \(\hbar^{-1}\), the \(\hbar^{-(d-1)}\) scaling at \(d=2\). Both statements hold
at the order kept and not beyond it. The perimeter has not left the problem, it has only fallen
below that order: differentiating the two-term law returns a correction to \(\bar\rho\) of
relative order \(L/Ak\), down by one power of the wavenumber, and
Exercise~\ref{ex:L3_timescales}(a) asks for the numerical factor and for the \(k\) past which
it stops mattering.

Notice what this last result implies. The smooth part of the spectral density sees the
domain only through its geometry, the area first and the perimeter next, and it is blind to
what the flow inside does. Given any stadium one can find an ellipse of the same area and
the same perimeter: caps of radius \(1\) joined by straight sections of length \(2\) give
\(A=\pi+4\) and \(L=2\pi+4\), matched by the ellipse with semi-axes \(2.1123\) and
\(1.0762\). Both Weyl terms then agree exactly, and so does the constant that follows them,
since both boundaries are smooth and simply connected. The ellipse is nonetheless integrable,
and the stadium is the standard example of hard chaos. A circle cannot play this role, which
is worth pausing on: at fixed area the circle has the smallest perimeter of any shape, so
nothing else matches it in both. Every trace of regular or chaotic dynamics must therefore
reside in the fluctuations about the Weyl terms rather than in the terms themselves. That is why
Lecture 4 has to unfold the spectrum before it can compare anything, an operation which
amounts to dividing out precisely the smooth density computed here.

What this lecture has established is that the opening question was the wrong one. The
subject took shape from a collision: classical mechanics had shown that deterministic systems
can be nonintegrable and unpredictable, while quantum mechanics had replaced trajectories with
wavefunctions, operators and spectra. Asking whether quantum mechanics can itself be chaotic
was the natural thing to ask and it leads nowhere. The productive question, and the one
answered here, is what is distinctive about the quantum Hamiltonian obtained by quantizing a
classical system that is chaotic. Three things are. Its spectrum shows level repulsion where
an integrable system's does not, a fact that grew out of Wigner's analysis of nuclear
resonances~\cite{Wigner1955,Wigner1958}. Its eigenstates equidistribute on the energy shell, scarring being the visible deviation at
finite \(\hbar\) rather than a counterexample, and the billiard numerics arrived between
the announcement of that theorem and its full
proofs~\cite{Shnirelman1974,ColindeVerdiere1985,Zelditch1987}. And its squared
commutator grows across the window \(t_E\ll t\ll t_H\). None of the three requires a
trajectory. The classical structure does survive quantization, but only after being rewritten
in the language of spectra, eigenstates and correlators, and it is that rewriting, rather than
any loss of information, that makes the subject subtle.

What the lecture has not done is measure any of it. Each of the three says what to look for
rather than supplying an instrument for looking. Lecture 4 builds the instruments for the
first two, starting from the trace formula that ties a spectrum to the classical periodic
orbits and ending at a statement about single eigenstates. The third has to wait, because the
squared commutator is not a functional of the spectrum at all and needs machinery of a
different kind.

\subsection{Exercises}

\begin{exercise}
\label{ex:L3_mechanisms}
A rectangular billiard and a stadium billiard share the same free Hamiltonian and differ only
in their boundaries. Describe qualitatively how a typical trajectory differs in the two cases.
Then supply what Sec.~\ref{sec:L3_billiards} asserts and does not derive. Writing \(\phi\) for
the position of a bounce on a circular boundary and \(\theta\) for the angle the outgoing
trajectory makes with the tangent there, show that \(\theta\) is the same at every bounce and
that the boundary map is the rigid rotation \((\phi,\theta)\mapsto(\phi+2\theta,\theta)\). A
rigid rotation has zero Lyapunov exponent for every \(\theta\), so the caps of a stadium cannot
by themselves be what makes it chaotic. Then test the explanation Sec.~\ref{sec:L3_billiards}
gives: build the boundary map of a stadium numerically, measure its largest Lyapunov exponent by
the recipe of Sec.~\ref{sec:L2_lyapunov}, and show that it falls to zero as the length of the
straight sections is taken to zero.
\end{exercise}

\begin{exercise}
\label{ex:L3_timescales}
A particle of mass \(m\) moves in a stadium billiard of area \(A\).
\begin{enumerate}
    \item[(a)] Use Weyl's law to show that the Heisenberg time is \(t_H=Am/\hbar\) to
    leading order, independent of energy.  The perimeter does appear at the next order: obtain
    the correction from the two-term law and say at what \(k\) it stops mattering.
    \item[(b)] Take a rubidium atom, \(m=1.4\times 10^{-25}\,\mathrm{kg}\), moving at
    \(v=1\,\mathrm{cm\,s^{-1}}\) in a stadium of area \(1\,\mathrm{cm^{2}}\). Evaluate
    \(t_H\). Then estimate the Ehrenfest time, using \(\lambda\approx v/\sqrt{A}\) for the
    Lyapunov exponent and \(\hbar_{\rm eff}\approx \ell_{\rm dB}/\sqrt{A}\) with
    \(\ell_{\rm dB}=\hbar/mv\).
    \item[(c)] Compare the two. How many decades wide is the window \(t_E\ll t\ll t_H\), and
    which of the two scales is the practical obstacle to seeing quantum chaos here?
\end{enumerate}
\end{exercise}

\begin{exercise}
\label{ex:L3_cat_state}
Compute the Wigner function of the superposition
\begin{equation}
    \psi(q)\propto e^{-(q-a)^{2}/4\sigma^{2}}+e^{-(q+a)^{2}/4\sigma^{2}},
    \qquad a\gg\sigma .
\end{equation}
Show that it consists of two positive Gaussian lobes centred at \(q=\pm a\) together with a
third contribution centred at \(q=0\) which oscillates in \(p\) as \(\cos(2ap/\hbar)\) and
therefore takes negative values, while the \(q\)-marginal remains non-negative everywhere.
Compare the period of the fringes with the momentum width \(\hbar/2\sigma\) of a single
lobe, and use Hudson's theorem to say why the negativity was unavoidable. Then show that smoothing over a coherent
state of width \(\sigma\) suppresses the fringes by \(e^{-a^{2}/4\sigma^{2}}\) rather than
removing them, and that the residual oscillates at half the frequency of the original.
\end{exercise}

\clearpage

\section{Lecture 4: Diagnostic tools of quantum chaos}

Lecture 3 closed on Weyl's law and on what follows from it: a billiard's smooth spectral
staircase is fixed by its area and its perimeter, so it cannot tell a stadium from an
ellipse built to match both, and any signature of chaos has to sit in the fluctuations
around it. Reaching those fluctuations
means removing the smooth part first. That operation, unfolding, is the business of
Sec.~\ref{sec:L4_unfolding}, and it comes second here rather than first because the trace
formula is what says why the fluctuations should be universal at all.

The question every instrument in this lecture answers can be put in a sentence. Given a quantum
system with a chaotic classical limit, or a strongly interacting many-body system expected to
thermalize, what measurable quantity separates it from an integrable one? Three answers read the
energies: the distribution of nearest-neighbour spacings, the ratio of consecutive spacings, and
the long-range measures of spectral rigidity together with their time-domain counterpart, the
spectral form factor. Random matrix theory supplies the benchmark curves for all three, and it
earns that role by asking what the symmetries of a Hamiltonian imply once its microscopic details
have been given up. A fourth answer reads the eigenstates instead of the energies, and ends in the
eigenstate thermalization hypothesis, the statement that a single eigenstate of a chaotic
many-body Hamiltonian already carries thermal expectation values for simple observables.

Each of these carries a condition. Spacing statistics say nothing until every exact symmetry has
been resolved; unfolding needs enough levels in a window for a smooth fit to be meaningful, and
the long-range statistics need more still; eigenstate thermalization holds for a restricted class
of observables, which has to be named before the claim can be tested. Their failure modes differ,
which is the reason for developing all four rather than settling on one.

\subsection{From periodic orbits to random matrices}
\label{sec:L4_trace_formula}

For an integrable system the route from classical mechanics to the spectrum is the EBK
quantization of Sec.~\ref{sec:L3_semiclassical}: each invariant torus satisfying
\begin{equation}
    \oint_{\gamma_i} p\cdot dq = 2\pi\hbar\left(n_i+\frac{\mu_i}{4}\right),
    \qquad i=1,\ldots,d,
\end{equation}
supplies one level, and the spectrum is the image of a \(d\)-dimensional lattice of quantum
numbers under a smooth map. Nothing in that construction correlates neighbouring levels, because
two levels adjacent in energy generally descend from distant lattice points, and that is where
the Poisson statistics of Sec.~\ref{sec:L4_poisson_wd} come from.

A chaotic system has no such tori, and the semiclassical organization of its spectrum runs
through periodic orbits instead. Split the density of states into a smooth part and a fluctuating
part,
\begin{equation}
    \rho(E)=\bar{\rho}(E)+\rho_{\mathrm{osc}}(E).
\end{equation}
Gutzwiller's trace formula expresses the second as a sum over the classical periodic orbits
\(p\), for a system whose periodic orbits are isolated and
unstable~\cite{Gutzwiller1971,Gutzwiller1990},
\begin{equation}
    \rho_{\mathrm{osc}}(E)
    \sim
    \sum_{p} A_p(E)\cos\left(\frac{S_p(E)}{\hbar}-\frac{\pi\mu_p}{2}\right),
    \qquad
    A_p \propto \frac{T_p}{\left|\det\left(1-M_p\right)\right|^{1/2}},
\end{equation}
where \(S_p\) is the classical action of the orbit, \(\mu_p\) a Maslov-type index, \(T_p\) the
period, and \(M_p\) the reduced monodromy matrix of
Sec.~\ref{sec:L2_poincare}. The hypothesis is what the denominator needs: an integrable
system carries its orbits in continuous families, \(\det(1-M_p)\) vanishes on every one of
them, and the amplitudes are infinite. The smooth part \(\bar\rho(E)\) is the Weyl term of
Lecture 3, so every dynamical distinction between the stadium and the ellipse matched in
Sec.~\ref{sec:L3_billiards} sits in \(\rho_{\mathrm{osc}}\), and which sum
\(\rho_{\mathrm{osc}}\) is written as is itself part of that distinction: the one above for
the stadium, the Berry--Tabor sum over tori for the
ellipse~\cite{BerryTabor1977}. At two degrees of freedom the torus sum is the larger of the
two by \(\hbar^{-1/2}\), which is the semiclassical face of the wider long-range fluctuations
an integrable spectrum turns out to have.

The formula reorganizes the problem without solving it. An unstable orbit in a system with two
degrees of freedom has monodromy eigenvalues \(e^{\pm\lambda_{p}T_p}\), in the notation
Sec.~\ref{sec:L2_poincare} fixed, so
\begin{equation}
    \left|\det\left(1-M_p\right)\right|
    =
    2\cosh\lambda_{p} T_p-2
    \simeq
    e^{\lambda_{p} T_p},
\end{equation}
and each term of the sum is damped as \(e^{-\lambda_{p}T_p/2}\). Against that, the number of
periodic orbits of period below \(T\) grows as \(e^{hT}\) up to algebraic factors, where \(h\) is
the topological entropy. That \(h\) is at least the Lyapunov exponent follows from the
variational principle together with Pesin's formula~\cite{Pesin1977}, the two ingredients
Sec.~\ref{sec:L1_stretchfold} named; the baker map worked out there is the case of equality,
\(2^{k}\) orbits of period \(k\) against an exponent \(\log2\), so \(h=\lambda=\log2\).
The two exponentials fail to cancel: writing \(\lambda\) for an exponent typical of the
orbits in the sum,
\begin{equation}
    \int^{T} dT'\, e^{(h-\lambda/2)T'}
\end{equation}
diverges. The orbit sum therefore converges at best conditionally, and recovering individual
levels from it requires resummation.

Random matrix theory takes the opposite route. It gives up the microscopic Hamiltonian entirely
and asks what spectral statistics follow from symmetry alone. The two routes meet. That same orbit sum, evaluated in the diagonal approximation in which each orbit is paired
with itself, reproduces the random-matrix form factor at short times~\cite{Berry1985}. That is where the
slope of the ramp in Sec.~\ref{sec:L4_sff} comes from.

\subsection{Unfolding the spectrum and the spacing ratio}
\label{sec:L4_unfolding}

Comparing a measured spectrum against any universal prediction needs a common scale, and the raw
levels supply none, since the mean density of states varies between systems and, within one
system, with energy. The spectral staircase
\begin{equation}
    N(E)=\sum_n \Theta(E-E_n),
    \qquad
    \rho(E)=\frac{dN(E)}{dE}=\sum_n \delta(E-E_n),
\end{equation}
is split into a smooth and a fluctuating part,
\begin{equation}
    N(E)=\bar{N}(E)+N_{\mathrm{fl}}(E),
\end{equation}
which is the split of Sec.~\ref{sec:L4_trace_formula} carried up one integration, and the
unfolded levels are the smooth part evaluated at the true ones,
\begin{equation}
    \varepsilon_n=\bar{N}(E_n),
    \qquad
    s_n=\varepsilon_{n+1}-\varepsilon_n .
\end{equation}
Then \(\langle s\rangle=1\) by construction. Unfolding is a change of units applied level by
level, chosen to make the mean spacing unity everywhere in the spectrum.

Section~\ref{sec:L3_billiards} computed \(\bar N\) for a planar billiard from Weyl's law, and
that calculation shows how much of the smooth part matters here. The area term sets the mean
density and with it the unit of \(s\); the perimeter term corrects it at relative order \(1/k\);
the constant term, which collects the boundary curvature and the corner angles, shifts every
\(\varepsilon_n\) by the same amount and cancels in \(s_n\). Two billiards of equal area and
perimeter, one chaotic and one integrable, therefore have smooth staircases differing only by
that constant, which the spacings never see. The whole distinction lives in
\(N_{\mathrm{fl}}\).

Two cautions attach to the procedure. Unfolding should remove the smooth part and nothing more:
a fit to \(\bar N\) with too many free parameters starts to follow the fluctuations themselves,
softening the long-range rigidity that Sec.~\ref{sec:L4_sff} sets out to measure, so the fit must
stay smooth over many mean spacings. And in many-body spectra the density of states rises steeply
and is roughly Gaussian in the middle of the band, with no analytic form available to subtract,
so one works inside a fixed symmetry sector, keeps away from the spectral edges, and fits
locally.

Where unfolding cannot be trusted it can be avoided. The ratio of consecutive
spacings~\cite{OganesyanHuse2007},
\begin{equation}
    r_n=\frac{E_{n+1}-E_n}{E_n-E_{n-1}},
    \qquad
    \tilde{r}_n=\min\left(r_n,\frac{1}{r_n}\right)\in[0,1],
\end{equation}
compares two adjacent gaps, so any smooth factor multiplying the local density cancels between
numerator and denominator. This is why the ratio is the statistic of choice for spin chains,
matrix models and truncated field theories.

Its benchmark values follow from short calculations. For an uncorrelated spectrum the two gaps
are independent exponentials, so \(r\) is distributed as \(P(r)=(1+r)^{-2}\) and
\begin{equation}
    \langle \tilde{r}\rangle_{\mathrm{P}}
    =
    2\int_0^1 \frac{r\,dr}{(1+r)^2}
    =
    2\ln 2-1
    \simeq
    0.386 .
\end{equation}
The chaotic benchmarks are quoted here and earned later: the three Gaussian ensembles and the
Dyson index \(\beta=1,2,4\) that labels them are built in Sec.~\ref{sec:L4_ensembles}, and the
Wigner surmise that the next formula generalizes is Sec.~\ref{sec:L4_poisson_wd}. That
analogue, drawn from \(3\times 3\) matrices rather than \(2\times 2\), is~\cite{Atas2013}
\begin{equation}
    P_\beta(r)
    =
    \frac{1}{Z_\beta}
    \frac{(r+r^2)^\beta}{(1+r+r^2)^{1+\frac{3\beta}{2}}},
    \qquad
    Z_1=\frac{8}{27},
    \qquad
    Z_2=\frac{4\pi}{81\sqrt3},
    \qquad
    Z_4=\frac{4\pi}{729\sqrt3},
\end{equation}
and integrating it gives
\begin{equation}
    \langle \tilde{r}\rangle_{\mathrm{GOE}}=4-2\sqrt{3}\simeq 0.536,
    \qquad
    \langle \tilde{r}\rangle_{\mathrm{GUE}}=\frac{2\sqrt{3}}{\pi}-\frac{1}{2}\simeq 0.603,
    \qquad
    \langle \tilde{r}\rangle_{\mathrm{GSE}}=\frac{32\sqrt{3}}{15\pi}-\frac{1}{2}\simeq 0.676,
\end{equation}
for \(\beta=1,2,4\) respectively. As with the Wigner surmise, these sit close to the values for
large matrices: \(0.5307\), \(0.5996\) and \(0.6744\), which the surmise overshoots by one
per cent, a half and a quarter respectively. That gap is far
smaller than the distance to the Poisson value, so the statistic separates the universality
classes long before the difference between surmise and asymptotics becomes relevant.

\subsection{Poisson versus Wigner--Dyson}
\label{sec:L4_poisson_wd}

A robust empirical rule organizes what unfolded spectra look like. Quantum systems whose
classical limit is integrable have levels that behave as though drawn independently, giving
Poisson statistics, while systems whose classical limit is chaotic have correlated levels obeying
the statistics of random matrices, once every exact symmetry has been resolved. The integrable
half is the Berry--Tabor statement met in Sec.~\ref{sec:L3_semiclassical}, with the genericity
qualification recorded there~\cite{BerryTabor1977}. The chaotic half is due to Bohigas, Giannoni
and Schmit~\cite{BohigasGiannoniSchmit1984}, advanced as a conjecture on numerical evidence and
since given semiclassical support for uniformly hyperbolic systems by summing correlated
pairs of periodic orbits, a construction introduced for the leading off-diagonal
term~\cite{SieberRichter2001} and carried to all orders in \(\tau\) for the spectral form
factor rather than for the spacing distribution itself~\cite{Mueller2004}, which is the orbit sum of
Sec.~\ref{sec:L4_trace_formula} evaluated beyond its diagonal terms.

\begin{figure}[!htbp]
	\centering
	\includegraphics[width=\textwidth]{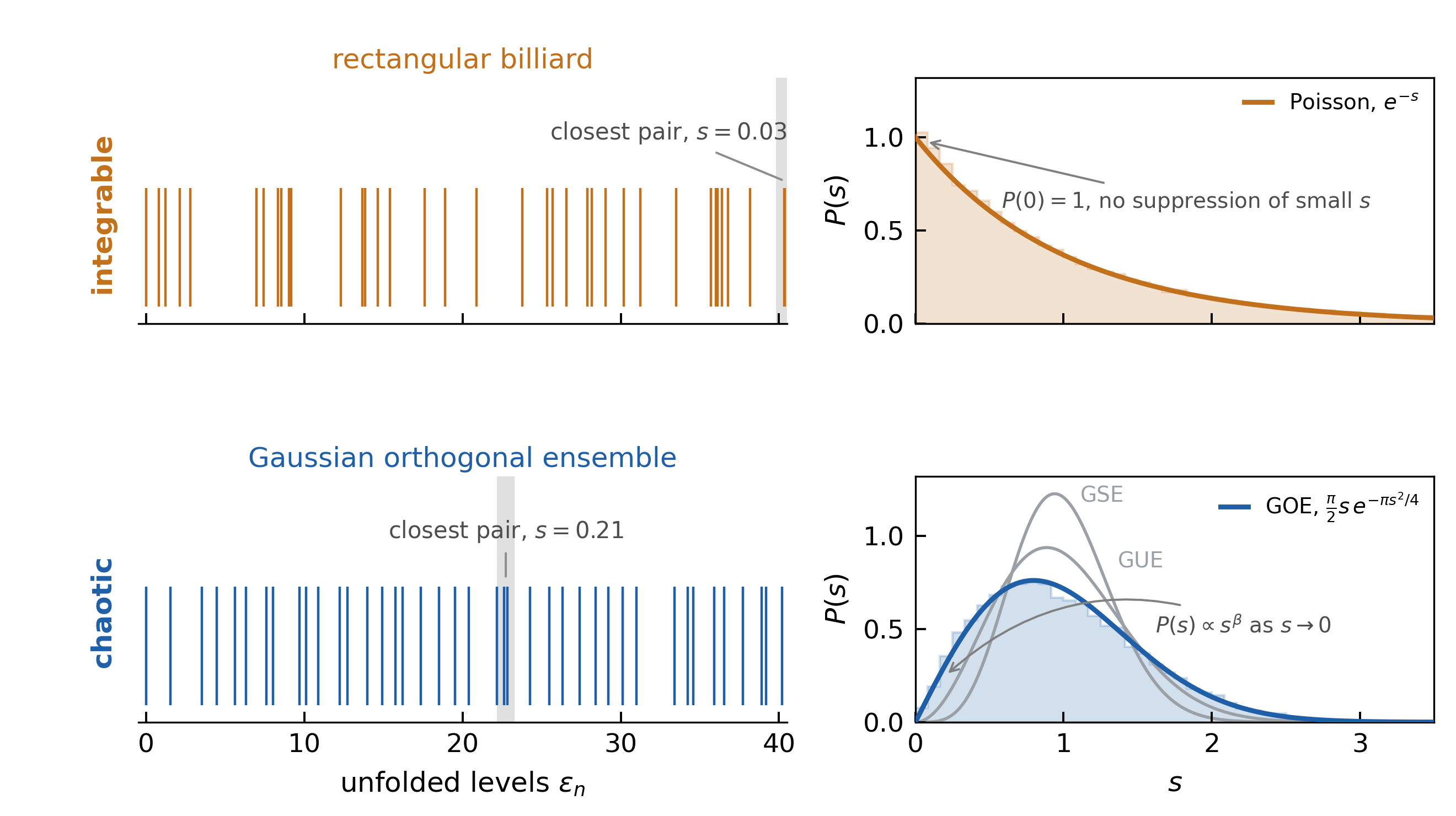}
	\caption{Level statistics of an integrable and a chaotic spectrum. Upper row: forty
	consecutive unfolded levels of a rectangular billiard of side ratio \((\sqrt5-1)/2\), and the
	nearest-neighbour spacing distribution of its spectrum, against \(e^{-s}\). Lower row: the
	same for the eigenvalues of a Gaussian orthogonal ensemble, against the Wigner surmise, with
	the GUE and GSE surmises drawn in grey. The rectangle's levels are exact and are unfolded with
	the two-term Weyl law; the GOE spectrum is unfolded with the semicircle law. Shaded bands mark
	the closest pair in each sequence.}
	\label{fig:poisson-wigner-dyson}
\end{figure}

Take the unfolded levels \(\varepsilon_n\) and spacings \(s_n\) of Sec.~\ref{sec:L4_unfolding}.
If the levels were independent points of unit mean density, the chance \(E(s)\) that an
interval of length \(s\), dropped at random rather than anchored to a level, contains none of
them would decay as \(e^{-s}\).  The spacing density is
\(P=\bar\rho\,E\) for independent points, and the unit mean density makes the two coincide,
\begin{equation}
    P_{\mathrm{P}}(s)=e^{-s},
    \qquad
    P_{\mathrm{P}}(0)=1 .
\end{equation}
The two are not the same object in general, and Exercise~\ref{ex:L4_superposition} needs the
harder relation \(P=E''\) that holds for any sequence.
Nothing suppresses small spacings, and levels that nearly coincide are common. The upper row of
Fig.~\ref{fig:poisson-wigner-dyson} shows this directly: the closest pair among forty consecutive
levels of a rectangular billiard is separated by three per cent of the mean spacing.

Chaotic spectra behave in the opposite way. For the Gaussian orthogonal ensemble the Wigner
surmise gives
\begin{equation}
    P_{\mathrm{GOE}}(s)
    =
    \frac{\pi}{2}s\exp\left(-\frac{\pi s^2}{4}\right),
    \qquad
    P_{\mathrm{GOE}}(s)\sim s
    \quad\text{as}\quad
    s\to 0,
\end{equation}
which vanishes linearly at the origin. The closest pair in the GOE sequence drawn beside it is
separated by a fifth of the mean spacing, which is on the generous side; over many realisations
of forty levels the median closest pair sits nearer a seventh. Either way the histogram is
pushed away from \(s=0\) altogether. Levels repel.

The strength of the repulsion is fixed by the symmetry class rather than by the system. The three
standard Dyson classes have
\begin{equation}
\begin{aligned}
	P_{\mathrm{GOE}}(s) &= \frac{\pi}{2}s\exp\left(-\frac{\pi s^2}{4}\right),\\[4pt]
	P_{\mathrm{GUE}}(s) &= \frac{32}{\pi^2}s^2\exp\left(-\frac{4s^2}{\pi}\right),\\[4pt]
	P_{\mathrm{GSE}}(s) &= \frac{2^{18}}{3^6\pi^3}s^4\exp\left(-\frac{64s^2}{9\pi}\right),
\end{aligned}
\end{equation}
all sharing the small-spacing form
\begin{equation}
    P(s)\sim s^\beta,
    \qquad
    \beta =
    \begin{cases}
        1, & \text{orthogonal class},\\
        2, & \text{unitary class},\\
        4, & \text{symplectic class}.
    \end{cases}
\end{equation}
The exponent \(\beta\) is the Dyson index. Section~\ref{sec:L4_repulsion} shows where it comes
from, by counting the conditions a Hamiltonian must satisfy for two of its levels to meet.

\subsection{Why levels repel}
\label{sec:L4_repulsion}

Level repulsion has a local origin visible in a \(2\times 2\) matrix. Take two states close in
energy that mix under a perturbation and write the Hamiltonian restricted to the space they span
as
\begin{equation}
    H=
    \begin{pmatrix}
        E_1 & H_{12}\\
        H_{12}^{*} & E_2
    \end{pmatrix},
    \qquad
    E_\pm=\frac{E_1+E_2}{2}\pm\frac{1}{2}\sqrt{(E_1-E_2)^2+4|H_{12}|^2}.
\end{equation}
The gap
\begin{equation}
    s=E_+-E_-=\sqrt{(E_1-E_2)^2+4|H_{12}|^2}
\end{equation}
closes only when \(E_1=E_2\) and \(H_{12}=0\) together.

Count those conditions class by class. For a real symmetric Hamiltonian \(H_{12}\) is a single
real number and degeneracy needs two conditions; for a complex Hermitian one it has a real and an
imaginary part and degeneracy needs three; for a quaternion self-dual one it has four real
components and degeneracy needs five. Each count is \(\beta+1\), where \(\beta\) is the number of
real components in an off-diagonal element. Degeneracy therefore has codimension \(\beta+1\) in
the space of Hamiltonians of the given class, and varying a single parameter cannot generically
reach it. This is the non-crossing rule~\cite{VonNeumannWigner1929,Haake2018}, and it is why avoided
crossings rather than crossings are what one sees.

The same count fixes the exponent. Treat the \(\beta+1\) quantities that must vanish as Cartesian
coordinates, in which the gap \(s\) is the distance from the origin. If their joint distribution
is smooth and does not vanish there, the probability of a gap \(s\) is that density multiplied by
the area of the sphere of radius \(s\), so
\begin{equation}
    P(s)\propto s^{\beta},
    \qquad s\to 0 .
\end{equation}
The Dyson index of Sec.~\ref{sec:L4_poisson_wd} is a statement about the codimension of
degeneracy.

The same picture gives the whole distribution. Basis invariance, imposed on the ensembles in
Sec.~\ref{sec:L4_ensembles}, forces the diagonal entries to carry twice the variance of each
off-diagonal component. Then \(E_1-E_2\) and the components of \(2H_{12}\) are independent
Gaussians of equal variance, no direction in the \((\beta+1)\)-dimensional space is
distinguished, and \(s\) is the length of an isotropic vector:
\begin{equation}
    P(s)\propto s^{\beta}e^{-as^{2}} .
\end{equation}
The width \(a\) is inherited from whatever sets the scale of the ensemble, and it does not have
to be tracked, since the two conditions below fix it in each class separately. Imposing \(\int_0^\infty P(s)\,ds=1\) and \(\int_0^\infty sP(s)\,ds=1\) recovers each of the three
distributions quoted in Sec.~\ref{sec:L4_poisson_wd}. This is why they are called surmises: they
are exact for two levels, and close to but not identical with the answer for large matrices, the
same relation the ratio surmise bears to its asymptotic value in Sec.~\ref{sec:L4_unfolding}.

The argument also says when levels fail to repel. If the two states carry different eigenvalues
of an operator commuting with \(H\), then \(H_{12}\) vanishes identically rather than
accidentally, the codimension collapses to one, and a single parameter suffices to bring the
levels together. An integrable system is cut into many such sectors by its conserved charges, so
its spectrum is an overlay of independent sequences whose members have no reason to avoid one
another. That is the algebraic counterpart of the lattice of quantum numbers in
Sec.~\ref{sec:L4_trace_formula}, and it is why
every exact symmetry must be resolved before any comparison with random matrix theory. It is not
on its own the reason an integrable spectrum is Poisson. Superposing \(m\) independent repelling
sequences leaves \(P(0)=1-1/m\), which reaches the Poisson value only as \(m\to\infty\), and
Exercise~\ref{ex:L4_superposition} works that out; the conserved quantities of an integrable
system are in any case continuous rather than a finite list of sectors. The statement that does
the work is Berry and Tabor's~\cite{BerryTabor1977}, quoted in
Sec.~\ref{sec:L4_trace_formula}.

\subsection{The Gaussian ensembles}
\label{sec:L4_ensembles}

The phrase ``random matrix theory'' is meant literally. A complicated Hamiltonian, that of an
excited heavy nucleus in Wigner's original setting~\cite{Wigner1955,Wigner1958}, cannot be
written down matrix element by matrix element. In place of it one asks which statistical properties follow if the Hamiltonian is
replaced by a matrix drawn at random, subject to the structural requirements of quantum mechanics
and to the symmetries of the system,
\begin{equation}
    H_{\mathrm{physical}}
    \quad\longrightarrow\quad
    H_{\mathrm{RMT}} .
\end{equation}
The claim being made is narrow. The physical Hamiltonian is fixed, and what the ensemble
reproduces is its local spectral statistics and nothing
else~\cite{Haake2018,Reichl2021}.

Three requirements determine the ensemble, two of them immediately. Hermiticity, \(H=H^\dagger\), makes the eigenvalues real
and the evolution unitary. Antiunitary symmetry then decides which matrices are allowed. If the
system has an antiunitary symmetry \(T\) with \(T^2=+1\), a basis exists in which \(H\) is real
symmetric and the ensemble is the Gaussian orthogonal ensemble. If no antiunitary symmetry
survives, as when a magnetic field is present, \(H\) is complex Hermitian with no real basis and
the ensemble is the Gaussian unitary ensemble. If there is an antiunitary symmetry with
\(T^2=-1\), the levels come in Kramers doublets, \(H\) may be written as a quaternion self-dual
matrix, and the ensemble is the Gaussian symplectic ensemble; half-integer spin with spin-orbit
coupling is the usual physical realization, although it is the symmetry rather than the spin that
is the criterion. Summarizing,
\begin{equation}
\begin{array}{c|c|c|c}
	\text{Ensemble} & \text{Matrix type} & \text{Antiunitary symmetry} & \beta \\ \hline
	\text{GOE} & \text{real symmetric} & T^2=+1 & 1\\
	\text{GUE} & \text{complex Hermitian} & \text{none} & 2\\
	\text{GSE} & \text{quaternion self-dual} & T^2=-1 & 4
\end{array}
\end{equation}
where \(\beta\) is the same Dyson index counted in Sec.~\ref{sec:L4_repulsion}, the number of
real components in an off-diagonal matrix element.

The Gaussian ensembles are the probability densities
\begin{equation}
    P(H)\,dH=\frac{1}{Z}\exp\left(-\frac{\beta}{4\sigma^2}\mathrm{Tr}\,H^2\right)dH ,
\end{equation}
with \(dH\) the flat measure over the independent matrix elements. Expanding the trace as
\(\mathrm{Tr}\,H^2=\sum_i H_{ii}^2+2\sum_{i<j}\sum_c H_{ij,c}^2\), the quaternion trace in
the symplectic class, half the ordinary trace of the \(2N\times2N\) complex representative one
would build numerically, where \(c\) runs over the
\(\beta\) real components of \(H_{ij}\), the diagonal entries carry variance \(2\sigma^2/\beta\)
and each off-diagonal component \(\sigma^2/\beta\), a ratio of two in every class. That ratio is
what makes the gap coordinates of Sec.~\ref{sec:L4_repulsion} isotropic, so the Wigner surmises
derived there follow from the measure written here rather than from an extra assumption.

The remaining requirement is invariance under the basis changes that preserve the symmetry class,
\begin{equation}
    H\to OHO^{T},\quad O\in O(N),
    \qquad
    H\to UHU^{\dagger},\quad U\in U(N),
\end{equation}
and the symplectic analogue, which says that no basis has been secretly built into the ensemble.
Hermiticity, the symmetry constraint and basis invariance are the three principles, and the
Gaussian weight is the simplest realization of them. The weight looks arbitrary, and in the
relevant sense it is: for the unitarily invariant ensembles the eigenvalue statistics on the
scale of a single
spacing are fixed by the symmetry class and the mean density of states
alone~\cite{PasturShcherbina1997}, with the orthogonal and symplectic classes settled later, and
unfolding divides out the second of these. Everything the
choice of weight controls is removed before the comparison is made.

What the randomness implies for the eigenvalues follows from a change of variables. Diagonalize
\(H=O\Lambda O^{T}\) with \(\Lambda=\mathrm{diag}(E_1,\ldots,E_N)\), so that
\begin{equation}
    dH=O\left(d\Lambda+[\Omega,\Lambda]\right)O^{T},
    \qquad
    \Omega=O^{T}dO
\end{equation}
with \(\Omega\) antisymmetric. Because \([\Omega,\Lambda]_{ij}=\Omega_{ij}(E_j-E_i)\) vanishes on
the diagonal, the cross terms drop out of the invariant line element and
\begin{equation}
    ds^2=\mathrm{Tr}\left(dH^2\right)
    =\sum_i dE_i^2+2\sum_{i<j}(E_i-E_j)^2\Omega_{ij}^2 .
\end{equation}
Two facts now combine. The line element \(\mathrm{Tr}(dH^{2})\) is Euclidean in the
independent matrix entries, so the flat measure \(dH\) is exactly the Riemannian volume
\(\sqrt{\det g}\) of the metric it defines; and in the coordinates \((E,\Omega)\) that metric
is diagonal. Computing its determinant therefore computes \(dH\),
\begin{equation}
    dH \propto \prod_{i<j}|E_i-E_j|\prod_i dE_i\,d\mu(O),
\end{equation}
with \(d\mu(O)\) the invariant measure on the group. In the other classes each pair \((i,j)\)
carries \(\beta\) independent components \(\Omega_{ij}\) in place of one, and the same computation
returns \(\prod_{i<j}|E_i-E_j|^{\beta}\). Integrating over the eigenvectors and using
\(\mathrm{Tr}\,H^2=\sum_i E_i^2\),
\begin{equation}
    P(E_1,\ldots,E_N)
    =
    \frac{1}{Z_N}
    \exp\left(-\frac{\beta}{4\sigma^2}\sum_i E_i^2\right)
    \prod_{i<j}|E_i-E_j|^{\beta}.
\end{equation}

These are three appearances of one number. In Sec.~\ref{sec:L4_repulsion} the index \(\beta\)
counted the real components of an off-diagonal element and, together with the diagonal
difference, gave codimension \(\beta+1\) for a degeneracy. Here the same components reappear as
the independent rotations \(\Omega_{ij}\) that mix a pair of eigenvectors, and each contributes
one power of \(|E_i-E_j|\) to the Jacobian. The exponent of the Vandermonde factor, the
codimension of a degeneracy and the small-spacing exponent are one fact counted three ways.

The joint distribution also has a mechanical reading. Writing
\begin{equation}
    P(E_1,\ldots,E_N)=\frac{1}{Z_N}e^{-\beta W},
    \qquad
    W=\frac{1}{4\sigma^2}\sum_i E_i^2-\sum_{i<j}\ln|E_i-E_j| ,
\end{equation}
the eigenvalues become charges confined to a line by a harmonic potential and repelling one
another logarithmically, at an inverse temperature numerically equal to the Dyson index. This is
Dyson's log-gas~\cite{Dyson1962a,Dyson1962b,Dyson1962c,Mehta2004}. The temperature is the
fictitious gas's own and has nothing to do with the temperature of any physical system, a
distinction that costs nothing until Sec.~\ref{sec:L4_sff} needs both at once. It is why
a random-matrix spectrum is rigid rather than merely irregular: the randomness sits in the matrix
entries, while the eigenvalues are strongly correlated.

The physical use of all this is a two-step statement. First one constructs an ensemble whose
entries are genuinely random, constrained only by Hermiticity and symmetry. Second, one asserts
that a single fixed chaotic Hamiltonian shares its local spectral statistics. The second step
does not follow from the first. It rests on evidence from nuclear spectra~\cite{HaqPandeyBohigas1982}, kicked
tops~\cite{HaakeKusScharf1987}, microwave cavities standing in for quantum
billiards~\cite{StockmannStein1990} and many-body systems, together with the semiclassical
derivation cited in Sec.~\ref{sec:L4_poisson_wd}.

This is also where symmetry resolution earns its place. Random matrices describe the statistics
of levels that are allowed to repel, and Sec.~\ref{sec:L4_repulsion} gave the mechanism: two
states carrying different eigenvalues of an operator commuting with \(H\) have a vanishing
off-diagonal element, so they do not repel. If \([H,Q]=0\) the Hilbert space decomposes as
\(\mathcal{H}=\bigoplus_q \mathcal{H}_q\) and the statistics must be computed inside a fixed
\(\mathcal{H}_q\). Combining sectors overlays independent sequences, and a superposition of
independent sequences drifts toward Poisson statistics~\cite{BerryRobnik1984}, so a chaotic
system whose sectors have been mixed can be mistaken for an integrable one~\cite{Lakshminarayan2018}. In practice one fixes
every exact quantum number first: parity, momentum, angular momentum, particle number, spin and
charge sectors, and any discrete symmetry. For a conformal field theory on a compact spatial
manifold this means organizing the spectrum by global charges, momentum, spin and parity before
asking whether the remaining levels repel.

For a periodically driven system the same logic applies to the Floquet operator over one drive
period \(T_{\rm d}\),
\begin{equation}
    U=\mathcal{T}\exp\left(-\frac{i}{\hbar}\int_0^{T_{\rm d}} H(t)\,dt\right),
    \qquad
    U|\phi_n\rangle=e^{-i\theta_n}|\phi_n\rangle ,
\end{equation}
rather than to a Hamiltonian. Four different objects wear the letter \(T\) in this subsection
and it is worth separating them once: the antiunitary symmetry above, the transpose in
\(OHO^{T}\), the drive period \(T_{\rm d}\), and the time ordering \(\mathcal{T}\), which is
the calligraphic one and appears only inside the exponential. Unitarity puts the spectrum of
\(U\) on the unit circle, so
the relevant ensembles are the circular ones, COE, CUE and CSE, defined by the same antiunitary
constraints as their Gaussian counterparts~\cite{Dyson1962a,Mehta2004}. Section~\ref{sec:L4_sff} uses the CUE result.

\subsection{Long-range correlations and the spectral form factor}
\label{sec:L4_sff}

The spacing distribution uses one gap at a time, and the rigidity that distinguishes a
random-matrix spectrum from a merely irregular one extends over many spacings. Two kinds of diagnostic
reach it. One counts levels in a window, the other watches the spectrum evolve in time, and they
carry the same two-level information in conjugate variables, that information being the
two-level cluster function \(Y_2\), the connected part of the two-point correlation of the
unfolded density, of which the two diagnostics below are two transforms.

Let \(n(L)\) be the number of unfolded levels in an interval of length \(L\), so that
\(\langle n(L)\rangle=L\). The number variance is
\begin{equation}
    \Sigma^2(L)=\left\langle \left(n(L)-L\right)^2\right\rangle .
\end{equation}
Independent levels give the Poisson result \(\Sigma^2_{\mathrm{P}}(L)=L\). The Gaussian ensembles
instead grow logarithmically,
\begin{equation}
    \Sigma^2_{\mathrm{GOE}}(L)\simeq\frac{2}{\pi^2}
    \left[\ln 2\pi L+\gamma+1-\frac{\pi^2}{8}\right],
    \qquad
    \Sigma^2_{\mathrm{GUE}}(L)\simeq\frac{1}{\pi^2}
    \left[\ln 2\pi L+\gamma+1\right],
\end{equation}
with \(\gamma\) Euler's constant~\cite{Haake2018}. The difference is not a matter of degree. Over thirty-two mean
spacings an uncorrelated spectrum fluctuates by about \(5.7\) levels and a GOE spectrum by about
\(1.1\), and the gap widens with \(L\). Once a level has been placed, its neighbours have nowhere
to go.

\begin{figure}[!htbp]
	\centering
	\includegraphics[width=\textwidth]{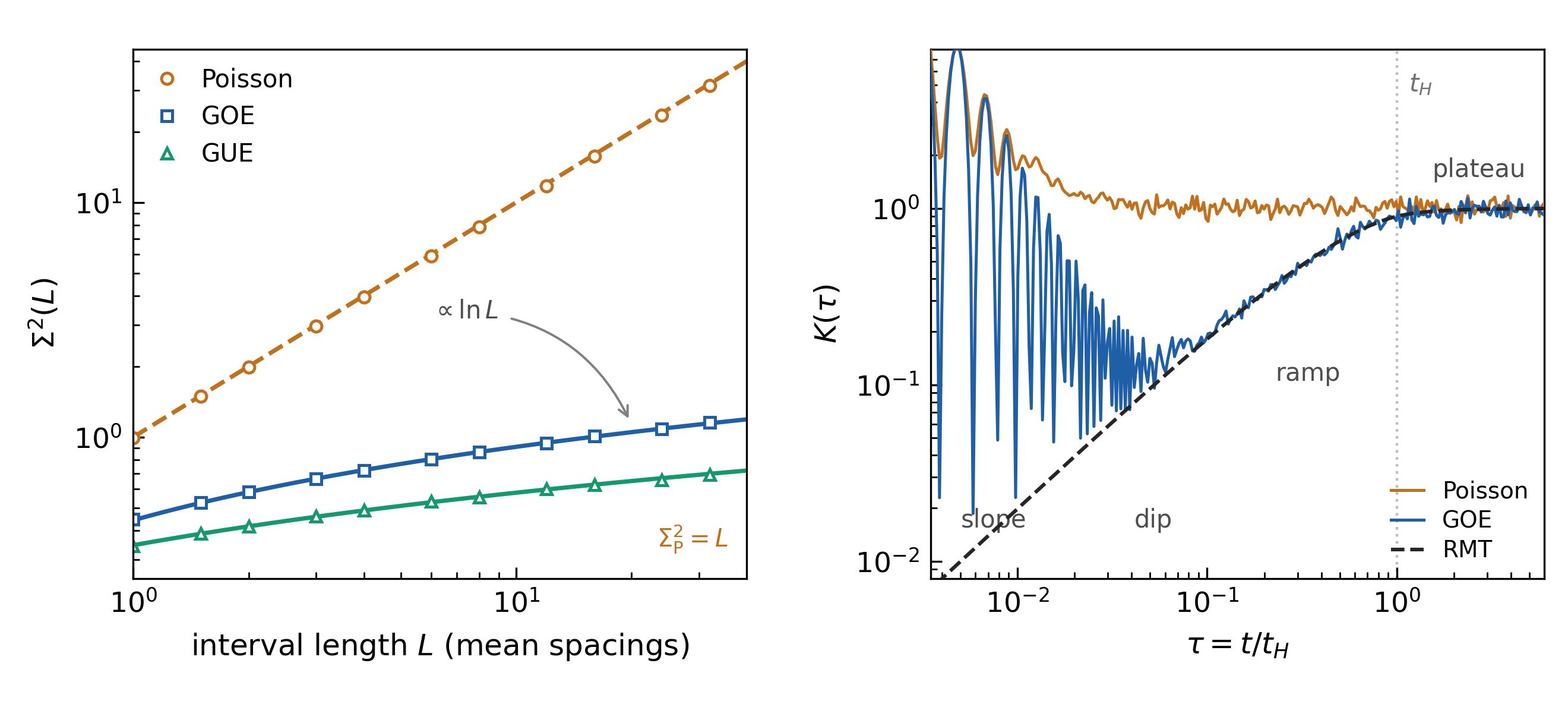}
	\caption{Long-range spectral rigidity and its counterpart in time. Left: number variance of
	unfolded spectra, with \(\Sigma^2=L\) for an uncorrelated sequence and logarithmic growth for
	the Gaussian ensembles; the points are computed from diagonalized matrices and the curves are
	the asymptotic formulas rather than fits. Right: spectral form factor against
	\(\tau=t/t_H\), averaged over 200 GOE matrices, showing slope, dip, ramp and plateau against
	the random-matrix result \(K(\tau)=2\tau-\tau\ln(1+2\tau)\), dashed. An uncorrelated spectrum
	of the same length, in orange, has no ramp. A hard spectral window is used, so the
	oscillations through the slope and the dip are the ringing of its edges rather than a property
	of the spectrum.}
	\label{fig:spectral-rigidity}
\end{figure}

A second energy-domain measure is the Dyson--Mehta statistic~\cite{Mehta2004}, which asks how
straight the unfolded staircase is over an interval,
\begin{equation}
    \Delta_3(L)
    =
    \left\langle
    \frac{1}{L}\min_{A,B}
    \int_{\varepsilon_0}^{\varepsilon_0+L}
    \left[N(\varepsilon)-A\varepsilon-B\right]^2 d\varepsilon
    \right\rangle_{\varepsilon_0}.
\end{equation}
The two statistics are not independent measurements. Carrying out the minimisation turns
\(\Delta_3\) into a smoothing of the number variance,
\begin{equation}
    \Delta_3(L)=\frac{2}{L^{4}}\int_{0}^{L}
    \left(L^{3}-2L^{2}r+r^{3}\right)\Sigma^{2}(r)\,dr ,
\end{equation}
from which the Poisson value follows in one line, \(\Sigma^{2}=L\) giving
\(\Delta_3(L)=L/15\), and the Gaussian ensembles inherit the logarithm with half the
coefficient,
\begin{equation}
    \Delta_3^{\mathrm{GOE}}(L)\simeq\frac{1}{\pi^{2}}
    \left[\ln 2\pi L+\gamma-\frac{5}{4}-\frac{\pi^{2}}{8}\right],
    \qquad
    \Delta_3^{\mathrm{GUE}}(L)\simeq\frac{1}{2\pi^{2}}
    \left[\ln 2\pi L+\gamma-\frac{5}{4}\right],
\end{equation}
which is why the two carry the same information~\cite{Mehta2004}. Over the same window of
thirty-two spacings used above, \(\Delta_3\) is \(2.13\) for an uncorrelated spectrum,
\(0.34\) for GOE and \(0.24\) for GUE. Both statistics need far more levels than the spacing
distribution does, which is the practical reason they are used less often. Figure~\ref{fig:spectral-rigidity} shows the
number variance for the three cases.

The time-domain version is the spectral form factor. From here \(\hbar=1\), so that time and
inverse energy are the same variable; it is restored in Lecture 5 wherever a classical limit is
being taken. Continuing the partition function to complex
inverse temperature,
\begin{equation}
    Z(\beta+it)=\sum_n e^{-(\beta+it)E_n},
\end{equation}
and pairing it with its conjugate,
\begin{equation}
    K(\beta,t)
    =
    \left\langle Z(\beta+it)Z(\beta-it)\right\rangle
    =
    \left\langle \sum_{m,n} e^{-\beta(E_m+E_n)}e^{-it(E_m-E_n)}\right\rangle ,
\end{equation}
where \(\beta\) is a physical inverse temperature. Both quantities are live from here on, so
the Dyson index is written \(\beta_{\rm D}\) for the rest of the lecture and \(\beta\) is kept
for the temperature. At \(\beta=0\) this reduces to
\begin{equation}
    K(0,t)=\left\langle \left|\sum_n e^{-iE_n t}\right|^2\right\rangle ,
\end{equation}
the Fourier transform of \(Y_2\). That is the second of the two transforms promised at the
head of this subsection, the first being the integral relating \(\Delta_3\) to
\(\Sigma^2\) above, and the conjugate variables are \(t\) and \(L\sim 1/t\).

The average is not optional. A single spectrum yields a form factor fluctuating by an amount
comparable to its own mean at every time, so the ramp described below is invisible in one sample~\cite{Prange1997}
and emerges only after averaging over an ensemble, over a disorder realization, over a spectral
window, or over a short interval of time.

A chaotic spectrum then shows four regimes: a slope, a dip, a ramp and a plateau. The slope is
the disconnected part \(|\langle Z\rangle|^2\), controlled by the smooth density of states and by
the spectral window rather than by any correlation between levels. It decays quickly, and the dip
is where it falls below the connected part. The ramp is that connected part, and is the
time-domain face of spectral rigidity. The plateau sets in once the evolution resolves individual
levels, at a time of order the Heisenberg time
\begin{equation}
    t_H=2\pi\hbar\bar{\rho}
\end{equation}
of Sec.~\ref{sec:L3_ehrenfest}, with \(\bar\rho\) the mean density of states in the window.
Of order, and not at: only in the unitary class does the form factor reach its plateau exactly at
\(\tau=1\). The orthogonal form below gives \(K_{\rm GOE}(1)=2-\ln3=0.9014\) and arrives only
asymptotically, \(0.978\) at \(\tau=2\) and \(0.9992\) at \(\tau=10\), while the symplectic
one plateaus at \(\tau=2\). The
time at which the connected part starts to follow the random-matrix curve is called the Thouless
time~\cite{Cotler2017,GharibyanEtAl2018}; below it the system has not yet explored the states available to it and universality has
not set in, so the universal
window is \(t_{\mathrm{Th}}\ll t\ll t_H\), alongside \(t_E\ll t\ll t_H\) from Lecture 3.
The two lower limits are set by different physics and do not track one another. The Ehrenfest
time grows like \(\log(1/\hbar)\) and knows about nothing but the Lyapunov exponent, while the
Thouless time is a relaxation time, and in a spatially extended system it is fixed by transport
across the sample and grows with its size. Which of them binds is a question about the system in
hand.

\begin{figure}[!htbp]
	\centering
	\includegraphics[width=\textwidth]{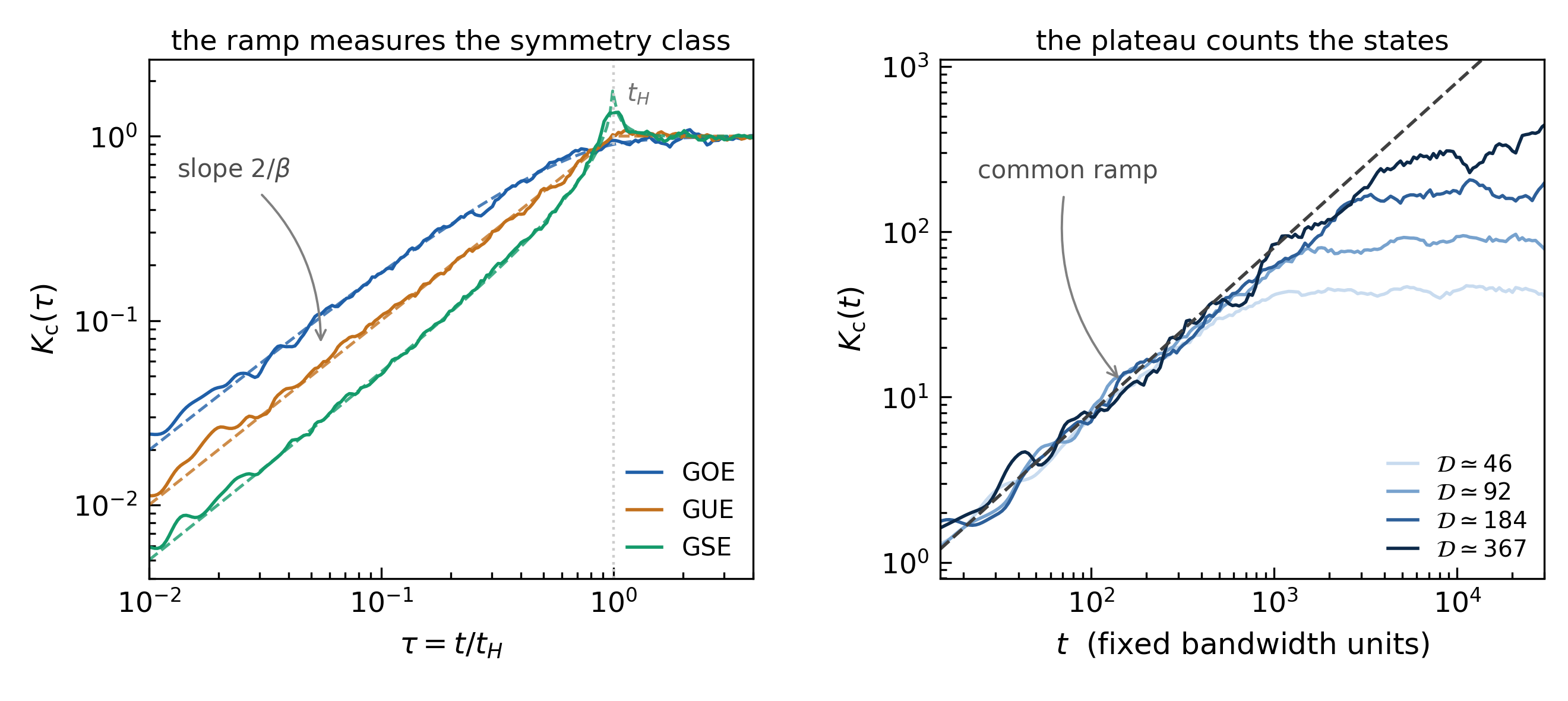}
	\caption{What the form factor measures. Left: connected spectral form factor of the three
	Gaussian ensembles, from ninety diagonalized matrices in each class unfolded with the
	semicircle law, against the analytic random-matrix results (dashed). The early ramps stand in
	the ratio \(2:1:\tfrac12\), which is \(2/\beta_{\rm D}\), measured here as
	\(K_{\mathrm{c}}/\tau=1.91\), \(1.08\) and \(0.53\); the rise in the symplectic curve near
	\(\tau=1\) is the logarithmic singularity of \(K_{\mathrm{GSE}}(\tau)=\tau/2-(\tau/4)\ln|1-\tau|\)
	and not a numerical artefact. Right: connected form factor of GOE spectra at fixed bandwidth,
	weighted by a Gaussian energy window of fixed width, so that the number of levels
	\(\mathcal{D}\) inside the window grows with the matrix dimension. All four follow one common
	ramp and leave it at a plateau whose height tracks \(\mathcal{D}\), measured at \(44\),
	\(88\), \(175\) and \(344\) against \(\mathcal{D}=46\), \(92\), \(184\) and
	\(367\), at a time proportional to \(\mathcal{D}\); the largest of them is still
	settling at the right-hand edge of the window. The disconnected part is subtracted in both panels, which is why neither shows
	the slope and dip of Fig.~\ref{fig:spectral-rigidity}.}
	\label{fig:sff-measures}
\end{figure}

The random-matrix prediction for the ramp is
\begin{equation}
    K(\tau)=2\tau-\tau\ln(1+2\tau),
    \qquad
    \tau=\frac{t}{t_H}\leq 1,
\end{equation}
with one convention to record, since three normalisations of \(K\) appear in this subsection.
The \(K(\beta,t)\) defined above plateaus at \(\mathcal{D}\), the number of levels in the
window, which is what the right-hand panel of Fig.~\ref{fig:sff-measures} measures. The
\(K(\tau)\) written here and the Floquet \(K(t)\) below are both divided by that number, so
their plateau is one, and it is the divided form that Exercise~\ref{ex:L4_sff} asks for.
With that fixed, \(K\simeq 2\tau\) at early times, and in general the slope of the ramp is
\(2/\beta_{\rm D}\). That factor is where the periodic orbits of
Sec.~\ref{sec:L4_trace_formula} return. In Berry's diagonal approximation each orbit in the trace
formula is paired with itself, and in a system with time-reversal symmetry also with its time
reverse, so the count doubles~\cite{Berry1985}. The ramp slope is level repulsion seen in the
time domain.

Figure~\ref{fig:sff-measures} separates the two things the form factor measures, the slope of
the ramp and the height of the plateau. Subtracting the disconnected part removes the slope and
the dip and exposes the ramp from early times. In the left panel the three
ensembles then run parallel, with slopes in the ratio \(2:1:\tfrac12\), so the ramp alone fixes
the symmetry class, exactly as the small-spacing exponent does in
Sec.~\ref{sec:L4_poisson_wd} and for the same underlying reason. In the right panel the ensemble
is held fixed and the number of levels inside a window of fixed width is increased instead: every
spectrum follows the same ramp and leaves it at a plateau whose height equals the number of states
and whose onset time grows in proportion. That is the sense in which the plateau counts states,
and it is why the form factor became a diagnostic for black holes, whose Hilbert space has
dimension \(e^{S}\). The later the plateau, the more microstates the spectrum contains.

For a periodically driven system one uses the Floquet operator of Sec.~\ref{sec:L4_ensembles} and
its eigenphases,
\begin{equation}
    K(t)=\frac{1}{N}\left\langle \left|\mathrm{Tr}\,U^{t}\right|^2\right\rangle .
\end{equation}
For the circular unitary ensemble this is exactly
\begin{equation}
    K_{\mathrm{CUE}}(t)=
    \begin{cases}
        t/N, & t<N,\\
        1, & t\geq N,
    \end{cases}
\end{equation}
the ramp and plateau in their cleanest form, with no slope and no dip because there is no smooth
density of states to produce them.

The form factor matters in holography because the ramp and the plateau are sensitive to the
discreteness of the spectrum and to correlations at times exponentially long in the entropy.
Ordinary semiclassical gravity saddles capture the early decay and not these late-time features,
which is why the form factor has become a diagnostic at the interface of quantum chaos, SYK and
low-dimensional gravity~\cite{Cotler2017}.

Neither feature has to be taken as an empirical fact about random matrices. Averaging over a
spectrum spontaneously breaks the symmetry that exchanges advanced and retarded energy
arguments, and the Goldstone modes of that breaking carry a nonlinear sigma model whose leading
saddle returns the sine kernel, and with it the ramp; a second saddle, reached from the first by
a Weyl reflection, supplies the plateau. The construction controls the level statistics to
order \(e^{-S}\), which is the mean spacing itself, and it has a bulk reading in which the
Goldstone modes are bound states of strings stretched between spectral
branes~\cite{AltlandSonner2021}. Adding operator insertions to the same effective theory
carries the result from the spectrum to correlation functions, and makes the Thouless time a
property of the operator rather than of the Hamiltonian
alone~\cite{AltlandBagretsNayakSonnerVielma2021}.

\subsection{Eigenstates and thermalization}
\label{sec:L4_eth}

The diagnostics so far have all read eigenvalues. Eigenvectors carry their own signature. In a chaotic
billiard a high-lying eigenfunction looks locally like a random superposition of plane waves of
fixed wavelength, which is the random-wave picture met in
Sec.~\ref{sec:L3_signatures}~\cite{Berry1977}. The random-matrix counterpart is that
eigenvectors are distributed uniformly on the unit sphere, so a typical component is
\(a_i\sim N^{-1/2}\) and the state spreads across the whole basis. Delocalization is measured by
the inverse participation ratio
\begin{equation}
    \mathrm{IPR}_n=\sum_i |\langle i|n\rangle|^4 ,
\end{equation}
equal to one for a state living on a single basis vector and of order \(1/N\) for one spread over
\(N\) of them. The measure is basis-dependent, and the basis has to be chosen by the physics:
position for a billiard, occupation number for a lattice model, a basis of states localized on
an energy shell semiclassically, and one labelled by gauge-invariant quantum numbers in a field
theory.

That picture is what turns eigenvector statistics into a statement about thermodynamics.
If the eigenfunctions of a classically chaotic system behave as Gaussian
random variables in this sense, then expectation values taken in a single eigenstate already
agree with the microcanonical ensemble. That is the observation of~\cite{Srednicki1994}, which
also supplied the name, and the same conclusion had been reached three years earlier by a
different route, perturbing an integrable Hamiltonian with a random matrix~\cite{Deutsch1991}.
The result is the eigenstate thermalization hypothesis, and it answers a question no spectral
diagnostic touches.

That question is how a closed system evolving unitarily from a pure state ever comes to look
thermal. Sec.~\ref{sec:L1_ergodic} settled the classical version and promised this one. There a
mixing flow spreads any initial region over the energy surface until the probability of finding
the trajectory in a given set is that set's share of the volume, so the initial condition is
forgotten and the equilibrium measure stops being a convenience. The quantum answer cannot take
that shape, and the reason is immediate. Literally a closed quantum system does not thermalize
at all. The state \(|\psi(t)\rangle=e^{-iHt}|\psi(0)\rangle\) remains pure
and its von Neumann entropy remains zero at all times. Thermalization has to mean something
weaker, that for a restricted class of simple observables the late-time expectation values become
indistinguishable from thermal ones. What thermalizes is what a simple probe can see of the
state.

Expand the initial state in the energy eigenbasis, \(|\psi(0)\rangle=\sum_n c_n|n\rangle\), so
that
\begin{equation}
    \langle O(t)\rangle=\sum_{m,n}c_m^{*}c_n e^{i(E_m-E_n)t}O_{mn},
    \qquad
    O_{mn}=\langle m|O|n\rangle .
\end{equation}
The diagonal terms are time-independent and fix the long-time average; the off-diagonal terms
oscillate and control both the relaxation and the residual fluctuations. Everything below
concerns one or the other.

If the spectrum is non-degenerate the phases dephase, and the infinite-time average is
\begin{equation}
    \overline{\langle O\rangle}
    =\sum_n |c_n|^2 O_{nn}
    =\mathrm{Tr}\left(\rho_{\mathrm{diag}}O\right),
    \qquad
    \rho_{\mathrm{diag}}=\sum_n |c_n|^2|n\rangle\langle n| ,
\end{equation}
the diagonal ensemble. This is not yet thermalization, since \(\rho_{\mathrm{diag}}\) still
remembers every \(|c_n|^2\). For the answer to be thermal the initial-state weights must drop
out, which happens when \(O_{nn}\) is a smooth function of energy across the occupied window.

That is the diagonal part of the hypothesis,
\begin{equation}
    O_{nn}=O(E_n)+\text{fluctuations},
    \qquad
    \langle n|O|n\rangle\simeq\langle O\rangle_{\mathrm{mc},\,E_n},
\end{equation}
with the fluctuations vanishing in the thermodynamic limit. A single stationary pure state
carries thermal expectation values. Plotting \(O_{nn}\) against \(E_n\) makes the distinction
visible: a chaotic system gives points hugging a smooth curve, an integrable one broad scatter,
because energy alone no longer determines the observable. The effect was demonstrated for five
hard-core bosons on a two-dimensional lattice, where the expectation value computed in one
many-body eigenstate reproduces the thermal average~\cite{Rigol2008}.  It has also been
checked in a model with a proposed black-hole dual: exact diagonalization of the
Sachdev--Ye--Kitaev model at up to seventeen sites finds the ansatz below satisfied, and finds
the out-of-time-order correlator of Lecture 5, evaluated in one energy eigenstate, reproducing
its thermal counterpart down to the scrambling time~\cite{SonnerVielma2017}.  In the conformal
sector of that model the same matrix-element structure follows analytically from three-point
coefficients rather than from diagonalization, at the price that the reparametrization mode,
and with it the gravitational degrees of freedom, has been removed~\cite{NayakSonnerVielma2019}. None of this conflicts with
unitarity. The eigenstate is pure, and what is thermal is the pair consisting of the state and
the class of observables.

Why should eigenstates behave this way? Take a microcanonical shell of dimension
\(\mathcal{D}\sim e^{S(E)}\) and a state drawn at random from it. Its expectation value of a
bounded simple operator has mean equal to the microcanonical average and variance of order
\(1/\mathcal{D}\), so the typical deviation is
\begin{equation}
    \delta O\sim \mathcal{D}^{-1/2}\sim e^{-S/2} .
\end{equation}
Almost every state in a large shell is thermal. The content of the hypothesis is the dynamical
claim that the eigenstates of a chaotic Hamiltonian are typical in this sense, which does not
follow from typicality alone and which fails for whole classes of Hamiltonians, and that is
exactly where chaos enters. Random matrix theory supplies the
statistical intuition; the dynamics decides whether it applies.

Controlling the fluctuations about the long-time average requires the off-diagonal elements, and
the two parts are usually written together as~\cite{Srednicki1999}
\begin{equation}
    O_{mn}=O(\bar E)\delta_{mn}+e^{-S(\bar E)/2}f_O(\bar E,\omega)R_{mn},
    \qquad
    \bar E=\frac{E_m+E_n}{2},
    \quad
    \omega=E_m-E_n ,
\end{equation}
with \(O(\bar E)\) and \(f_O\) smooth and \(R_{mn}\) a fluctuating quantity of zero mean and unit
variance, constrained by Hermiticity of \(O\) to satisfy \(R_{mn}=R_{nm}^{*}\). The exponential prefactor is the
typicality estimate above, \(e^{-S/2}\sim\mathcal{D}^{-1/2}\). The envelope is not random, and
what it fixes is worth deriving rather than quoting. Write
\begin{equation}
    \mathcal{C}(\omega)=\frac{1}{2}\int dt\,e^{i\omega t}
    \bigl\langle\{O(t),O(0)\}\bigr\rangle_\beta ,
    \qquad
    \chi''(\omega)=\frac{1}{2}\int dt\,e^{i\omega t}
    \bigl\langle[O(t),O(0)]\bigr\rangle_\beta
\end{equation}
for the symmetrized thermal correlation function and the dissipative response. Inserting the
ansatz and turning the sum over \(m\) into an integral weighted by the density of final states,
\(\sum_m\to\int d\omega\,e^{S(\bar E+\omega/2)}\), each transform picks up
\(e^{S(\bar E+\omega/2)-S(\bar E)}\simeq e^{\beta\omega/2}\), since
\(\partial S/\partial E=\beta\). The two orderings differ only in the sign that
\(\omega\to-\omega\) carries, so the even and odd combinations are
\begin{equation}
    \mathcal{C}(\omega)=\cosh\left(\frac{\beta\omega}{2}\right)\left|f_O\right|^{2},
    \qquad
    \chi''(\omega)=2\sinh\left(\frac{\beta\omega}{2}\right)\left|f_O\right|^{2},
\end{equation}
so a single envelope fixes both the thermal fluctuations and the dissipative response, and their
ratio \(\chi''/\mathcal{C}=2\tanh(\beta\omega/2)\) is the fluctuation-dissipation
relation~\cite{DAlessio2016}. Only the variation from one
matrix element to the next is left structureless, and that last step is an assumption rather
than a result.

Structureless means independent, and independence means Gaussian: every higher moment would
follow from the variance by Wick contraction, so every model's higher correlation functions
would be one universal functional of its two-point function. Lecture 5 shows what that costs.
The squared commutator is a sum over four energy indices, and Wick contraction keeps only the
terms carrying a coincident pair, which return products of two-point functions and no growth at
all. The piece with all four indices distinct is the sole source of an exponential, and it is
exactly the piece independence sets to zero, so a positive Lyapunov exponent requires
correlations between matrix elements that no symmetry enforces. The ansatz has to be extended
to a hierarchy of smooth higher moments, the \(n\)-index one suppressed by \(e^{-(n-1)S}\)
against \(n-1\) free index sums each worth \(e^{S}\), so that the suppression cancels and
the correlations survive at leading order~\cite{FoiniKurchan2019}. Read as free cumulants of
the operator, those moments organize the multi-time correlators over non-crossing
partitions~\cite{PappalardiFoiniKurchan2022}. The extension has a holographic home as well: it
is the matrix-element statistics of a two-matrix model in the Hamiltonian and the operator,
which cannot be Gaussian if the thermal four-point function is to be crossing
symmetric~\cite{JafferisKolchmeyerMukhametzhanovSonner2023a}, and whose double-scaled limit
reproduces Jackiw--Teitelboim gravity coupled to a massive
scalar~\cite{JafferisKolchmeyerMukhametzhanovSonner2023b}.

Together these give the mechanism. Dephasing reduces the long-time average to the diagonal
ensemble; diagonal ETH makes \(O_{nn}\) a smooth function of energy, so a narrow initial energy
distribution yields a single answer; off-diagonal ETH makes the temporal fluctuations of order
\(e^{-S/2}\), so the expectation value stays near that answer at almost all late times.

\subsection{Typicality, the Page curve, and the limits of the hypothesis}
\label{sec:L4_typicality}

Typicality extends from expectation values to entanglement, and in a direction that matters for
black holes. Split the Hilbert space as \(\mathcal{H}_A\otimes\mathcal{H}_B\) with
\(d_A\leq d_B\), draw a pure state at random, and reduce to \(A\). The mean von Neumann entropy
of \(\rho_A\) is known in closed form,
\begin{equation}
    \langle S_A\rangle
    =
    \sum_{k=d_B+1}^{d_A d_B}\frac{1}{k}-\frac{d_A-1}{2d_B}
    \simeq
    \ln d_A-\frac{d_A}{2d_B},
\end{equation}
conjectured by Page and proved soon afterwards~\cite{Page1993,FoongKanno1994,Sen1996}. The largest value
available to \(S_A\) is \(\ln d_A\), so a random state falls short of maximal entanglement by
\(d_A/2d_B\), which is exponentially small in the number of degrees of freedom once \(B\) is
larger than \(A\) by an extensive amount, the deficit being \(\tfrac12 e^{-(n_B-n_A)}\) in the
counts of degrees of freedom on the two sides. Larger by a hair will not do. At \(d_A=d_B\) the
deficit is half a nat, which Page's exact expression puts at \(0.4639\) for
\(d_A=d_B=4\) and \(0.5000\) at \(1024\), against \(0.0073\) for \(d_A=4\),
\(d_B=256\). A random pure state is very nearly maximally entangled across any small cut.

Read as a function of subsystem size, this is the Page curve, and it is what makes the purity of
the global state compatible with the thermality of its parts. A region small compared with its
complement is almost maximally entangled and therefore indistinguishable from thermal to anything
confined to it, while the state of the whole remains pure and carries all of the information.
Applied to an evaporating black hole, with \(A\) the emitted radiation and \(B\) what remains of
the hole, the same counting is the origin of the expectation that the entropy of the radiation
turns over rather than growing without bound~\cite{PageBH1993}. What follows is the eigenstate
version of the same statement.

For field theory and holography the natural statement concerns reduced density matrices. Split
the system into a region \(A\) and its complement and write
\(\rho_A^{(n)}=\mathrm{Tr}_{\bar A}|n\rangle\langle n|\). Subsystem ETH asserts that
\begin{equation}
    \rho_A^{(n)}\simeq\rho_A^{\mathrm{thermal}}(E_n)
\end{equation}
for high-energy chaotic eigenstates and subregions small compared with the whole, with the
comparison made in trace distance~\cite{Dymarsky2018}. The full state is pure, and the subregion
is thermal because it is entangled with the rest. In holographic language many black-hole
microstates look alike to simple boundary probes, which see only coarse data fixed by the energy
and the conserved charges. The statement needs qualification, since protected sectors,
supersymmetry, global charges, large-\(N\) factorization and finite-volume effects all modify it.

Several misreadings are worth blocking. The hypothesis is not about every observable: given the
eigenbasis one can always construct an operator that separates eigenstates, and the claim is
restricted to simple ones, meaning local, few-body or low-complexity probes. It is not about
every system: integrable models, many-body localized phases, protected sectors and quantum
many-body scars all violate it, the last of these named by analogy with the single-particle scars
of Lecture 3~\cite{Serbyn2021}. It does not say that the pure state becomes mixed. It extends
random matrix theory from eigenvalues to matrix elements rather than replacing it. And it does
not remove the need to resolve symmetries, since in the presence of conserved charges it must be
formulated within a fixed sector. Integrability is the sharpest failure. With many conserved
quantities two eigenstates of nearly equal energy can differ in their other charges, so
\(O_{nn}\) need not collapse onto a function of energy, and after a quench the late-time state is
described by a generalized Gibbs ensemble~\cite{RigolEtAl2007}
\begin{equation}
    \rho_{\mathrm{GGE}}=\frac{1}{Z}\exp\left(-\sum_i \mu_i Q_i\right),
    \qquad
    \langle Q_i\rangle_{\mathrm{initial}}=\mathrm{Tr}\left(\rho_{\mathrm{GGE}}Q_i\right),
\end{equation}
with the multipliers \(\mu_i\) fixed by the initial charge expectation
values~\cite{DAlessio2016}. In a
chaotic system the energy and the global charges determine what simple observables see; in an
integrable one many more numbers are needed.

The diagnostics of this lecture are built from a single object, the spectrum of \(H\) together
with its eigenvectors. Level spacings, spacing ratios, the number variance, the form factor and
the matrix elements entering the ETH are all functionals of that data, and random matrix theory
supplies the benchmark for each. For a holographic audience the payoff is direct: in AdS/CFT a
thermal state of the boundary theory can be dual to a black hole in the bulk, so if black holes
are strongly chaotic then level statistics, the spectral form factor and eigenstate
thermalization become boundary diagnostics of bulk gravitational physics.

Three caveats travel with that statement. The strict large-\(N\) limit produces an effectively
continuous spectrum, and large degeneracies can hide level repulsion entirely, so random-matrix
universality is expected only once finite-\(N\) effects, a finite volume and a finite energy
window are in play. And not every state in a holographic CFT is a black hole,
since low-energy states, protected states and perturbative multi-particle states need not share
the statistics of the black-hole microstates. The working picture is that random matrix theory
captures a universal late-time sector of chaotic dynamics while gravity supplies the
semiclassical account of earlier collective behaviour, and reconciling the two is one motivation
for studying SYK, Jackiw--Teitelboim gravity and the spectral form factor
together~\cite{Cotler2017}.

There is a limitation of a different kind, worth stating plainly because it is what makes a
further lecture necessary. Unfolding was introduced in Sec.~\ref{sec:L4_unfolding} to strip a
spectrum of everything system-specific, and Sec.~\ref{sec:L4_ensembles} recorded that what
survives depends only on the symmetry class. The universal part therefore yields a classification
and a count: the symmetry class, read from the spacing distribution or equally from the slope of
the ramp, and the number of accessible states, read from the plateau. Spectral statistics do
carry one system-specific scale, the Thouless time, which marks where universality sets in and
which in a spatially extended system is governed by transport and grows with the system size. But
a relaxation rate is not a Lyapunov exponent. The quantity measuring how fast neighbouring
initial conditions separate, which carried the whole of Sec.~\ref{sec:L2_lyapunov}, appears in
this lecture in exactly one place, Sec.~\ref{sec:L4_trace_formula}, and there only as a classical
input to the convergence of an orbit sum. Level repulsion places a system in a chaotic
universality class without saying how chaotic it is.

Recovering a rate therefore needs a quantity that is not a functional of the spectrum alone. It
has to involve operators and the order in which they act, and it has to be sensitive to the early
part of the window \(t_E\ll t\ll t_H\) opened in Lecture 3, rather than to the neighbourhood of
the Heisenberg time where the ramp gives way to the plateau. The squared commutator of two
operators separated in time is such a quantity. Lecture 5 builds it and follows its growth
rate out to a bound.

\subsection{Exercises}

\begin{exercise}
\label{ex:L4_superposition}
Superpose \(m\) independent spectra, each drawn from the Gaussian orthogonal ensemble and
rescaled to mean spacing \(m\), so that the combined sequence again has unit mean spacing. Write
\(E_1(s)\) for the gap probability of a single Gaussian orthogonal sequence at \emph{unit}
mean spacing, so that one of the rescaled sequences has gap probability \(E_1(s/m)\).
\begin{enumerate}
    \item[(a)] Argue that the gap probability of the combined sequence is \(E(s)=E_1(s/m)^m\),
    and that the nearest-neighbour spacing distribution is \(P(s)=E''(s)\).
    \item[(b)] Using \(E_1(s)=1-s+O(s^3)\), which follows from \(P_{\rm GOE}(s)\sim s\), show that
    \begin{equation}
        P(0)=1-\frac{1}{m},
    \end{equation}
    so the superposition interpolates between the Wigner surmise at \(m=1\) and the Poisson value
    \(P(0)=1\) as \(m\to\infty\).
    \item[(c)] Verify the result numerically for \(m=2\) and \(m=3\) by diagonalizing GOE
    matrices and unfolding with the semicircle law.
    \item[(d)] Explain, using the argument of Sec.~\ref{sec:L4_repulsion}, why analysing a
    Hamiltonian with an unresolved parity symmetry produces exactly the case \(m=2\).
\end{enumerate}
\end{exercise}

\begin{exercise}
\label{ex:L4_ratios}
Choose a quantum system whose spectrum can be computed numerically, for example a rectangular
billiard, a kicked top, or a finite spin chain. Working inside one fixed symmetry sector, compute
the ordered levels \(E_n\) and the ratios
\begin{equation}
    \tilde{r}_n=\min\left(\frac{E_{n+1}-E_n}{E_n-E_{n-1}},\frac{E_n-E_{n-1}}{E_{n+1}-E_n}\right).
\end{equation}
\begin{enumerate}
    \item[(a)] Evaluate \(\langle\tilde r\rangle\) from the raw levels and again from the
    unfolded ones, and confirm that the two agree. This insensitivity to the smooth density is
    the property that makes the statistic worth having.
    \item[(b)] Compare with \(2\ln 2-1\) for an uncorrelated spectrum and with the surmise values
    of Sec.~\ref{sec:L4_unfolding}. For the chaotic case, roughly how many levels are needed
    before the one per cent difference between the surmise value \(4-2\sqrt3\) and the
    large-matrix value can be resolved?
\end{enumerate}
\end{exercise}

\begin{exercise}
\label{ex:L4_eth}
Take a finite nonintegrable spin chain and a simple local observable, for example
\(O=\sigma^z_i\) or \(O=\sigma^z_i\sigma^z_{i+1}\). Diagonalize the Hamiltonian in a fixed
symmetry sector.
\begin{enumerate}
    \item[(a)] Plot the diagonal matrix elements \(O_{nn}\) against \(E_n\), and repeat for an
    integrable limit of the same model. Account for the difference in the two pictures.
    \item[(b)] In a narrow window at the centre of the spectrum, compute the variance of the
    off-diagonal elements \(O_{mn}\) and check that it falls as \(1/\mathcal{D}\) with the
    dimension of the window as the chain is lengthened. This is the \(e^{-S}\) of the ansatz in
    Sec.~\ref{sec:L4_eth}.
\end{enumerate}
\end{exercise}

\begin{exercise}
\label{ex:L4_sff}
Compute the spectral form factor \(K(\tau)\) of a Gaussian orthogonal ensemble and of an
uncorrelated spectrum of the same length, reproducing
Fig.~\ref{fig:spectral-rigidity}. Average over realizations, and note how the result looks before
you do.
\begin{enumerate}
    \item[(a)] Identify the slope, the dip, the ramp and the plateau, and locate the Heisenberg
    time on your plot.
    \item[(b)] Check that the early ramp is \(K\simeq2\tau\) by comparing with
    \(K(\tau)=2\tau-\tau\ln(1+2\tau)\), then repeat the calculation for the unitary ensemble
    and confirm that it becomes \(K\simeq\tau\), as the factor \(2/\beta_{\rm D}\)
    requires.
    \item[(c)] Replace the hard spectral window by a smooth one and explain what happens to the
    oscillations.
\end{enumerate}
\end{exercise}

\begin{exercise}
\label{ex:L4_surmise}
Sec.~\ref{sec:L4_poisson_wd} quotes all three Wigner surmises and derives none, although
Sec.~\ref{sec:L4_ensembles} has already built the measure they follow from. Derive them.
\begin{enumerate}
    \item[(a)] Take the \(2\times2\) real symmetric case with density proportional to
    \(e^{-\operatorname{Tr}H^{2}/4\sigma^{2}}\). Change variables to the two eigenvalues and the
    rotation angle, using the Vandermonde Jacobian of Sec.~\ref{sec:L4_ensembles}, and integrate
    out the angle and the centroid \(\tfrac12(E_1+E_2)\). Show that what is left is
    \(P(s)\propto s\,e^{-as^{2}}\) for some \(a>0\).
    \item[(b)] Fix \(a\) by \(\langle s\rangle=1\) rather than by \(\sigma\), which is what
    unfolding does, and recover \(P_{\rm GOE}(s)=\tfrac{\pi}{2}s\,e^{-\pi s^{2}/4}\). Note that
    \(\sigma\) has dropped out, so the surmise is a statement about the symmetry class and not
    about the ensemble's width.
    \item[(c)] Repeat for \(\beta=2\) and \(\beta=4\), where the Jacobian carries
    \(|E_1-E_2|^{\beta}\), and confirm the two remaining surmises. Check in each case that
    \(P\) integrates to one and has unit mean.
    \item[(d)] The result is exact for \(2\times2\) matrices and only close for large ones.
    Diagonalize \(N\times N\) GOE matrices for \(N=2,4,8,64\), unfold, and watch
    \(\langle\tilde r\rangle\) move from the surmise value toward the large-matrix one quoted in
    Sec.~\ref{sec:L4_unfolding}.
\end{enumerate}
\end{exercise}

\clearpage

\part{Scrambling and Holography}

\section{Lecture 5: Scrambling, OTOCs, and the route to holography}

The previous lecture closed by naming a quantity rather than evaluating one.  The squared
commutator
\begin{equation}
    C(t)=-\big\langle [W(t),V(0)]^{2}\big\rangle_\beta
\end{equation}
introduced in Sec.~\ref{sec:L3_signatures} is the simplest object answering to the description
it gave.  This lecture
asks what it measures, when it grows exponentially, and why its growth rate turns out to be
bounded.

Three things have to be settled before \(C(t)\) can be trusted.  The first is what its growth
corresponds to physically, which is the spreading of an initially simple operator through the
operator algebra of a many-body system until no small set of degrees of freedom carries the
perturbation any longer.  The second is why the ordering matters, since a time-ordered
four-point function assembled from the same operators decays for reasons that have nothing to
do with chaos.  The third is usually passed over quickly.  The semiclassical argument that
suggests \(C(t)\sim\hbar^{2}e^{2\lambda t}\) is formal, and a quantized chaotic billiard fails
to obey it; a clean exponential window is a property of systems with many degrees of freedom
and a parametric separation of timescales, not a property of chaos by itself.

Each of these brings in a scale.  Operator growth introduces the scrambling time \(t_*\), at
which an initially local perturbation has spread over everything available to it, and in
systems whose interactions are all-to-all \(t_*\) grows only as the logarithm of the number of
degrees of freedom.  Spatial locality introduces a second scale, the butterfly velocity
\(v_B\), which sets how fast the growing operator advances through space and gives the squared
commutator a front resembling a light cone.  Between microscopic dissipation and \(t_*\) lies
the only window in which an exponent can be read off at all.

The destination is gravitational.  A black hole in anti-de Sitter space is dual to a thermal
state of a strongly coupled field theory, and the bulk process computing the boundary
correlator is the scattering of one infalling excitation from the shock wave left by another.
Near-horizon time translations act as boosts, boosts grow exponentially, and the exponent
supplied by Einstein gravity is \(\lambda_L=2\pi/\beta\).  The same number bounds \(\lambda_L\)
from above for any thermal quantum system meeting a short list of assumptions, which places
black holes at the top of the allowed range.  Lecture 6 constructs the gravitational side; the
aim here is to reach the bound and to be clear about what it bounds.

\subsection{Scrambling as operator growth}
\label{sec:L5_growth}

Scrambling is the process by which information initially accessible in a simple part of a
quantum system becomes encoded in correlations among many of its degrees of freedom.  The
evolution stays unitary and the information survives; what grows is the complexity of the
observable needed to retrieve it.  A perturbation that was simple at \(t=0\) becomes
invisible to simple late-time probes once its weight has spread across a large operator
subspace.

Two claims in that paragraph can be made exact rather than left as pictures, and both are
easier to recognize once stated in words.  The first is that unitary evolution conserves the
total weight of an operator and only moves it onto more complicated ones, so the survival of
the information is a matter of bookkeeping rather than a figure of speech.  The second is
that the squared commutator of \(W(t)\) with a simple operator at a given site measures, up
to a fixed constant, the fraction of that weight sitting on terms which act at the site.  An
out-of-time-order correlator is in that sense a measurement of operator size.  The rest of
this subsection establishes both, and a reader willing to grant them can pass to
Sec.~\ref{sec:L5_bracket} without losing the thread.

Take \(N\) qubits, with
\begin{equation}
    \mathcal H=\mathcal H_1\otimes \mathcal H_2\otimes\cdots\otimes \mathcal H_N ,
\end{equation}
and expand operators in the basis of Pauli strings
\begin{equation}
    \mathcal{S}=\sigma^{a_1}_1\sigma^{a_2}_2\cdots\sigma^{a_N}_N ,
    \qquad a_i\in\{0,x,y,z\},\quad \sigma^0=\mathbb{I} .
\end{equation}
The \(4^N\) strings are orthogonal under \(\langle A,B\rangle=\operatorname{Tr}(A^\dagger
B)/2^N\), so every Heisenberg operator has a unique expansion
\begin{equation}
    W(t)=\sum_{\mathcal{S}} c_{\mathcal{S}}(t)\,\mathcal{S} ,
    \qquad
    c_{\mathcal{S}}(t)=\frac{1}{2^N}\operatorname{Tr}\bigl[\mathcal{S}^\dagger W(t)\bigr].
\end{equation}
Call the \emph{size} of a string the number of sites on which it acts nontrivially.  The
numbers \(|c_{\mathcal{S}}(t)|^2\) then define a distribution over sizes, and the motion of that
distribution is what the rest of this lecture tracks.

Unitarity constrains the motion.  Since \(\operatorname{Tr}[W^\dagger(t)W(t)]\) is conserved,
\begin{equation}
    \sum_{\mathcal{S}} |c_{\mathcal{S}}(t)|^2 = \text{const},
\end{equation}
so weight is neither created nor lost, only moved between strings.  Start from
\(W(0)=\sigma^z_1\), a single string of size one holding all of it.  Interactions transfer
weight to longer strings, and in a chaotic chain it finishes spread over the exponentially
many strings of size of order \(N\).  A measurement of any few sites reads only the
small-string part of the distribution, which by then holds almost none of the weight.

The squared commutator measures this migration exactly.  Let \(V=\sigma^a_i\) act on site
\(i\), write \(\langle A\rangle_{0}=\operatorname{Tr}A/2^N\) for the infinite-temperature
expectation value, and average over the three orientations at that site:
\begin{equation}
    \bar C_i(t)=\frac{1}{3}\sum_{a=x,y,z}\frac{1}{2}
    \Bigl\langle\bigl[W(t),\sigma^a_i\bigr]^\dagger
    \bigl[W(t),\sigma^a_i\bigr]\Bigr\rangle_{0}.
\end{equation}
Each string commutes or anticommutes with \(\sigma^a_i\), commuting when it acts at site
\(i\) either as the identity or as \(\sigma^a\) itself.  For an anticommuting string \([\mathcal{S},\sigma^a_i]=2\mathcal{S}\sigma^a_i\), the
cross terms vanish by orthogonality, and each such string contributes \(4|c_{\mathcal{S}}|^2\).  A string
acting nontrivially at site \(i\) anticommutes with two of the three orientations, so the
factor of one half and the average over orientations leave
\begin{equation}
    \bar C_i(t)=\frac{4}{3}\sum_{\mathcal{S}\,:\,\mathcal{S}_i\neq\mathbb{I}}\bigl|c_{\mathcal{S}}(t)\bigr|^{2},
\end{equation}
which is the \(q=2\) case of the relation \(\bar C=q^2(q^2-1)^{-1}\mu\) between the averaged
squared commutator and the operator weight \(\mu\) at a site, for spins of local dimension
\(q\)~\cite{Nahum2018,vonKeyserlingk2018}.  The counting above does not generalize as it
stands, since the clock-and-shift operators that replace the Pauli matrices at \(q>2\) are
neither Hermitian nor simply anticommuting.  What replaces it is the twirl
\(q^{-2}\sum_a T_a^{\dagger}XT_a\), which projects any operator onto its identity component,
and averaging the squared commutator over the \(q^2\) generators \(T_a\) returns the same
relation with the combinatorial factor \(q^{2}/(q^{2}-1)\) in place of \(4/3\).  Summing over \(i\) returns the mean operator
size up to that same factor, \(\sum_i\bar C_i=\tfrac43\sum_i\mu_i\) for spins.  This is the sense in which an out-of-time-order correlator counts how large an
operator has become, using nothing but two operators and a thermal trace.

Two aspects of the growth then separate.  One is its rate.  The time at which the weight has
spread over everything available to it is the scrambling time \(t_*\), whose parametric form
follows from the growth law itself in Sec.~\ref{sec:L5_otoc}.  The other is its geometry.
Where the interactions are all-to-all, a perturbation reaches a finite fraction of the
degrees of freedom in one step and the size distribution simply advances.  In a chain, weight
must also be carried through space: strings grow at their ends, and \(\bar C_i(t)\) stays
exponentially small until the front arrives at site \(i\), at a speed called the butterfly
velocity \(v_B\) and taken up in Sec.~\ref{sec:L5_vb}.  Random circuits of Haar-distributed
gates make both statements sharp, the coarse-grained weight obeying a biased diffusion whose
drift is \(v_B=(q^2-1)/(q^2+1)\), strictly below the circuit light-cone speed, and whose
front broadens as \(\sqrt t\) rather than staying
sharp~\cite{Nahum2018,vonKeyserlingk2018}.  In the SYK model, which has no space at all, the
mean size instead grows exponentially at early times at the infinite-temperature chaos
exponent~\cite{RobertsStanfordStreicher2018}.

This is what separates scrambling from thermalization, and the two are often run together.  A
local observable can relax to its thermal value while the operator remains small, because
relaxation is the decay of the few small-string coefficients that the observable happens to
sample.  Scrambling is the arrival of weight on strings of size of order \(N\), and it takes
longer.  Ref.~\cite{SekinoSusskind2008} conjectured that the fastest systems reach it in a time growing
only as the logarithm of the number of degrees of freedom, and that black holes are the
fastest scramblers there are.  Granting that, information dropped
into an old black hole reappears in its radiation almost as soon as it has been
scrambled~\cite{HaydenPreskill2007}, which is the reason this timescale became a question for
gravity rather than only for condensed matter.

\subsection{From Poisson brackets to squared commutators}
\label{sec:L5_bracket}

Sec.~\ref{sec:L5_growth} reached the squared commutator by counting operator weight.  It can
be reached from the other end, starting from the classical definition of chaos, and the two
routes answer different questions: the first says what the object measures in a many-body
system, the second says where the exponential was supposed to come from and why anyone wrote
the thing down.  Sec.~\ref{sec:L3_signatures} asserted the second route's conclusion and left
it at that; what follows is where it comes from.

Classically, sensitivity to initial conditions is a derivative.  In one degree of freedom the
response of the coordinate at time \(t\) to a displacement of the initial coordinate is
\(\partial q(t)/\partial q(0)\), and the variational equation \eqref{eq:L2_variational}
makes it grow as \(e^{\lambda t}\) along a chaotic orbit, with \(\lambda\) the largest
Lyapunov exponent.  That derivative is already a Poisson bracket.  Taking the bracket in the
initial variables \((q_0,p_0)\),
\begin{equation}
    \{A,B\}_{\rm PB}=\frac{\partial A}{\partial q_0}\frac{\partial B}{\partial p_0}
    -\frac{\partial A}{\partial p_0}\frac{\partial B}{\partial q_0},
\end{equation}
and setting \(A=q(t;q_0,p_0)\) and \(B=p_0\), the second term vanishes and
\(\partial B/\partial p_0=1\), which leaves
\begin{equation}
    \{q(t),p(0)\}_{\rm PB}=\frac{\partial q(t)}{\partial q(0)}\sim e^{\lambda t}.
\end{equation}

Quantization sends \(\{A,B\}_{\rm PB}\to(i\hbar)^{-1}[\hat A,\hat B]\), so the quantum
counterpart of the classical response derivative is \([\hat q(t),\hat p(0)]\), of magnitude
\(\hbar e^{\lambda t}\), and the positive square of it is
\begin{equation}
    -\bigl[\hat q(t),\hat p(0)\bigr]^{2}\sim \hbar^{2}e^{2\lambda t}.
\end{equation}
This is where the object came from.  It was written down in 1969, in the course of asking when a semiclassical treatment of
electron scattering in a superconductor stays valid~\cite{LarkinOvchinnikov1969}, and it sat
unused until it was revived in 2015 as a diagnostic for black holes~\cite{Kitaev2015}.  A semiclassical treatment confirms the form:
the leading contribution to the correlator is of order \(\hbar^{2}\) and grows exponentially
over an intermediate window~\cite{JalabertGarciaMataWisniacki2018}.

The rate in that window deserves care, because it is not the Lyapunov exponent.  The
classical response derivative fluctuates from one initial condition to another, so writing
\(\partial q(t)/\partial q(0)=e^{\lambda_T t}\) with \(\lambda_T\) the finite-time exponent
\eqref{eq:L2_finite_time}, evaluated on the orbit through that point over a window equal to
the elapsed time, the correlator involves
\begin{equation}
    C(t)\sim\hbar^{2}\bigl\langle e^{2\lambda_T t}\bigr\rangle
    \;\geq\;
    \hbar^{2}e^{2\langle\lambda_T\rangle t}=\hbar^{2}e^{2\lambda t},
\end{equation}
by convexity of the exponential, with equality only if \(\lambda_T\) does not fluctuate at
all.  The averaged correlator is therefore controlled by the fastest-separating trajectories
rather than by the typical ones.  Writing the variance of \(\lambda_T\) in its generic
large-deviation form \(D/t\), the Gaussian average is one line,
\begin{equation}
    \bigl\langle e^{2\lambda_T t}\bigr\rangle
    =
    \exp\!\left[2\lambda t+\tfrac12(2t)^{2}\frac{D}{t}\right]
    =
    e^{2(\lambda+D)t},
    \label{eq:L5_gaussian_rate}
\end{equation}
so \(C\) grows at \(\lambda_L=2(\lambda+D)\).  The classical exponent it reports is
\(\lambda+D\), above \(\lambda\) by an amount fixed by the spread of finite-time
exponents.  Both numbers are worth keeping straight, because the factor of two between them is
the same one the front matter records.  Numerically the
excess is large: in the quantum kicked rotor the rate extracted from the correlator exceeds
the Lyapunov exponent at every nonzero kicking strength, and by a wide margin where the
chaos is weak~\cite{RozenbaumGaneshanGalitski2017}.  The systematic statement is that the
correlator measures a \emph{generalized} Lyapunov exponent of order two rather than the
ordinary one, the ordinary exponent being recovered only in the limit of vanishing
order~\cite{PappalardiKurchan2023}.  This also resolves the factor of two between
conventions: with \(C(t)\sim e^{\lambda_L t}\), as is standard in the holographic literature,
the semiclassical estimate above corresponds to \(\lambda_L=2\lambda\), which is the case of
no fluctuations.  The contrast is visible in the numerical recipe
\eqref{eq:L2_renormalization}, which averages the logarithm of the growth factors.  The correlator averages their product
instead, and the two agree only when the \(a_k\) do not fluctuate.

There is a second restriction on the window, already established.  The prefactor
\(\hbar^{2}\) is what gives the exponential room to run, and the amount of room is the
logarithm of that small number, which is the Ehrenfest time
\(t_E\sim\lambda^{-1}\log(1/\hbar_{\rm eff})\) of Sec.~\ref{sec:L3_ehrenfest}.  Past it the
spectrum is discrete, the squared commutator is an almost periodic bounded function, and the
limits \(\hbar\to0\) and \(t\to\infty\) fail to commute.  Whether an exponential window is
visible at all in a system with few degrees of freedom is a further question again, taken up
in Sec.~\ref{sec:L5_billiards}.

Why square, rather than study the commutator itself?  There are three answers and the third
is the strongest.  The first is that \(\langle[W(t),V(0)]\rangle_\beta\) is not positive, and
can vanish by symmetry or through phase cancellation in situations where the operator has
spread perfectly well~\cite{GarciaMata2023}.  The classical signed response derivative
behaves the same way, taking both signs across an ensemble of initial conditions while its
magnitude climbs steadily.  The second is that the unsquared object is nothing new: up to a
factor and a step function it is the Kubo response function for the response of \(W\) to a
perturbation coupling to \(V\), and in a system that thermalizes such a response decays on
the dissipation time.  Linear response theory has always had it, and it has never diagnosed
chaos.  The third answer is already proved, for a particular version of the object.  At infinite
temperature and averaged over the three orientations at a site,
Sec.~\ref{sec:L5_growth} showed the squared
commutator to be a sum of \(|c_{\mathcal{S}}|^{2}\) over strings acting there, manifestly
non-negative and directly readable as operator weight.  The \(\langle\cdot\rangle_\beta\) with
one fixed \(V\) used here is not literally that object; non-negativity survives the move, since
\([W,V]\) is anti-Hermitian for Hermitian \(W\) and \(V\), while the reading as a weight is
what the restricted setting bought.  The unsquared expectation value
admits no such reading, its coefficients entering linearly and with signs.  Squaring is what
turns a response function into a census.

Squaring also changes what kind of correlator is involved.  For Hermitian \(W\) and \(V\),
expanding the commutator gives
\begin{equation}
\begin{aligned}
    -\bigl\langle[W(t),V(0)]^{2}\bigr\rangle_\beta
    &= \bigl\langle W(t)V(0)V(0)W(t)\bigr\rangle_\beta
    + \bigl\langle V(0)W(t)W(t)V(0)\bigr\rangle_\beta \\
    &\quad - 2\,\mathrm{Re}\bigl\langle W(t)V(0)W(t)V(0)\bigr\rangle_\beta .
\end{aligned}
\end{equation}
In the first two terms each operator stands beside its own partner, and both approach
constants fixed by the thermal two-point functions.  If \(W\) and \(V\) are chosen unitary as
well as Hermitian, so that \(W^{2}=V^{2}=\mathbb{I}\), those terms equal one exactly and
\begin{equation}
    C(t)=2-2\,\mathrm{Re}\,F(t),
    \qquad
    F(t)=\bigl\langle W(t)V(0)W(t)V(0)\bigr\rangle_\beta .
    \label{eq:L5_otoc_def}
\end{equation}
Every trace of chaotic growth therefore sits in the single term whose time arguments
alternate.  That term is the out-of-time-order correlator, and Sec.~\ref{sec:L5_otoc} takes
it up.

\subsection{The out-of-time-order correlator}
\label{sec:L5_otoc}

Sec.~\ref{sec:L5_bracket} isolated the one term in the squared commutator that carries the
growth.  This subsection is about that term, and about the window in which it grows.

Begin with a property usually stated as a sign convention.  For Hermitian \(W\) and \(V\) the
commutator is anti-Hermitian, \([W,V]^{\dagger}=-[W,V]\), so
\begin{equation}
    C(t)=-\bigl\langle[W(t),V(0)]^{2}\bigr\rangle_\beta
    =\bigl\langle[W(t),V(0)]^{\dagger}[W(t),V(0)]\bigr\rangle_\beta \;\geq\; 0
\end{equation}
in any state whatever.  The minus sign is what makes \(C\) a norm.

The four-point function \(F(t)\) of \eqref{eq:L5_otoc_def} takes its name from the
alternation of its time arguments, late, early, late, early.  The
contrast with a monotonically ordered arrangement is sharper than it first looks.  If \(W\)
and \(V\) are unitary as well as Hermitian, then \(W(t)^{2}=\mathbb{I}\) and
\begin{equation}
    \bigl\langle W(t)W(t)V(0)V(0)\bigr\rangle_\beta=1
\end{equation}
identically, at every time and every temperature.  The ordered arrangement of the same four
operators carries no information at all, while \(F(t)\) runs from one down to zero for \(W\) and
\(V\) far enough apart to commute at \(t=0\). That proviso is not idle. At the same point it
can fail outright: \(W=\sigma^{x}\) and \(V=\sigma^{z}\) on one site give \(F(0)=-1\) and
\(C(0)=4\), and Sec.~\ref{sec:L5_energy} takes \(W=x\), \(V=p\) at one point, where
\([x,p]=i\hbar\) puts \(C(0)=\hbar^{2}\) rather than zero.  Whatever
the correlator knows about the dynamics, it knows because of the ordering.

That ordering has a direct reading.  Applied to a state, \(W(t)V(0)\) means: perturb with
\(V\), evolve forward for a time \(t\), perturb with \(W\), evolve back.  The reversed
product \(V(0)W(t)\) means the same two perturbations performed in the opposite order, and
\(F(t)\) is the overlap of the two results.  Where the early perturbation leaves no lasting
mark the two operations commute in effect and the overlap stays near one.  In a chaotic
system the early perturbation changes what the later one finds, the two states drift apart,
and \(F\) decays.  This is the butterfly effect written as an inner product, and it is why
the correlator folds the time contour twice rather than once.

Growth of \(C\) and decay of \(F\) are thus the same statement, made exact for involutions by
\(C=2-2\,\mathrm{Re}\,F\) in Sec.~\ref{sec:L5_bracket}.

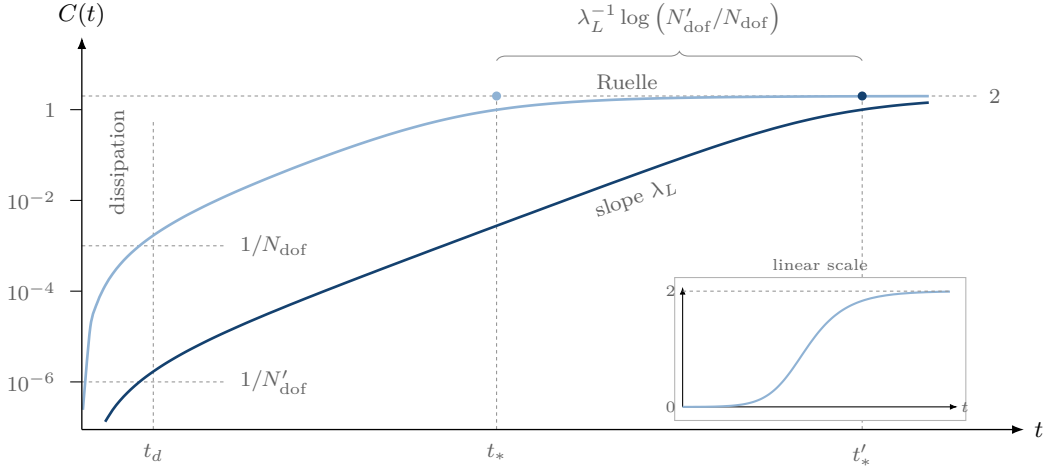
\begin{figure}[!htbp]
\centering
\def\otocCurveA{%
    (0.020,-6.615) (0.160,-4.792) (0.300,-4.229) (0.441,-3.878) (0.581,-3.620) (0.721,-3.413) (0.861,-3.238)
    (1.001,-3.087) (1.141,-2.951) (1.282,-2.828) (1.422,-2.715) (1.562,-2.610) (1.702,-2.511) (1.842,-2.417)
    (1.982,-2.327) (2.123,-2.242) (2.263,-2.159) (2.403,-2.079) (2.543,-2.002) (2.683,-1.926) (2.824,-1.852)
    (2.964,-1.780) (3.104,-1.710) (3.244,-1.641) (3.384,-1.573) (3.524,-1.506) (3.665,-1.440) (3.805,-1.375)
    (3.945,-1.310) (4.085,-1.247) (4.225,-1.185) (4.365,-1.123) (4.506,-1.063) (4.646,-1.003) (4.786,-0.944)
    (4.926,-0.886) (5.066,-0.829) (5.206,-0.773) (5.347,-0.718) (5.487,-0.664) (5.627,-0.612) (5.767,-0.560)
    (5.907,-0.511) (6.048,-0.462) (6.188,-0.416) (6.328,-0.371) (6.468,-0.327) (6.608,-0.286) (6.748,-0.246)
    (6.889,-0.208) (7.029,-0.172) (7.169,-0.138) (7.309,-0.106) (7.449,-0.076) (7.589,-0.048) (7.730,-0.021)
    (7.870,0.003) (8.010,0.026) (8.150,0.048) (8.290,0.068) (8.431,0.086) (8.571,0.103) (8.711,0.118)
    (8.851,0.133) (8.991,0.146) (9.131,0.158) (9.272,0.170) (9.412,0.180) (9.552,0.190) (9.692,0.198)
    (9.832,0.206) (9.972,0.214) (10.113,0.221) (10.253,0.227) (10.393,0.233) (10.533,0.238) (10.673,0.243)
    (10.814,0.247) (10.954,0.252) (11.094,0.255) (11.234,0.259) (11.374,0.262) (11.514,0.265) (11.655,0.268)
    (11.795,0.270) (11.935,0.273) (12.075,0.275) (12.215,0.277) (12.355,0.279) (12.496,0.280) (12.636,0.282)
    (12.776,0.283) (12.916,0.285) (13.056,0.286) (13.196,0.287) (13.337,0.288) (13.477,0.289) (13.617,0.290)
    (13.757,0.291) (13.897,0.292) (14.038,0.292) (14.178,0.293) (14.318,0.294) (14.458,0.294) (14.598,0.295)
    (14.738,0.295) (14.879,0.296) (15.019,0.296) (15.159,0.296) (15.299,0.297) (15.439,0.297) (15.579,0.297)
    (15.720,0.298) (15.860,0.298) (16.000,0.298)}
\def\otocCurveB{%
    (0.441,-6.878) (0.581,-6.619) (0.721,-6.412) (0.861,-6.238) (1.001,-6.086) (1.141,-5.950) (1.282,-5.827)
    (1.422,-5.714) (1.562,-5.608) (1.702,-5.509) (1.842,-5.415) (1.982,-5.325) (2.123,-5.239) (2.263,-5.156)
    (2.403,-5.075) (2.543,-4.997) (2.683,-4.921) (2.824,-4.847) (2.964,-4.774) (3.104,-4.702) (3.244,-4.632)
    (3.384,-4.563) (3.524,-4.494) (3.665,-4.427) (3.805,-4.360) (3.945,-4.293) (4.085,-4.228) (4.225,-4.162)
    (4.365,-4.098) (4.506,-4.033) (4.646,-3.969) (4.786,-3.906) (4.926,-3.842) (5.066,-3.779) (5.206,-3.716)
    (5.347,-3.653) (5.487,-3.591) (5.627,-3.528) (5.767,-3.466) (5.907,-3.404) (6.048,-3.342) (6.188,-3.280)
    (6.328,-3.219) (6.468,-3.157) (6.608,-3.095) (6.748,-3.034) (6.889,-2.973) (7.029,-2.911) (7.169,-2.850)
    (7.309,-2.789) (7.449,-2.727) (7.589,-2.666) (7.730,-2.605) (7.870,-2.544) (8.010,-2.483) (8.150,-2.422)
    (8.290,-2.361) (8.431,-2.301) (8.571,-2.240) (8.711,-2.179) (8.851,-2.118) (8.991,-2.058) (9.131,-1.997)
    (9.272,-1.936) (9.412,-1.876) (9.552,-1.816) (9.692,-1.755) (9.832,-1.695) (9.972,-1.635) (10.113,-1.575)
    (10.253,-1.516) (10.393,-1.456) (10.533,-1.397) (10.673,-1.338) (10.814,-1.279) (10.954,-1.220) (11.094,-1.162)
    (11.234,-1.105) (11.374,-1.047) (11.514,-0.990) (11.655,-0.934) (11.795,-0.878) (11.935,-0.823) (12.075,-0.769)
    (12.215,-0.716) (12.355,-0.664) (12.496,-0.612) (12.636,-0.562) (12.776,-0.513) (12.916,-0.466) (13.056,-0.419)
    (13.196,-0.375) (13.337,-0.332) (13.477,-0.290) (13.617,-0.251) (13.757,-0.213) (13.897,-0.177) (14.038,-0.143)
    (14.178,-0.111) (14.318,-0.081) (14.458,-0.052) (14.598,-0.026) (14.738,-0.001) (14.879,0.023) (15.019,0.044)
    (15.159,0.064) (15.299,0.083) (15.439,0.100) (15.579,0.116) (15.720,0.130) (15.860,0.144) (16.000,0.156)}
\def\otocInset{%
    (11.350,-6.550) (11.396,-6.550) (11.443,-6.550) (11.489,-6.550) (11.535,-6.550) (11.582,-6.549)
    (11.628,-6.549) (11.674,-6.549) (11.721,-6.548) (11.767,-6.548) (11.813,-6.547) (11.860,-6.546)
    (11.906,-6.545) (11.952,-6.544) (11.999,-6.542) (12.045,-6.540) (12.091,-6.538) (12.138,-6.535)
    (12.184,-6.532) (12.230,-6.529) (12.277,-6.525) (12.323,-6.520) (12.369,-6.514) (12.416,-6.507)
    (12.462,-6.500) (12.508,-6.491) (12.555,-6.480) (12.601,-6.468) (12.647,-6.454) (12.694,-6.438)
    (12.740,-6.419) (12.786,-6.398) (12.833,-6.374) (12.879,-6.346) (12.925,-6.315) (12.972,-6.280)
    (13.018,-6.241) (13.064,-6.198) (13.111,-6.150) (13.157,-6.098) (13.203,-6.041) (13.250,-5.980)
    (13.296,-5.915) (13.342,-5.846) (13.389,-5.773) (13.435,-5.698) (13.481,-5.620) (13.528,-5.541)
    (13.574,-5.461) (13.620,-5.381) (13.667,-5.301) (13.713,-5.223) (13.759,-5.146) (13.806,-5.072)
    (13.852,-5.000) (13.898,-4.931) (13.944,-4.866) (13.991,-4.803) (14.037,-4.744) (14.083,-4.689)
    (14.130,-4.637) (14.176,-4.588) (14.222,-4.543) (14.269,-4.501) (14.315,-4.461) (14.361,-4.425)
    (14.408,-4.391) (14.454,-4.360) (14.500,-4.331) (14.547,-4.304) (14.593,-4.280) (14.639,-4.257)
    (14.686,-4.236) (14.732,-4.217) (14.778,-4.199) (14.825,-4.183) (14.871,-4.168) (14.917,-4.154)
    (14.964,-4.142) (15.010,-4.130) (15.056,-4.119) (15.103,-4.110) (15.149,-4.101) (15.195,-4.092)
    (15.242,-4.085) (15.288,-4.078) (15.334,-4.071) (15.381,-4.066) (15.427,-4.060) (15.473,-4.055)
    (15.520,-4.051) (15.566,-4.047) (15.612,-4.043) (15.659,-4.039) (15.705,-4.036) (15.751,-4.033)
    (15.798,-4.030) (15.844,-4.028) (15.890,-4.026) (15.937,-4.023) (15.983,-4.021) (16.029,-4.020)
    (16.076,-4.018) (16.122,-4.017) (16.168,-4.015) (16.215,-4.014) (16.261,-4.013) (16.307,-4.012)
    (16.354,-4.011) (16.400,-4.010)}
\def\otocTsA{7.835}
\def\otocTsB{14.742}
\begin{tikzpicture}[x=0.70cm,y=0.60cm,
    curve/.style={line width=1.0pt},
    guide/.style={black!42,line width=0.4pt,dash pattern=on 1.3pt off 1.3pt},
    slbl/.style={font=\scriptsize,black!62}]
\draw[guide] (0,0.301) -- (16.9,0.301);
\draw[guide] (1.35,-7.05) -- (1.35,-0.2);
\draw[guide] (\otocTsA,-7.05) -- (\otocTsA,0.301);
\draw[guide] (\otocTsB,-7.05) -- (\otocTsB,0.301);
\draw[guide] (0,-3) -- (2.7,-3);
\draw[guide] (0,-6) -- (2.7,-6);
\node[right,slbl] at (2.8,-3) {$1/N_{\rm dof}$};
\node[right,slbl] at (2.8,-6) {$1/N_{\rm dof}'$};
\draw[curve,otocLight] plot[smooth] coordinates {\otocCurveA};
\draw[curve,otocDark]  plot[smooth] coordinates {\otocCurveB};
\draw[-{Latex[length=4.5pt]},line width=0.5pt] (0,-7.05) -- (17.8,-7.05) node[right,font=\footnotesize] {$t$};
\draw[-{Latex[length=4.5pt]},line width=0.5pt] (0,-7.05) -- (0,1.6) node[above,font=\footnotesize] {$C(t)$};
\foreach \y in {0,-2,-4,-6}{\draw[line width=0.4pt] (0,\y)--(-0.22,\y);}
\node[left,slbl] at (-0.3,0)  {$1$};
\node[left,slbl] at (-0.3,-2) {$10^{-2}$};
\node[left,slbl] at (-0.3,-4) {$10^{-4}$};
\node[left,slbl] at (-0.3,-6) {$10^{-6}$};
\node[right,slbl] at (16.95,0.301) {$2$};
\fill[otocLight] (\otocTsA,0.301) circle (1.7pt);
\fill[otocDark]  (\otocTsB,0.301) circle (1.7pt);
\node[below,slbl] at (1.35,-7.15) {$t_d$};
\node[below,slbl] at (\otocTsA,-7.15) {$t_*$};
\node[below,slbl] at (\otocTsB,-7.15) {$t_*'$};
\node[slbl,rotate=90] at (0.70,-1.15) {dissipation};
\node[slbl,rotate=19,anchor=south] at (10.6,-2.35) {slope $\lambda_L$};
\node[slbl] at (10.3,0.62) {Ruelle};
\draw[decorate,decoration={brace,amplitude=3.5pt,raise=1pt},black!45,line width=0.4pt]
      (\otocTsA,1.02) -- (\otocTsB,1.02);
\node[above,slbl] at (11.29,1.42) {$\lambda_L^{-1}\log\big(N_{\rm dof}'/N_{\rm dof}\big)$};
\fill[white] (11.20,-6.85) rectangle (16.70,-3.70);
\draw[black!30,line width=0.4pt] (11.20,-6.85) rectangle (16.70,-3.70);
\draw[-{Latex[length=3pt]},line width=0.4pt] (11.35,-6.55) -- (16.55,-6.55) node[right,font=\tiny,black!62,inner sep=1pt] {$t$};
\draw[-{Latex[length=3pt]},line width=0.4pt] (11.35,-6.55) -- (11.35,-3.90);
\draw[curve,otocLight,line width=0.8pt] plot[smooth] coordinates {\otocInset};
\draw[guide] (11.35,-4.00) -- (16.40,-4.00);
\node[left,font=\tiny,black!62,inner sep=1.5pt] at (11.30,-4.00) {$2$};
\node[left,font=\tiny,black!62,inner sep=1.5pt] at (11.30,-6.55) {$0$};
\node[font=\tiny,black!62] at (13.95,-3.42) {linear scale};
\end{tikzpicture}
\caption{The squared commutator against time, on a logarithmic vertical scale, drawn for two
values of \(N_{\rm dof}\).  Each curve leaves zero on the dissipation time \(t_d\), climbs at
rate \(\lambda_L\) from a height set by \(1/N_{\rm dof}\), and approaches its saturation value
of two.  Raising \(N_{\rm dof}\) lowers the start of the ramp and delays the saturation while
leaving the slope untouched, so the separation of the two marked times is exactly
\(\lambda_L^{-1}\log(N_{\rm dof}'/N_{\rm dof})\).  The inset redraws the lighter curve on a
linear scale, the form in which the object is usually displayed; the exponential regime is
invisible there, compressed against the axis, which is why the logarithmic scale is used for
the main panel.  The curve drawn is the interpolating form
\(C/2=(1-e^{-t/t_d})^{2}\,[\,1-(1+\epsilon e^{\lambda_L t})^{-\mu/\lambda_L}]\) with
\(\epsilon=1/N_{\rm dof}\), drawn in units where \(\lambda_L=1\) with \(\mu=0.55\) and
\(t_d=1.35\), so the three rates enter only through those parameters.  Two
features are conventions rather than results.  The take-off is drawn as \(C\sim t^{2}\), the
perturbative onset for operators whose commutator vanishes at \(t=0\); the power is
model-dependent, and a spin chain probed by two diagonal operators a distance \(r\) apart
gives \(t^{4r+2}\) instead.  And a computed curve fluctuates about the plateau rather than
approaching it smoothly, since the correlator of a single spectrum does not self-average.  The layout follows Fig.~2
of~\cite{Jahnke2019}, with the two values of \(N_{\rm dof}\), the interpolating form, the
linear-scale inset and the take-off added here.}
\label{fig:otoc_regimes}
\end{figure}

For a system with many degrees of freedom and a clean separation of scales, the behaviour at
intermediate times is
\begin{equation}
    C(t)\sim\frac{1}{N_{\rm dof}}e^{\lambda_L t},
    \qquad t_d\ll t\ll t_* ,
\end{equation}
with \(t_d\) the dissipation time on which ordinary two-point functions
decay~\cite{MaldacenaShenkerStanford2016}.  The prefactor does the work \(\hbar^{2}\) did in
Sec.~\ref{sec:L5_bracket}: it is small, so the exponential has room to run, and the room is
its logarithm.  Setting \(C(t_*)\sim 1\) gives
\begin{equation}
    t_*\sim\frac{1}{\lambda_L}\log N_{\rm dof},
\end{equation}
the scrambling time of Sec.~\ref{sec:L5_growth}, derived now rather than quoted.
Fig.~\ref{fig:otoc_regimes} shows the shape and shows the logarithm at work: raising
\(N_{\rm dof}\) lowers the start of the ramp and pushes \(t_*\) out in proportion to
\(\log N_{\rm dof}\), leaving the slope untouched.

Three cautions travel with this form.  It presumes \(t_d\ll t_*\), which is a statement about
having many degrees of freedom rather than about being chaotic, so a system without that
separation offers no window to fit an exponential into.  The exponent is the quantum growth
rate of Sec.~\ref{sec:L5_bracket} rather than the classical Lyapunov exponent.  And the form
describes only the window in which \(C\) is still small, which leaves both of its edges to be
accounted for.

Those edges turn out to be the same physics.  Past \(t_*\) the correlator approaches its
saturation value exponentially,
\begin{equation}
    2-C(t)\sim e^{-\mu t},
\end{equation}
at a rate \(\mu\) fixed by the smallest Ruelle resonance of the system.  The dissipation time
bounding the window from below is set by the same family of resonances, since Ruelle
resonances are what make ordinary two-point functions decay~\cite{Jahnke2019}, so \(t_d\) and
\(\mu^{-1}\) are of the same order without being the same number; the curve of
Fig.~\ref{fig:otoc_regimes} is drawn with \(\mu t_d=0.74\).  The
window is therefore bounded below by relaxation and above by the size of \(N_{\rm dof}\), and
those are different things rather than one thing seen twice: raising \(N_{\rm dof}\) by three
decades pushes \(t_*\) out by \(6.9/\lambda_L\) and leaves \(t_d\) and \(\mu\) exactly
where they were, which is what the two curves of Fig.~\ref{fig:otoc_regimes} show. A system
needs fast relaxation and many degrees of freedom together before there is any window at all.  In a holographic theory those resonances are the quasinormal
modes of the black hole, so the rate at which the boundary correlator settles onto its plateau
is the rate at which the bulk geometry rings down after being disturbed.  Lecture 6 constructs
that correspondence.  Whether the growth in between is visible at all in a system with few
degrees of freedom is taken up in Sec.~\ref{sec:L5_billiards}.

\subsection{OTOCs in the energy basis}
\label{sec:L5_energy}

The definitions so far have been formal.  Computing an OTOC for a given system needs its
spectrum and its matrix elements, and writing the correlator in the energy basis shows which
of the two does the work.  The development in this subsection and the examples in
Sec.~\ref{sec:L5_billiards} follow~\cite{HashimotoMurataYoshii2017}.

Take a time-independent Hamiltonian, \(H|n\rangle=E_n|n\rangle\), and choose the two
operators to be \(x\) and \(p\), the simplest pair with a nonvanishing commutator.  The
thermal state is diagonal in energy, so the thermal correlator is a Boltzmann average of
microcanonical ones,
\begin{equation}
    C_T(t)=\frac{1}{Z}\sum_n e^{-\beta E_n}C_n(t),
    \qquad
    C_n(t)=-\langle n|[x(t),p(0)]^{2}|n\rangle ,
\end{equation}
which keeps two questions apart: how one energy eigenstate responds to operator growth, and
how the ensemble weights those responses.

The microcanonical object has structure of its own.  Writing
\begin{equation}
    b_{nm}(t)=-i\langle n|[x(t),p(0)]|m\rangle
\end{equation}
and inserting a complete set of states,
\begin{equation}
    C_n(t)=\sum_m \bigl|b_{nm}(t)\bigr|^{2},
\end{equation}
which follows because the commutator of two Hermitian operators is anti-Hermitian, so
\(-A^{2}=A^{\dagger}A\) as in Sec.~\ref{sec:L5_otoc}.  The microcanonical OTOC is therefore a
census of the amplitudes with which the commutator connects \(|n\rangle\) to every other
eigenstate, which is the energy-basis form of the operator-weight statement of
Sec.~\ref{sec:L5_growth}.

Heisenberg evolution supplies the time dependence immediately.  With \(E_{nm}=E_n-E_m\) and
\(x_{nm}=\langle n|x|m\rangle\), the evolved matrix elements are
\(x_{nk}(t)=e^{iE_{nk}t}x_{nk}\), so
\begin{equation}
    b_{nm}(t)=-i\sum_k\Bigl(e^{iE_{nk}t}x_{nk}p_{km}-e^{iE_{km}t}p_{nk}x_{km}\Bigr).
\end{equation}
For a Hamiltonian of the form
\begin{equation}
    H=p^{2}+U(x),
\end{equation}
where units have been chosen so that \(\hbar=k_B=2m=1\), one finds \([H,x]=-2ip\).  The
identity is worth a second look for the box below, where \(p\) is not self-adjoint on the
natural domain: it survives because every eigenfunction vanishes at both walls, so the boundary
terms that the integration by parts produces cancel.  Hence
\begin{equation}
    p_{nm}=\frac{i}{2}E_{nm}x_{nm},
\end{equation}
so the momentum matrix elements carry no information beyond the position ones.  Substituting,
\begin{equation}
    b_{nm}(t)=\frac{1}{2}\sum_k x_{nk}x_{km}
    \Bigl(E_{km}e^{iE_{nk}t}-E_{nk}e^{iE_{km}t}\Bigr).
\end{equation}
Every quantity on the right is a level or a position matrix element.  This is what makes
single-particle OTOCs cheap to compute: diagonalize once, keep \(E_n\) and \(x_{nm}\), and the
entire time dependence follows from a pair of matrix products at each time.

The formula also settles what kind of diagnostic an OTOC is.  The tools of Lecture 4 divide
cleanly, spacing statistics and the form factor being functionals of the eigenvalues alone
while the eigenstate thermalization hypothesis of Sec.~\ref{sec:L4_eth} is an ansatz for
matrix elements.  The squared commutator requires both at once, with the phases
\(e^{iE_{nk}t}\) reweighting the matrix elements as time advances.  That combination is what
allows it to carry a rate where the spectral diagnostics of Lecture 4 could carry only a
classification and a count.  It needs more of the ansatz than its standard form supplies,
though.  The rate lives in the correlations between matrix elements that
Sec.~\ref{sec:L4_eth} had to add, and a reader who takes the \(R_{mn}\) there to be
independent will find the sums above pairing off into two-point functions with no exponential
anywhere.

\subsection{Integrable benchmarks and the billiard warning}
\label{sec:L5_billiards}

Before trusting a diagnostic on a system whose behaviour is in question, it is worth running
it on systems whose answer is known.  Two integrable examples do that, and they fail to grow
for different reasons.

The harmonic oscillator is the degenerate case.  With
\begin{equation}
    H=p^{2}+\frac{\omega^{2}}{4}x^{2}
\end{equation}
in the units of Sec.~\ref{sec:L5_energy}, the Heisenberg equations close on themselves,
\begin{equation}
    x(t)=x(0)\cos\omega t+\frac{2}{\omega}p(0)\sin\omega t ,
\end{equation}
so that
\begin{equation}
    [x(t),p(0)]=i\cos\omega t .
\end{equation}
The commutator is a c-number.  It therefore takes the same value in every state, and
\begin{equation}
    C_n(t)=C_T(t)=\cos^{2}\omega t
\end{equation}
for every eigenstate and every temperature.  In the language of Sec.~\ref{sec:L5_growth} the
operator does not grow at all: its weight stays on the two elements it started on, and
nothing about the thermal ensemble can change that.

The particle in a box is more instructive.  With
\begin{equation}
    H=p^{2}+U_{\rm box}(x),
    \qquad
    U_{\rm box}(x)=\begin{cases} 0, & 0<x<1,\\ \infty, & \text{otherwise},\end{cases}
\end{equation}
the eigenfunctions and levels are
\begin{equation}
    \psi_n(x)=\sqrt{2}\sin(\pi n x),
    \qquad
    E_n=\pi^{2}n^{2},
\end{equation}
and the position matrix elements follow in closed form~\cite{HashimotoMurataYoshii2017},
\begin{equation}
    x_{nm}=\frac{1-(-1)^{n+m}}{\pi^{2}}
    \left[\frac{1}{(n+m)^{2}}-\frac{1}{(n-m)^{2}}\right],
    \qquad x_{nn}=\frac{1}{2}.
\end{equation}
The correlator built from these does grow, by an order of magnitude, and then returns exactly.
Its scale is not the one of Sec.~\ref{sec:L5_otoc}: with \(W=x\) and \(V=p\) rather than
involutions, \(C\) starts at \(\hbar^{2}\) rather than zero and has no ceiling at two, so
``grows by an order of magnitude'' and ``saturates at two'' are statements in two different
normalizations.
The recurrence is usually attributed to the commensurability of the
spectrum~\cite{HashimotoMurataYoshii2017}, and commensurability is indeed enough to give a
period: every \(E_{nm}\) is an integer multiple of \(\pi^{2}\), so every phase in
\(b_{nm}(t)\) repeats after \(2\pi/\pi^{2}=2/\pi\).  The true period is half of that.  The cited reference gives the correct value; what is added
here is the reason, and it comes from the matrix elements rather than from the levels.  Measure position from the
centre of the box first, replacing \(x\) by \(x-\tfrac12\).  A constant shift commutes with
\(p\), so it changes neither the commutator nor the correlator, and it removes the diagonal
elements \(x_{nn}=\tfrac12\), which are the only ones with \(n+m\) even.  What is left vanishes
unless \(n+m\) is odd, so in the sum \(\sum_k x_{nk}x_{km}\) of Sec.~\ref{sec:L5_energy} both
\(n+k\) and \(k+m\) are odd, which is to say \(b_{nm}\) is supported on \(n+m\) even, and
therefore
\begin{equation}
    \frac{E_{nk}}{\pi^{2}}=(n-k)(n+k)
\end{equation}
is a product of two odd numbers.  Every energy difference that contributes is an \emph{odd}
multiple of \(\pi^{2}\), so at \(t=1/\pi\) every phase equals \(e^{i\pi(\rm odd)}=-1\)
together.  Hence
\begin{equation}
    b_{nm}\!\left(t+\tfrac{1}{\pi}\right)=-\,b_{nm}(t),
\end{equation}
and since \(C_n=\sum_m|b_{nm}|^{2}\) the sign cancels: the amplitude is antiperiodic with
period \(1/\pi\) while the correlator is periodic with it.  The recurrence time is
\begin{equation}
    T_{\rm rec}=\frac{1}{\pi},
\end{equation}
and it is exact rather than approximate, at every temperature.

\begin{figure}[!htbp]
    \centering
    \includegraphics[width=\textwidth]{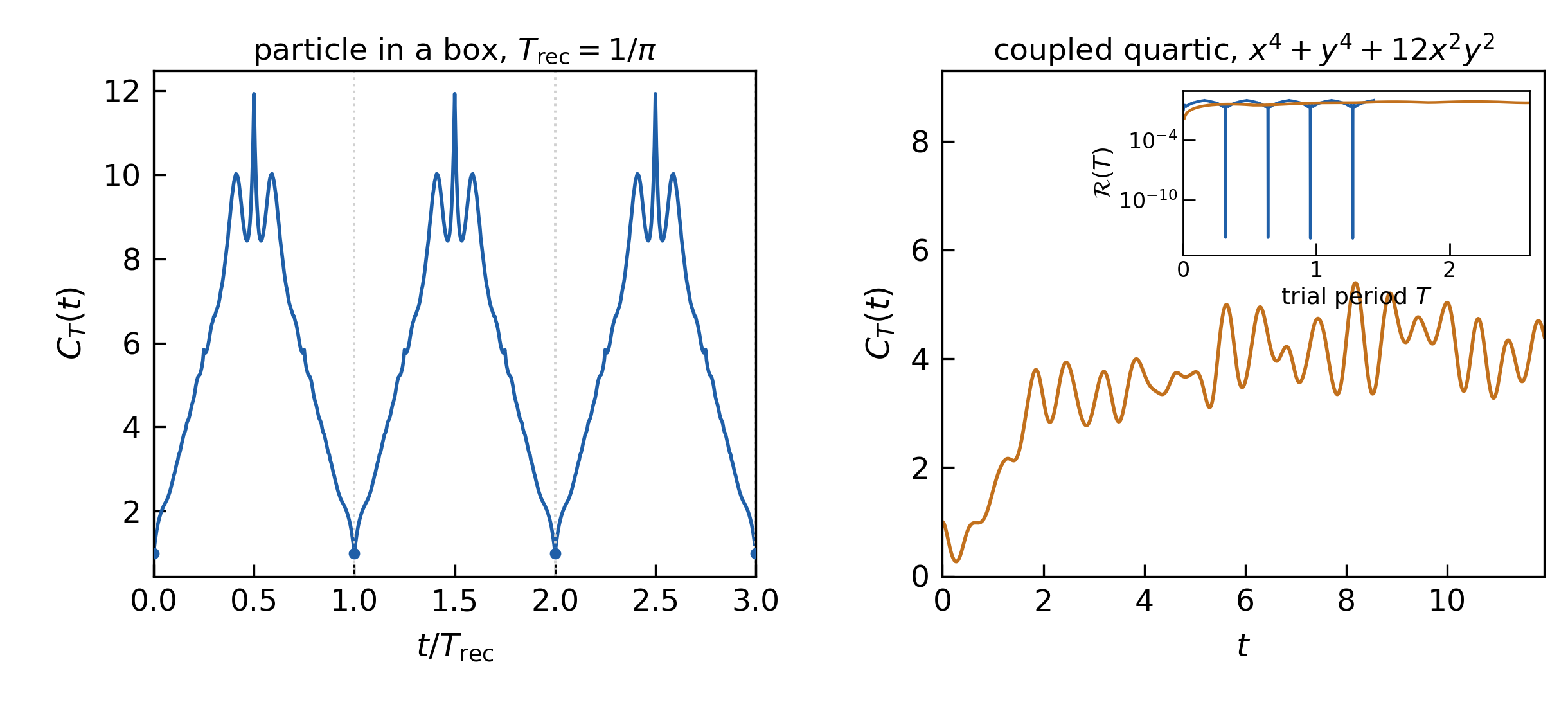}
    \caption{Left: the thermal correlator of a particle in a box, over three recurrence
    periods, returning exactly to its earlier values at each multiple of
    \(T_{\rm rec}=1/\pi\) (marked).  Right: the same quantity for the coupled quartic
    oscillator \(H=p_x^{2}+p_y^{2}+x^{4}+y^{4}+12x^{2}y^{2}\), whose phase space is
    predominantly chaotic, which grows and then fluctuates without ever repeating.  The two are drawn on their own
    time axes, since the quartic evolves over \(t\sim1\) to \(10\) and would be featureless
    on the scale of the box's period.  Inset: the recurrence deficit
    \(\mathcal{R}(T)=\max_t|C(t+T)-C(t)|/\max_t C(t)\), a direct test for a period.  For the box it
    falls to \(1.5\times10^{-14}\) at every multiple of \(T_{\rm rec}\); for the quartic it
    never falls below \(0.27\) at any trial period.  The coupling coefficient is not
    incidental: the symmetric ratio \(1\!:\!6\!:\!1\) is integrable, with second integral
    \(I=p_xp_y+2xy(x^{2}+y^{2})\), and \(\alpha=2\) is integrable for a
    different reason, the potential becoming the isotropic \((x^{2}+y^{2})^{2}\) with
    \(L_z\) conserved.  Between those values the phase space is mixed rather than either one
    thing or the other, so a Lyapunov scan at generic \(\alpha\lesssim 8\) returns regular
    and chaotic regions side by side rather than a single number, and it is only well beyond
    that the typical orbit is chaotic.
    At \(\alpha=12\) the measured \(\lambda/E^{1/4}\) is \(0.55\), \(0.51\) and \(0.57\) at
    \(E=5,20,80\), constant as the homogeneity of the potential requires.  Both correlators come from the
    energy-basis formula of Sec.~\ref{sec:L5_energy}, which needs only \([H,x]=-2ip_x\) and so
    applies unchanged in two dimensions.  The quartic is diagonalized in a scaled harmonic
    basis; its OTOC converges considerably more slowly than its spectrum does, since the sum
    over \(k\) in \(b_{nm}\) reaches states the eigenvalues do not, and the temperature used
    here is one at which dropping twelve converged levels moves the result by under one per
    cent.}
    \label{fig:otoc_box_quartic}
\end{figure}

Fig.~\ref{fig:otoc_box_quartic} sets that against a chaotic system, the quartic oscillator
\(H=p_x^{2}+p_y^{2}+x^{4}+y^{4}+\alpha x^{2}y^{2}\) at \(\alpha=12\).  The choice of
\(\alpha\) carries a warning of its own, since the symmetric case \(\alpha=6\) is integrable
despite looking no less coupled than any other, and a numerical Lyapunov exponent is the only
reliable way to tell one from the other.  The first lesson follows.  Growth of a commutator is not by itself evidence of chaos, since the box grows
substantially and then unwinds completely.  What separates the two cases is arithmetic: a
spectrum that is rationally related throughout forces the phases back into alignment at a
fixed time, and chaos destroys the relation, after which the phases never realign and the
correlator has nowhere to return to.

That suggests an obvious next test, and it is the one that disappoints.  A stadium billiard,
introduced in Sec.~\ref{sec:L3_billiards}, is classically chaotic by Bunimovich's defocusing
mechanism, so one expects the quantized version to show an exponentially growing correlator.
It was computed in~\cite{HashimotoMurataYoshii2017} and comes out otherwise: the correlator
saturates to an
approximately constant value with no clean exponential regime, and the saturation value grows
linearly with temperature, as \(C_T\sim mT L^{2}\) with \(L\) a typical system size, the
mass restored here rather than held at the
\(2m=1\) of the units used elsewhere in this subsection.

The reason was assembled in Sec.~\ref{sec:L5_bracket}.  The semiclassical bridge
\begin{equation}
    [x(t),p(0)]\;\sim\; i\hbar\{x(t),p(0)\}_{\rm PB}
\end{equation}
holds inside a window that closes at the Ehrenfest time.  For a single particle there is no
large \(N_{\rm dof}\) to open a parametric gap between the dissipation time and the scrambling
time, so there may be no interval into which an exponential can be fitted at all.  Wave-packet
spreading, interference, the boundary, and truncation of the Hilbert space each shorten what
remains.  Sec.~\ref{sec:L5_bracket} added a caution of a different kind: even where a rate can
be extracted, it is a generalized Lyapunov exponent rather than \(\lambda\) itself.

Both examples fail in the direction one expects, showing no growth where there is no chaos.
The failure that costs more runs the other way. A stationary point at which the classical
motion is unstable produces exponential growth in the correlator whether or not the system is
chaotic, because a local instability is all the growth requires. The point is made in~\cite{XuScaffidiCao2020} for the
Lipkin--Meshkov--Glick model, which is classically integrable and still carries a
saddle in its phase space, where the correlator's rate is bounded below by the unstable
exponent of that saddle alone. Ref.~\cite{HashimotoHuhKimWatanabe2020} reaches
it in the barest setting available, an inverted harmonic oscillator, where above a temperature
threshold the correlator grows at a rate of order the curvature at the maximum and there is no
chaos anywhere in the problem. An exponential in the
correlator therefore certifies a local instability rather than a positive Lyapunov exponent,
and the two agree only when the instability is spread through the accessible phase space
instead of sitting at one point in it.

The conclusion to draw is narrower than the slogan.  An out-of-time-order correlator is a
diagnostic of operator growth and scrambling, and a good one.  Its relation to the classical
Lyapunov exponent is a separate question whose answer depends on the model and on the time
window, and in few-body quantum mechanics that answer is often that no such relation is
visible.  The clean exponential regime of Sec.~\ref{sec:L5_otoc} belongs to systems with a
parametrically large number of degrees of freedom, which is why the subject moved to
holography rather than staying with billiards.

\subsection{The butterfly velocity}
\label{sec:L5_vb}

Separating the two operators in space turns the squared commutator into a function of
distance as well as time,
\begin{equation}
    C(t,\mathbf{x})=-\bigl\langle[V(0,\mathbf{0}),W(t,\mathbf{x})]^{2}\bigr\rangle_\beta ,
\end{equation}
which asks whether an operator that started at \(\mathbf{x}\) has grown enough to stop
commuting with one at the origin.  Sec.~\ref{sec:L5_growth} gave the answer in terms of
operator weight: \(C\) at a site is the fraction of \(W(t)\) sitting on strings that touch it,
so the question is when the growing operator arrives.

For chaotic systems with many degrees of freedom the intermediate-time behaviour takes the
form
\begin{equation}
    C(t,\mathbf{x})\sim
    \exp\left[\lambda_L\left(t-t_*-\frac{|\mathbf{x}|}{v_B}\right)\right],
\end{equation}
in which \(\lambda_L\) sets the growth in time and \(|\mathbf{x}|/v_B\) is a delay: the
perturbation has to travel before it can be felt.  The velocity \(v_B\) is called the
butterfly velocity, and it defines a cone,
\begin{equation}
    t-t_*\gtrsim\frac{|\mathbf{x}|}{v_B}\ \Longrightarrow\ C\sim O(1),
    \qquad
    t-t_*<\frac{|\mathbf{x}|}{v_B}\ \Longrightarrow\ C\ll1 ,
\end{equation}
outside which one operator still cannot detect the other.  This is the many-body version of
the butterfly effect: the amplification of Lecture 1 acquires a speed.

\begin{figure}[!htbp]
    \centering
    \includegraphics[width=\textwidth]{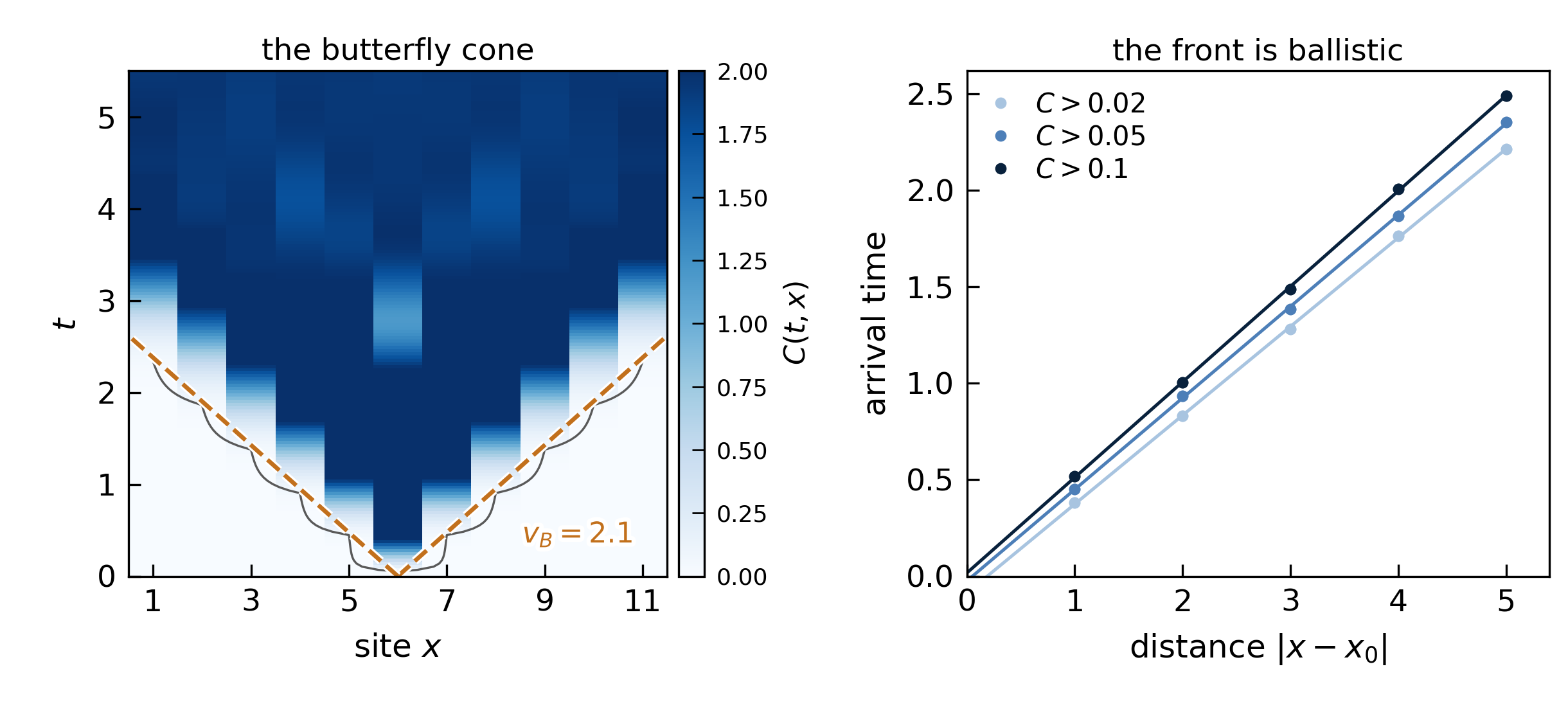}
    \caption{The cone, computed rather than sketched.  Left: the squared commutator between
    \(\sigma^z\) at the centre of an eleven-site tilted-field Ising chain,
    \(H=\sum_i Z_iZ_{i+1}+1.05\sum_i X_i+0.5\sum_i Z_i\), and \(\sigma^z\) at each other
    site, at infinite temperature.  The dashed lines are the measured cone
    \(x=x_0\pm v_Bt\) with \(v_B=2.1\) obtained from the fit at right, not a guide to the
    eye.  Right: the time at which \(C\) first exceeds a threshold, against distance, for
    three thresholds.  The linearity is what defines a velocity; the offset between the three
    lines is the finite width of the front, which is why the cone has a blurred edge rather
    than a sharp one.  The horizontal blockiness is the lattice, eleven sites and no more;
    interpolating across them would invent structure that is not there, whereas time is
    sampled finely enough to be smooth.  The contour marks \(C=0.05\).  Evaluating \(C\) at every site costs nothing extra here: both
    operators are diagonal in the computational basis, so the commutator is an elementwise
    mask and \(C_x=2(\|W\|_F^{2}-\eta^{\mathsf T}|W|^{2}\eta)/\mathcal{D}\) with \(\eta=\pm1\)
    the eigenvalues of \(\sigma^z_x\) and \(\mathcal{D}=2^{L}\) the dimension of the Hilbert
    space, leaving two matrix products per time step however many sites are probed.  The
    factor of two is the one that makes \(C\) saturate at \(2\), which is its value once
    \(W(t)\) and \(\sigma^z_x\) have stopped being correlated.}
    \label{fig:butterfly_cone}
\end{figure}

Fig.~\ref{fig:butterfly_cone} shows the cone for a spin chain small enough to diagonalize.
Two features of it are worth separating from the schematic form above.  The front is
ballistic, so a velocity exists at all, and this is the content of the linear fit rather than
an assumption.  But the front also has a width: the arrival time depends on the threshold
used to define it, and the random-circuit models of Sec.~\ref{sec:L5_growth} show why: the
front broadens as \(\sqrt{t}\)~\cite{Nahum2018,vonKeyserlingk2018}.  That exponent is
one-dimensional.  Above one dimension the front is a growing interface in the
Kardar--Parisi--Zhang class instead, of width \(t^{1/3}\) in \(2+1\) dimensions and
\(t^{0.240}\) in \(3+1\).  A cone with a
sharp edge is an idealization of the large-system limit.

The butterfly velocity is not the Lieb--Robinson velocity, and the difference is more than
one of tightness.  A Lieb--Robinson bound is a rigorous statement that commutators are
exponentially small outside a cone~\cite{LiebRobinson1972}, so a front defined by any fixed threshold cannot outrun
\(v_{\rm LR}\); but \(v_{\rm LR}\) is a property of the Hamiltonian alone and knows nothing
about the state.  The butterfly velocity is the speed the front actually moves, and it does
depend on the state, and is characterized in~\cite{RobertsSwingle2016} as a state-dependent
effective Lieb--Robinson velocity, fixed by infrared data such as the temperature rather than by
microscopic couplings.  How far that dependence can go is worth one example.  Take a
two-dimensional holographic theory driven out of equilibrium by a sudden injection of energy,
and insert one operator pair before the injection and the other after.  The cone then opens
transiently to \(v_B=T_f/T_i\), the ratio of the final temperature to the initial one, which is
greater than the equilibrium value and so, in two dimensions, faster than
light~\cite{BalasubramanianCrapsDeClerckNguyen2019}.  Nothing is broken by that.  The squared
commutator still vanishes identically outside the light cone, and what has moved is the front
defined by a threshold, not the support of the commutator.  Those are exactly the two objects
this paragraph has been separating: the Lieb--Robinson bound constrains the support, and the
butterfly velocity measures the front.  In a holographic theory with dynamical
exponent \(z\) it scales there as \(v_B\propto T^{1-1/z}\); requiring that it stay finite as
\(T\to0\) then forces \(z\geq1\), which is an inference drawn here rather than the route by
which that reference arrives at the same condition.  Fig.~\ref{fig:butterfly_cone} returns
\(v_B=2.1\) for the chain used there, in units of the coupling, with a spread of \(0.15\)
across the three thresholds; that spread is the width of the front, not an uncertainty in the
velocity.

In holographic theories \(v_B\) is fixed by the geometry near the horizon rather than by any
lattice scale, and it emerges from the same shock-wave calculation that supplies
\(\lambda_L\)~\cite{RobertsStanfordSusskind2015}.  That calculation is the subject of
Sec.~\ref{sec:L5_shock} and of Lecture 6, and it is where the two quantities in the exponent
above acquire a common gravitational origin: \(\lambda_L\) from the near-horizon boost, and
\(v_B\) from how far the shock wave spreads transversely before it reaches the boundary.

\subsection{The route to holography}
\label{sec:L5_shock}

Black holes behave as thermal systems, with a temperature, an entropy and a relaxation rate,
and gauge/gravity duality sharpens that into an identification: a black hole in
asymptotically anti-de Sitter spacetime is a thermal state of a strongly coupled field theory
on the boundary~\cite{Maldacena1998}.  If that boundary theory is chaotic then its squared commutator grows, and
the question is which bulk process computes it.

The setting has to be stated precisely, because the calculation depends on it.  Take the
eternal AdS black hole, which has two asymptotic boundaries and is dual to the thermofield
double state~\cite{Maldacena2003}, a particular entangled state of two copies of the field theory at inverse
temperature \(\beta\).  Perturb one boundary at a time \(t\) in the past by adding a few
quanta and follow them in~\cite{ShenkerStanford2014}.

The perturbation falls toward the horizon and the geometry does the rest.  Boundary time
translation acts near the horizon as a boost, so a quantum released with boundary energy
\(E\) at time \(t\) in the past crosses the central slice with proper energy
\begin{equation}
    E_p \sim E\,e^{2\pi t/\beta},
\end{equation}
the blueshift growing exponentially with how long ago it was released.  A second excitation
crossing the horizon later meets this one at an exponentially large centre-of-mass energy,
and gravitational scattering at large centre-of-mass energy is strong.  That is the whole
mechanism: the exponential in the correlator is the exponential of the near-horizon
blueshift, and \(2\pi/\beta\) is the rate at which a boost accumulates at a horizon of
temperature \(1/\beta\).  Reading off the exponent,
\begin{equation}
    \lambda_L=\frac{2\pi}{\beta},
\end{equation}
a property of the geometry rather than of the matter falling into it.

The scrambling time follows from the same estimate rather than needing a separate argument.
Scattering becomes strong, and the perturbation stops being a small correction, once
\(G_N E_p\sim1\).  Since the black hole entropy scales as \(S\sim1/G_N\), that happens at
\begin{equation}
    t_*\sim\frac{\beta}{2\pi}\log\frac{S}{E},
\end{equation}
so up to the logarithm of the injected energy \(t_*\sim(\beta/2\pi)\log S\): the fast
scrambling conjectured on general grounds in Sec.~\ref{sec:L5_growth}, here obtained from a
geometry~\cite{ShenkerStanford2014}.

Assembling the pieces gives the holographic form of the correlator,
\begin{equation}
    F(t,\mathbf{x})\approx 1-\frac{c}{N_{\rm dof}}
    \exp\left[\frac{2\pi}{\beta}\left(t-\frac{|\mathbf{x}|}{v_B}\right)\right],
    \qquad
    C(t,\mathbf{x})\sim\frac{1}{N_{\rm dof}}
    \exp\left[\frac{2\pi}{\beta}\left(t-\frac{|\mathbf{x}|}{v_B}\right)\right],
\end{equation}
valid while the correction is still small, with \(c\) a positive number of order one and both
quantities in the exponent now geometrical: \(\lambda_L\) is the boost rate at the horizon and \(v_B\) measures how far the
shock spreads along it.

The velocity does not stay abstract.  Localizing the perturbation rather than spreading it
over the horizon turns the shock into one that grows outward, and for a planar
AdS--Schwarzschild brane the transverse spread gives
\begin{equation}
    v_B=\sqrt{\frac{d}{2(d-1)}},
\end{equation}
with \(d\) the spacetime dimension of the boundary theory, the perturbed operator filling a
region of radius \(r\approx v_B(t-t_*)\)~\cite{RobertsStanfordSusskind2015}.  The number is
worth evaluating.  It equals one for \(d=2\), where the front moves exactly at the speed of
light; it falls to \(\sqrt{2/3}\approx0.82\) on the four-dimensional boundary of
\(\mathrm{AdS}_5\); and it approaches \(1/\sqrt2\) from above as \(d\) grows.  So chaos in a
holographic theory spreads strictly more slowly than light in every boundary dimension above
two, with no bound imposed by hand, which is the holographic counterpart of the
state-dependence discussed in Sec.~\ref{sec:L5_vb}.

\subsection{The chaos bound}
\label{sec:L5_bound}

The exponent obtained in Sec.~\ref{sec:L5_shock} turns out to be the largest one available.
Maldacena, Shenker and Stanford showed that in a thermal quantum system at inverse temperature
\(\beta\) the squared commutator can grow no faster than
\begin{equation}
    \lambda_L\leq\frac{2\pi}{\beta}=\frac{2\pi k_BT}{\hbar},
\end{equation}
so temperature alone sets a ceiling on how fast chaos can
develop~\cite{MaldacenaShenkerStanford2016}.  Black holes described by classical Einstein
gravity sit exactly at that ceiling, which is what calling them maximally chaotic means.

The assumptions deserve naming, since the bound is often quoted as though it governed any
exponential anywhere.  The state is thermal and the system has many degrees of freedom.  The
operators \(V\) and \(W\) are simple and Hermitian, each built from \(O(1)\) degrees of
freedom, and have vanishing thermal one-point functions.  And the growth must have a window to
occupy, which requires the dissipation time to be far shorter than the scrambling time.  That
last condition is the substantive one: in a strongly coupled system \(t_d\sim\beta\), so the
hierarchy is a statement about having many degrees of freedom rather than about being chaotic,
and a system lacking it has no exponential to bound in the first place.

The argument also needs a change of object, and it is easy to miss. The theorem is about a
\emph{regularized} correlator, in which the four operators are spaced evenly around the
Euclidean circle rather than inserted together against a single thermal factor,
\begin{equation}
    F_{\rm reg}(t)=\operatorname{tr}\!\left[y\,V(0)\,y\,W(t)\,y\,V(0)\,y\,W(t)\right],
    \qquad
    y^{4}=\frac{1}{Z}e^{-\beta H},
\end{equation}
and not the \(F(t)\) of Sec.~\ref{sec:L5_otoc}, which is what a lattice calculation or an
experiment returns. The distinction is worth keeping: the argument constrains
\(F_{\rm reg}\), and carrying the conclusion across to the unregularized correlator is a
further step.

What the regularization buys is a domain. \(F_{\rm reg}(t+i\tau)\) is analytic in the
half-strip \(t>0\), \(|\tau|\leq\beta/4\), and unitarity keeps it bounded there once \(t\)
exceeds the dissipation time, since the correlator has by then factorized. That is the whole
input. A strip of that width measures \(\beta/2\) across, so the map taking it to a half-plane
is \(z\mapsto e^{2\pi z/\beta}\), which is the Schwarz--Pick lemma in the form Hadamard's
three-lines theorem takes on a strip: a bounded analytic function cannot outrun the conformal
factor that carries it there. The \(2\pi/\beta\) in the bound is that factor and nothing else,
which is why the conclusion is one about quantum mechanics at finite temperature rather than
about gravity. Sec.~\ref{sec:L6_tfd} shows where the even spacing comes from: it is what the
thermofield double gives once the two boundaries sit at antipodal points of the thermal
circle.

Which exponent is bounded takes some care, and Sec.~\ref{sec:L5_bracket} has already laid the
groundwork.  The \(\lambda_L\) above is the growth rate of the squared commutator, which is the
generalized Lyapunov exponent of order two rather than the ordinary one.  That point is made explicitly in~\cite{PappalardiKurchan2023}, which extends the bound to the
whole family, proving
\begin{equation}
    \frac{L_{2q}}{2q}\leq\frac{\pi}{\beta\hbar}
\end{equation}
for every \(q\), with the Maldacena--Shenker--Stanford result recovered at
\(q=1\).  Because \(L_{2q}/2q\) is the growth rate of an \(L^{2q}\) norm and so cannot decrease
with \(q\), while the typical exponent is its \(q\to0\) limit, the typical rate satisfies
\(\lambda_1\leq\pi/\beta\), half the headline figure. That \(\lambda_1\) is the smallest
member of the generalized family, the rate a typical trajectory shows, and not the largest of a
Lyapunov spectrum, which is what \(\lambda_1\) means in Sec.~\ref{sec:L2_lyapunov}.  The famous bound constrains the annealed rate, the one dominated by
the fastest-separating trajectories, and the quantity a classical simulation would actually
report is held more tightly still.

Saturation carries two consequences.  The first fixes what a fast scrambler is.  With
\(\lambda_L\) as large as it may be, the scrambling time
\(t_*\sim\lambda_L^{-1}\log N_{\rm dof}\) is as short as it may be, so the conjecture recorded
in Sec.~\ref{sec:L5_growth} acquires a reason rather than remaining an observation about black
holes.

The second picks out Einstein gravity, and it pays to be exact about which departures from
Einstein gravity matter.  A local higher-derivative term, one built from finitely many
derivatives, does not move the exponent at all.  It leaves the exchanged graviton at spin two,
and the exponent depends on nothing else, so what such a term moves is \(v_B\) and not
\(\lambda_L\)~\cite{RobertsSwingle2016}.  Changing the exponent means changing the spin of
what is exchanged, and string theory supplies the change.  The graviton is the leading member
of a Regge trajectory, and the correlator grows at a rate set by the effective spin of whatever
is exchanged,
\begin{equation}
    \lambda_L=\frac{2\pi}{\beta}\bigl(j_{\rm eff}-1\bigr),
\end{equation}
which returns \(2\pi/\beta\) for a graviton at \(j=2\).  A finite string length lowers the
effective spin, to leading order
\begin{equation}
    j_{\rm eff}=2-\frac{d(d-1)}{4}\frac{\ell_s^{2}}{\ell_{\rm AdS}^{2}},
\end{equation}
and the exponent drops below the bound in
proportion~\cite{ShenkerStanfordStringy2015,Jahnke2019}.  That shift cannot be seen at any
finite order in a local \(\alpha'\) expansion, which is the other half of the distinction
drawn above: the correlator probes a boost \(s\sim e^{2\pi t/\beta}\), at which
\(\alpha'\log s\) is of order one and every member of the tower contributes equally, so only
the resummed trajectory sees it.  Measuring how far below is therefore
a measurement of \(\ell_s/\ell_{\rm AdS}\), which is what turns the squared commutator from a
diagnostic of chaos into an instrument for reading the structure of the dual.

What this lecture has established is a rate, a speed, and an account of where they come from.
The squared commutator measures operator size; its growth in time defines \(\lambda_L\), its
spatial front defines \(v_B\), and in a holographic theory both are read off a horizon, the
first from the rate at which boundary time accumulates boost and the second from how far a
shock spreads along the horizon before returning to the boundary.  The bound then says that
the boost rate is the fastest any thermal system can manage.

What the lecture has not done is build the geometry.  The shock wave was described rather than
solved for, the eternal black hole was invoked rather than drawn, and the statement that
boundary time acts near the horizon as a boost was used without ever writing the coordinates in
which it becomes obvious.  Lecture 6 supplies those: the Kruskal extension in which both
boundaries and the horizon are visible at once, the backreacted metric of a null shell crossing
it, and the calculation that converts the blueshift into a correlator.  It also turns the same
diagnostics on systems nearer to experiment, where the dual is a QCD-like theory rather than a
conformal one, and where a classical string in the bulk can be chaotic in its own right.

\subsection{Exercises}

\begin{exercise}
\label{ex:L5_bracket}
Take a Hamiltonian system with \(d\) degrees of freedom.
\begin{enumerate}
    \item[(a)] Sec.~\ref{sec:L5_bracket} takes the bracket in one degree of freedom. Extend it:
    working in the initial variables, show that
    \(\{q_i(t),p_j(0)\}_{\rm PB}=\partial q_i(t)/\partial q_j(0)\), so the bracket is one
    block of the tangent map rather than an analogue of it. Identify the brackets that carry the
    other three blocks, and conclude that the whole Lyapunov spectrum of
    Sec.~\ref{sec:L2_lyapunov}, and not only its largest member, is encoded in objects of this
    kind.
    \item[(b)] Explain why the thermal expectation \(\langle[W(t),V(0)]\rangle_\beta\), which
    quantization suggests first, is the Kubo response function for \(W\) under a perturbation
    coupling to \(V\) up to a factor of \(i/\hbar\) and a step function \(\theta(t)\), and why it therefore decays on the dissipation time in any system that
    thermalizes.
    \item[(c)] Show that at infinite temperature \(\langle[W(t),V(0)]\rangle\) vanishes
    identically for \emph{every} pair of operators, however much weight \(W(t)\) has
    acquired, and say in one line why the squared commutator escapes the same argument. The
    proof is shorter than the sentence stating it.
\end{enumerate}
\end{exercise}

\begin{exercise}
\label{ex:L5_benchmarks}
Sec.~\ref{sec:L5_energy} works both integrable benchmarks out. This exercise asks for what it
does not.
\begin{enumerate}
    \item[(a)] For the harmonic oscillator the text finds \(C=\cos^{2}\omega t\) with \(W=x\)
    and \(V=p\). Show that this is not a lucky choice of operators: for a harmonic oscillator
    \emph{no} pair of polynomial operators makes \(C\) grow, because Heisenberg evolution sends
    a polynomial of degree \(n\) in \((x,p)\) to a polynomial of the same degree with bounded
    coefficients. Check the first nontrivial case, \(W=x^{2}\), explicitly. Then say what that
    implies for the operator-growth reading of Sec.~\ref{sec:L5_growth}.
    \item[(b)] Verify the box's period \(1/\pi\) numerically, sampling on a grid commensurate
    with it. Then restore the diagonal matrix elements and make them unequal. The
    antiperiodicity of \(b_{nm}\) fails, yet the period of \(C_T\) survives exactly. Find what
    protects it.
    \item[(c)] Compute the recurrence deficit \(\mathcal{R}(T)\) of
    Fig.~\ref{fig:otoc_box_quartic} for both systems. The quartic's floor of \(0.27\) rules out
    an exact period; say whether it is also evidence of chaos, and if not, what would be.
\end{enumerate}
\end{exercise}

\begin{exercise}
\label{ex:L5_generalized}
This exercise establishes that the correlator measures the wrong Lyapunov exponent, in a
precise sense.  Write \(\partial q(t)/\partial q(0)=e^{\lambda_T t}\) with \(\lambda_T\) the
finite-time exponent \eqref{eq:L2_finite_time}, so that
\(C(t)\sim\hbar^{2}\langle e^{2\lambda_T t}\rangle\).
\begin{enumerate}
    \item[(a)] Let the finite-time exponents obey a large-deviation law, so that the density
    of \(\lambda_T\) at time \(t\) behaves as \(e^{-tI(\lambda)}\) with \(I\geq0\) vanishing
    at \(\lambda=\langle\lambda_T\rangle\). Show by steepest descent that \(C\) grows at the rate
    \(\sup_\lambda\left(2\lambda-I(\lambda)\right)\), that this is at least
    \(2\langle\lambda_T\rangle\) because \(I\) vanishes there, and that it is strictly
    larger unless \(\lambda_T\) does not fluctuate at all. The convexity argument of Sec.~\ref{sec:L5_bracket} is the special case
    that keeps only the inequality.
    \item[(b)] Take \(\lambda_T\) Gaussian with variance \(D/t\), the generic large-deviation
    scaling, and show that \(C\) then grows at exactly \(2(\lambda+D)\), so that the
    classical exponent it reports is \(\lambda+D\) and not \(\lambda\).  Verify by
    sampling.
    \item[(c)] Show that the generalized exponents are the Legendre transform of the same rate
    function, \(L_{2q}=\sup_\lambda\left(2q\lambda-I(\lambda)\right)\), so that the family
    \(\{L_{2q}\}\) and the distribution of finite-time exponents carry the same information.
    Deduce that \(\lambda_1=\lim_{q\to0}L_{2q}/2q\) is \(\langle\lambda_T\rangle\), and hence
    what the bound of Sec.~\ref{sec:L5_bound} on \(L_2\) implies for it. Here \(q\) is the
    order, not the position of parts (a) and (b).
\end{enumerate}
\end{exercise}

\begin{exercise}
\label{ex:L5_cone}
Reproduce Fig.~\ref{fig:butterfly_cone}.  Take a nonintegrable spin chain of ten or eleven
sites, place \(\sigma^z\) at the centre, and compute \(C(t,x)\) at infinite temperature.
\begin{enumerate}
    \item[(a)] The probe \(\sigma^z_x\) is diagonal in the computational basis, which is all
    that is needed, so
    \([W(t),\sigma^z_x]_{ab}=W(t)_{ab}(\eta_b-\eta_a)\) with \(\eta=\pm1\).  Use this to show
    that all sites can be probed for the cost of evolving one operator, and be careful with the
    constant: \((\eta_b-\eta_a)^{2}\) is \(4\) on a differing pair while
    \(1-\eta_a\eta_b\) is \(2\).  Check the resulting expression against a direct commutator
    on a smaller chain, which is what fixes the constant.
    \item[(b)] Extract \(v_B\) from the time at which \(C\) first exceeds a threshold, as a
    function of distance.  Eleven sites with the probe at the centre give five distances on each
    side, which is too few to test linearity; say what the fit is actually good for, and what
    would be needed to test the linearity itself.
    \item[(c)] Repeat for two other thresholds.  The velocity should shift slightly; explain
    what that shift measures, and why it does not indicate an error in the fit.
\end{enumerate}
\end{exercise}

\clearpage
\section{Lecture 6: Chaos, black holes, and holography}

Lecture 5 ended with a rate, a speed, and a list of things it had not done.  The rate
\(\lambda_L=2\pi/\beta\) and the speed \(v_B=\sqrt{d/2(d-1)}\) were both read off a horizon,
but the geometry they were read off was described rather than constructed.  Boundary time was
said to act near the horizon as a boost, without ever writing the coordinates in which that is
visible.  The shock wave entered as something an infalling excitation scatters from and as something
that spreads along the horizon, with the geometry producing it never written down.  And
the eternal black hole, the two-sided spacetime the entire calculation takes place in, was
named in one sentence and left undrawn.

Sections~\ref{sec:L6_tfd} to~\ref{sec:L6_correlator} build the three of them.  The answers do
not change.  What changes is that \(2\pi/\beta\) becomes a surface gravity, \(v_B\) becomes an
inverse screening length, and the exponential in the correlator becomes the parameter of a
boost.

The rest of the lecture gives that universality up on purpose.
Sec.~\ref{sec:L6_tfd} shows that \(\lambda_L\) is a surface gravity and nothing besides, so in
two-derivative gravity it returns \(2\pi/\beta\) whatever \(f(r)\) happens to be.  A quantity
that takes the same value on every such geometry cannot tell one from another.  The quantities
that discriminate belong to probes rather than to the horizon.
Sections~\ref{sec:L6_thermo} to~\ref{sec:L6_probe} follow an unstable circular geodesic and
find its Lyapunov exponent tracking the thermodynamic phase structure of the background: it
turns multivalued in temperature exactly where the free energy develops a swallowtail, and the
branches merge with a critical exponent of one half.  Sections~\ref{sec:L6_hqcd}
to~\ref{sec:L6_bounds} trade the geodesic for a string, in a bulk geometry built to imitate QCD
rather than to be conformal, and ask what a magnetic field and a chemical potential do to the
chaos of a heavy quark pair.

\subsection{The eternal black hole and the thermofield double}
\label{sec:L6_tfd}

Take a static, asymptotically anti-de Sitter black hole,
\begin{equation}
    ds^{2}=-f(r)\,dt^{2}+\frac{dr^{2}}{f(r)}+r^{2}d\Omega_{d-1}^{2},
\end{equation}
with a single non-degenerate horizon at \(f(r_h)=0\), \(f'(r_h)>0\).  The restriction is not
cosmetic: everything below is built on the surface gravity of that one horizon, and for an
\(f\) carrying an inner horizon as well, as the charged solutions of
Sec.~\ref{sec:L6_thermo} do, the chart fails there and the maximal extension is an infinite
tower rather than the four regions found here.  Those solutions are used later only for their
exterior thermodynamics.  Two quantities follow from \(f\) alone: the
surface gravity \(\kappa=\tfrac12 f'(r_h)\), and, by regularity of the Euclidean section, the
temperature \(T=\kappa/2\pi\), so that \(\beta=4\pi/f'(r_h)\).  The chart itself covers the
exterior and stops there, since \(f\) vanishes at \(r_h\) and \(g_{rr}\) diverges.  The
curvature invariants stay finite, so what fails is the coordinate system.

Repair it by exponentiating the null coordinates at the rate the horizon itself supplies.  With
the tortoise coordinate \(r_*=\int^{r}dr'/f(r')\), set
\begin{equation}
    u=-e^{-\kappa(t-r_*)},\qquad v=e^{\kappa(t+r_*)},
\end{equation}
so that
\begin{equation}
    ds^{2}=-\frac{f(r)}{\kappa^{2}}\,e^{-2\kappa r_*}\,du\,dv+r^{2}d\Omega_{d-1}^{2},
    \qquad uv=-e^{2\kappa r_*(r)} .
\end{equation}
The second relation defines \(r\) in the exterior, where \(f>0\), and continues across the
horizon to cover the rest.  The prefactor only looks singular.  Near the horizon
\(f\simeq2\kappa(r-r_h)\) while \(r_*\simeq(2\kappa)^{-1}\log(r-r_h)\), so
\(e^{-2\kappa r_*}\) diverges at exactly the rate at which \(f\) vanishes, and the product
tends to a finite nonzero constant.  Nothing in the metric then distinguishes \(u=0\) or
\(v=0\) from any other null line, the chart continues through them, and the maximal extension
is the eternal black hole.

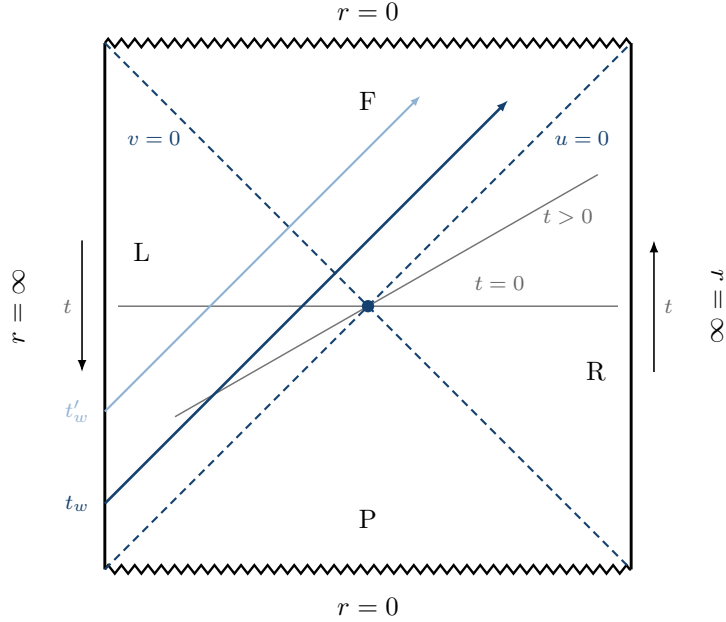
\begin{figure}[!htbp]
\centering
\begin{tikzpicture}[scale=1.16,
    bdy/.style={line width=1.1pt},
    sing/.style={line width=0.9pt,decorate,decoration={zigzag,segment length=5pt,amplitude=1.6pt}},
    hor/.style={otocDark,line width=0.8pt,dash pattern=on 3pt off 2pt},
    slice/.style={black!55,line width=0.55pt},
    ray/.style={otocDark,line width=1.0pt},
    rayb/.style={otocLight,line width=0.8pt},
    lbl/.style={font=\small},
    slbl/.style={font=\scriptsize,black!60}]

  \draw[sing] (-3,3) -- (3,3);
  \draw[sing] (-3,-3) -- (3,-3);
  \node[lbl] at (0,3.36) {\(r=0\)};
  \node[lbl] at (0,-3.42) {\(r=0\)};

  \draw[bdy] (3,-3) -- (3,3);
  \draw[bdy] (-3,-3) -- (-3,3);
  \node[lbl,rotate=-90] at (3.98,0) {\(r=\infty\)};
  \node[lbl,rotate=90]  at (-3.98,0) {\(r=\infty\)};

  \draw[-{Latex[length=4.5pt]},line width=0.6pt] (3.26,-0.75) -- (3.26,0.75)
        node[midway,right,slbl] {\(t\)};
  \draw[-{Latex[length=4.5pt]},line width=0.6pt] (-3.26,0.75) -- (-3.26,-0.75)
        node[midway,left,slbl] {\(t\)};

  \draw[hor] (-3,-3) -- (3,3);
  \draw[hor] (-3,3) -- (3,-3);
  \node[slbl,otocDark] at (2.44,1.92) {\(u=0\)};
  \node[slbl,otocDark] at (-2.44,1.92) {\(v=0\)};

  \draw[slice] (-2.85,0) -- (2.85,0);
  \draw[slice] (-2.20,-1.26) -- (2.62,1.50);
  \node[slbl] at (2.28,1.02) {\(t>0\)};
  \node[slbl] at (1.50,0.26) {\(t=0\)};

  \fill[otocDark] (0,0) circle (2.0pt);

  \node[lbl] at (2.60,-0.74) {R};
  \node[lbl] at (-2.58,0.62) {L};
  \node[lbl] at (0,2.34) {F};
  \node[lbl] at (0,-2.42) {P};

  \draw[rayb,-{Latex[length=4.0pt]}] (-3,-1.20) -- (0.60,2.40);
  \node[slbl,otocLight,anchor=east] at (-3.06,-1.20) {\(t_w'\)};
  \draw[ray,-{Latex[length=4.5pt]}]  (-3,-2.25) -- (1.60,2.35);
  \node[slbl,otocDark,anchor=east]  at (-3.06,-2.25) {\(t_w\)};
\end{tikzpicture}
\caption{The maximally extended eternal anti-de Sitter black hole in Kruskal coordinates
\((u,v)\).  The two vertical edges are the asymptotic boundaries, each carrying a copy of the
field theory, and the arrows give the direction of increasing Killing time, future-directed on
the right and past-directed on the left.  Dashed diagonals are the horizons \(u=0\) and
\(v=0\), meeting at the bifurcation surface (dot); the zigzag edges are the curvature
singularities.  Surfaces of constant \(t\) are straight lines through the bifurcation surface,
which is what makes time translation a boost.  The two solid rays are null quanta released
from the left boundary at Killing times \(t_w\) and \(t_w'<t_w\) in the past, travelling at
constant \(u=u_w\sim e^{-2\pi t_w/\beta}\).  The earlier release runs closer to the horizon and
crosses it nearer the bifurcation surface, approaching it exponentially in \(t_w\) without ever
arriving, which is the geometric content of \(p^{v}\sim e^{2\pi t_w/\beta}\).  The square shape
is particular to anti-de Sitter space, where \(r_*\) converges at large \(r\) so that the
boundaries are timelike at finite \(uv\).  The arrangement of the two released quanta
follows~\cite{ShenkerStanford2014}.}
\label{fig:kruskal}
\end{figure}

That extension has four regions, drawn in Fig.~\ref{fig:kruskal}.  The original exterior is
\(u<0,\,v>0\), a second exterior sits at \(u>0,\,v<0\), and the two interiors have \(uv>0\),
bounded by the curvature singularities.  Because \(f\) grows like \(r^{2}\) in anti-de Sitter
space, \(r_*\) converges as \(r\to\infty\), so each boundary sits at a finite negative value of
\(uv\) and is timelike; choosing \(r_*(\infty)=0\) puts them at \(uv=-1\).  The diagram is
therefore a square with two vertical edges, rather than reaching out to the null infinities of
the asymptotically flat case, and those two edges carry two copies of the field theory.

The statement Lecture 5 used without proving is now one line.  A Killing time translation
\(t\to t+t_0\) acts on the Kruskal coordinates as
\begin{equation}
    u\to e^{-\kappa t_0}\,u,\qquad v\to e^{\kappa t_0}\,v,
\end{equation}
leaving \(uv\), and therefore \(r\), untouched.  This is a boost in the \((u,v)\) plane, exact
and global, rather than an approximation valid close to the horizon.  Surfaces of constant
\(t\) are the straight lines through the origin in Fig.~\ref{fig:kruskal}, and time evolution
slides them into one another.  The fixed point \(u=v=0\) is the bifurcation surface, where the
two horizons meet.  One asymmetry matters later: the Killing vector is future-directed in the
right exterior and past-directed in the left, so if time is taken to run upward on both
boundaries then the isometry generates \(H_{\rm R}-H_{\rm L}\).

The exponential follows immediately.  Following~\cite{ShenkerStanford2014}, add the perturbation on
the left boundary at Killing time \(t_w\) in the past.  It travels
on a surface of constant \(u\), and since \(u=e^{-\kappa(t-r_{*})}\) in the left exterior, later
release means smaller \(u\): pushing the release further back into the past drives that surface
toward the horizon as
\begin{equation}
    u_w\sim e^{-2\pi t_w/\beta},
\end{equation}
so the quantum runs ever closer to \(u=0\) without reaching it, as in Fig.~\ref{fig:kruskal}.
Its Killing energy \(E\) is fixed; what grows is the momentum measured by a frame regular at
the horizon.  The two ingredients are the Killing vector in Kruskal coordinates,
\(\xi=\kappa\left(v\partial_v-u\partial_u\right)\) with \(\kappa=2\pi/\beta\), and the fact
that a ray released at constant \(u\) carries \(p_v=0\).  Then
\(E=-\xi\cdot p=\kappa\left(up_u-vp_v\right)=\kappa\,u_w\,p_u\), and the metric's off-diagonal
\(uv\) structure turns \(p_u\) into \(p^{v}\), so
\begin{equation}
    p^{v}\sim\frac{E}{u_w}\sim E\,e^{2\pi t_w/\beta} .
\end{equation}
This is the estimate quoted in Sec.~\ref{sec:L5_shock} with its origin now visible: the
exponent is a boost parameter, and the rate \(2\pi/\beta\) is the surface gravity.

Which state does a geometry with two boundaries compute?  The Euclidean section is a cigar of
circumference \(\beta\), and cutting it in half along the Euclidean time circle leaves a path
integral with two boundaries, preparing a state on two copies of the Hilbert space,
\begin{equation}
    |\mathrm{TFD}\rangle=\frac{1}{\sqrt{Z(\beta)}}\sum_{n}e^{-\beta E_{n}/2}\,
    |\bar n\rangle_{\rm L}\otimes|n\rangle_{\rm R} ,
\end{equation}
the thermofield double~\cite{Israel1976,Maldacena2003}.  Here \(|\bar n\rangle\) is the CPT
conjugate of \(|n\rangle\), which is what the Euclidean path integral produces on the left
copy; The state is written this way in~\cite{MaldacenaShenkerStanford2016}, while the earlier
references leave the conjugation implicit.  Tracing out either side returns \(\rho_{\rm R}=e^{-\beta H_{\rm R}}/Z(\beta)\), so an
observer confined to one boundary sees exactly the thermal ensemble in which every correlator
of Lecture 5 was defined, and the second copy is its purification.  The geometry's staticity is
visible in the state, since \(|\mathrm{TFD}\rangle\) is annihilated by
\(H_{\rm R}-H_{\rm L}\), the generator of the boost written above.

The second boundary earns its keep in the correlators.  Continuing an insertion around the
Euclidean circle to the other side amounts to shifting its time argument by \(i\beta/2\), so
the two boundaries sit at antipodal points of the thermal circle and a two-sided correlator is
a single-sided thermal one at complex time.  That is the origin Sec.~\ref{sec:L5_bound}
promised and did not supply.  The even \(\beta/4\) spacing of \(F_{\rm reg}\) is not a
technical device chosen to make an estimate go through; it is what a two-sided geometry looks
like written in boundary variables, one insertion for each quarter turn between the two
boundaries.  The half-strip of width \(\beta/4\) on which the bound is proved is the interval
separating them.  The same object is read in~\cite{MaldacenaShenkerStanford2016} as a two-sided
overlap \(\langle\Psi|V_{\rm L}V_{\rm R}|\Psi\rangle\) in a thermofield double perturbed by
\(W\), which is the form the gravity calculation of Sec.~\ref{sec:L6_shock} uses.

All of this concerns the unperturbed geometry, in which the infalling quantum is a test
particle.  It cannot remain one.  Its momentum at the crossing grows without bound as \(t_w\)
grows, and once \(G_N p^{v}\) reaches order unity its own gravitational field stops being a
small correction.  Sec.~\ref{sec:L6_shock} works out what that field is.

\subsection{Shock waves and the backreacted geometry}
\label{sec:L6_shock}

Sec.~\ref{sec:L6_tfd} left the infalling quantum with a momentum growing like
\(e^{2\pi t_w/\beta}\) and with the observation that it cannot stay a test particle.  What does
a null quantum of very large momentum, travelling essentially along the horizon, do to the
geometry?  The answer predates holography and is exact rather than perturbative.

Write the unperturbed metric in Kruskal form,
\begin{equation}
    ds^{2}=\ell_{\rm AdS}^{2}\left[-\mathcal{A}(uv)\,du\,dv+\mathcal{B}(uv)\,dx_{i}dx^{i}\right],
\end{equation}
with \(x^{i}\) the \(d-1\) directions along the horizon. The script letters are deliberate:
\(A\) and \(B\) are already spoken for later in this lecture, as the warp factor and the
magnetic field of Sec.~\ref{sec:L6_hqcd}.  Cut this spacetime along \(u=0\) and
glue the two halves back together with a relative displacement in \(v\), so that a curve
crossing the cut finds its \(v\) coordinate advanced by \(h(\mathbf{x})\). Writing that as one
chart means expressing the old coordinate through a new one that is continuous across \(u=0\),
\begin{equation}
    v_{\rm old}=v-\theta(u)\,h(\mathbf{x}) ,
\end{equation}
with the step function carrying the sign. Substituting \(v\to v+h\) without it flips the sign
of the term below, and a reader who does that concludes the two boundaries have moved closer
together.
The result solves Einstein's equations with a null stress tensor supported on \(u=0\), and in
coordinates that make the displacement explicit the metric picks up exactly one new term,
\begin{equation}
    ds^{2}=\ell_{\rm AdS}^{2}\left[-\mathcal{A}(uv)\,du\,dv+\mathcal{B}(uv)\,dx_{i}dx^{i}
    +\mathcal{A}(uv)\,\delta(u)\,h(\mathbf{x})\,du^{2}\right].
\end{equation}

\begin{figure}[!htbp]
\centering
\begin{tikzpicture}[scale=1.16,
    bdy/.style={line width=1.1pt},
    sing/.style={line width=0.9pt,decorate,decoration={zigzag,segment length=5pt,amplitude=1.6pt}},
    hor/.style={black!45,line width=0.7pt,dash pattern=on 3pt off 2pt},
    shock/.style={otocDark,line width=1.2pt},
    slice/.style={otocDark,line width=0.8pt},
    ghost/.style={black!45,line width=0.6pt,dash pattern=on 2pt off 2pt},
    lbl/.style={font=\small},
    slbl/.style={font=\scriptsize,black!60}]

  \draw[sing] (-3,3) -- (3,3);
  \draw[sing] (-3,-3) -- (3,-3);
  \node[lbl] at (0,3.34) {\(r=0\)};
  \node[lbl] at (0,-3.40) {\(r=0\)};

  \draw[bdy] (3,-3) -- (3,3);
  \draw[bdy] (-3,-3) -- (-3,3);
  \node[lbl,rotate=-90] at (3.42,0) {\(r=\infty\)};
  \node[lbl,rotate=90]  at (-3.42,0) {\(r=\infty\)};

  \draw[hor] (-3,-3) -- (3,3);
  \draw[hor] (-3,3) -- (3,-3);

  \draw[shock] (-3,-2.7) -- (2.7,3)
        node[pos=0.72,above,sloped,slbl,otocDark] {shock at \(u=u_w\)};
  \node[slbl,black!55] at (2.52,2.10) {\(u=0\)};

  \draw[slice] (3,0) -- (-0.3,0);
  \draw[slice] (0.20,0.5) -- (-3,0.5);
  \draw[slice] (-0.3,0) -- (0.20,0.5);
  \draw[ghost] (-0.3,0) -- (-3,0);
  \node[slbl,anchor=north] at (-1.75,-0.07) {without the shock};

  \draw[-{Latex[length=4pt]},otocDark,line width=0.5pt] (0.62,0.12) -- (0.30,0.44);
  \node[slbl,otocDark,anchor=west] at (0.60,0.06) {\(h(\mathbf{x})\)};

  \node[lbl] at (2.15,-1.05) {R};
  \node[lbl] at (-2.15,-1.05) {L};
  \node[lbl] at (-1.95,2.35) {F};
\end{tikzpicture}
\caption{The backreacted geometry.  A null quantum released early from the left boundary
travels essentially along the horizon, at \(u=u_{w}\) with \(u_{w}\sim e^{-2\pi t_w/\beta}\),
and the spacetime on the two sides of it is glued with a relative displacement
\(h(\mathbf{x})\) in \(v\).  The consequence is drawn on a slice sent in from the right
boundary: on crossing the shell it is displaced along the null direction by \(h\) and emerges
where the dashed line shows it would not have, so the two boundaries are further apart than in
the unperturbed spacetime of Fig.~\ref{fig:kruskal}.  The shell is drawn at a visible distance
from \(u=0\) for legibility; in the regime of interest \(u_{w}\) is exponentially small.}
\label{fig:shock}
\end{figure}
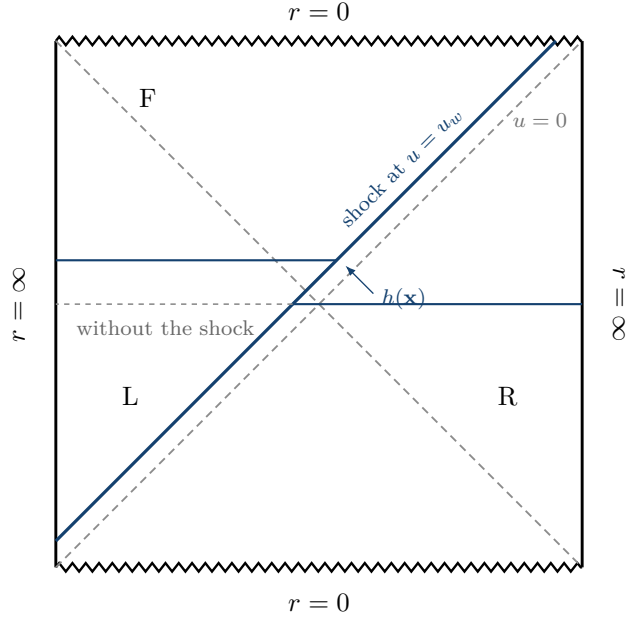

The form of the answer carries the physics.  The quantum does not curve the spacetime around
itself in the way a static mass does.  A geodesic crossing \(u=0\) resumes its journey
unchanged except that its \(v\) coordinate has jumped by \(h(\mathbf{x})\), so a signal sent
from one boundary toward the other finds that much further to go, as in
Fig.~\ref{fig:shock}.  That is exactly what~\cite{ShenkerStanford2014} measures: the shift increases the
distance through the interior, and driving it up is what eventually sends the mutual
information between the two boundaries to zero; stacking several shocks drives it
further~\cite{ShenkerStanford2014b}.  The construction
is due to Dray and 't Hooft, for a massless particle travelling along the Schwarzschild
horizon and for shells falling onto it~\cite{DrayTHooft1985a,DrayTHooft1985b}, and it recovers
as its flat-space case the field of an infinitely boosted particle found by Aichelburg and
Sexl~\cite{AichelburgSexl1971}, which is the quickest way to see why a shock is the right
object: boost a gravitational field hard enough and all that survives is a transverse profile
on a null plane.  The vacuum construction was generalized in~\cite{Sfetsos1995} to backgrounds carrying matter
and a cosmological constant, anti-de Sitter black holes among them.

Einstein's equations reduce, for this ansatz, to one linear equation for \(h\).  With the
perturbation carrying boundary energy \(E\), released at \(t_w\), and with transverse profile
\(a_{0}(\mathbf{x})\) normalized to unit integral,
\begin{equation}
    \left(-\partial_{i}\partial^{i}+m_{\rm scr}^{2}\right)h(\mathbf{x})
    =\frac{16\pi G_{N}}{\mathcal{A}(0)\,\ell_{\rm AdS}^{d-1}}\,E\,e^{2\pi t_w/\beta}\,a_{0}(\mathbf{x}),
    \qquad
    m_{\rm scr}^{2}=\frac{d(d-1)}{2},
\end{equation}
for the \(\mathrm{AdS}_{d+1}\)-Schwarzschild brane with the horizon radius set to
one~\cite{RobertsStanfordSusskind2015}, who call \(m_{\rm scr}\) simply
\(\mu\).  This is a screened Poisson equation: the shift is the potential of a source placed
where the perturbation entered.  The screening belongs to the transverse geometry rather than
to the perturbation, which~\cite{RobertsSwingle2016} makes explicit by writing the mass through the
metric functions, \(m_{\rm scr}^{2}\propto \mathcal{B}'(0)/\mathcal{A}(0)\).  The
exponential in \(t_w\) rides through untouched from Sec.~\ref{sec:L6_tfd}, having entered only
through the source.

Solving it for a perturbation localized at the origin gives, at large separation,
\begin{equation}
    h(\mathbf{x})\;\simeq\;
    \frac{\exp\left[\frac{2\pi}{\beta}\left(t_w-t_*\right)-m_{\rm scr}|\mathbf{x}|\right]}
         {|\mathbf{x}|^{(d-2)/2}} ,
    \qquad
    t_*=\frac{\beta}{2\pi}\log N^{2},
\end{equation}
with \(N\) the rank of the boundary gauge group, so that \(h\) reaches order unity at
\(t_w=t_*\) for a probe sitting within an AdS radius of the
insertion~\cite{RobertsStanfordSusskind2015}.  Further out the same condition is met later,
by the transverse term, which is the content of the second point below.  Two things follow.

The scrambling time has become geometric.  Sec.~\ref{sec:L5_shock} obtained \(t_*\) by asking
when \(G_{N}E_{p}\) reaches order unity and found \(t_*\sim(\beta/2\pi)\log S\); here the same
time is where the shift stops being a small deformation of the spacetime.  The two agree
because \(S\sim N^{2}\sim1/G_{N}\) counts the same degrees of freedom, and the second is a
solution where the first was an estimate.

And the two exponents combine into a cone.  Writing the exponential as
\begin{equation}
    h\;\sim\;\exp\left[\frac{2\pi}{\beta}
    \left(t_w-\frac{|\mathbf{x}|}{v_B}\right)\right],
    \qquad
    v_{B}=\frac{2\pi}{\beta\,m_{\rm scr}},
\end{equation}
identifies the butterfly velocity as the boost rate divided by the screening mass.  For the
brane, \(f(r)=r^{2}-r^{2-d}\) with horizon at \(r=1\) gives \(f'(1)=d\), hence
\(2\pi/\beta=d/2\), and
\begin{equation}
    v_{B}=\frac{d/2}{\sqrt{d(d-1)/2}}=\sqrt{\frac{d}{2(d-1)}} ,
\end{equation}
the value quoted without derivation in Sec.~\ref{sec:L5_shock}, now with a screening length
behind it.

One caution for anyone comparing sources.  Ref.~\cite{RobertsSwingle2016} writes these results
with \(d\)
counting the boundary \emph{spatial} dimensions, so that \(v_{B}=\sqrt{(d+1)/2d}\) and the
prefactor reads \(|\mathbf{x}|^{-(d-1)/2}\).  The conventions differ
by one unit in \(d\) and the two sets of formulas are identical; this lecture follows
Sec.~\ref{sec:L5_shock}, where \(d\) is the boundary spacetime dimension.

The power of \(|\mathbf{x}|\) in the denominator is worth keeping.  It makes the front a
screened potential rather than a step, so for \(d>2\) the position at which \(h\) crosses any
fixed threshold trails \(v_{B}t\) by a term growing like \(\log t\).  Reading that off the
profile above takes two lines and is worth doing.
The asymptotic velocity is exactly \(v_{B}\).  A velocity fitted as position over elapsed time
falls short of it by a term of order \(\log t/t\), with a coefficient
\(\beta v_{B}(d-2)/4\pi\) that is the same whatever threshold is used, so asymptotically the fit
undershoots. It need not do so at any finite time. The threshold enters one order later, in a
term of order \(1/t\) whose coefficient carries the logarithm of the threshold and is positive
for any threshold below the amplitude, and at accessible times that term is often the larger of
the two, so a threshold-fitted velocity overshoots as readily as it falls short.
Exercise~\ref{ex:L6_cone}(c) works the expansion out.
The restriction to \(d>2\) is not decorative: at \(d=2\) the power vanishes, the profile is a
pure exponential, and there is no drift to fit at all.
Exercise~\ref{ex:L5_cone} asked what the drift in a threshold-fitted \(v_{B}\) measures; this
is that effect in the geometry, and Exercise~\ref{ex:L6_cone} works it out.

The bulk solution is now completely specified.  Turning it into a correlator, which is what
the boundary theory actually measures, is the subject of Sec.~\ref{sec:L6_correlator}.

\subsection{From the shock to the correlator}
\label{sec:L6_correlator}

Sec.~\ref{sec:L6_shock} produced a geometry.  The boundary theory measures a correlator, and
the step between the two is a scattering problem near the horizon.

Start from the form the correlator took in Sec.~\ref{sec:L6_tfd}.  The out-of-time-order
correlator is an inner product between two states, each the thermofield double acted on by
\(V\) and \(W\) in one of the two orders,
\begin{equation}
    |\Psi\rangle=W(t)^{\dagger}V(0)^{\dagger}|\mathrm{TFD}\rangle,
    \qquad
    |\Psi'\rangle=V(0)\,W(t)\,|\mathrm{TFD}\rangle,
    \qquad
    F(t)=\langle\Psi|\Psi'\rangle ,
\end{equation}
so the question is how far apart the two orderings push the
state~\cite{ShenkerStanfordStringy2015}.  Regularizing means giving the four insertions the
imaginary-time offsets of Sec.~\ref{sec:L6_tfd}.  They are left implicit below, but they are
not cosmetic: they make the momenta complex, and that is what turns the \(i\chi\) of the
next display into a real decrease rather than a pure phase.
Exercise~\ref{ex:L6_spin}(a) works it out.  In the bulk each operator creates a quantum.  The
\(W\) inserted a time \(t\) in the past makes the shell of Sec.~\ref{sec:L6_shock}, sitting at
\(u=u_w\) and carrying \(p^{v}\sim e^{2\pi t/\beta}\).  The \(V\) inserted near \(t=0\) makes a
quantum falling in at fixed \(v\) with \(p^{u}\) of order the temperature, and it has to cross
the shell to get to the other side.

The kinematics are what make this tractable.  The two quanta meet near the bifurcation surface
with a squared centre-of-mass energy proportional to \(p^{u}p^{v}\), growing like
\(e^{2\pi t/\beta}\), while their transverse separation \(|\mathbf{x}|\) stays fixed.  Large
centre-of-mass energy at fixed impact parameter is the one regime in which gravitational
scattering is simple: single graviton exchange dominates and the amplitude exponentiates.

Restricted to the two-particle subspace the S-matrix is then multiplication by a
phase~\cite{ShenkerStanfordStringy2015},
\begin{equation}
    S=e^{i\chi},\qquad \chi=p_{v}\,h(\mathbf{x}),
\end{equation}
with \(h\) exactly the shift computed in Sec.~\ref{sec:L6_shock}.  That identity is a
restatement rather than a coincidence.  Crossing the shell translates the infalling quantum's
\(v\) coordinate by \(h\); translating a wavefunction multiplies it by the phase conjugate to
the translation; and the conjugate momentum is the one that quantum carries.  The geometry of
Sec.~\ref{sec:L6_shock} and the amplitude here are one calculation written twice.  Since
\(h\propto G_{N}p^{v}\), and since the Kruskal metric written above is off-diagonal in \(u\)
and \(v\), so that \(p_{v}=g_{vu}p^{u}\) is proportional to \(p^{u}\),
\begin{equation}
    \chi\;\propto\;G_{N}\,p^{u}p^{v}\;\propto\;G_{N}\,e^{2\pi t/\beta} .
\end{equation}

Averaging \(e^{i\chi}\) against the wavefunctions of the two quanta and expanding for small
\(\chi\) gives
\begin{equation}
    F(t)\;\approx\;1-\frac{c}{N_{\rm dof}}\,e^{2\pi t/\beta},
    \qquad
    N_{\rm dof}\sim\frac{1}{G_{N}}\sim N^{2},
\end{equation}
with \(N^{2}\) counting the degrees of freedom of that gauge group, which is what
Sec.~\ref{sec:L5_shock} wrote down without
deriving~\cite{MaldacenaShenkerStanford2016}.  The prefactor now has an identity.  The small
parameter suppressing the exponential is Newton's constant, so the coupling that produced the
shift is the coupling that keeps the correction small, and at strictly infinite \(N\) the
perturbation has no gravitational field and there is no chaos to see.

The expansion also shows where it ends.  The linear term is the first in a series in \(\chi\),
and \(\chi\) reaches order unity at \(t\sim t_*\).  The eikonal survives that: the full
\(\langle e^{i\chi}\rangle\) is an oscillatory average that decays rather than growing, so the
same amplitude produces both the exponential and the saturation of \(C\) described in
Sec.~\ref{sec:L5_otoc}.  What expires at \(t_*\) is the linearization, not the approximation.

Written this way the chaos bound acquires a bulk reading.  Exchange of a particle of spin
\(J\) gives a phase growing like \((p^{u}p^{v})^{J-1}\), so
\(\lambda_{L}=(2\pi/\beta)(J-1)\).  The graviton has \(J=2\) and saturates, and anything of
higher spin would run faster.  What Sec.~\ref{sec:L5_bound} proves from analyticity and
unitarity of a thermal correlator becomes, on this side of the duality, a statement about what
the bulk may contain.  Ref.~\cite{MaldacenaShenkerStanford2016} draws it carefully: the bound rules out a
weakly coupled, large-radius bulk carrying a \emph{finite} number of light species of spin
above two.  An infinite tower is a different matter, and
the next paragraph is about the one string theory supplies.

Finite string length is that statement one step further.  The graviton becomes the leading
member of a Regge trajectory and the effective spin drops below two.  Sec.~\ref{sec:L5_bound}
recorded the leading correction as
\(j_{\rm eff}=2-d(d-1)\ell_s^{2}/4\ell_{\rm AdS}^{2}\), and in the language of
Sec.~\ref{sec:L6_shock} that coefficient is half the squared screening mass,
\begin{equation}
    \frac{d(d-1)}{4}=\frac{m_{\rm scr}^{2}}{2},
\end{equation}
so a single number governs both how far the shock spreads along the horizon and how far the
exponent falls below the bound.  The scrambling time inherits it,
\begin{equation}
    t_*=\frac{\beta}{2\pi}
    \left(1+\frac{m_{\rm scr}^{2}\ell_s^{2}}{2\ell_{\rm AdS}^{2}}+\cdots\right)\log S ,
\end{equation}
so a fatter string scrambles more slowly~\cite{ShenkerStanfordStringy2015}.

That closes the account Lecture 5 left open.  The exponent is a surface gravity, the velocity
an inverse screening length, the prefactor Newton's constant, and the correlator carrying them
a two-particle amplitude at a horizon.  One of those four is the same for every black hole of
two-derivative gravity: the exponent, which is \(2\pi/\beta\) whatever the metric, because a
horizon has only one rate to offer.  That is at once why the holographic answer is universal
and why it identifies nothing.

Universal, but local and instantaneous, and two limits are worth knowing because in both the
formula returns something short of the whole story.  A state driven far from equilibrium by a
sudden injection of energy has no single exponent at all: the correlator accumulates growth at
whichever temperature is current, each stage saturating its own \(2\pi/\beta\), so what the
late-time answer records is a product of exponentials rather than
one~\cite{BalasubramanianCrapsDeClerckNguyen2019}.  At extremality the formula instead returns
zero, which is true and useless.  The eikonal phase there grows as \(t^{3}\), making the
scrambling time a power of the central charge rather than its logarithm, and superposed on that
power law are brief exponential bursts, growing at the rate that the right-moving temperature
sets and not the Hawking one, since the first stays finite at extremality where the second does
not~\cite{CrapsDeClerckHackerNguyenRabideau2021}.  Scrambling survives the vanishing of
\(2\pi/\beta\); maximal chaos does not.  Neither result is reachable by the argument given
above, which puts the shock exactly on the horizon, an approximation the blueshift justifies at
finite temperature and nothing justifies at zero.  Both were obtained in position space
instead, from the geodesic saddle of each propagator, with the phase built from one geodesic's
stress tensor against the other's shock wave.

The other three do carry information about the background, the
velocity most visibly, since the screening length it divides is a property of the transverse
geometry.  The rest of the lecture changes the question, asking what a probe
moving in the bulk sees rather than what a boundary operator measures, and what survives when
the geometry is built to imitate QCD instead of to be conformal.  Isotropy is the first
casualty: a background magnetic field picks out a direction, so nothing carrying a direction
need be the same along it as across it. Sec.~\ref{sec:L6_aniso} finds exactly that for the
exponent of a suspended string. The butterfly velocity of such a background would be
anisotropic for the same reason, though it is not computed here.  A different pair has to be
kept apart, and Sec.~\ref{sec:L6_bounds} is where that is done: the \(\lambda_L\) the theorem
of Sec.~\ref{sec:L5_bound} constrains belongs to the horizon, while what a geodesic or a
suspended string returns is the instability rate of a probe.

\subsection{Black-hole thermodynamics and the swallowtail}
\label{sec:L6_thermo}

The question therefore has to change, and what follows changes it to one about probes,
meaning an object moving in the bulk rather than an operator inserted on the boundary.  A probe samples
the geometry where it happens to sit rather than at the horizon, so its Lyapunov exponent has
no reason to be universal, and the useful question becomes what it can tell apart.  This
subsection supplies something worth telling apart.

Take the dyonic black hole in four-dimensional anti-de Sitter space, the simplest
generalization of Reissner--Nordstr\"om--AdS carrying both an electric and a magnetic
charge~\cite{ShuklaDasDudalMahapatra2024},
\begin{equation}
    ds^{2}=-f(r)dt^{2}+\frac{dr^{2}}{f(r)}
    +r^{2}\left(d\theta^{2}+\sin^{2}\theta\,d\varphi^{2}\right),
    \qquad
    f(r)=1+\frac{r^{2}}{\ell_{\rm AdS}^{2}}-\frac{2M}{r}
    +\frac{q_{e}^{2}+q_{m}^{2}}{r^{2}} ,
\end{equation}
with \(\ell_{\rm AdS}\) the anti-de Sitter radius, written as it was in
Sec.~\ref{sec:L6_shock} rather than as a bare \(l\), and gauge field \(A=q_{e}\left(1/r_{h}-1/r\right)dt+q_{m}\cos\theta\,d\varphi\).  The
magnetic charge earns its place twice over.  It enriches the phase diagram, and it gives the
dual theory a background magnetic field, which is what makes these solutions useful for
holographic treatments of the Hall effect and of
ferromagnetism~\cite{HartnollKovtun2007}.

Fix the electric potential \(\Phi_{e}=q_{e}/r_{h}\) and the magnetic charge \(q_{m}\), with
\(G_{N}=1\).  The ensemble is mixed, grand canonical in the electric sector and canonical in
the magnetic one, and the mixture is not incidental: a purely electric black hole at fixed
\(\Phi_{e}\) has no critical point at all~\cite{ChamblinEmparanJohnsonMyers1999b}, so everything below is
driven by the charge that is being held fixed rather than by the potential that is not.  The horizon condition \(f(r_{h})=0\) fixes \(M\), and
\begin{equation}
    S=\pi r_{h}^{2},
    \qquad
    T=\frac{f'(r_{h})}{4\pi}
    =\frac{1}{4\pi r_{h}^{3}}
     \left(\frac{3r_{h}^{4}}{\ell_{\rm AdS}^{2}}+r_{h}^{2}\bigl(1-\Phi_{e}^{2}\bigr)
     -q_{m}^{2}\right),
\end{equation}
\begin{equation}
    G=M-TS-\Phi_{e}q_{e}
    =\frac{1}{4}\left(r_{h}\bigl(1-\Phi_{e}^{2}\bigr)-\frac{r_{h}^{3}}{\ell_{\rm AdS}^{2}}
    +\frac{3q_{m}^{2}}{r_{h}}\right).
\end{equation}
One may instead treat the cosmological constant as a pressure, \(P=-\Lambda/8\pi G_{N}\), and
with it the mass as an enthalpy rather than an
energy~\cite{KastorRayTraschen2009,KubiznakMann2012}; nothing in the dyonic analysis below
changes under that reading, so we stay in the standard ensemble.

Whether the ensemble holds one black hole or three at a given temperature is settled by
whether \(T(r_{h})\) is monotonic, and that is a short calculation.  Setting
\(\ell_{\rm AdS}=1\),
\begin{equation}
    \frac{dT}{dr_{h}}=\frac{1}{4\pi}
    \left(3-\frac{1-\Phi_{e}^{2}}{r_{h}^{2}}+\frac{3q_{m}^{2}}{r_{h}^{4}}\right),
\end{equation}
which vanishes when \(3r_{h}^{4}-(1-\Phi_{e}^{2})r_{h}^{2}+3q_{m}^{2}=0\).  That is a
quadratic in \(r_{h}^{2}\) whose roots multiply to \(q_{m}^{2}\), so both are positive when the
discriminant is positive, \((1-\Phi_{e}^{2})^{2}>36\,q_{m}^{2}\), and the magnetic charge does
not vanish.  The second condition is easy to pass over and it carries the physics: at
\(q_{m}=0\) one root falls to the origin, one turning point is lost, and the swallowtail with
it.  The temperature is non-monotonic for any \(q_{m}\) below \(q_{mc}\), including \(q_{m}=0\),
where it falls to a single minimum and rises again.  What a nonzero magnetic charge adds is the
second turning point, and with it a window of temperatures over which three black holes coexist,
which happens precisely for
\begin{equation}
    0<q_{m}<q_{mc}=\frac{\ell_{\rm AdS}}{6}\left(1-\Phi_{e}^{2}\right),
\end{equation}
the two turning points merging at
\begin{equation}
    r_{hc}=\frac{\ell_{\rm AdS}\sqrt{1-\Phi_{e}^{2}}}{\sqrt{6}}
\end{equation}
when the discriminant vanishes.  This is the same point the inflection conditions
\(\partial_{r_{h}}T=\partial_{r_{h}}^{2}T=0\) pick
out~\cite{ShuklaDasDudalMahapatra2024}, and both expressions require \(|\Phi_{e}|<1\), above
which there is no transition to look for.

\begin{figure}[!htbp]
\centering
\includegraphics[width=\textwidth]{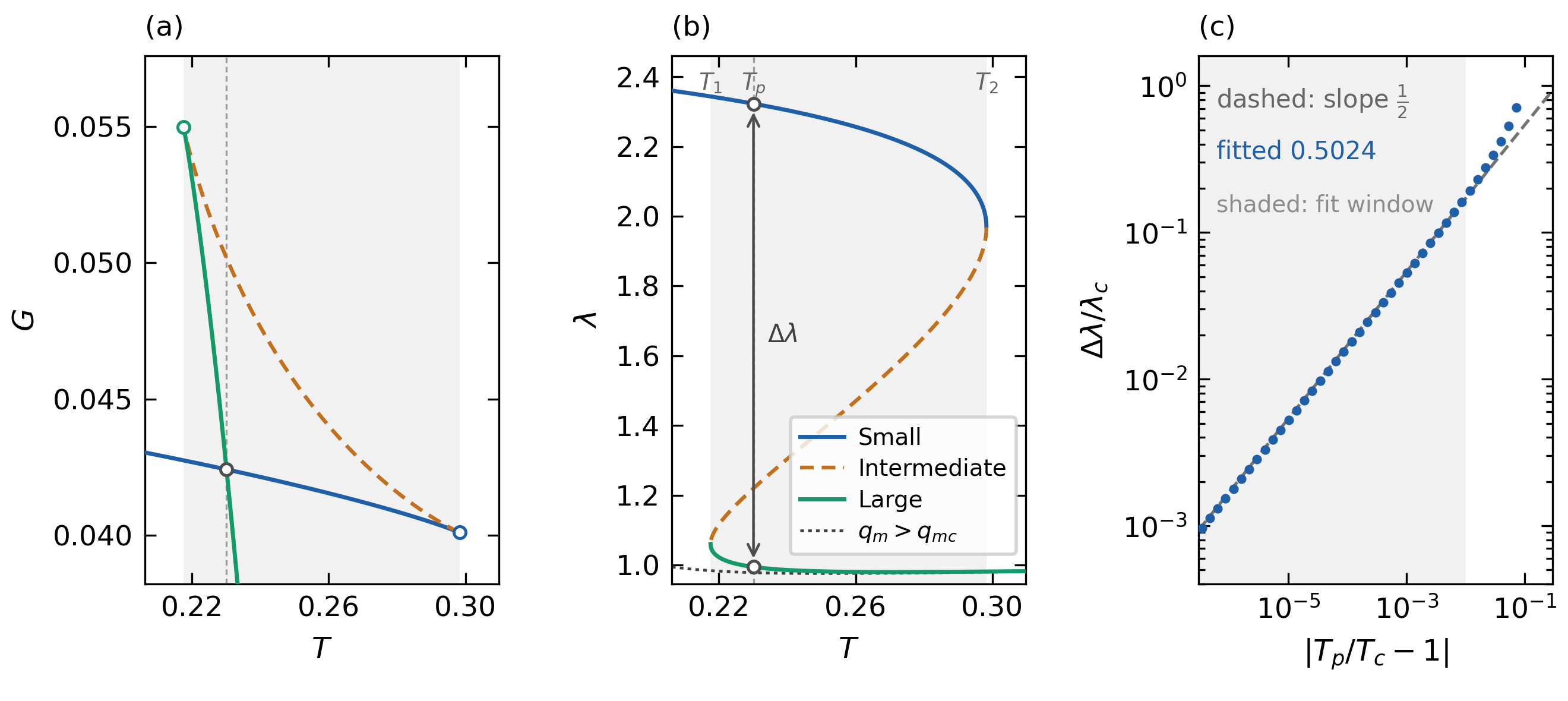}
\caption{The small/large transition of a dyonic anti-de Sitter black hole at
\(\Phi_{e}=0.6\), with \(\ell_{\rm AdS}=G_{N}=1\), and \(q_{m}=0.55\,q_{mc}\) in the first
two panels.
(a) Gibbs free energy against temperature, showing the swallowtail.  Open circles mark the two
cusps at \(T_{1}\) and \(T_{2}\) and the crossing at \(T_{p}\) where the small and large
branches exchange dominance; the shaded band is the temperature range over which all three
branches coexist.  (b) The Lyapunov exponent of the unstable circular null geodesic in the
same background, computed as in Sec.~\ref{sec:L6_geodesic}.  Its folds sit at the same two
temperatures as the cusps in (a), which is the sense in which a probe sees the thermodynamics,
and \(\Delta\lambda\) is the gap between the two stable branches at \(T_{p}\).  The dotted
curve is a supercritical black hole, which has no fold.  (c) That gap against the reduced
transition temperature as \(q_{m}\) is swept up towards \(q_{mc}\), which is a different sweep
from the fixed-\(q_{m}\) one of the first two panels and has to be, since the gap closes at
\(q_{mc}\) itself. The shading here is the fit window, four and a half decades wide in
\(|T_{p}/T_{c}-1|\), over which the fitted exponent is \(0.5024\)
against the value \(\tfrac12\) derived in Sec.~\ref{sec:L6_probe}.  The colour code for the
three branches follows Ref.~\cite{ShuklaDasDudalMahapatra2024}.}
\label{fig:lyap_phase}
\end{figure}

The three branches are the familiar ones.  The small and large black holes are
thermodynamically stable, the intermediate branch has negative heat capacity and never
dominates, and the free energy traces the swallowtail of Fig.~\ref{fig:lyap_phase}(a), with
small and large exchanging dominance at a temperature \(T_{p}\) where a first-order transition
occurs, the subscript here standing for the transition rather than for an orbit period as it did
in Sec.~\ref{sec:L4_trace_formula}.  Raising \(q_{m}\) shrinks the swallowtail until it closes at \(q_{mc}\), so
\(q_{mc}\) is a second-order critical point terminating a line of first-order transitions,
which is the pattern of the van der Waals liquid--gas system.  That pattern, swallowtail included, was found for the electrically charged black hole at
fixed \emph{charge} in~\cite{ChamblinEmparanJohnsonMyers1999}; the role \(q_{e}\) plays there is
played by \(q_{m}\) here, which is why the fixed potential does no harm.  Setting \(q_{m}=0\)
removes the structure entirely for any \(|\Phi_{e}|<1\), leaving the Hawking--Page transition
between a black hole and thermal anti-de Sitter space~\cite{HawkingPage1983}.

Three other families of asymptotically anti-de Sitter black hole, Bardeen, Gauss--Bonnet and
Lorentz-violating massive gravity, carry the same qualitative structure, and the critical
exponent derived in Sec.~\ref{sec:L6_probe} comes out the same for all
four~\cite{ShuklaDasDudalMahapatra2024}.  The details differ, and the last two are handled
there in the extended phase space rather than the ensemble used here, so what transfers is the
exponent and not every statement along the way.  Working through the dyonic case keeps the
formulas short.  What the thermodynamics cannot say is
whether any of it is visible to a probe, and building that probe is the subject of
Sec.~\ref{sec:L6_geodesic}.

\subsection{Lyapunov exponents of unstable circular geodesics}
\label{sec:L6_geodesic}

The probe is the simplest one available: a particle on a circular orbit that happens to sit at
a maximum of its effective potential rather than a minimum.  Displace it radially and the
displacement grows, at a rate the geometry fixes.

Restrict to the equatorial plane of a static, spherically symmetric black hole.  The Killing
vectors give a conserved energy \(E\) and angular momentum \(L\), and the radial equation
reduces to
\begin{equation}
    \dot r^{2}+V_{\rm eff}(r)=E^{2},
    \qquad
    V_{\rm eff}(r)=f(r)\left(\frac{L^{2}}{r^{2}}-\epsilon\right),
\end{equation}
with \(\epsilon=0\) for null geodesics and \(\epsilon=-1\) for timelike
ones~\cite{ShuklaDasDudalMahapatra2024}.  The overdot is differentiation with respect to the
parameter along the geodesic, normalized by \(g_{MN}\dot x^{M}\dot x^{N}=\epsilon\): proper
time in the timelike case, and in the null case an affine parameter, since proper time vanishes
identically there and cannot be used.  A circular orbit sits at \(V_{\rm eff}'(r_{0})=0\),
and it is unstable when \(V_{\rm eff}''(r_{0})<0\), which is the case of interest.

Linearizing the equations of motion in \((\delta r,\delta p_{r})\) about such an orbit gives
\(\dot X=KX\) for \(X=(\delta r,\delta p_{r})\), with \(K\) off-diagonal, \(K_{12}=K_{1}\)
and \(K_{21}=K_{2}\),
\begin{equation}
    K_{1}=\frac{f(r_{0})}{\dot t},
    \qquad
    K_{2}=-\frac{V_{\rm eff}''(r_{0})}{2f(r_{0})\,\dot t} ,
\end{equation}
whose eigenvalues are \(\pm\sqrt{K_{1}K_{2}}\).  Taking the growing one,
\begin{equation}
    \lambda=\sqrt{K_{1}K_{2}}=\sqrt{-\frac{V_{\rm eff}''(r_{0})}{2\dot t^{\,2}}} .
\end{equation}
The factor of \(\dot t\) is doing real work.  It converts a rate per unit of the geodesic's own
parameter into a rate per unit coordinate time, which is the time of the observer at infinity,
and in a holographic setting the time of the boundary theory.  That is what will make \(\lambda\)
comparable with \(2\pi T\) in Sec.~\ref{sec:L6_bounds}.

Since \(\dot t\) depends on the kind of geodesic, the two cases separate here.  For null
geodesics \(\dot t=L/\bigl(r_{0}\sqrt{f(r_{0})}\bigr)\), giving
\begin{equation}
    \lambda=\sqrt{-\frac{r_{0}^{2}f(r_{0})}{2L^{2}}\,V_{\rm eff}''(r_{0})} ,
\end{equation}
and because \(V_{\rm eff}\) is proportional to \(L^{2}\) when \(\epsilon=0\), the angular
momentum cancels: the exponent is a property of the orbit and not of the particle on it.  For
timelike geodesics \(\dot t^{\,2}=2/[2f(r_{0})-r_{0}f'(r_{0})]\), giving
\begin{equation}
    \lambda=\tfrac12\sqrt{\bigl(r_{0}f'(r_{0})-2f(r_{0})\bigr)V_{\rm eff}''(r_{0})} ,
\end{equation}
where both factors under the root are negative on an unstable
orbit~\cite{ShuklaDasDudalMahapatra2024}.

The null case is worth following through, because it has no free parameters.  With
\(V_{\rm eff}=f(r)L^{2}/r^{2}\), the condition \(V_{\rm eff}'(r_{0})=0\) is
\begin{equation}
    r_{0}f'(r_{0})=2f(r_{0}),
\end{equation}
which locates the photon sphere.  For Schwarzschild, \(f=1-2M/r\) puts it at \(r_{0}=3M\) and
returns
\begin{equation}
    \lambda=\frac{1}{3\sqrt3\,M} .
\end{equation}
The same orbit has coordinate angular velocity
\(\Omega_{c}=\sqrt{f(r_{0})}/r_{0}=\sqrt{f'(r_{0})/(2r_{0})}\), and in four dimensions that
evaluates to \(1/3\sqrt3 M\) as well, so the photon sphere of a Schwarzschild black hole
becomes unstable at exactly the rate at which it orbits.

That coincidence is not what makes the exponent interesting.  What does is that in one setting
it is measurable.  In the eikonal limit
the quasinormal spectrum of a static, spherically symmetric, asymptotically flat black hole is
\begin{equation}
    \omega_{\rm QNM}=\Omega_{c}\,\ell-i\left(n+\tfrac12\right)|\lambda| ,
\end{equation}
with \(\ell\) the angular momentum of the perturbation and \(n\) the
overtone~\cite{CardosoEtAl2009}.  The real part is the orbital frequency at the photon sphere
and the imaginary part is set by its instability rate, the fundamental mode decaying at
\(|\lambda|/2\) and each overtone adding another \(|\lambda|\).  So what a perturbed black hole
does after it is struck is controlled by how quickly light peels away from the circular orbit.
Their dimensionless ratio \(\gamma=\Omega_{c}/2\pi\lambda\) equals \(1/2\pi\sqrt{d-3}\) for a
Schwarzschild--Tangherlini black hole in \(d\) dimensions, so an orbit becomes more unstable
relative to its own period as the dimension grows, and \(d=4\) is the case where the two rates
coincide.

The hypotheses of that theorem are worth reading twice, because the rest of this lecture sits
outside them in two separate ways.  Asymptotic flatness is one of them, and Ref.~\cite{CardosoEtAl2009} says
plainly that the argument does not seem to extend to anti-de Sitter spacetimes; every background in Sec.~\ref{sec:L6_probe} is
asymptotically anti-de Sitter.  Being a test field is the other.  Konoplya and Stuchl\'ik
exhibit gravitational perturbations in Einstein--Lovelock gravity whose eikonal frequencies are
not those of the null orbit, and map out where the identification survives, which is for test
fields and not for the metric's own~\cite{KonoplyaStuchlik2017}.  Neither restriction touches
\(\lambda\) itself, which is defined by the geodesic and needs no wave equation.  What they
touch is the claim that \(\lambda\) is something a distant observer could measure, and that
claim is not available for the black holes of Sec.~\ref{sec:L6_probe}.  What the exponent does
there it does as a property of an orbit.

Two cautions, both easy to trip over.  The first is a convention.  Ref.~\cite{CardosoEtAl2009} writes
\(\lambda=\sqrt{V_{r}''/2\dot t^{\,2}}\) with \(\dot r^{2}=V_{r}\), so that \(V_{r}\) is
\(E^{2}-V_{\rm eff}\) and its second derivative is minus the one used here.  The two
expressions agree; the two conventions sit side by side in the literature.

The second is about the word.  The definition \eqref{eq:L2_lyapunov_def} makes a Lyapunov
exponent a long-time average along a trajectory, characterizing a flow.  The \(\lambda\) here is a local
linear instability rate at one orbit, and geodesic motion in a static spherically symmetric
spacetime is completely integrable, so the flow it belongs to has no positive Lyapunov
exponent in that earlier sense at all.  The situation is the pendulum's: an integrable system
with a perfectly well-defined instability rate at its unstable equilibrium.  The quantity
below is that rate, and the name it carries is conventional rather than a claim that the
geodesics are chaotic.

The timelike formula is worth one sanity check, since it is otherwise a result to be taken on
trust.  For Schwarzschild it returns a real \(\lambda\) throughout \(3M<r_{0}<6M\), falling to
zero exactly at \(r_{0}=6M\) where \(L^{2}=12M^{2}\), and admits no unstable circular orbit
beyond.  It therefore reproduces the innermost stable circular orbit as the radius at which
the instability rate vanishes, which is what a formula for the instability rate should do.

Nothing so far involves the phase structure of Sec.~\ref{sec:L6_thermo}.  The geometry enters
\(\lambda\) only through \(f(r)\), and the thermodynamic branches differ only in \(f(r)\), so
the two are linked whether or not the link is useful.  Sec.~\ref{sec:L6_probe} asks how much
of the thermodynamics survives the trip.

\subsection{The Lyapunov exponent as a probe of the transition}
\label{sec:L6_probe}

Feed the branches of Sec.~\ref{sec:L6_thermo} into the machinery of
Sec.~\ref{sec:L6_geodesic} and the answer is immediate.  At a temperature where three black
holes coexist there are three horizon radii, hence three blackening functions, hence three
photon spheres and three exponents.  So \(\lambda(T)\) is multivalued over exactly the
temperature range where \(G(T)\) is, and folds at exactly the two temperatures where the free
energy has its cusps, as Fig.~\ref{fig:lyap_phase} shows in its first two panels.  Above
\(q_{mc}\) there is one branch and the exponent is a single-valued function of temperature.

It is worth being clear about how much that observation contains on its own.  Not a great
deal.  At fixed charges the only thing that moves along a branch is \(r_{h}\), so \(\lambda\)
is a smooth function of the horizon radius and would fold wherever \(r_{h}(T)\) folds whatever
smooth function it happened to be.  What \(\lambda\) adds is the provenance rather than the
folding.  It is a property of a probe orbit rather than of a thermodynamic potential, computed
from the geodesic equation without reference to any free energy.  Whether it is also something
an observer could measure is a separate question, and Sec.~\ref{sec:L6_geodesic} answered it
in the negative for backgrounds of this kind: the quasinormal reading of \(\lambda\) rests on
asymptotic flatness, which these black holes do not have.

The sharper statement concerns the first-order transition.  Define
\begin{equation}
    \Delta\lambda=\lambda_{\rm small}-\lambda_{\rm large}
\end{equation}
at \(T=T_{p}\), the gap between the two thermodynamically stable branches at the transition,
drawn as the arrow in Fig.~\ref{fig:lyap_phase}(b).  It is nonzero below \(q_{mc}\) and
vanishes as \(q_{m}\to q_{mc}\), which is what is asked of an order parameter, and its
approach to zero has an exponent.

Getting that exponent takes two observations.  The first is about the geometry near the
critical point.  Both \(\partial_{r_{h}}T\) and \(\partial_{r_{h}}^{2}T\) vanish there, by
definition, so \(T-T_{c}\) begins at cubic order in \(r_{h}-r_{c}\); moving the charge slightly
below its critical value adds a term linear in \(r_{h}-r_{c}\) with coefficient proportional
to \(q_{mc}-q_{m}\), together with a term independent of \(r_{h}\).  The second is Maxwell's
construction: equality of the free energies places the two coexisting radii symmetrically
about \(r_{c}\) to leading order, the asymmetry being of order \(q_{mc}-q_{m}\) against a
separation of order \(\sqrt{q_{mc}-q_{m}}\) and so dropping out of what follows.  Imposing that \(T\) take the same value at \(r_{c}\pm(r_{h}-r_{c})\) then
cancels the even part and leaves
\begin{equation}
    (r_{h}-r_{c})^{2}\;\propto\;q_{mc}-q_{m},
\end{equation}
while substituting back shows the transition temperature itself moves linearly,
\(T_{p}-T_{c}\propto q_{mc}-q_{m}\).  Eliminating the charge between the two,
\begin{equation}
    r_{\rm large}-r_{\rm small}\;\propto\;\left|T_{p}-T_{c}\right|^{1/2}.
\end{equation}
This is the van der Waals order parameter and its mean-field exponent, reached by the van der
Waals argument.

The exponent of \(\Delta\lambda\) follows without further work.  Near the critical point
\(\lambda\) is a smooth function of \(r_{h}\), so to leading order
\begin{equation}
    \Delta\lambda\simeq\left(\frac{\partial\lambda}{\partial r_{h}}\right)_{c}
    \bigl(r_{\rm small}-r_{\rm large}\bigr)
    \qquad\Longrightarrow\qquad
    \frac{\Delta\lambda}{\lambda_{c}}\;\propto\;
    \left|\frac{T_{p}}{T_{c}}-1\right|^{1/2},
\end{equation}
recovering the exponent reported in Ref.~\cite{ShuklaDasDudalMahapatra2024}.  The route taken
here, through the vanishing of the second derivative and a symmetric Maxwell construction, is
the one that survives that vanishing; an expansion stopped at quadratic order has nothing left
to work with.  Fig.~\ref{fig:lyap_phase}(c) measures the exponent for the dyonic case over four and
a half decades in the reduced transition temperature and returns \(0.5024\).

Two features of that argument deserve notice.  Nothing in it used \(f(r)\).  It used only that
a critical point is where the first two derivatives of the temperature vanish, and that
\(\lambda\) is a smooth and generically nondegenerate function of the horizon radius.  That is
why the same exponent appears for the dyonic, Bardeen, Gauss--Bonnet and massive-gravity
families alike~\cite{ShuklaDasDudalMahapatra2024}, and why it takes the mean-field value: an
analytic equation of state, degenerate at cubic order, with a symmetric Maxwell construction,
is precisely the van der Waals situation.

The other feature is a limitation, and stating it is what keeps the result honest.
\(\Delta\lambda\) inherits its exponent rather than possessing one.  The order parameter of
this transition is the horizon-radius gap, and near the critical point \(\Delta\lambda\) is a
linear image of it, so the value \(\tfrac12\) is not independent information about the
criticality.  What it does establish is that the map from thermodynamic branch to probe orbit
is smooth and nondegenerate, and that is a statement worth checking rather than assuming: were
\((\partial\lambda/\partial r_{h})_{c}\) to vanish in some background, the exponent would
change.

It does not deliver \(T_{p}\) either, which still comes from equating free energies.  What
the exponent does is show a purely geometrical property of
a probe orbit changing discontinuously across a thermodynamic transition of the background it
moves in, and doing so with the mean-field exponent.  The construction has since been carried to rotating
black holes, where the spherical symmetry assumed in Sec.~\ref{sec:L6_geodesic} is
absent~\cite{ChenYangLiu2025}.

That exhausts what a geodesic can say.  The remainder of the lecture keeps the idea of a probe
and changes both the probe and the background: a string rather than a particle, in a geometry
built to imitate QCD rather than to be conformal.

\subsection{Holographic QCD and the anisotropy of chaos}
\label{sec:L6_hqcd}
\label{sec:L6_aniso}

The black holes of Sec.~\ref{sec:L6_thermo} solve Einstein--Maxwell in anti-de Sitter space,
so their duals are conformal theories carrying a charge.  QCD is not conformal, it confines,
and the physics one wants sits near its deconfinement crossover, where the coupling is strong
and real-time quantities are out of reach of the lattice.  A bottom-up holographic model
replaces the problem of solving QCD with the problem of engineering a bulk geometry whose
thermodynamics reproduces it, and then computes real-time quantities there.

The model used here is five-dimensional Einstein--Maxwell--dilaton gravity carrying two gauge
fields, one supplying a chemical potential \(\mu\) and one a background magnetic field
\(B\)~\cite{ShuklaNongmaithemDudalMahapatra2025}, the two-field extension of the purely
magnetic background of Ref.~\cite{ShuklaDudalMahapatra2023}.  Its metric,
\begin{equation}
    ds^{2}=\frac{e^{2A(z)}}{z^{2}}
    \left[-g(z)dt^{2}+\frac{dz^{2}}{g(z)}+dx_{1}^{2}
    +e^{B^{2}z^{2}}\left(dx_{2}^{2}+dx_{3}^{2}\right)\right],
    \qquad
    F^{(2)}=B\,dx_{2}\wedge dx_{3},
\end{equation}
puts the field along \(x_{1}\) and warps the plane transverse to it, breaking the boundary
\(SO(3)\) to \(SO(2)\) and recovering it as \(B\to0\).  That broken symmetry is the origin of
everything anisotropic below.  The holographic coordinate is \(z=1/r\), running from the
boundary at \(z=0\) to the horizon at \(z=z_{h}\), so \(z\) and \(r\) increase in opposite
directions; the string calculation is written in \(r\) from here on with \(r_{h}=z_{h}=1\), so
tip positions above \(1\) mean tips outside the horizon.

Rather than choosing a dilaton potential and solving, one chooses the warp factor
\(A(z)=-az^{2}\) and reconstructs \(V\) from the equations of motion, a strategy that predates
the calculations quoted here~\cite{GubserNellore2008,DudalMahapatra2017,BohraDudalHajilouMahapatra2021}
and closes the system in closed form, the second gauge kinetic function
included~\cite{ShuklaNongmaithemDudalMahapatra2025}, which is why it never has to be specified.
Two parameters remain and neither is free: the \(a\) of the warp factor, fixed at
\(0.15\,\mathrm{GeV}^{2}\) by a deconfinement temperature near \(270\) MeV at vanishing \(\mu\)
and \(B\), and the constant \(c\) in the first gauge kinetic function
\(f_{1}(z)=e^{-A(z)-B^{2}z^{2}-cz^{2}}\), fixed at \(1.16\,\mathrm{GeV}^{2}\) by the lowest
heavy-meson states.  Demanding that the dilaton stay real then imposes \(B^{4}\leq6a^{2}\), so
the model carries its own ceiling \(B\lesssim0.61\,\mathrm{GeV}\) rather than admitting
arbitrarily strong fields.

One point of bookkeeping before any string is put on the geometry.  The solution is written in
the Einstein frame, but a string does not couple to that metric; it couples to the string-frame
one, \(g^{(s)}_{MN}=e^{\sqrt{2/3}\,\phi}g_{MN}\), and everything below is computed there for
that reason.  The temperature is indifferent to the distinction: writing the metric in the
radial variable as \(-\mathcal{F}\,dt^{2}+dr^{2}/\mathcal{G}\), in script letters because
\(F\) is a field strength and \(H\) a Hamiltonian everywhere else in these notes, the
conformal factor appears in \(\mathcal{F}'(r_{h})\) and its inverse in \(\mathcal{G}'(r_{h})\),
so it cancels from \(T=\sqrt{\mathcal{F}'\mathcal{G}'}/4\pi\) and \(2\pi T\) is unambiguous
when Sec.~\ref{sec:L6_bounds} comes to compare against it.

\begin{figure}[!htbp]
\centering
\includegraphics[width=\textwidth]{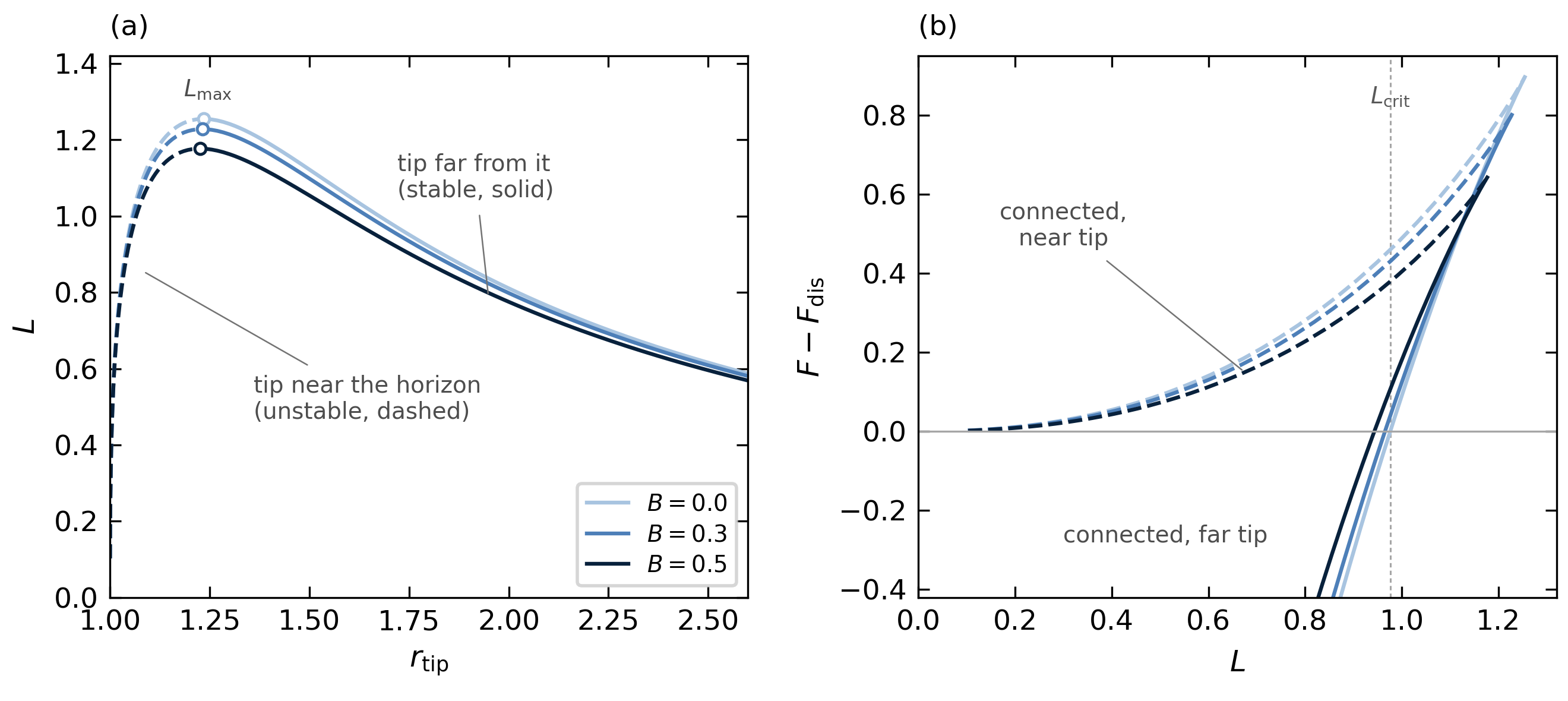}
\caption{The suspended string in the magnetised Einstein--Maxwell--dilaton model, in the string
frame at vanishing chemical potential with \(r_{h}=1\), for a string aligned with the
magnetic field.  (a) The separation \(L\) of the endpoints against the position of the string tip.  It
is not monotonic, so below the maximum \(L_{\max}\), marked by open circles, two connected
strings share the same separation: one with its tip near the horizon (dashed) and one far from
it (solid).  (b) The free energy of those solutions relative to the disconnected pair of
straight strings, so that the horizontal axis at zero is the disconnected configuration
itself.  The far-tip branch is the ground state up to \(L_{\rm crit}=0.976\), drawn here
for \(B=0\), beyond which the disconnected pair takes over, while the near-tip branch lies
above both at every separation.  Raising the magnetic field lowers \(L_{\max}\), from
\(1.255\) at \(B=0\) to \(1.228\) and \(1.177\) at \(B=0.3\) and \(0.5\); at fixed separation
it lifts the tip away from the horizon, the tip at \(L=1.1\) moving from \(r_{\rm tip}=1.083\)
to \(1.089\) and \(1.107\) over the same range.  Computed from the closed-form background, and
reproducing the tip positions tabulated in
Ref.~\cite{ShuklaNongmaithemDudalMahapatra2025} to two parts in \(10^{4}\).}
\label{fig:qcd_string}
\end{figure}

The probe is an open string with both endpoints on the boundary, dual to a heavy
quark-antiquark pair, or equivalently to a Wilson loop.  Integrating its Nambu--Goto profile
gives the separation \(L\) of the endpoints as a function of the tip position \(r_{\rm tip}\),
and that function is not monotonic~\cite{ShuklaNongmaithemDudalMahapatra2025}.  It rises from
zero as the tip lifts off the horizon, reaches a maximum \(L_{\max}\), and falls again, so
below \(L_{\max}\) two connected strings share the same separation, one with its tip near the
horizon and one well away from it, alongside the disconnected configuration of two straight
strings.  Comparing free energies settles which is which, Fig.~\ref{fig:qcd_string}: the
far-tip solution has the lowest free energy up to a critical separation, beyond which the
disconnected pair takes over, while the near-tip solution lies above both everywhere.  Chaos
appears on that energetically disfavoured branch, the one reaching closest to the horizon.  On
the far-tip branch it is not merely weaker, it is absent, the sections there showing closed
orbits and no scatter at any \(B\) or \(\mu\).

Perturbing the static profile in the direction normal to itself and reducing the quadratic
action leaves a Sturm--Liouville problem,
\begin{equation}
    \partial_{\sigma}\left(C_{\sigma\sigma}\xi'\right)-C_{00}\,\xi=\omega^{2}C_{tt}\,\xi ,
\end{equation}
with coefficients fixed by the background and lowest eigenvalue negative on the near-horizon
branch for every \(B\) and \(\mu\), which is what it means for that configuration to be
unstable.  Keeping the two lowest modes and carrying the expansion one order further leaves a
mechanical system in two degrees of freedom with a four-dimensional phase space, which is
precisely the setting of Lecture 2, so Poincar\'e sections and Lyapunov exponents apply to it
unchanged.  Three of the cubic terms carry velocities, so the kinetic energy depends on the
coordinates rather than being a fixed quadratic form and a near-identity redefinition of the
mode amplitudes is needed to restore positivity; it leaves the trajectories, and the exponents
read off them, unchanged at the order kept.  None of that construction is new here.  Perturbing
a suspended string normal to itself, truncating to two modes and expanding to cubic order is
the method of~\cite{HashimotoMurataTanahashi2018}, applied to a confining background
in~\cite{AkutagawaHashimotoMurataOta2019}, and it belongs to the older line of work on chaotic
semiclassical strings in holographic
geometries~\cite{PandoZayasTerreroEscalante2010,BasuPandoZayas2011a}.  What
Ref.~\cite{ShuklaNongmaithemDudalMahapatra2025} supplies is the magnetised background and the
numbers.

At the saddle of the reduced potential the four Lyapunov exponents converge to
\(\bigl(\pm\sqrt{-\omega_{0}^{2}},0,0\bigr)\), so \(\lambda_{\max}=\sqrt{-\omega_{0}^{2}}\).
The largest exponent at the unstable fixed point is exactly the growth rate of the unstable
normal mode, so the linear stability analysis and the full nonlinear evolution return the same
number, and the exponents summing to zero confirms that the integration has preserved the
conservative structure.  The caveat entered in Sec.~\ref{sec:L6_geodesic} applies here too and
for the same reason: this is a local instability rate at one fixed point rather than a
long-time average along a generic orbit.  It is quoted anyway, because it is the quantity the
bound of Sec.~\ref{sec:L6_bounds} is stated for; what shows the system to be chaotic rather
than merely unstable is the structure of its Poincar\'e sections, which scatter near the origin
and resolve into curves as \(\mu\) is raised, the transverse configuration ordering ahead of
the longitudinal one.

The background carries two knobs beyond temperature, and both act the same way, by lifting the
string out of the region where the horizon can reach it.  They are not equivalent otherwise:
the chemical potential preserves the symmetry of the boundary theory and the magnetic field
does not, breaking the rotation group to \(SO(2)\) about its own direction, so a quark pair
aligned with the field and one laid across it are physically distinct configurations rather
than two descriptions of the same one.  Fix the quark separation at
\(L=1.1\,\mathrm{GeV}^{-1}\), about \(0.22\) fm.  The tip then moves from
\(r_{\rm tip}=1.0832\) at vanishing field and chemical potential to \(1.1120\) along the field
and \(1.1355\) across it at the largest value of each, the field lifting the transverse string
further than the longitudinal one and the two coinciding exactly at \(B=0\) as the unbroken
\(SO(3)\) requires.  The exponent falls with both, from \(1.72\) GeV to \(1.14\) GeV over the
same range, a reduction of about a third.  The fall is monotonic in the field except at the two
largest chemical potentials, where the smallest nonzero field lifts the longitudinal exponent
slightly before the decrease resumes.  Tables I and II of
Ref.~\cite{ShuklaNongmaithemDudalMahapatra2025} give the full grid, \(B=0\) to \(0.5\) GeV and
\(\mu=0\) to \(1.2\) GeV for both orientations, and are worth having open alongside this; the
numbers quoted here are its corners rather than a summary of it.

The anisotropy is the part that is difficult to obtain any other way.  At every nonzero field
and every chemical potential the exponent for a string laid across the magnetic field is
smaller than for one aligned with it, the two coinciding only at \(B=0\) where there is no
direction to be laid across.  In the string frame a magnetic field makes a quark pair less
chaotic, and it makes the transverse pair less chaotic than the longitudinal one, with the gap
widening as the field grows.  That is the broken \(SO(3)\) above printing itself onto a
real-time observable.

The frame qualifier is load-bearing.  Repeat the calculation in the Einstein frame and the
signs come out differently: there the magnetic field still suppresses the exponent for the
transverse string but raises it for the longitudinal one, and the chemical potential raises it
for both orientations.  That disagreement is the main result of
Refs.~\cite{ShuklaDudalMahapatra2023,ShuklaNongmaithemDudalMahapatra2025} rather than a side
effect of them, and it is not resolved here.  Nothing in this subsection should be read as a
frame-independent statement about QCD-like matter.

One further piece of bookkeeping matters for Sec.~\ref{sec:L6_bounds}.  The grid is taken at
fixed horizon radius rather than at fixed temperature, and \(T=|g'(z_{h})|/4\pi\) depends on
both knobs, so the exponent and the temperature move across it together.  A ratio
\(\lambda_{\max}/2\pi T\) therefore need not follow \(\lambda_{\max}\) itself, which is why the
largest ratio quoted there is not reached where the exponent is largest.

Whether any of this reaches QCD is a fair question, and the honest answer is that it is
suggestive rather than settled, and suggestive in a way that cuts against the grain.  If chaos
in the confining string is tied to the onset of deconfinement, then two parameters that both
suppress it in the string frame point toward deconfinement being harder to reach at larger
field, an inference the Einstein-frame signs would run backwards.  The lattice finds the
opposite for the magnetic field: inverse magnetic catalysis is the observation that the
transition temperature \emph{falls} as \(B\) grows~\cite{BaliEtAl2012}, so deconfinement gets
easier rather than harder.  Ref.~\cite{ShuklaNongmaithemDudalMahapatra2025} raises the
connection without claiming agreement, and that is the right level of confidence, since the
link between string chaos and the transition is itself a conjecture and either end of it could
be what fails.  The orientation dependence is on firmer ground: the transverse direction
softening faster than the longitudinal one parallels the anisotropy the lattice finds in the
quark-antiquark potential at nonzero magnetic field~\cite{BonatiEtAl2014}.  Neither connection
is established here.  What is established is narrower and still worth having: in a background
tuned to reproduce QCD thermodynamics, the real-time chaos of a heavy quark pair is measurably
anisotropic, and both of the parameters that a heavy-ion collision actually supplies act to
suppress it in the frame the string couples to.

What remains is to ask whether any of these exponents come close to the ceiling that
Sec.~\ref{sec:L5_bound} placed on thermal chaos.

\subsection{Probe exponents and the chaos bound}
\label{sec:L6_bounds}

Sections~\ref{sec:L6_geodesic} to~\ref{sec:L6_aniso} have produced classical Lyapunov exponents
for probes moving in black-hole backgrounds, and the temptation now is to hold them against
\(2\pi T\).  The comparison is worth making, but three different statements travel under the
name of the chaos bound, and separating them is most of the work.

The first is the theorem of Sec.~\ref{sec:L5_bound}.  The Maldacena--Shenker--Stanford bound
constrains the
growth rate of a regularized thermal out-of-time-order correlator in a quantum system with many
degrees of freedom, and its proof runs on analyticity and unitarity.  It is a statement about a
squared commutator, and Sec.~\ref{sec:L5_bound} sharpened which exponent it constrains: the
generalized exponent of order two, with the typical exponent obeying the tighter
\(\lambda_{1}\leq\pi/\beta\).

The second is a classical statement about horizons, and it is also a theorem, though of a much
narrower kind.  A particle held near a non-extremal horizon
of a static, spherically symmetric black hole by an external force, electromagnetic or scalar,
sits at an unstable maximum of its effective potential whose instability rate is the surface
gravity itself,
\begin{equation}
    \lambda=\kappa=2\pi T_{H},
\end{equation}
independently of the force used to hold it there, of the particle's mass, and of where along
the near-horizon region the maximum happens to fall~\cite{HashimotoTanahashi2017}.  Moving the
maximum itself out of the near-horizon region turns the equality into the bound
\(\lambda\leq2\pi T_{H}\).  Nothing quantum enters the derivation.  The bound
coincides numerically with the first, and that coincidence is what makes the phrase
\emph{classical analogue of the chaos bound} reasonable; the two are nonetheless established by
unrelated arguments about unrelated quantities.

The third is not a statement but a habit: comparing any classical exponent computed in a
black-hole background against \(2\pi T\) because both carry units of inverse time.

The exponents of this lecture belong to the second category at best.  The geodesic exponent of
Sec.~\ref{sec:L6_geodesic} is the instability rate of a circular orbit, the string exponent of
Sec.~\ref{sec:L6_hqcd} the instability rate of a normal mode.  Neither is the growth rate of a
squared commutator, so neither is constrained by the theorem of Sec.~\ref{sec:L5_bound}, and a
reader who finds one of them above \(2\pi T\) has not falsified anything.

For the confining string the comparison can at least be made cleanly, since
Sec.~\ref{sec:L6_hqcd} established that the temperature is indifferent to the conformal factor
and \(2\pi T\) is therefore unambiguous.  Across every combination of magnetic field, chemical
potential and orientation entering Sec.~\ref{sec:L6_aniso}, the exponent stays below the bound
without exception, and it does not come close to saturating it: the largest ratio anywhere in
that range is
\begin{equation}
    \frac{\lambda_{\max}}{2\pi T}=0.785 ,
\end{equation}
reached at the smallest nonzero field and the largest chemical potential.  The confining string
is at most four-fifths as chaotic as a thermal system at the same temperature is permitted to
be.

The geodesic side is less well behaved, and that is the more interesting half.  The classical
statement holds in the setup it was proved in and not beyond it, and the literature reporting
exponents above \(2\pi T_{H}\) is worth reading with that in mind.  Ref.~\cite{GaoChenYuWang2022} finds
the bound violated by charged particles around a charged Kiselev black hole at a finite distance
from the horizon, and explicitly no violation in the near-horizon region once the charge to mass
ratio is fixed, which places those violations outside the regime the
theorem covers rather than against it.  Ref.~\cite{TargemaEtAl2026} returns to the same black
hole and finds a second way for a violation to be an artefact: the angular momentum of the
orbiting particle is fixed by the circular-orbit conditions and is not free to be varied
alongside them, and some of the reported violations do not survive imposing that.  What does
survive in that analysis appears only once higher-curvature terms are added to the geometry,
at large charge-to-mass ratio, which makes it a statement about the curvature rather than
about the orbit.  A classical exponent above \(2\pi T_{H}\) is therefore
not automatically a counterexample.  It may just as easily be a statement about a geometry, or
about a regime, lying outside the hypotheses the theorem was proved under, and telling the two
apart takes work.

There is a further question about which temperature the comparison should even use.  The
exponent of Sec.~\ref{sec:L6_geodesic} belongs to an orbit at \(r_{0}=3M\) in the
Schwarzschild case, which is nowhere near the horizon, while \(T_{H}\) is a property of the
horizon.  Setting one against the other compares quantities defined at different places, and a
ratio above one carries less weight than it looks to.  One way out is to supply a temperature
that belongs to the orbit rather than to the horizon.  Ref.~\cite{GiataganasEtAl2026} does
this with a string probe, whose induced worldsheet metric is Rindler and carries a surface
gravity of its own, and reports that on equatorial photon-ring orbits of a broad class of
Kerr-like geometries the exponent of Sec.~\ref{sec:L6_geodesic} equals \(2\pi\) times that
temperature exactly.  The same exponent is recovered there from an out-of-time-order
correlator in the near-ring region, which is the one place in this discussion where a geodesic
exponent and a squared commutator are made to meet.

What survives is worth stating compactly.  Surface gravity sets the natural instability scale
at a horizon, which is why \(2\pi T\) keeps appearing.  The classical bound is a theorem for
particles held against a horizon and not a general law.  The quantum bound governs a different
object entirely, and no calculation in this lecture tests it.  And a reported violation deserves reading closely before it is taken as a counterexample,
since the exponent and the temperature it is measured against are frequently defined at
different places in the geometry.

Lecture 6 opened by building the geometry that Lecture 5 had only described, and closes by
computing exponents for objects moving in geometries built to resemble QCD.  The two halves
answer different questions.  The first says what chaos looks like when the dual is Einstein
gravity, and the answer is entirely universal and therefore says nothing about which black hole
is being looked at.  The second gives that universality up deliberately, and what it buys is
discrimination: an exponent that tracks a thermodynamic phase transition with a mean-field
exponent, and one that resolves the direction of a magnetic field in a confining background.
Neither is the quantum Lyapunov exponent that Lecture 5 bounded, and keeping the two apart is
the last thing this lecture has to insist on.

\subsection{Exercises}

\begin{exercise}
\label{ex:L6_cone}
The shift of Sec.~\ref{sec:L6_shock} is the potential of a screened point source in the
\(d-1\) transverse directions, and the butterfly cone is its behaviour far from that source.
\begin{enumerate}
    \item[(a)] Solve \(\left(-\partial_{i}\partial^{i}+m_{\rm scr}^{2}\right)h
    =A\,\delta^{(d-1)}(\mathbf{x})\) and show that at large \(|\mathbf{x}|\) the solution
    behaves as \(e^{-m_{\rm scr}|\mathbf{x}|}/|\mathbf{x}|^{(d-2)/2}\).  The overall constant
    is not the point; the power is, and it is the one quoted in Sec.~\ref{sec:L6_shock}.
    \item[(b)] Restore the source and read off the butterfly velocity,
    \(v_{B}=2\pi/\beta m_{\rm scr}\).  Sec.~\ref{sec:L5_vb} quoted the numbers this returns;
    the point of getting them this way is that a speed has come out of a screening length in
    the transverse geometry and of nothing else.  Say which feature of the brane fixes
    \(m_{\rm scr}\), and show that the large-\(d\) limit \(1/\sqrt2\) is approached from
    above.
    \item[(c)] Define the front by \(h=\varepsilon\) at fixed small \(\varepsilon\) and solve
    for its position \(x_{c}(t)\).  Show that
    \begin{equation}
        x_{c}(t)=v_{B}t-\frac{\beta v_{B}(d-2)}{4\pi}\log\left(v_{B}t\right)
        +\frac{\beta v_{B}}{2\pi}\log\frac{C}{\varepsilon}+\cdots ,
    \end{equation}
    so a velocity fitted as \(x_{c}/t\) sits below \(v_{B}\) by a term of order
    \(\log t/t\), while the next term is of order \(1/t\) and carries \(\log\varepsilon\).
    Conclude that the asymptotic velocity is exactly \(v_{B}\) and that at accessible times
    the choice of threshold can dominate the discrepancy.
    Exercise~\ref{ex:L5_cone} asked what that shift measures in a spin chain; this is the same
    effect where the profile is known in closed form.
\end{enumerate}
\end{exercise}

\begin{exercise}
\label{ex:L6_spin}
The eikonal phase of Sec.~\ref{sec:L6_correlator} carries the bound as a statement about spin.
\begin{enumerate}
    \item[(a)] The two quanta meet with squared centre-of-mass energy \(s\propto p^{u}p^{v}\)
    at fixed transverse separation.  Using \(\chi=p_{v}h\) and \(h\propto G_{N}p^{v}\), show
    \(\chi\propto G_{N}s\propto G_{N}e^{2\pi t/\beta}\).  Expanding \(e^{i\chi}\) to first
    order then returns \(F(t)\approx1-(c/N^{2})e^{2\pi t/\beta}\), but only once the
    \(\beta/4\) imaginary-time offsets of Sec.~\ref{sec:L5_bound} are restored, which is what
    makes the momenta complex and turns \(i\chi\) into a real decrease rather than a phase.
    Show that it is a phase without them.
    \item[(b)] Exchange of a particle of spin \(J\) contributes a phase growing like
    \((p^{u}p^{v})^{J-1}\).  Deduce \(\lambda_{L}=(2\pi/\beta)(J-1)\), and say in one sentence
    why the graviton saturates the bound of Sec.~\ref{sec:L5_bound} while nothing consistent
    exceeds it.
    \item[(c)] With \(j_{\rm eff}=2-d(d-1)\ell_{s}^{2}/4\ell_{\rm AdS}^{2}\), verify
    \(d(d-1)/4=m_{\rm scr}^{2}/2\) and recover the corrected scrambling time of
    Sec.~\ref{sec:L6_correlator}.  A single screening mass therefore fixes both how far the
    shock spreads and how far the exponent falls short.  Show that as \(d\) grows at fixed
    \(\ell_{s}/\ell_{\rm AdS}\) the first of those effects saturates while the second grows
    like \(d^{2}\), and say which of the two the bulk calculation is trustworthy for.
\end{enumerate}
\end{exercise}

\begin{exercise}
\label{ex:L6_photon}
Circular orbits beyond the case worked in Sec.~\ref{sec:L6_geodesic}.
\begin{enumerate}
    \item[(a)] For Schwarzschild--Tangherlini in \(d\) spacetime dimensions,
    \(f(r)=1-(r_{s}/r)^{d-3}\), show that the circular null orbit sits at
    \(r_{0}^{d-3}=\tfrac12(d-1)r_{s}^{d-3}\) and that \(f(r_{0})=(d-3)/(d-1)\).
    \item[(b)] Show that \(\Omega_{c}/\lambda=1/\sqrt{d-3}\), hence
    \(\gamma=1/2\pi\sqrt{d-3}\), and say what is special about \(d=4\) and what happens to
    the orbit's stability relative to its own period as the dimension rises.
    \item[(c)] For four-dimensional Schwarzschild with \(T_{H}=1/8\pi M\), show
    \(\lambda/2\pi T_{H}=4\sqrt3/9\simeq0.770\).  Then say where the shortfall lives: in the
    exponent, in the temperature, or in the fact that \(r_{0}=3M\) is nowhere near the horizon.
    \item[(d)] Take the timelike case, \(V_{\rm eff}=f(r)\left(L^{2}/r^{2}+1\right)\), and fix
    \(L\) by \(V_{\rm eff}'(r_{0})=0\).  Show \(L^{2}=Mr_{0}^{2}/(r_{0}-3M)\) and
    \begin{equation}
        \lambda^{2}=\frac{M(6M-r_{0})}{r_{0}^{4}} .
    \end{equation}
    Read off the range of \(r_{0}\) on which the exponent is real, and say which familiar
    radius the upper end of that range is.
\end{enumerate}
\end{exercise}

\begin{exercise}
\label{ex:L6_swallowtail}
The critical exponent of Sec.~\ref{sec:L6_probe} is short enough to check from scratch. Work at
\(\ell_{\rm AdS}=G_N=1\) and fixed \(\Phi_e\), with \(M\) eliminated by \(f(r_h)=0\).
\begin{enumerate}
    \item[(a)] From \(T(r_h)\) and \(G(r_h)\) of Sec.~\ref{sec:L6_thermo}, confirm
    \(q_{mc}=\tfrac16(1-\Phi_e^{2})\) and \(r_{hc}=\sqrt{1-\Phi_e^{2}}/\sqrt6\), and notice that
    they satisfy \(r_{hc}^{2}=q_{mc}\).
    \item[(b)] For \(q_m<q_{mc}\), find the first-order transition temperature \(T_p\) by
    equating the free energies of the small and large branches, and let \(r_a\) and \(r_b\) be
    their horizon radii there. Show numerically that
    \begin{equation}
        \left(r_b-r_a\right)^{2}=6\left(q_{mc}-q_m\right)
    \end{equation}
    and check that this is exact rather than asymptotic, by testing it far from the critical
    point as well as near it.
    \item[(c)] Show that \(T_p-T_c\) is linear in \(q_{mc}-q_m\) as the critical point is
    approached, and combine with (b) to predict the exponent in
    \(\Delta\lambda\propto|T_p/T_c-1|^{\,\theta}\) without computing a single Lyapunov exponent.
    \item[(d)] Now compute them. Take the unstable circular null geodesic on each branch at
    \(T_p\), evaluate \(\lambda\) from Sec.~\ref{sec:L6_geodesic}, and fit \(\theta\). The
    approach is slow: the local slope is near \(0.60\) a per cent from criticality and reaches
    \(0.50\) only in the last decade, which is worth seeing, since a fit over too wide a range
    would report the wrong exponent.
\end{enumerate}
\end{exercise}

\section*{Outlook}
\addcontentsline{toc}{section}{Outlook}

These lectures followed one arc. Lecture 1 defined classical chaos geometrically and Lecture 2 built the
instruments that measure it, every one of which reads a phase-space trajectory. Quantization removes that
trajectory, so Lectures 3 and 4 rebuilt the instruments from spectra and eigenstates instead, and what they
return is a classification and a count rather than a rate. Lecture 5 recovered the rate from the squared
commutator, followed it out to the butterfly velocity and to a ceiling on how fast chaos can develop, and
Lecture 6 built the geometry behind that ceiling before turning the same machinery on backgrounds where an
exponent discriminates instead of saturating: one tracking a thermodynamic phase transition, the other
resolving an anisotropy induced by a background field. The endpoint, the chaos bound, is not the end of the
story.

Its saturation is special to theories with a classical Einstein gravity dual. Away from that limit, at finite
coupling, at finite $N$, or at finite $\alpha'$ where an infinite tower of higher-spin states contributes,
the Lyapunov exponent sits strictly below $2\pi T$, and how far below measures the departure from Einstein gravity
rather than any weakness in the bound. Exponents reported above $2\pi T$ are classical quantities of the
kind computed in Sec.~\ref{sec:L6_geodesic}, which Sec.~\ref{sec:L6_bounds} argues is not the object the
theorem constrains. The geodesic and string calculations of Lecture 6 are worked examples
rather than results about the bound, and the same machinery applies to any new bulk geometry a reader
wants to test.

One direction these notes have not taken is worth naming, because a good deal of the recent work
sits there. The late-time physics of Lecture 4, the ramp and the plateau, has been pushed much
further in two dimensions, where the Sachdev--Ye--Kitaev model and Jackiw--Teitelboim gravity
supply a boundary theory and a bulk that can be matched at the scale of individual level spacings
rather than only semiclassically. A recent review sets that programme out from the
beginning~\cite{AltlandSonner2026}, and it complements this course rather than extending it: the
form factor, which occupies one subsection here, is most of its subject, while the
out-of-time-order correlator, which occupies two lectures here, is one subsection of it.

The tools built in Lectures 3 and 4, random matrix statistics and the eigenstate thermalization hypothesis,
came from outside holography and are still the standard toolkit in condensed matter
and atomic physics: many-body localization, disordered spin chains, and quantum simulator experiments all
use them. A researcher who has followed these two lectures already has what is needed to read that
literature as well.

\section*{Acknowledgments}
\addcontentsline{toc}{section}{Acknowledgments}

It is a pleasure to thank the organizers of the ST4 Workshop at the Chennai Mathematical Institute for the
invitation to deliver this lecture series and for their hospitality, and the participants for their questions,
which shaped the presentation of several sections.  I am grateful to Julian Sonner for generous and
detailed correspondence on eigenstate thermalization and on spectral statistics in holography,
which improved this version in several places.

\section*{Author Contributions}
\addcontentsline{toc}{section}{Author Contributions}

The author delivered the lecture course, set its scope and sequence, and reviewed and revised the
material presented here.  No scientific result in these notes originates with the assistant. Claude Opus 5, an AI research assistant developed by Anthropic, worked on the
notes under that direction: expanding the lecture material into continuous prose, writing the Python
scripts that generate the figures, checking algebra and cross-references, and preparing the manuscript.
The work was carried out in Cowork, a feature of Anthropic's Claude application. The author is fully responsible for the scientific content
and integrity of these notes.

\bibliographystyle{JHEP}
\bibliography{references}

\end{document}